SVEUČILIŠTE U ZAGREBU

HRVATSKI STUDIJI

Mladen Domazet

# OBJAŠNJENJA IZ SUVREMENIH KVANTNIH TEORIJA: NEKE ONTOLOŠKE KARAKTERISTIKE

DOKTORSKI RAD

UNIVERSITY OF ZAGREB

STUDIA CROATICA

MLADEN DOMAZET

# EXPLANATIONS FROM CONTEMPORARY QUANTUM THEORIES: SOME ONTOLOGICAL CHARACTERISTICS

DOCTORAL THESIS

Supervisor:

Dr Stipe Kutleša

Zagreb, 2009

University of Zagreb

Studia Croatica

Mladen Domazet

# *Explanations from contemporary quantum theories: some ontological characteristics*

Doctoral Thesis

Zagreb, 2009.

Sveučilište u Zagrebu

Hrvatski studiji

Mladen Domazet

# *Objašnjenja iz suvremenih kvantnih teorija: neke ontološke karakteristike*

Doktorska disertacija

Zagreb, 2009.

Mentor:

dr. sc. Stipe Kutleša

## Acknowledgements

Through the years the construction of this dissertation has been helped by numerous mentors and friends, especially Stipe, Harvey, Rada, Žarko, MikiTrasi, Prle, Rajko, Ksenija, Goran, Sanja, Željka, Dnjepar, Branislava and the Centre, Karin, Tvrtko, Kruno, Pavel, Will, James, Lana, Sian, Marko, Silvija, Domagoj, Sue, Anthony, Elisabeth, Dennis, Chris, Boris, Simon, Balázs, Tim, Dunja, Nenad, Filip, Danijela, Tihomir and all those who offered a hand or a shoulder.

Ana has made this project possible.

I wish to sincerely thank the Institute for Social Research Zagreb, Centre for Educational Research and Development, National Foundation for Science, Higher Education and Technological Development of the Republic of Croatia, Ministry of Science, Education and Sports of the Republic of Croatia, Central European University, and University College, Oxford for the material and financial support throughout the development of this dissertation.

## Zahvale

Kroz godine su izradu ovog rada pomogli brojni mentori i prijatelji, posebno Stipe, Harvey, Rada, Žarko, MikiTrasi, Prle, Rajko, Ksenija, Goran, Sanja, Željka, Dnjepar, Branislava i Centar, Karin, Tvrtko, Kruno, Pavel, Will, James, Lana, Sian, Marko, Silvija, Domagoj, Sue, Anthony, Elisabeth, Dennis, Chris, Boris, Simon, Balázs, Tim, Dunja, Nenad, Filip, Danijela, Tihomir te svi koji su pružili ruku ili rame.

Bez Ane ovaj projekt ne bi bio moguć.

Iskreno se zahvaljujem Institutu za društvena istraživanja Zagreb, Centru za istraživanje i razvoj obrazovanja, Nacionalnoj zakladi za znanost, visoko školstvo i tehnologijski razvoj RH, Ministarstvu znanosti, obrazovanja i športa RH, Central European University i University College, Oxford za materijalnu i financijsku podršku izradi rada.

## Sažetak

Početna je pozicija ovoga rada da je znanstveno znanje nepotpuno bez objašnjenja, gdje se razvojem novih teorija to znanje proširuje i produbljuje (tako što temeljne teorije objašnjavaju širi spektar pojava i postaju općenitije primjenjive). Povijesno gledano, kvantna je teorija (početkom 20. st.), ili inicijalno 'kvantna mehanika', konačno potkopala navodni neograničeni uspjeh redukcionističke mehanistične eksplanatorne filozofije (uzevši u obzir i konceptualni dodatak koji čini Maxwellova konceptualizacija polja). Tako je ponovo otvoren put skepticizmu prema eksplanatornim ciljevima znanosti. U radu mu se suprotstavlja obrazloženje temeljne uloge kvantne teorije, bilo kao dijela fundamentalne potpune teorije ili iznova osmišljene u okvirima ograničenja prikupljanja informacija o osjetilno nedostupnoj ontologiji fizikalnog svijeta.

Izlaganje počinje pregledom povijesnih stajališta u različitim pokušajima razumijevanja materijalnog svijeta od uspona novovjekovne znanosti, sa posebnim naglaskom na ulogu kartezijanskih primarnih kvaliteta na eksplanatornu konceptualizaciju. Naglašava se i da iako su eksplanatorni narativi u osnovi epistemološke konstrukcije, oni zahtijevaju metafizičku podlogu kroz prihvaćanje referencijalnosti pojmova od kojih se sastoje. Nadalje se predstavljaju dvije metodološke perspektive na konstrukciju znanstvenih teorija, slijedeći Einsteinovu podjelu na principne i konstruktivne teorije. One su prikazane i obzirom na njihove metafizičke i eksplanatorne karakteristike i oblikovane u istraživački instrument kroz koji se razmatraju pojedine 'studije slučaja' eksplanatorne rekonstrukcije kvantne teorije. Specifična strategija zagovaranja znanstvenog realizma (nazvana 'jednostavnom transcendentalnom strategijom') se izlaže i povezuje s izazovima koje određene pojave (EPR korelacije i 'teleportacija') iz domene suvremene kvantne teorije postavljaju pred nju. Prvo se poglavlje zaključuje pregledom općih modela objašnjenja i njihovom pozicioniranjem u odnosu na gore spomenutu dihotomiju principno-konstruktivno.

U konačnici se zagovara eksplanatorni model koji uz protegnute materijalne entitete sadrži i dodatni temeljni ontološki element: sveobuhvatni primitivni prirodni zakon. U skladu s time, u svrhu odgovora skepticizmu o eksplanatornim ciljevima znanosti, preporuča se i u 'svakodnevnom pojmovnom okviru' zamijeniti trenutno stanje protegnute materijalne strukture, kao temeljne jedinice ontologije realizma, 'poopćenim predmetima' kao invarijantama re-identifikacije u procesu promjena.

## Summary

The starting position of this dissertation is that scientific knowledge is incomplete without explanations, whereupon with the development of new theories our knowledge both broadens and deepens (as fundamental theories explain more and become more general). Historically, it has been quantum theory (early 20th century), or initially quantum mechanics, that finally undermined the supposed runaway success of reductionist mechanistic philosophy (modulo Maxwellian updating), re-opening the door for scepticism about the explanatory aims of science. However, recent years have seen a revival of the belief in some version of quantum theory, either as part of a fundamental complete theory or as reinvented in terms of constraints on information gathering about the underlying unobservable ontology of the physical world.

We begin by surveying the historical positions in different attempts to understand the material world since the rise of modern science, with specific focus on the role of Cartesian primary qualities in explanatory conceptualisation. Moreover the opening chapter argues that although explanatory narratives are essentially epistemological constructions, they require a general metaphysical backing through the explainer's and explainee's commitments to take the concepts and higher structures composed of them as directly referential. Two methodological perspectives on theory construction, Einstein's division into principle and constructive theories, are then delineated along the lines of their metaphysical and explanatory potential, and presented as the research instrument with which to approach the specific-case-study instances of quantum theory reconstruction. A specific strategy of arguing for scientific realism ('the simple transcendental strategy') is then presented and connected to the challenges that the phenomena (EPR correlations and 'teleportation') from the domain of contemporary quantum theory pose for it. The opening chapter concludes with a survey of the general models of explanation and their position with respect to the principle-constructive dichotomy introduced above.

Finally, an explanatory model containing extended material-like entities along with further primitives, universal laws of temporal evolution, is argued for. Consequently, in the interest of a wholesome realistic response to the sceptical challenge about the explanatory aims of science, we recommend that even the common sense conceptual framework replaces the extended material structure with 'generalised things' as re-identifiable invariants through change as the fundamental unit of the realist ontology.

**Keywords:** explanation, realism, quantum theory, principle and constructive theories, information, law of nature

Contents

# INTRODUCTION

There is a plausible view according to which scientific knowledge consists primarily of explanations, whereupon with the development of new theories our knowledge both broadens and deepens (as fundamental theories explain more and become more general). One might claim that science is, then, aiming at an integrated understanding of reality that consists "not only of reductionist ingredients such as space, time and subatomic particles, but also, for example, of life, thought and computation" (D. Deutsch, The Fabric of Reality, 1997). In this context, the thesis aims to contribute to the general considerations concerning structure and ontological commitments of scientific explanation capable of including the specific case-study instances.

Historically, it has been quantum theory (early 20th century), or initially quantum mechanics, that finally undermined the supposed runaway success of reductionist mechanistic philosophy (modulo Maxwellian updating), re-opening the door for scepticism about the explanatory aims of science (2nd half of the 20th century). However, recent years have seen a revival of the belief in some version of quantum theory, as part of a fundamental complete theory, as well as (alternatively) its 'reinvention' as a weak-realist (in some instances non-physical) theory that delineates the constraints of information gathering about the underlying unobservable ontology of the physical world (end of 20th and early 21st century).

Aside from numerous philosophical perplexities associated with interpretations and re-formulations of the theoretical framework behind the empirical success of quantum mechanics, the recent developments named are also interesting for their approaches to scientific explanation of the physical phenomena. Due to the potential status of quantum mechanics as a fundamental theory, it is important for any scientific explanation, to investigate the constraints it imposes on the explanatory aim of science, as well as any departure it requires from the basic explanatory construct of matter evolving on the space and time stage. In this thesis, two broad perspectives on the integrated understating of reality will be delineated: a principle and a constructive one, and these will be applied as criteria in a comparative analysis of specific interpretations of contemporary quantum theory.

The 'principle' and 'constructive' perspectives are formed on the basis of the following (broad) criteria: methodological approach to the development of new theories about segments of physical reality (principle vs. constructive in the narrow sense), metaphysical attitude towards existence of the unobservable theoretical entities (agnostic weak-realism vs. simple realism), and the method of providing an explanation (unification-type vs. causal). With regard to the former, from which the two theoretical perspectives draw their names, constructive theories attempt to build a picture of the more complex phenomena out of the relatively simple ontology from which they conceptually start out. Principle theories, on the other hand, employ the analytic, not the synthetic, method. The elements that form their basis and conceptual starting point are not hypothetically constructed, but empirically discovered, general characteristics of natural processes. In other words, the fundamental task of principle theories is the analysis of principles, with the aim of arriving at certain necessary conditions or constraints on observed phenomena; the phenomena that underwrite and reconcile these empirical principles. On the other hand, it has long been received knowledge in the philosophy of physics that when we say we have succeeded in understanding a group of natural processes, we invariably mean that a constructive theory has been found which covers the processes in question.

Regarding the second criterion, the metaphysical attitude towards existence of the unobservable theoretical entities (this need not be just the dimensionally 'small' things) a basic realist account of the constructive approach accepts that some mind-independent referents, or tokens, of most currently observable common-sense and physical types (constituting our known world) objectively exist independently of the mental. The general weak-realist stance of the principle approach claims that the independent reality is beyond the reach of our knowledge and language (but not that it does not exist), and that the known world is partly constructed by the human imposition of concepts. All the worlds defined by such concepts differ according to the social group that introduced them, and thus exist only relative to the (mental) imposition of concepts. The thesis investigates the concurrence of the metaphysical commitments of either account with the simple transcendental strategy for realism. Namely, that the concepts employed in an account of everyday experience can have a philosophical foundation in the physical constraints imposed by quantum theory.

Finally, in terms of explanation, explanations aiming at the unification conception of understanding (those of our principle approach) primarily focus on uncovering the unity that underlies the apparent diversity of the observed phenomena, without particular reliance on causality. Explanations in the manner of the causal conception of understanding (our constructive approach) highlight the structural mechanisms that cause the observed phenomena. In that they can be seen as a subset of the unification-type if the causal picture is presented as the unity behind diverse phenomena, but needn't in those cases where the structural mechanisms, characterised as fundamental for other reasons, break the unity and only partially account for the diverse set of phenomena. It is generally thought that unification-type explanations lag behind the causal ones in stopping the regress of explanation, since with the causal explanation (as with realist metaphysics) the explanatory regress stops with the bare fact of how things are in the world.

Historical analysis, though, places the unification-type explanations as a starting point for the development of causal ones (as a specifically motivated special case), possibly justifying the viewing of explanatory success (of any workable kind) as more fundamental than causal relatedness. Direct comparison of the two approaches over case-study instances and their framing in the general considerations of deeper explanations invites far-reaching consequences for the application of the common sense conceptual framework as the starting point of the realist strategy. These call for the abandonment of the instantaneous state of the extended material structure as the fundamental unit of the realist ontology, and its replacement with 'generalised things' partly defined as objects of fundamental laws.

The structure of the thesis is as follows. The opening chapter outlines the details of the proposed methodological instrument and justifies its construction, as well as its application to theories and associated world-views from the history of science. It draws conclusions for the proposed instrument from an in-depth analysis of the research context (including contemporary analyses of the history of science) and most notably the proposal of the simple transcendental strategy for realism. The latter suggests that it is most rational to assume the validity of the conceptual scheme that contains objects existing independently from us in an objective framework of space and time, a simple unpacking of the conceptual commitments of the everyday language.

Chapter 2 introduces the main variants of the principle approach as a case-study instance, and applies the principle side of the methodological instrument to the explanatorily troublesome phenomena from the domain of quantum information theory. The motivation for the principle approach, the nature of explanation it is able to provide, as well as the extent of its metaphysical commitment, is distilled in the conclusion to this chapter.

Chapter 3 explores the constructive side of the methodological instrument and aligns it with a case-study instance of contemporary Bohmian theory. Furthermore, the chapter explores the metaphysical (though still constructive) expansion and alteration of the simple constructive scheme, as through the introduction of primitive laws of temporal evolution This is explored as the desired connection between the requirements of quantum phenomena and the construction of explanatory narratives along realist lines.

In the final, fourth, chapter, principle and constructive perspectives as instantiated in the case-study instances are brought face to face in comparative analysis, against the theoretical accounts of deeper explanations. As a result, suggestions for an altered view of primary qualities and immediate objects of experience, with respect to the entrenched nature of the basic physical concepts of most human languages and the fundamental scientific role of quantum mechanics, is offered. It is argued that constructive approaches along the Bohmian lines, even with the modifications of the everyday conceptual framework, offer a deeper explanation of the paradoxical phenomena, whilst still respecting the simple transcendental strategy for preferring the realist worldview.

# 1. SPATIAL EXTENSION, NONLOCALITY, EXPLANATION

> Thus science seems to be at war with itself: when it most means to be objective, it finds itself plunged into subjectivity against its will. Naive realism leads to physics, and physics, if true, shows that naive realism is false. Therefore naive realism, if true, is false; therefore it is false. (Russell, 1940, p. 15)

## 1.1 Understanding the material world

### Philosophy and a physical problem

In the simplest of terms, this thesis takes it as given that contemporary physics is at an impasse concerning the empirical equivalence of formalised quantum theories. In other words, science has come up against the wall of empirical equivalence of different formal approaches to the problems to be elaborated below, but these approaches carry widely differing associated metaphysics. Empirical investigations cannot decide between them. This might immediately suggest that we are dealing with a pseudo-problem, something to be rejected altogether and replaced by a fresh perspective (such examples have been known in the history of science). Scientifically, no such perspective has been offered so far, at least not sufficiently overarching so as not to be just another pseudo-solution for the pseudo-problem. This, on the other hand, might suggest that we need to at least look at the problem more closely using the existing paradigms only in ‘new hands’. The ‘new hands’ are to be provided by philosophy. The aim is to help science explain.

> So many people today – and even professional scientists – seem to me like somebody who has seen thousands of trees but has never seen a forest. A knowledge of the historic and philosophical background gives that kind of independence from prejudices of his generation from which most scientists are suffering. (Einstein to Thornton, 7th December 1944, indexed in the Einstein Archive as 61-574; as quoted in (Howard, 2004))

> Concepts that have proven useful in ordering things easily achieve such an authority over us that we forget their earthly origins and accept them as unalterable givens. Thus they come to be stamped as "necessities of thought," "a priori givens," etc. The path of scientific advance is often made impassable for a long time through such errors. For that reason, it is by no means an idle game if we become practiced in analyzing the long commonplace concepts and exhibiting those circumstances upon which their justification and usefulness depend, how they have grown up, individually, out of the givens of experience. By this means, their all-too-great authority will be broken. They will be removed if they cannot be properly legitimated, corrected if their correlation with given things be far too superfluous, replaced by others if a new system can be established that we prefer for whatever reason. ((Einstein, 1916, p. 102); as cited in (Howard, 2004))

At the beginning of the 20th century Pierre Duhem famously claimed that physics and science were not expected to provide explanations, but merely descriptions. However, explanations remained in the domain of philosophy (which, concerning quantum theory, was not expected to be separated from physics before 20th century). A simple illustration from Hitchcock (2004) will help us set the stage for the type of explanation we are concerned with (as opposed to those that we are not, though will be often skirting them).

> This banishment of explanation from science seems to rest on a confusion, however. If we ask “Why did the space shuttle Challenger explode?”, we might mean something like “Why do such horrible things happen to such brave and noble individuals?”. That is certainly a question for religion or philosophy, rather than science. But we might instead mean

> "What were the events leading up to the explosion, and the scientific principles connecting those events with the explosion?". It seems entirely appropriate that science [and, by extension also philosophy of science] should attempt to answer that sort of question. (Hitchcock, 2004, p. 8)

But one might object that all the effort expended over the following three chapters in comparing the depth and width of proposed explanations is a consequence of a stubborn refusal to accept Kuhn's view of scientific paradigms. Briefly, in such a view what we are dealing with here are two paradigms, concerning the same scientific project, and depending on which paradigm wins over the physics community (given the empirical equivalence), we will have our problem resolved one way or another. Though frivolously sketched here, this issue can be easily dismissed by pointing out that we are dealing with a problem that has to be fitted into a larger framework (cf. separability violations), and that therefore paradigm shifts would involve more than just the narrow community of specialists. An even simpler, but as effective, answer is that we are dealing with a philosophical question of general preferability for structures of explanation, and that the community decisions in one historical instance do not bear on such matters however powerful they may appear in a given social setting or historical context.

In terms of explanatory ontology, our central problem is whether "there is a genuine **nonlocality** in the workings of nature, *however* we attempt to describe it" (Albert, 1992 , p. 70, my bold script), or not. To answer the question affirmatively is to be committed to 'hardcore' ontological scientific realism and whatever theoretical models it has to carry in tow (only one of which we shall investigate as a case study instance). To answer it negatively is to seek an explanatory model based on weaker realism (cf. section 1.4. below). But crucially we must bear in mind that the latter position is not to be agnostic about nonlocality, on the contrary it is to strongly deny it. Yet, to position the debate in terms of nonlocality rather than specific physical entities, is to move to a different level of the realism debate. It is to rise away from peculiarities of the details of different ontological postulates to the issue of overall conceptualisation of the world through physical theory (the task of 'descrying the world in physics').

Given empirical equivalence of the theoretical, physical[1] approaches to the supposedly locality-violating phenomena (i.e. lack of prediction of empirical, observable difference between the phenomena as predicted by one or the other physical formalisation) what is expected from the more general philosophical considerations of explanation? Philosophy, done in the wake of Wittgenstein, teaches us to look again, and look hard, at the most obvious aspects of the problem before us, because the real solution is hidden behind the simplicity and familiarity. The phenomena of teleportation, EPR-style correlations and the like are hardly familiar to many people, but their problem-generating aspects such as spatial separation, propagation of causal influences, individuation of objects etc. are. It is those familiar aspects, such as the conceptualisation of the world founded on geometrical permanence of primary qualities at every level of detail, i.e. the conceptual ontological foundation of all of the material world on primary qualities, that we need to keep an eye on, most notably when describing the unfamiliar phenomena in a language employing a pre-existing conceptual scheme.

As will become very clear from the exposition below, and will be explicitly addressed in several more technical instances in what follows, this thesis proposes to look at the explanatory structure as it can be distilled from some quantum theories with a slant on its ontological characteristics. Some might object that explanations are essentially epistemic constructions and that any ontology tied with their particular instances is added at a later stage or stems from some requirements that are extraneous to explanation itself. In considering a possible realist strategy of response to numerous (for our purposes collected and simplified here) ontology-agnostic or explicitly anti-realist philosophies garnered by postmodernist movements in general philosophy, it will be of importance to focus in the analysis that follows on those explanations that are taken to be of the ontological, or the ontic, type and then finally the specific ontological characteristics they display. That this should not be an impossible strategy even from the general philosophical perspective can be glimpsed from e.g. recurrent theme in Ruben (1990) that explanation is an epistemological concept, that requires a general metaphysical (and this includes a more specific ontological) backing. Our transcendental strategy, to be introduced in section 1. 4 below, explicitly requires that we look into the commitments that stand behind (as a 'backing' of) the concepts we employ even in everyday communication.

[1] In the spirit of the opening paragraphs we might say 'scientific' here, although the distinction between scientific and philosophical aspects of the discussion will increasingly be blurred below.

In general it might be assumed, though, that through focus on ontological features of explanations we are giving precedence to a particular type of explanation, and with it a general scientific world-view, and thus prejudicing the question to be settled through a more detailed consideration of the case-study instances of quantum theories below. The supposedly preferred type of explanation is the causal-mechanical type (see section 1. 6 below for a more detailed exposition), as suggested by Salmon (1984, p. 81): "to explain an event is to exhibit it as a occupying its (nomologically necessary) place in the discernible pattern of the world". But we shall be interested in leaving an option of ontic explanations more widely accessible, as generally requiring of an explanation that it is about some real worldly feature, relation or something else (cf. Ruben (1993, p. 5)). Such further relevant concepts might be given by Kim's (1974) considerations of various determinative or dependency relations, of which causal relations are only a smaller sub-kind. This opens up other determinative relations (e.g. 'Cambridge dependency', supervenience, relation between actions, relation between a disposition and its structural basis and the like) that pertain to essential links within the observed general conditions and the phenomena to be explained, but are short of identity, to be used in ontic explanations by our case-study instances. Whatever the general conclusion of these metaphysical considerations it lays sufficient ground for our considerations of the ontological characteristics of explanations.

## Explanations in the philosophy of science from a historical perspective

We shall try to make at least a partial break away from the tradition in the 20$^{th}$ -century-philosophy-of-science analyses of scientific explanations. Though perhaps the most natural reading of the problem we are addressing in terms of explanation would be to consider all approaches to the 'troublesome' phenomena from the deductive-nomological paradigm (Hempel, 1965) with some aspects of inductive-statistical model[2] thrown in, we shall not go down that route. The primary reason is that it does not provide enough ground to distinguish between the two approaches in our case-study instances below. Furthermore, such models by and large tend to be anti-metaphysical (Bird, 2005) trying not to squabble over the details of ontology behind the phenomena at all, but to merely present the syntactic deduction of the formal description of the phenomena as resulting from the formal description of the initially observed conditions and the codification of laws. They are thus not suited for investigating the ontological characteristics of different accounts and their agreement with an overall worldview.

[2] Basically, we could deduce the phenomena from the formalism of the theory, allowing for the statistical aspects in where the predictions are chancy and our ontology (if we specify it in enough detail) permits the introduction of objective statistical elements.

When viewing explanations in a different way, more suited to scientific realism, classifications of explanations that differentiate between our two case-study approaches open up. This different way connects the phenomenon and the background theory through semantic entailment (thus saving it from obvious problems faced by the traditional models, such as the flagpole-shadow example; for further examples cf. Bromberger (1966)). It is too early to get into more detail concerning models of explanation at this stage, but we ought to make a note that the search for an explanation with satisfactory ontological characteristics will have to take into account more than mere deducibility of phenomena from the theory, it will have to show what such deduction would *mean* for the real world. This will of course be of importance when considering the acceptability of the violations of separability, through the phenomena exhibiting nonlocal characteristics.

Most recently (from the historical perspective of this section) Woodward and Hitchcock (2003) develop a model of explanation from an argument that to explain why some phenomenon occurs is to show what (e.g. other phenomena, presence or not of entities etc.) that phenomenon depends upon. Showing the latter satisfactorily is not to play with general counterfactual situations based on the phenomenon to be explained, but only with those that consider variations in what would happen under *interventions on the 'system at hand'*. Thus on their account the choice of basic ontology precedes the attempts of explanations, but explanations will be more or less successful based on the success of this prior choice. Of course, identifying the system at hand may not be so difficult when dealing with macroscopic objects, so that may be a good place to start for both our approaches, though in the end some sort of reduction to less obvious ontology may be required.

It is worth adding a warning though, even before we properly discuss the various possible ontological aspects of the problem in the case-study instances, that the success of explanatory models will not only depend on the choice of ontology, but also on its epistemic accessibility. Though subscribing to the overall realist perspective, our approaches are empirically equivalent and we have no recourse to the all-knowing arbiter to tell us which of them gets closer to the truth. So it is important to limit the explanatory 'buck-passing' that is characteristic of the *hidden structure strategy* (Woodward, 2003), and thus limit the pitfalls of excessively speculative metaphysics. Genuine candidates for explanation will have to identify epistemically accessible, non-hidden features *in virtue of which* they are explanatory.

On the basis of this some say that explanation in general is impossible in quantum theory (Salmon, 2002), whilst others take comfort in the fact that quantum theory can be formulated on the basis of a small number of highly general principles, and that it is universally applicable as a theory of material phenomena (the essence of the principle approach to be outlined in 1. 6. and Chapter 2). For the latter, it is acceptable that quantum theory provides unification/type explanations, whilst not providing those of the causal-mechanical sort.[3] On the other hand, Chapter 3 will illustrate that the causal, even mechanical, explanations can be constructed, at the price of giving up on locality. The deadlock situation brings quantum theory, and with it fundamental physics, close to the more contestable special sciences where we can also provide functional explanations of the phenomena without the possibility of constructing the causal mechanism behind them. This is why some of our considerations will apply more generally, beyond the narrow scope of a few 'troublesome' phenomena in contemporary physics. We shall return to the issues of use and depth of explanation in the final chapter.

## 1. 2 Historical background of explanatory conceptualisation of the world

### Quantum theory and everyday intuitions

We can thus expect the possibility of theoretical justification for locating the explanatory power in physical sciences in ontology, i.e. the primary entities assumed to exist in the domain under investigation and producing the observable phenomena through the specificities of their interaction (Cao, 2004). As Cao says, "primary entities are those from which all appearances (other entities, events, processes, and regularities) are derivable as consequences of their properties and behaviour; these primary entities display regularities and obey laws, the so-called fundamental laws in the domain covered" (Cao, 2004, p. 175). Yet it is precisely this common-sensically sound view that runs into trouble in providing explanations based on quantum theory.

Why should quantum theory be special, as opposed to genetics or meteorology? After all, Cao (2004) does not advocate a simple reduction of the observable phenomena to the primary entities (as if zooming in with a microscope), nor does he seem to warrant the possibility of explanation of all phenomena solely in terms of the entities open to direct observation and experimental experience. What he in fact advocates is the reliance on metaphor, a metaphor that allows for change of the primary actors with the adherence to the overall structure. To understand the meaning of a phenomenon as presented through an explanation in a specific scientific domain, we must provide a chain of metaphors from such fundamental explanation to everyday life reliant on the structural similarity possessed by each link of the chain. And a great number of these metaphors are historically developed, not created on the spot for the purposes of explaining away troublesome phenomena.

[3] For the differences between these types cf. section 1.6. below.

What Cao in (2004) seems to advocate then is to start up with seemingly intuitive understanding of the most basic mechanics of the directly observable phenomena, motion of human sized objects in the Euclidian space of our visual field and from it link up structurally sound metaphors to the supposed existents in the less accessible domains. Yet the less accessible domains should also contain entities with properties whose structure of interaction we can link (though the chain of metaphors) to our intuitive understanding of the macroscopic world around us. Even if we were to accept the existence of such intuitive understanding, quantum mechanics is still capable of denying the tenability of this strategy.

This is because such quantum existents seem to resist consistent ascription of a factual property before its status has been measured. This applies also to the processes following the measurement of such property that involve further interactions between existents; they in a way lose the firm property until we can establish it by measurement again. Furthermore, this instability of property ascription can be taken to the very existence of the entities (i.e. treating existence as a property), especially if the latter is characterised by continuous occupation of the space-time points (i.e. something like a space-time trajectory). This furthermore threatens the construction of a continuous causal process, where the power of the cause reaches from one end to the other of the causal chain. Finally, there appears to be an inherent randomness in the evolution of causal processes threatening the account of singular causality. What we effectively have is the abstract mathematical formalism that expresses general laws and principles such that they cannot be taken as representing physical processes visualizable in spatial-temporal terms.

## Brief history of primary qualities: how we got where we are now

Again, there is no room here to properly lay out the historical role played by space and primary qualities (susceptible to mechanical treatment) in development of scientific explanations, but a brief outline of the general idea is in order. This can prove illuminating due to the importance of something like the primary qualities view in the common sense contemporary conceptualisation of the world, as well the preference for causal-mechanical explanation in contemporary philosophy of science. Some criticisms of historical development of the view popular today may help us open doors to their revision that at first glance appeared too radical to muster.

### *Ontology*

In classical times two major explanatory worldviews can be contrasted. The perversely compounded[4] Aristotelian-Platonic view construed the everyday world as a confused reflection of an underlying reality. In Aristotle's view this reality is given by the necessary relation between the universals, of which the observed individuals were combined instantiations. Explicating the universals instantiated in them is the necessary step in understanding the world, for once a given universal is highlighted the understanding follows. In the Platonic view, the true reality is merely more perfect, but not structurally radically different from the one we observe. In fact, a relationship between a universal and individual could be shown to be of importance here as well. But in both we have reality and common-sense (and scientific) conception of it as an original and its imperfect copy (similar in every respect, only of poorer quality).

A radical discontinuity between the observed and the real is suggested by the atomists Democritus and Leucippus (Losee, 1993), because we can no longer view the everyday and the real as the original and an imperfect copy. The reality was for them different in kind from the world known by the senses. It consisted of the motion of atoms through void (space), and these motions and various combinations resulting from them gave rise to the experiences such as colours, odours and tastes. But the real existents, the atoms, only bore the properties of size, shape, impenetrability and the propensity to enter into various associations. Thus they did not themselves bear all the properties they gave rise to, such as colour.

What is crucial here for explanatory methodology is the notion that observed changes can be explained by reference to systematically fundamental processes occurring at a more elementary level of organization (Losee, 1993). Seventeenth century philosopher-scientists readily adopted this view. In itself this was not a result of fashion or revolutionary feeling, but of observation that it is in fact impossible to adequately explain the qualities and processes at one level by the same qualities and processes at a deeper level.[5] The worry is, though, whether this replacement of properties can go too far. Before considering that question, let us see a further strength of the atomistic explanation. Namely, the atomists suggested the replacement of qualitative changes at the level of observation by the quantitative (i.e. mathematically formalizable) changes at the atomic (fundamental) level. This was in line with the Pythagorean notion that scientific explanations ought to be given in terms of geometrical and numerical relationships (Losee, 1993).

---

[4] I am not aware of literature that provides such unification of the two dominant classical views. I do not even wish to claim that such unification can get far off the ground as a theory in history of philosophy. My main purpose is to contrast it with the atomist view.

Yet one difficulty of the atomistic explanations was apparent from the outset: they could not be verified by direct observation. Moreover, from the outset they were plagued by some ad-hoc replacements for the lack of contemporary experimental and observational precision. As Losee (1993) illustrates, the atomists could not explain why salt dissolves in water whereas sand doesn't, other than stating that the salt atoms are such as to produce the phenomenon of dissolution whereas the sand ones aren't.

Descartes (and his immediate predecessors and contemporaries also, to a varying degree) took the atomistic worldview further, and linked it inextricably to space in proclaiming spatial extension as a necessary characteristic of any fundamental physical ontology. To do this Descartes sought to extricate what is 'clear and distinct' about all physical objects, and deduced that it must be spatial extension (coupled with impenetrability). Thus he distinguished between the primary qualities that all bodies must possess in order to be material bodies, and secondary qualities that exist only in the perceptual experience of those bodies and phenomena that they are a part of.[6]

In summation, primary qualities were those that really belonged to the material objects, whilst the secondary qualities were derived from (i.e. explained by) the state of the objects' primary qualities. The primary caused and explained the secondary (Shapin, 1996, p. 53).

---

[5] It can be argued that development of optics, particularly rudimentary microscopy, opened the door to radically new structures behind the everyday observable phenomena.

[6] Though, of course Descartes was not the first to introduce the distinction, its elements can be traced back to the early atomists, and its first clear seventeenth century articulation is attributed to Galileo (Shapin, 1996, p. 52). But more interestingly for us, Descartes' approach seems to follow the principle paradigm in that he did not speculate (in deriving the primacy of extension as a quality) about the detailed structure of the construction of material existents, but followed a general rule seeking 'clear and distinct' perceptions of properties. Moreover, he directly diverged from the atomists over the existence of empty space: in principle for him all space had to be filled by matter, i.e. effectively equated with matter. Yet, it can be argued, his physics contained manifestations of practical commitment to vacuum and absolute space (Losee, 1993; Huggett, 1999).

Yet as the corpuscular explanations of the phenomena became more technical the gap between the philosophically legitimate account and common sense widened, so that increasingly the sensory experience offered no reliable guide to how the world really was. Economising on an extended debate over the details of this picture, it suffices to say that the corpuscular mechanical explanations were providing a successful alternative to the Aristotelian doctrine of "substantial forms" (i.e. abstract and non-quantitative real qualities). The "substantial forms" were a product of rational examination of relationships in reality, and were ostensively as inaccessible as the atomic corpuscles. But the 'mechanical philosophers' (Shapin, 1996) claimed their explanations were more intelligible, or in our terms had greater explanatory power. In Lipton's (2004) terms they embody a powerful combination of unification and causation (by reducing the phenomena to mechanical processes) styles of explanation, and avoid the need to introduce a gratuitous multiplicity of explanatory principles (Della Rocca, 2002).

Though unification is undoubtedly their great strength, such reductions to supposed underlying mechanism have been known to be pushed too far in an attempt to explain *all* encountered physical phenomena. Thus objections to their historical success have recently been raised, suggesting that they may not have universally relied on greater intelligibility, but on philosopher-scientists' agreement that this simply is the right explanatory paradigm to follow (Shapin, 1996, p. 57). We come to notice a 'circle' in that the phenomena to be explained were caused by the entities whose structure was such that they caused the phenomena (Gabbey, 1985). It has been suggested that the reasons for success of the mechanical explanations ought to be sought as much in historical circumstances (such as increasing practical success of mechanical machinery (Marsden, 2004)) as in their philosophical plausibility.

### *Space*

Though the investigation of space has perhaps been the most fruitful interaction between physics and philosophy historically, its main debate concentrated on the metaphysical status of space: whether it is something *absolute* (endowed with existence independent of all things material[7]) or a construct of *relations* between other existents (namely, material bodies). Though we will primarily be concerned with the explanations that rely on the reduction to the microscopic, we can assume, as is generally done in contemporary physics, that ‘space’ is the same concept presupposed by motion (spatial change) of all bodies, from tiniest particles, through human-sized bodies to the whole universe. The main debate between the absolute and relative views of space will not be our concern here. What is of interest to us is the nature of influences, or the forming of correlations, between the changes in objects that are not spatially contiguous. Whether there is absolute space between them, or instance of formal relation functionally indistinguishable from absolute space, will not influence the outcome of our discussion.

[7] We can, for the purposes of the discussion that is to develop subsequently, ignore the relativistic (i.e. pertaining to Relativity Theory) interaction between matter and space. The characteristics of space that concern us will not be affected by its 'bending' by mass of material existents.

This is because, despite being omnipresent, space in physics (and even relativistic space can be shown to fall into this category)[8] is exceptionally inert. It does not even have the indirect causal effect such as we attribute to the supposed unobservable material existents. As shall be explained in more detail later, for our purposes, space acts as a barrier, a constriction on the proposed explanatory models. The problem is that without this barrier we are unable to do structured physics the way we have been used to doing even from classical antiquity. Abandoning space, thus, may be too high a price to pay, one we shall not be risking here. Yet, we will expect of our barrier to not act in a haphazard way: standing up or falling down randomly. This consistency is something easily visualizable from everyday life: separations are sturdy and we do not expect them to expand, shrink or disappear at whim. This does not make them impenetrable, but merely penetrable according to consistent ‘laws’: separated things can influence each other, but they have to do so by transmitting ‘the influence’ through every bit of space between them. This can be formalised even if ‘space’ does not exist, but is a mere relation between the bodies. This relation is consistently systematic.

But we cannot completely ignore issues of space in the history of physics, because somewhat like unobservable microscopic entities, space has been employed in physics to provide better explanations. And this use was then backed up by metaphysical speculations about its nature. So we have to be aware of the ground the concept stands on physically, when employing it in the discussion to come. The other reason is that in the metaphysical model founded on primary qualities as measurable, and thus real and firm, properties of the foundational physical ontology, space plays an undeniable role. It shares the same essence with all matter (according to some interpretations, it is a part of the essence of matter): extension. On the other hand, in the very formalism of quantum mechanics, space does not appear as a fundamental element of the theory or a fundamental observable. But, when combined with macroscopically observable phenomena it has to be accounted for, as space is an essential part of the conceptual scheme at that level. Effectively, we want macroscopically spatially separated objects not to be conjoined, contiguous or interwoven at the microscopic level as that produces problems in the structural isomorphism between the observable phenomena and their explanatory reduction. And the isomorphism, easily formalizable through geometry, was one of the strong reasons for choosing this particular aspect to be fundamental (rather than, say, colour, scent or rate of vibration). Einstein can be interpreted as saying as much (cf. (Born, 1971) and quotes below) when claiming that the whole of physics as we know it depends on it.

[8] There have been suggestions to exploit extreme bends, shortcuts in space-time, known as wormholes, to explain the apparent connection between otherwise spatially separated objects in quantum mechanics. But as Maudlin (2002) elucidates, this is not a promising route to take, as the wormholes would have to have strange choice of appearance, as well as allowing the hypothesised ‘information’ to pass between the objects, but not the objects themselves, or their radiation or massive parts. Most importantly, if wormholes are indeed a part of the game, then one ought to be able to use them to send superluminal signals, which is not the case in the ‘troublesome’ situations we are dealing with.

### *Method*

The ways to deal with the problem then, require ontology of explanations that either does not need space such as it had been historically presented (including the properties of matter that are associated with it: namely the fundamental role of the primary qualities) or that introduces ontological elements that are independent of space. Historically, that calls for the mystical substance of mind, but we shall not go down that route. We can introduce completely new ontologies that do not rest on extension. The interesting issue, of course, is to see how those figments of imagination can be made to fit with the rest of the standard conceptual scheme so as to save most of our appearances and not call for a single-sweep and all-pervading replacement of the world-view. What we need is a change of paradigm, such that it replaces the problematic parts, whilst keeping the rest of the picture as much like the old one as possible. The question is whether the explanations based on primary qualities can be simply augmented, or whether we will, in the end, be forced to abandon them. If the latter is the case what can come to replace them, given their deep entrenchment in the ordinary conceptual scheme?

But there are historical precursors to our predicament, in for example Kepler's approach to the empirical equivalence of the contemporary competing 'astronomical hypotheses'. Predictive success of either could not help choose between them, and Kepler had to resort to other means to achieve, as he termed it 'change of syllogistic context'.[9] Kepler terms all the problems that result from empirical equivalence pseudo-problems, and advocates changing the syllogistic context so that the competing hypotheses no longer display empirical equivalence and thus the impasse of the pseudo-problem is overcome. So far, this is what most science textbooks advocate also, one must find the means by which to falsify some hypotheses and corroborate others. But of course, there are real experimental situations in physics in which this can't easily be done. And, history teaches us, this is where we step outside the realm of pure physics, into philosophical, even aesthetical, speculation. What Kepler did was to look into physical plausibility (above mere calculational adequacy) of a mechanical model that was to support the observed phenomena on either hypothesis. Nothing revolutionary by today's standards (e.g. choose the simplest hypothesis), but an important historical precursor nonetheless, because it indicates that in search for a better explanation we must consider the wider picture (without prejudicing the choice between causal and unificatory explanation-types here, cf. section 1.6. below, both can provide the fitting into the wider picture). But in Kepler's case there is a much more elaborate justification for an appeal to simplicity, namely as an understandable geometrical order underlying apparently diverse phenomena. This was not a mere appeal for a search for the grand unifying theory no matter how crazy it may be (for example a numerological explanation of the planetary distances), but also a call for further-reaching testing opportunities[10], and avoidance of ad hoc modifications (Martens, 1999). And the unification in Kepler's style, as Martens argues, leads to a wider explanation of the very different phenomena, i.e. points to the truly fundamental elements of explanation, including the ontological ones. The second example of the escape from impasse based on the simple foundational principles is the famous one of Einstein' s Special Theory of Relativity, which is to be recounted in greater detail below (section 1.3. and Chapter 2).

### *Quantum theory in the historical narrative*

The twentieth century produced two radical revisions of the physical worldview – relativity and quantum mechanics. Although it is the theory of relativity that has more deeply pervaded the public consciousness, in many ways quantum mechanics

[9] I am indebted to Rhonda Martens for useful pointers on this issue.

[10] As testing on isolated samples affects the understanding of the whole, requiring a single cause for all the diverse phenomena, or at least a single principle behind the causes of the diverse phenomena.

> represented the more radical change. Relativity required its own accommodations, but at least it still allowed the retention of classical views of determinism and local causality, as well as the conceptual separation of the experimental object from the measuring apparatus. (Evans, 2007, p. 1)

This supposed rejection of the classical worldview was received with different attitudes amongst the developers of the theory in the first part of the twentieth century. Whilst some, most notably Werner Heisenberg welcomed it, others, such as Albert Einstein, Erwin Schrödinger and Louis de Broglie worried about its implications, with Einstein steadfastly rejecting their metaphysical side. Niels Bohr seemed to make peace with a necessary cut between the classical conceptualisation of our everyday physical experience, that of the macroscopic objects, and the novel, strange but orderly non-classicality of the microscopic entities described by quantum mechanics. As Evans (2007) points out, this divide between the microscopic and the macroscopic along the lines of quantum and classical was (or is) no less drastic than the Aristotelian separation between the celestial and sublunar realm, or Descartes' division between the substances of matter and spirit.

By and large, the 'troublesome' aspects of the theory hinge on the notion of entanglement:

> When two systems, of which we know the states by their respective representatives, enter into temporary physical interaction due to known forces between them, and when after a time of mutual influence the systems separate again, then they can no longer be described in the same way as before, viz. by endowing each of them with a representative of its own. **I would not call that *one* but rather *the* characteristic trait of quantum mechanics, the one that enforces its entire departure from classical lines of thought.** By the interaction the two representatives [the quantum states] have become entangled. (Schrödinger, 1935, p. 555) (my bold typeface)

Soon enough further, formally justifiable, conceptual problems had arisen out of this, most notably with the EPR situation. Einstein, Podolsky and Rosen claimed as early as 1935 (Einstein, Podolsky, & Rosen, 1935) that the theoretical formalism predicts the occurrence of certain phenomena that go against the grain of both common sense and classical-physical conception of reality, and thus the formalism must be incomplete and in need of further development (i.e. better alignment with what is really going on in the physical world). Einstein saw the realistic interpretation of the quantum formalism to be attacking the important principle of separability, the one he claimed the whole of physics (and we might project even further: the whole of common sense conceptual scheme) rested on.

His argument rests on the situation in which a pair physical systems A and B, jointly described in the language of quantum theoretical formalism by an entangled (joint quantum) state, which does not tell us anything about the *individual* properties of the systems become functionally spatially separate (i.e. become operationally distinct). When a measurement of a certain property is performed on the system A, the outcome of the measurement together with the laws of the formalism, immediately assigns a new state to the distant system B. Subsequent measurement can confirm the correctness of this ascription in accordance with the standard rule for ascription of states in quantum formalism. As our conceptual framework, and the description of this hypothetical situation, makes the system sufficiently separated to bar physical influence propagating between them[11], we must conclude that no physical change has occurred with the ascription of the new state to the system B. But if there had been no change, that means that the system B already had the contested property at the outset, before the measurement on system A. This leads Einstein et al. to conclude that the quantum theoretic descriptions of the world (most commonly those that hinge on entangled states, but not necessarily cf. Horodecki et al. (1999)) are just not complete.

For some time the foundational problems had been swept under the proverbial carpet, due, in part, to great practical success of the theory, but also the belief that the divide is benign. Though the quantum world of the small was conceptually threatening it seemed to remain contained (*pace* Schrödinger's' cat's ill fate) behind the said divide, not endangering tables, chairs and cannon balls. In the 1960s, influenced by the work of John Bell, even physicists began to take the foundational issues, those of the theory's place in the overall worldview, seriously once again. Most of the 'troublesome' phenomena (such as macroscopic exploitations of the supposed entanglement of the microscopic objects, or the demonstration of their teleportation) that will be the focus of so much of the discussion to come are the recent theoretical and experimental breakthrough stemming from that reawakening.[12]

[11] Or at least, the separation is such to make any known physical influence (such as an electromagnetic signal or alteration in potential energy in the relationship of the pair) at least detectable if not downright impossible.

Subsequently, this led to the advances in what is today an independent field of research, the Quantum Information Theory. The work in that field that is of interest to us because the occurrence of some of the 'troublesome' phenomena rests on the technologically exploitable non-local correlations among macroscopically observed phenomena: theoretical formalism predicts that in certain situations the outcomes of interactions with matter conducted very far from each other are coordinated, and this is empirically confirmed and cannot be explained by any local theory. Cushing (1991) says that in the realm of quantum phenomena the "apparently nonlocal nature of the effects" goes over and above the irreducible mystery (the regress of the 'why' question) contained in any explanation. He claims that the importance of locality for explanations is that local interactions allow one to follow the time evolution of the physical processes 'in the mind's eye', which again follow from the deep-seated (though, possibly unjustified) expectations we have of the physical world. The problem arises when nonlocal phenomena clash with those expectations (cf. sections 1.4. and 1.5.).

It is suggested that nonlocal phenomena, even before the appearance of those resulting from the Quantum Information Theory, mandate the modification of at least some of the assumptions that are part and parcel of the core of traditional scientific metaphysics. Yet, one might say, we have been here before, action-at-a-distance (or at least passion-at-a-distance) has always been a problem in scientific metaphysics, the best known example being one of Newton's gravitational interaction. Yet, there are differences between the two situations taken as indicative of further complications in the case of quantum theory. In the quantum case, unlike the one of gravitation, the mysterious interaction is fully instantaneous and does not weaken with spatial distance; it in fact exhibits a complete disregard for the 'quantity of space'. Also, it is limited only to the physical systems from the initial pre-separation set-up (as if a private connection of its own), regardless of how many systems of the same type there are in the surrounding space (Maudlin, 2002).

At the expense of repeating the central tenet of this thesis, two ways out of this predicament take centre stage in our case studies (Chapters 2 and 3). One is to attempt to sever the 'metaphysical' link between the underlying structure of reality and the interpretation of the phenomena as currently available to us: principle approaches holding firm to the epistemic interpretation of the elements of quantum formalism that give rise to the ‘troublesome’ phenomena. The other, to hold fast to the ‘metaphysical’ link and claim that the phenomena are an empirical proof that our hitherto (traditional, standard, classical, everyday) conception of reality is mistaken. The mysterious connection is real and must be accounted for in explanation.

[12] For a more detailed timeline, for which there is no room here, cf. (Evans, 2007, pp. 2-7).

## 1. 3 The research instrument: principle and constructive approaches

### What is a principle theory?

There are probably as many motivations for the principle approach as there are different adherents of it, or at least as many as different versions of the approach, but the drop that started the overflow seems to be the exploitation of the theoretical notion of entanglement in Quantum Information Theory. Once entanglement came to be viewed as a tool in technologically valuable processes a new perspective on its ‘troublesome’ consequences developed.

> After decades in which everyone talked about entanglement but no one did anything about it, physicists have begun to *do* things with entanglement. (Popescu & Rohrlich, 1998, p. introduction)

Though the principle/constructive theories distinction appeared before Einstein (Howard, 2004) he brought it into a sharper focus in his philosophy of science, particularly his justifications of the methodology used in the derivation of the Special Theory of Relativity. Most theories in physics are constructive theories, theories that go hand-in-hand with reductive explanations of observed phenomena in terms of causal interactions between foundational entities. In Einstein’s own words, constructive theories attempt to “build up a picture of the more complex phenomena out of the materials of a relatively simple formal scheme from which they start out” (Einstein, 1954, p. 228). Einstein calls upon a model of kinetic theory of gases which reduces the mechanical, thermal and heat-diffusion processes to movements of molecules, i.e. reconstructs those processes on the hypothesis of motion of the constituents of the gases described.

Principle theories, on the other hand, use the analytic, not the synthetic, method. The elements that form their starting point are general characteristics of the observed phenomena, formulated as mathematical criteria (constrictions) which the phenomena or their theoretical representations have to satisfy. The example Einstein uses here is thermodynamics which seeks to describe (explain) the behaviour of gases without speculating about their constituent elements, but by simply constraining it by the universal principles derived from the experienced fact that perpetual motion is impossible.

Bub (2000) summarises the difference thus. A constructive theory begins with certain hypothetical elements, the elementary entities in terms of which it attempts to construct models of more complex processes representing the phenomena that we directly observe. The fundamental problem for such a theory is how to synthesize the complex processes out of the hypothesized fundamental entities, i.e. how to reduce the complex phenomena to the properties and interactions of those entities. The starting point of a principle theory is a set of empirical 'laws' or principles which provide unexceptionable generalizations of the directly observable properties of the experienced phenomena. The fundamental theoretical task for such theories is to derive a set of formally expressed necessary conditions or constraints on events (events covered by the theoretical framework) that can be seen as fundamental laws behind the observed empirical generalizations. It aims to explain what the world must be like, what the necessary constraints on events must be, if certain empirical laws are to hold (i.e. if observed generalizations are to be recognised as 'laws of nature').

There are a number of problems with the clear cut division presented above, and it is to be used as a guiding model, but one that we needn't adhere to literally at every step. First of all, as later discussions will show there is a clear popular preference for constructive theories in the philosophy of science. We could, in fact, view the foundations of modern science as shaped in terms of constructive theories based on material existents endowed with primary qualities. Einstein himself states that in terms of explanation nothing beats constructive theories:

> When we say we have succeeded in understanding a group of natural processes, we invariably mean that a constructive theory has been found which covers the processes in question. (Einstein, 1954, p. 228)

Yet he is also reported to have added (Howard, 2004) that progress in theory construction (and subsequent explanation provision) is often impeded by premature attempts to develop constructive theories in the absence of sufficient constraints. That is, we get wildly speculative about the nature of the elementary entities running into the danger of 'creating' entities with no more reality than a disposition to fit into the explanatory models we have constructed for them top down, eventually sliding into the danger of the so-called generalization of secondary qualities (cf. Chapters 3 and 4 ). Howard interprets Einstein as advocating reliance on principle theories as a first step in progress to complete understanding of the phenomena in question. Ergo, his derivation of the Special Theory of Relativity as an intermediate step towards the General Theory. In a situation characterised by long-standing lack of explanation (cf. (Cushing, 1991), (Reutsche, 2002), (Maudlin, 2002), (Putnam, 2005)) straightforwardly unifiable with the common sense conception of the material world, and the explanatory constructions of other physical theories, this need not be seen as an unnecessarily complicated strategy.

There is however a further objection that such an idealisation into a two-step conceptually clear process will simply not work. That is, Brown and Pooley (2001) claim that Einstein's own derivation of Special Theory of Relativity does not adhere sufficiently to the principle theory model. Namely, they show that in the said derivation Einstein makes implicit assumptions about the dynamical behaviour of the rods and clocks (material objects) used to define the reference frames in relative motion. Even though he claims to make no assumptions about the nature of the underlying entities out of which material objects in motion are constructed, his second application of the Principle of Relativity in derivation of kinematical transformations rests on the assumption that motion has no absolute effect on the microstructure of the objects used to define the reference frames. This is certainly not an explicit description of the elementary entities out of which the observable measuring rods and clocks are constructed, but is a step towards listing their properties that is not explicated as the universal constraint from empirical generalisation.[13] Though Einstein nowhere exhibits awareness of this non-principle step he is clearly uneasy about the special status accorded to measuring rods and clocks in the Special Theory (Brown & Pooley, 2001).

[13] It is important to bear in mind the difference between dynamics and kinematics here. Einstein's derivations concern kinematical transformations, observable macroscopic effects of motion, but make no explicit claims (and indicate no interest in making them) about dynamics, about forces acting on or within the moving bodies.

> […] strictly speaking measuring rods and clocks would have to be represented as solutions of the basic equations (objects consisting of moving atomic configurations), not as it were, as theoretically self-sufficient entities. (Einstein, 1951, pp. 59, 61)

Yet, it is also obvious that although a deviation from the principle theory ideal, this is by no means its utter falsification. The measuring rods and clocks hold a special status, but only as 'special' entities anyway as they are used to conceptualise the reference frames not provide real-life measurements. The assumption about absence of effects of motion on the microstructure is seen as even less worrying once we adopt Einstein's denigration of the absolute rest frame (aether, absolute space or some such) as then the rods and clocks are properly speaking 'at rest' in their rest-frame and in the absence of the dynamical interaction between rest frames in relative motion there is no reason to suppose anything but the principle of relativity holds for their microstructure as well. Nonetheless, it is a deviation from the principle ideal that makes no speculations about the microstructure except for the explicitly stated constraining principles.

Finally, it is worth briefly surveying the objection that principles in 'principle theories' should have the status of axioms and should not be derivable from the completed formal expression of the theory. If the latter were the case they would be theorems not foundational principles (axioms) upon which the theory is built. Hilgevoord and Uffink (2006) argue that though this is a fine logical requirement, it fails to be satisfied even by Einstein's exemplary principle theory: thermodynamics. Namely, once the theory of thermodynamics is formalised (or at least formulated as clearly as possible), one can derive the impossibility of various kinds of perpetual motion (from the violation of the laws of energy conservation and entropy increase). Likewise, once we have the formal apparatus of Special Theory of Relativity, we can prove the validity of the light postulate and the Principle of Relativity in formal notation. But this does not deny them the status of the foundational principles because in their non-formal expression they did not rely on the theoretical concepts (such as entropy and energy) for their meaning. That is, the 'rule of thumb' says that foundational principles ought to be understood without the introduction of any new special concepts inimical to the theory being developed, i.e. the concepts assigned hypothetical status such as the entities and their properties bear in the constructive theories.

It may seem a lot of concern is placed here on the principle theories, without additional discussions concerning the constructive ones. The reason for this is that constructive theories are more familiar, more common, whilst principle theories are rare, problematic in the sense of explanatory models offered above, and certainly mysterious about the characteristics of ontology they rely on. At first glance they actually say nothing about the ontology behind the phenomena, but it would be a mistake to assume them to be purely instrumentalist. They merely refrain from the speculations about the various details of the entities, even about their most essential (in some cases we might call these 'primary') qualities, over and above what can be gleaned from the constraints imposed by the natural understanding of the foundational principles. But we shall discover more about the principle/constructive distinction as we work through the case-study instances in the subsequent chapters.

### Non-methodological aspects of the principle-constructive dichotomy

Before introducing those instances, something more has to be said about the goggles through which they will be viewed and, finally, compared; the so-called research instrument. The primary dichotomy in the research instrument is one of the principle or constructive approach and follows closely the methodological dichotomy outlined above. It is not freely selected here, but is adopted from the authors of the case-study instance formulations of quantum theory (introduced in the subsequent chapters). Yet, for the purposes of comparing them along the lines of explanation, our research instrument has to explicate divisions between the two approaches that go beyond methodology of theory-construction. We need to glance at most natural explanatory models to associate with the given methodology, as well as the metaphysical status of the theoretical concepts, or more precisely the ontological entities assumed to be the building blocks of the objects participating in the processes the phenomena to be explained consist of.

Chapter 2 presents the principle approach to the phenomena to be explained. Methodologically it relies on the formal expression and subsequent formalised theory construction of the general constraints observed in the phenomena. It is not anti-realist in the sense of making the theory a mere instrument for outcome prediction, as that would not lay sufficient grounds for physical *explanation* of the phenomena. It is anti-realist though in the sense of being agnostic about the nature and mechanical construction of the unobservable entities supposed to produce the phenomena. Its own version of realism gains strong foothold in adherence to separability as the crucial criterion for reality of all physical entities including the possible microstructure behind the phenomena. Real individual entities must for certain experimental purposes be isolated from the rest of the physical universe, or sufficiently isolated so that the effects of their connection to the rest of the universe can be ignored. Hypothetical entities that cannot satisfy this requirement cannot, on this view, be considered real. Through this insistence on separability (to be reviewed in more detail further in the subsequent section of this chapter) the principle approach of Chapter 2 subscribes to the unification model of explanation, as the separability foothold provides for the explanatory terms sufficiently clear from other physical theories and the common-sense worldview. They basically say they don't know the detailed structure that brings about the phenomena, but they know what the real elements of the structure must carry.

The constructive approach, presented in Chapter 3, poses explicit hypotheses about the nature of the entities out of which the explanation of the phenomena can be built. It is realist in the strong sense of taking the unobservable entities as true constituents of the material reality, with properties such that they can give rise to the observed phenomena. They are unashamed of the potential conflict the entities with such properties may have with the common-sense view, most notably the requirement for separability. In their view if explanation of phenomena requires entities that violate separability then we must get used to living in the world in which the fundamental entities are not separable in a way required by Einstein (in (Born, 1971, pp. 170-171)). Obviously this kind of explanation is closer to the causal-mechanical model in which the understanding is provided by detailing the causal interactions between the structural elements. As such, it adheres to the preferred model of theory construction and explanation at the possible expense of having to revise much of the common-sense worldview and the unification of physical explanations.

Modulo potential overlaps between the given idealisations, about which we shall aim to be as explicit as possible, our stage is set to search for the preferred approach to satisfy our explanatory hunger, given the starting point of common-sense conceptualisation of the material world in terms of primary qualities. Our research task is to lay pointers for preferring either approach with a minimal expense to what we already take as understood, most notably the status traditional primary qualities have in the conceptualisation of the isomorphism between the explanatory ontology and the observable characteristics of the phenomena. However, the approaches provision of explanation that we shall survey all rest on the work-in-progress advances in physical sciences and will in some cases not be able to present definitive conclusions as yet. In that case we shall have to do with having pointed out the problems clearly enough.

## 1. 4 Philosophy and the two approaches

In connecting the explanatory strategies of the case-study instances with the wider philosophical world-views concerning status of knowledge, truth and reality in science and scientific practice two philosophical traditions most readily stand out. Even though the principle and constructive approaches presented above will focus on a narrow specialised issue, in a highly theoretical domain of physics, if the conclusions reached are to have a wider application they will touch upon the issues of epistemological status of science as a whole. That is, issues of scientific explanation, whichever narrow domain of science they may originate from, will come across the postmodernist anti-realist criticism. In that respect it is worth positioning the key players in that overarching debate, as well as be aware of the points of contact between any of the overarching schools and the case-study instances of explanatory frameworks presented in the following chapters (primarily, Chapters 2 and 3).

Thus we have scientific realism (for more see below), a doctrine that spans the empiricist and rationalist epistemologies, and maintains that there is an absolute reality beyond the experimenters' consciousness and interpretative alteration. Such reality is translatable and explainable under the employment of prearranged (most notably, objective) method of investigation. The much more heterogeneous doctrine of postmodernism, roughly a continuant of the historical philosophical doctrines of idealism and nominalism, denies it is possible to ever ground knowledge in some absolutist or naturalistic view of reality, guaranteed by firm methodological procedures of investigation. All knowledge, whatever its content and however it may have been arrived at, is forever mediated by language and interpretation (Ward, 1996). The third possible doctrine, though some may see it as part of the overall postmodern critique, social realism, will not be further elaborated on here, as it more properly belongs to sociological analysis of science in the footsteps of Thomas Kuhn, and as stated above there is no room here for a sociological analysis.[14]

The 'postmodernists' (henceforth addressed as antirealists, focusing on that aspect of their position, as broadly illustrated in the positions of (Rorty, 1980); (Putnam, 1981); more recently (Pettit, Realism and Resposnse-Dependence, 1991); (Pettit, 1998)) may raise a challenge that both case-study approaches have little or nothing to do with reality (especially as they deal with such a fringe segment of contemporary physics) and that we are, again, deciding between two world-views preferred by two social groups (perhaps directly competing for power). In the least case, antirealists may claim that neither approach can guarantee the access to the “cosmic register of truths” (Luntley, 1995) which would demonstrate that one worldview, however myopic due to limitations of human perception and conceptualization, is on the right track (i.e. closer to truth than the others). Though aiming to respect (as far as that is possible in the details of individual theoretical speculations) the abolishment of the dichotomy between the reality and the conceptual framework we describe it in, “giving up dependence on the concept of uninterrupted reality, something outside all schemes and science” (Davidson, On the Very Idea of a Conceptual Scheme, 1974), most of the work done here will precisely concern the modifications of the overall conceptual framework so that it may exemplify greater internal coherence in the absence of the precise empirical reference fixing.[15] The latter is not a consequence of the ‘metaphysical’ holism, such as is advocated by Davidson and Quine ( (Davidson, 1977); (Quine, 1969)), though it falls under their general theoretical framework, but of the scientifically ascertainable empirical adequacy of both case-study instances under consideration. It is the leitmotif of this entire work to evaluate under explanation what cannot be adjudicated between with respect to truth (usual standard of comparison of holistic frameworks), with the hope that some overarching conclusions can be drawn as lessons useful even for the ‘bigger picture’.

[14] As our conceptual frameworks shape our record of observations as well, the observable (empirical) aspects of the two approaches have to largely agree on conceptual frameworks in order to be comparable at all.

The general discussion concerning scientific realism (cf. (Gutting, 1982); (Boyd, 2002)) suggests the following starting point for a minimal realist ontological requirement. Both the ‘hardcore’ realist and the constructive empiricist (a softer version of our antirealists above) agree on the coarse ontological requirements of the everyday conceptual framework (tables and chairs, Sellars’ “manifest image” (Sellars, 1963)). The stronger realist sees the need to go beyond that in describing and explaining real phenomena. The weaker (i.e. closer to constructive empiricist) denies this need, i.e. claims that anything beyond this common ground is speculation. Useful speculation, but speculation nonetheless. Manifest image, and more importantly only its coarse version,[16] is the minimal requirement both will agree on.

It is easily acceptable that from a historical perspective science has made an enormous progress in explanation, prediction and subsequent control of the material reality we find ourselves a part of. In this case we shall focus only on the explanation aspect, thus circumnavigating the objections to the consequences of its other two interactions with material reality as given above, cf. (Luntley, 1995, pp. 45-47) . In terms of explanations we expect science to rely on the conceptual framework that is capable of describing the world independently of the dispositional aspects that we find peculiar to our particular position (be it 'human' position, the vaguely 'macroscopic' position, a 'provincial' galactic position, or some such). This is another way of requiring objectivity in the explanatory reports, i.e. excluding from them all aspects dependent on the peculiarities of individual viewpoints. It is very tempting therefore to argue in the modernist fashion that the scientific explanatory conceptual framework rests on the privileged link to what Luntely (1995) terms the 'cosmic register of truths'. Such conception immediately brings with it the notion of a language, as a system of concepts, that can be understood by any creature regardless of how it was constructed or what its spatiotemporal relation to the rest of the universe was, what kind of mind or perception it had or what its history and culture was (Luntley, 1995, p. 48). And we standardly assume that the language of mathematical physics provides just such foundation and it therefore affords us the most fundamental explanations of the world as it is independent of our individual perception of it, as well as the explanation of how our individual perception arises.

[15] And this, on the face of it, seems to be pushing towards the unificatory model of explanation, but a more explicit argument is needed to labour that point. On the other hand it should not be seen as pushing for a specific type of realist argument based on internal coherence of a realist world-view alone.

[16] Coarse because there are details of the manifest image itself which are unobservable, such as unobservable properties of observable entities.

Several problems arise for this picture that are relevant for this thesis, but we cannot go into all of them to the same degree. We have to take as more or less given that the postmodern criticism is capable of challenging the above presupposition of the primacy of link between the scientific conceptualisation of the world and the 'cosmic register of truths' in general. Luntley (1995) can be taken to provide a good introductory summary of the postmodern arguments in this vein (for more detailed accounts and different strategies see for example (Ward, 1996) and (Goldman, 1999)). What is particularly interesting in our case, and something that we shall dedicate more time to is that the case-study instances of quantum theory that we consider in this thesis seem to add grist to the post-modernist mill though both are well versed in the vagaries of mathematical physics and contain elaborate formal accounts of how to address the phenomena we deem 'troublesome'. This is because we take them to be formally equally empirically adequate with respect to providing predictive accounts of what takes place in the 'troublesome' phenomena. Now antirealists have something to point to and claim that mathematical physics itself has through the 'troublesome' phenomena in quantum theory hit the wall of relativism of metaphysical explanations and cannot employ its own supposedly superior methods to get out of the dire predicament.

The antirealist points out that the history of science shows that no set of agreed observations can of its own accord falsify a theoretical conceptual framework, that all of the latter can always be made observationally compatible with the agreed upon set of data. Such sloppiness is defended against in the philosophy of science by abhorrence of the ad hoc additions to a theory and general pursuit of both unification and simplicity. Yet, the antirealists may challenge, even with rigour imposed by the philosophy of science in the case of quantum theories you have a clear case in point, rigorous and formally well supported interpretations are to a large extent conceptually at odds with each other concerning what the minimal metaphysical requirements of the world-interpretation (or explanation of the material processes we encounter or engender) are. They conclude that there is no purely rational procedure (even when enshrined in the theoretical formalism) that can take us from an account of experience to a decision as to which of the two competing theoretical frameworks is true (Luntley, 1995, p. 80).

They can then generalize this to a conclusion that given that all experience is based on interpretation (as presumably the competing conceptual frameworks differ precisely in interpretation, and cannot rest on concepts rooted in experience that would be guaranteed to be free from it), and that there are no other more secure foundations of knowledge (such as Descartes found in the epistemological protection provided by the benevolent deity), there can be no single conceptual framework suitable for reporting majority of what we say about the world (Luntley, 1995). So, from the perspective of explanation there is no need to even burden ourselves with the heavy conceptual framework of the contemporary science, as that is explanatorily as valid as any other 'wish-wash' narrative one cares to produce, provided it can account for the experience of the human subjects (the explainee). This conclusion can be reached by other anti-realist routes (cf. (van Fraassen, 1980) on the pragmatic, not epistemic utility of explanations), but this is a particularly interesting one for our purposes. Precipitating a more detailed exposition in Chapter 3 such arguments suggest that "*everything* we say about an object is of the form: it is such as to affect *us* in such-and-such way. *Nothing at all* we say about any object describes the objects as it is in itself, independently of its effects on *us*" (Putnam, 1981, p. 61). This, however, is a highly impractical position to take, the one that does not allow any realist background against which details of competing explanations can be checked, whilst still asking for some hint of an explanation as to why a particular account is one way and not the other. Even though there is no direct answer to such scepticism, there is a simple strategy that we shall follow below: to ask for a minimal set of 'typings of objects' (Devitt, 1997) that are not dependent on human conceptualisation to explain the experiences they produce. An anti-realist position such as Putnam advocates above has not got such a minimal set to even begin to explain anything.

This is a strategy similar to Descartes' original search for the escape from doubt (though without the role for the deity). Namely, a bit of a transcendental argument and some common sense can help anyone who wants to be helped to escape the antirealist doubt. What even the staunch antirealists have to agree to is that there are external limitations to what we can and cannot do in life, to what it is and is not sensible to believe (cf. (Devitt, 2006) and section 3. 2 below). Even the antirealists don't go jumping off buildings expecting to defy gravity nor do they tend to stop eating upon discovering the underdetermination of the theories of nourishment.

Now this is not to argue that all worries about the reliability and utility of our conceptual framework and the accompanying explanations are just academic exercises, in positing worries as much as in refuting them. What we are counting on, following Luntley (1995, pp. 110-115) is the fact that acceptance of even those basic limitations to our acting and thinking commits us to the sensibility of the notion of things as they are independently of our thinking about them. That is we seem to hold some elements of the conceptual framework to be non-dispositional. As the experience of and interactions with the material objects form one of our most basic such non-phantasmal experiences (i.e. experiences characterized by seemingly externally imposed limitations), Luntley proposes a transcendental argument[17] that it is most rational to assume the conceptual scheme that contains objects existing independently from us in an objective framework of space and time (Luntley, 1995, p. 111).[18] Yet to differentiate it from philosophically burdened traditional form of transcendental argument that proposes as necessary condition in the transcendental step a conceptual background of acceptance of some starting position, whereas all we require is the unpacking of conceptual commitments, we shall henceforth call it the transcendental strategy. That is, given that even the antirealists (of the 'postmodern kind' as suggested above) are committed to thoughts about such objects, Luntley argues that it is more rationally prudent to take them to be originating in some way from the objects themselves, rather than just seeming to us that they do. In a similar vein one might put it to the antirealist that he does not doubt the reality of past events, even though they are not directly empirically accessible, but can be reasonably reconstructed from the present evidence. This of course is a summary of the age old argument for simple realism, but toned here to serve a particular purpose. A very strong argument for accepting the given conceptual scheme, the conceptual scheme of objects in space-time, in just such a way is that it plays a vital role in almost every language known to us and is capable of generating an extensively rich set of beliefs about the world. It is so wide-spread and strong that even the antirealists use it when they go about their daily activities. Luntley argues that they must accept it even at an academically more serious level, and even proposes ways for them to accommodate it deeper into their own particular modifications of the worldview.

[17] Though it may be objected that the 'transcendental argument' is a misnomer in this case, from the perspective of the more famous forms of such arguments, we shall adhere to using the terms for the following reasons. 'Inference to the best explanation' is a much used term in philosophy of science and carries a lot of philosophical baggage which there is no room to get into here. Though our transcendental argument could be seen as an instance of inference to the best explanation, for reasons of generality the former term is preferred. It is also not a form of the general transcendental argument that relies on necessity of some step to push for the conclusion. We merely aim to argue, following Luntley and Devitt, for the sensibility of application of the transcendental step: it is not necessary to see the common-sense conceptual framework as originating in the realist ontology, but it is sensible to do so when explanations of the experienced phenomena are sought. As Luntley puts it, an understanding of the concepts of experience commits us to a belief in the external world, rather than showing the external world to be a necessary condition for the possibility of experience. As to the related objection that transcendental steps are not fully justified and can still lead to errors, this is acceptable from the simple realist position that Luntley (1995) advocates. For the rest of the discussion to make sense we do not require that inferences based on the transcendental step be certain beyond all doubt, but merely that they be seen as sensible enough in search for an explanation. Again, if this brings us back to the 'inference to the best explanation', so be it, but it is illuminating to arrive at it via a different route which does not presuppose the familiarity with much of the existing debate in the philosophy of science; a fresh approach of sorts.

Yet, we shall soon (and more extensively in Chapter 4) be forced to argue that science forces us to accepts modifications of the said conceptual scheme, both in adding to and in changing some of its more central aspects, and that may seem to jeopardize its validity in this thesis again. The saving grace is to make (along with the ancient atomists, and in modern times Descartes and Locke for example) some aspects of it more foundational and unchallengeable and other subject to gradual change under the increase of empirical knowledge. As the changes potentially go astray it is always possible to fall back on the foundational elements. The foundational element is provided, loosely speaking, by the geometrical isomorphism of extension as essential constituent of all material objects, regardless of how large or small they are compared to us. This is the well known story of the primacy of extension, of considering extension and its modes as primary qualities of everything material. With particular reference to our case-study instances, this seems to be the aspect of material reality that neither of them can deny. What is more they must find a way to include it in the construction of their explanations of the troublesome phenomena.

[18] We have to be careful to note here that requiring the conceptual foundation of explanation routed in the unambiguous description of definite objects with definite properties is not identical to Bohrian demands for necessary use of classical concepts in providing objective descriptions of all physical phenomena. We shall delve more into the Bohrian world-view in the following chapter (Chapter 2).

And this is where we come to the final problem for the primacy of the scientific explanatory framework of material world, as suggested above. Quantum theory introduces some phenomena that require a careful selection of the agreed upon set of characteristics so as to construct explanations that respect the essential elements of the common-sense conceptual framework. For, at first glance, and we shall look into this in more detail below, these very phenomena seem to again provide the postmodern-style critic with material to claim the whole scientific conceptual framework has run into serious conceptual difficulties and not only can it not find a way out of an impasse of the empirical equivalence of different interpretations of the formalism (that, we might argue is very specific and academic), but calls for explanatory conceptualizations that do not share the widespread and foundationally firm minimal conceptual framework of objects in space and time. And they do this by supposedly violating separability.

Briefly (as we shall look into this in more detail in section 1. 5 and Chapter 4), violations of separability threaten to knock-down the whole house of cards defence from postmodernism as given above by denying the sensibility of the foundations of the common-sense conceptual scheme. As the following section shows, the idea of physical things existing and arranged into “a space-time continuum” (Einstein, 1948, p. 321) requires that they can “claim an existence independent of one another, insofar as these things “lie in different parts of space”” (Einstein, 1948, p. 321). In other words these objects arranged in space, as required by the core elements of our foundational conceptual scheme, ought to have an intrinsic thisness[19], i.e. whether they are interacting or not they should have separate intrinsic states (Howard, 1994, p. 206). The states can change as a result of interactions, but those interactions can be accounted for again in terms of the extension through the space-time continuum and, provided that the interaction is epistemically accessible in the given small region of space the object occupies, it is always to be separately definable. Furthermore, all composite objects acquire all their properties from the constituents' intrinsic states and locally intrinsic interactions.

[19] This should not be confounded with the notion of primitive thisness and identity as championed most notably in the works of R. M Adams. It allocates a foundational identity, for want of a better term an ‘itness’ (as suggested by D. Lehmkuhl in private correspondence), to the elements of reality but not one they retain independent of their potential for interaction with other elements of reality. At this stage we have to contend with an intuitive understanding of this term, given the proviso that it is not the technical term as advocated by Adams. For our purposes it suffices at this stage to allocate intrinsic states to elements of reality that are not wholly dependent on their ocurrent interactions with other such elements, i.e. not requiring an ontological holism in accounts of the material constituents of reality.

And, as our troublesome phenomena will purport to illustrate, quantum formalism seems to deny this property to the objects in its domain. The fundamental formal difference is that classical formalism allows for the lack of definite separable (formally factorizable) descriptions of the phenomena as ignorance, i.e. enables us to claim that the participating objects are properly separable only we don't have enough information to formally represent that; whereas quantum theory formally precludes such interpretation of the situation (by precluding the aforementioned factorizability).[20] This means that either quantum theory is not a fundamental physical theory and is not concerned with fundamental scientific explanatory ontology, or that we have to find some way of explaining how such separability violations are either benign (to our fundamental conceptual scheme) or just an illusion that does not actually affect the fundamental common sense explanatory conceptualization based on the notion of primary qualities (as sketched above). We have to bear in mind that at least for some properties (and the crucial question is whether for those we are most interested in: the traditional primary qualities) separability allows us to say that this definite object possesses this definite property (Howard, 1994, p. 209), and also to account for the changes of that property through the processes that foundationally rely on the primacy of extension in material world. The depth of explanation accounts (cf. Chapter 4) tend to require conceptualization of manipulations of definite object properties. It will then be our task to investigate what that provision does for the construction of explanatory accounts of the material processes, especially those involved in the troublesome phenomena themselves. Before that we will have to see just how each of our case-study instances proposes to deal with possible separability violations, as well as whether we can find a way of understanding separability so that the proposed violations are not damaging to the foundational aspects of the conceptual scheme.

So what remains of our conceptual scheme and the transcendental strategy, if separability is violated? Howard (1989) interprets Einstein as claiming that separability is the only conceivable objective criterion for ascription of intrinsic 'thisness' to elements of reality, to their objective (and this is important in our transcendental strategy) individuation. This rests on an even deeper metaphysical assumption that spatiotemporal separation is the only conceivable[21] objective criterion of individuation and definition of the foundational ontology. Philosophically this is not an entirely pedestrian observation, as Strawson's (1959) theory of the role of the concept of material object in the conceptual scheme in terms of which we think (and talk) about particulars illustrates. The particulars, along the lines of 'local beables' above and historically exemplified by the macroscopic objects in space and time, form the foundation of our most universal conceptual scheme. In other words they form the core element of every conceptual scheme as they are particulars that can be identified and re-identified without reference to the particulars of a different sort; they are ontologically foundational.

[20] Winsberg and Fine (2003) argue that metaphysical separability does not imply the factorizability of the formal functions associated with the phenomena, but their argument poses further difficulties for the aims of our transcendental argument. We shall return to those issues in more detail in the middle sections of Chapter 4.

We might wonder what the role of the space and time is then. The objective particulars (the 'local beables') serve as our empirical access point to the conception of space and time, as they are three-dimensional (or spatially extended in our terminology above) and enduring through time (allowing not only for identification, but also for re-identification). At the bottom of this conceptual scheme lies a conception of separable (i.e. locally completely definable) space (or space-time) providing for unique objective relations between material particulars and all conscious (and this presumes: linguistically capable) agents. An important aspect of Strawson's ontological foundation for the conceptual scheme must be noted, especially in the light of the forthcoming 'troublesome' phenomena[22]: the elements of ontology (the particulars) that provide the foundation of the conceptual framework must be taken to exist *continuously* through changes of place and time, so that we could re-identify them and thus rely on unique conceptualization for all conscious agents. The question arises what happens if the assumption of the continuous existence is threatened, not haphazardly but in a formal and systematic way. Can we still maintain the necessary re-identification and thus a simple rational assumption of the independent existence of the said 'particulars' when no conscious agent is performing the identification, nor is even suitably disposed to in-principle perform it?

[21] But, Howard (1989, p. 243) notes, we must distinguish this from possible in the sense of either logical or physical as expressed though theory formalisms. In fact, 'conceivable' here marks out precisely what our transcendental argument needs so as to work on the postmodernist as well: that which is conditioned by objective and historical factors, the models with which "we have been outfitted".

[22] What we shall be concerned with in the following chapters is the possibility of granting the existence of the spatially extended basic particulars, but not necessarily their continuous endurance, i.e. we might have to try to contend with them making 'jumps' in identity, if possible.

## 1. 5 Conceptual problems and quantum 'troublesome' phenomena

Separability is the principle behind classical physical explanations of the world, and states that material (include fields here as well) occupants of any two parts of space sufficiently distant from one another[23] must be considered separate in a sense that they each have their own definite set of qualities and that their joint set of qualities is wholly determined by these separate sets (Maudlin, 2002, p. 97). An immediate dynamical consequence of such an assumption is known as the principle of *locality*: an event sufficiently separated (spacelike separated in the language of Special Theory of Relativity) from a given small region cannot influence the physical state assigned to that region. But Bell's theorem shows that quantum theory cannot conform to this picture ((Bell, On the Einstein-Podolsky-Rosen Paradox, 1964); (Bell, Speakable and Unspeakable in Quantum Mechanics, 1987); (Maudlin, 2002)). It accounts for the occurrence of phenomena in which some behaviour of separated pairs of objects (physical systems) cannot be explained by any local physical theory (be it current quantum theory or some general theory that might replace it) without including some non-local interaction between the objects.

Yet, it must be stressed that the nonlocality as implied by the quantum theory is subtle, and despite providing for some further interesting phenomena in the Quantum Information Theory, it does not allow for unpalatable science-fiction-style phenomena akin to telepathy (distant communication without use of classical communication channels) or 'quantum' jumps (non-classically-assisted modifications of properties of distant objects). In summary the said nonlocality (Maudlin, 2002) **does not** require nor mandate:

1. superluminal exchange of matter or energy,
2. superluminal signalling,

but **does** require:

3. superluminal causal connections, or
4. superluminal information transmission.[24]

---

[23] Of course, this needs in fact to be supplemented with a more complete account of physical isolation, including isolating/individuating effects achieved in some other way, e.g. boxes or other barriers. But even those are describable in terms of properties based on extension.

[24] This does not contradict the above anti-telepathy claim, unless one takes information to be necessarily exchanged between human sender and receiver. But in parts of this thesis information transfer is a necessary prerequisite of superluminal causal connections and does not necessarily involve human subjects, but can be

Nonetheless, from the simple explanatory perspective, separability cannot be upheld, as despite of what probabilistic predictions we can make about the distant objects, the explanation of the changes they undergo will require some account of the characteristics of the situation which arises holistically over and above what we know about each separated object individually. Moreover, some of these characteristics will only be available to some experimenters in special circumstances (i.e. will not seem to objective relations established between objects and available to every investigator). We get a feeling that given the connections established between distant objects, perhaps they are not distinct objects or do not really occupy the different regions of space. But this options should not be so lightly accepted for we shall investigate below whether Einstein's expectations of a stable reality arise from their 'thisness' being fully independently specifiable (Maudlin, 1998, p. 54)).

The discussion about the subtle nature of these phenomena is wide ranging, but for the time being it suffices to illustrate how it clashes with the standard explanatory world-view, without committing to the technical details. Namely, traditional folk (everyday) and physical (technical, scientific) conceptual construction of the material world couples the assumption of individual 'thisness' with the principle of separability, to provide an account of individuation (as a basis for *inter*action) of material objects (our physical systems). Howard (1989, p. 244) says that separability is the physical necessity for any account of extension (understood as a sufficient criterion of metaphysical individuation, cf. also (Howard, 1994)), as to make explanatory sense of it we need the extension to come in discrete individuated packets (this is not a claim for necessity of atoms, but for a necessity of provisionally individuatable parcels of matter smaller than the totality of the matter in the universe; in fact small enough to fit on the table top and be susceptible to experiments). A theory that denies separability, such as quantum theory, jeopardises explanations built on this scheme by making the properties of some parcel of the extended stuff depend on something other than the properties of (surrounding) local extension (shape, position, motion or field-based local interactions) alone. The mysterious holistic connection provides for changes in the separated, thus individuated, parcels of the extended stuff, such that they cannot in principle be accounted for by the (known) physical interaction (i.e. by energy, signals or matter; arising from the locally constructed account of the extended stuff) and the properties of the individual parcels themselves. In formal terms: classical phase space built on the notion of extension as primary

---

assumed exchanged between inanimate physical systems. Though, how much this characterization will help us with the final explanatory project remains to be seen.

is expressed in terms of position and momentum. The quantum phase space is different, and it seems that this will need to be reflected in the metaphysics and the explanation of the phenomena.

The separability principle is, according to Howard (1989), tacitly behind the ascription of primary qualities as the only objective qualities of material existents in Newtonian physics, and their further gradual reduction to *position* as the sole objective criterion in distinguishing elements of material reality subject to formal theoretical description. This is of course supplemented by the divisibility of material objects along the lines of extension down to point particles, and finally with the need to explain interaction between the fundamental existents by spatial influences other than perfectly elastic contact action. Thus, all on tacit assumption of separability, we historically build up a half-scientific half-lay conceptual scheme of objects interacting along identifiable continuous 'lines' in space time. This conceptual scheme (for reasons logical or historical is not of utmost importance to us) provides a smooth transition between the explanations resulting from formal physical theories to the common-sense world-view of objects existing outside ourselves and in physical interaction with our material aspect. To abandon this tradition, claims Howard (1989, p. 244), is possibly to go along the lines of Leibnizian metaphysics which (however potentially philosophically complex and sound) was never a widespread foundation for the explanation of the real phenomena, nor was it easily accommodated with the wide-spread (so as to include the antirealist, as well) everyday conceptual scheme.

Dickson (1998, p. 156) objects to the tenability of holism alone as a scientific, and especially as an explanatory doctrine. Holistic metaphysics allows for no individuation of objects that can be said to be in an interaction, nor for their re-identification across space and time. In that sense it is robbing us of the core of our conceptual scheme, its essential part needed to construct an explanation of the phenomena. Also, its connection to the concepts of everyday parlance, all of them structured on objects with intrinsic 'thisness' would be difficult to construct in a manageable number of steps. Namely, permitting the holistic aspect to theoretical metaphysics leaves the generation of the everyday conceptual framework out of the theoretical conceptual framework as essentially unexplainable, bluntly postulated and required but not counterfactually manipulable. We then seem to be back to the knuckles of the early measurement problem: "[...] in what sense and with what objects have we [brought about the occurrence of our 'troublesome' phenomena]? And how are [the phenomena that really occurred] related to the phenomena we thought [we observed]?" (Dickson, 1998, p.

156). And Dickson is quick to point just how a simple resignation to holism does not help remove the worry that the 'troublesome' phenomena raised for the possibility of explanations from physics. For whether we call the correlations formally apparent in the 'troublesome' phenomena results of action at a distance, or the observant correlations between the two parts of the same objects, we still have to explain how the correlations of the space-like separated events come to be formally established and empirically verified.

One possibility is to distinguish separable and non-separable aspects of ontology, maintaining that the link between separability and the core of the conceptual scheme can be achieved solely through the separable part. Thus, Maudlin (2007b, p. 3158) argues that for the conceptual connection between the contemporary physical theories and the common-sense to hold, only some of its foundational elements need to be local (i.e. conform to the requirements of separability), whilst the separability violating segments can be relegated squarely to the section of ontology, different in kind, that is non-local. In Maudlin's words: we can have local beables and the non-local laws.[25] He says this is actually the case in that classical beacon, Newtonian mechanics. One could not get a complete picture of the physical phenomena in the theory solely from the observation of the isolated region of space, as the objects there might behave as if caused to do so from outside the region. That is, a more satisfactory, from a unification of phenomena point of view, explanation is achieved if it is observed that the local objects can change their behaviour under influences from outside the region that are not evident on the local picture. (Of course if we posited the existence of some causal mechanism that governs the troublesome Newtonian action-at-a-distance, such as the exchange of force particles then we could localize all dynamical phenomena in the region.) In Newtonian mechanics, as it is most commonly understood, a change in a distant gravitating body can bring about a change in the local body in the proximal region. To account for that the explanatory conceptualization that includes Newtonian mechanics and the common sense experience posits the existence of local ontology of objects and the non-locality of laws governing change in those objects.

The other is to try to diffuse the potential effects of the separability-violating phenomena as either illusions arising from an ontological mis-ascription of the elements of quantum formalism to the elements of fundamental ontology, or to show them to be constricted by limitations so as not to endanger our everyday conceptualization (something along the lines of: our fundamental building blocks are non-local, but only on occasions in which they are not providing the function we crucially expect from them, i.e. playing the role of the fundamental building blocks in the phenomena that feature in our experiences). From explanatory perspective and the requirement to relate the elements of the common-sense conceptual scheme to those of quantum theory, we must then either show how the mis-ascription arises or how what was intended as fundamental theory manages to produce so radically different common sense concepts. This is to admit that there can be no conceptually foundational connection made between the common sense and the contemporary theories. It then leaves an open question in science, but also a task in philosophy, of explaining how come quantum systems are so radically different, given that they are expected to be the building blocks of all other objects in the physical world (Wessels, 1989, p. 96).

[25] To be precise Maudlin does not attribute the laws to ontological postulations in this text, and in fact talks about the local ontology *and* non-local laws. With foresight to the discussions in the following chapters we can call them both elements of the explanatory ontology here.

### Quantum teleportation

A further, and for present purposes more interesting 'troublesome' phenomenon, is provided by the so-called teleportation protocol. In the protocol the sender and receiver again separate each with one end of the entangled physical system A and B, respectively. For sake of clarity, let us assume each of the systems A and B is a photon, and the photon-pair starts off in a state 'described' by the entangled quantum state. The sender has in possession another photon in some unknown state of polarisation, $u$. She then performs local operations on two photons in her possession, so that the formalism predicts that the distant (receiver's) photon will be disentangled and the sender's two photons will become entangled. But the receiver's photon is not simply left in any odd state, but is steered by the 'disentanglement' procedure into a state $u^*$, which is related to state $u$ in a definite way (Bub, 2007). After the sender then communicates the outcome of her operations (i.e. the result of the measurement on her two photons) to the receiver through a classical communication channel, he knows that his photon is either in a state $u^*=u$ or how to transform $u^*$ to $u$ by a local operation at his end.

To hammer this point home, consider what Bub (Bub, 2006) says about the density of coding (the quantity of information) employed in this transfer, by the sender and receiver he calls Alice and Bob.

> What is extraordinary about this phenomenon is that Alice and Bob have managed to use their shared entangled state as a quantum communication

> channel to destroy the state *u* of a photon in Alice's part of the universe and recreate it in Bob's part of the universe. Since the state of a photon requires specifying a direction in space (essentially the value of an angle that can vary continuously), without a shared entangled state Alice would have to convey an infinite amount of classical information to Bob for Bob to be able to reconstruct the state *u* precisely. [...This is because] to specify the value of an arbitrary angle variable requires an infinite number of bits. To specify the outcome of Alice's operation, which has four possible outcomes, with equal *a priori* probabilities, requires two bits of classical information. Remarkably, Bob can reconstruct the state *u* on the basis of just two bits of classical information communicated by Alice, apparently by exploiting the entangled state as a quantum communication channel to transfer the remaining information. (Bub, 2006)

"The state has 'disappeared' from Alice's region and 'reappeared' in Bob's, hence the use of the of the term *teleportation* for this phenomenon" (Timpson, 2004, p. 66). Of course, a lot of detail is missing from this introductory presentation and will be furnished when revisiting it in the sections below (alternatively, sufficiently detailed presentation can be found in (Timpson, 2004), and a more precise technical exposition in e.g. (Diosi, 2007)). For present purposes suffices to say that the phenomenon is 'troublesome' because nothing like that is possible in classical physical theories, however imprecise the discussion of information theory associated with the situation (i.e. whatever one's views of information-ontologies) may be. It is instructive, though, that it is the information transfer and not the matter or energy transfer that creates the puzzling effects here, perhaps another hint as to what direction to look in for the constraining principle of nature. The receiver has not created a photon out of nothing, but has merely transformed his existing photon into the distant one, without knowing exactly what the distant one was like in the first place. In fact no one knew exactly what the transmitted photon looked like before it was sent, not even the sender, no one had the infinite information. Unless a mysterious connection between all provisionally distant objects in the universe is postulated, we are 'troubled' by trying to explain what goes on here. Similarly in the 'dense coding' situation to be presented in the following chapter, the classical analogue requires that the separated communicators know in advance what the distant half of the coded message says (which is *ex hypothesi* impossible) in order to recreate the coded messages that can arise through manipulations of the quantum formalism and the attendant elements of material reality.

But stepping back from 'information-speak', that is to be more thoroughly analysed below, the teleportation phenomenon is still puzzling from the perspective of the potential for construction of the transcendental strategy of section 1. 4. Namely, it seems to deny an individuating 'thisness' to the supposed fundamental objects behind the phenomena by actively reducing their continuous space-time existence to the formal manipulations by experimenters. It illustrates most forcefully how the properties of the fundamental objects are dependent on the proscriptions from the formalism, and thus non-separably manipulable, rather than intrinsically inherent in the objects themselves. The experimenter that is able to more closely read the proscriptions of the wavefunction can come to know more about the distant object than the experimenter in possession of the object. The question then arises what other characteristics, other than being-thus, our fundamental objects have, and whether their location is a sufficient conceptual foundation to be connected with the common-sense conceptual framework. Teleportation is just a vivid illustration of how the fundamental objects are rid of all but their point positions.[26] Is that enough to reconstruct the phenomena of everyday experience?

Maudlin (2007a) argues that for the proposed transcendental account to go through the conceptual connection between the contemporary physical theories and common-sense must have at least some "local beables".[27] This is not to say that it can't postulate any non-local such beable, but merely that for the connection to be established in the most straightforward way it must contain at least some. “We take the world to contain localized objects (of unknown composition) in a certain disposition that changes through time. These are the sorts of beliefs we *begin with.*” (Maudlin, 2007a: 3160). In principle a theory without local beables could also account for these beliefs, but the construction of explanation from such a theory would prove a much harder task and one ridden with many more frailties, claims Maudlin. And the role of “local beables” is similar to that required of the material structure described essentially in terms of primary qualities, for they allow for a most direct connection between the experience of the phenomena and the ontology that explanatorily accounts for them by providing a most commonly agreeable vocabulary, a conceptual framework, through which to account for that connection (Maudlin, 2007a, p. 3160). The question that the teleportation, as the key ‘troublesome’ phenomenon, raises is: given how much of the conceptual framework is relegated to the non-local beable, are the local beables conceptually strong enough to uphold the simple transcendental strategy?

[26] For a detailed exposition of similar experimental situations that illustrate the qualitative paucity of the localized fundamental objects cf. (Brown, Elby, & Weingard, 1996).

[27] This is a terminology introduced in Bell (1987), where a 'beable' is a speculative piece of ontology, something that a theory postulates as being physically real. It is the foundational stone of our constructive approaches, the very construct that the explanation along the causal-mechanical lines rests on. Beables are the physical ontology that a theory postulates to exist. (These will be further explicated in the forthcoming sections.) 'Local beables', on the other hand, "do not merely exist: they exist *somewhere*." (Maudlin, 2007a, p. 3157). If local beables are all there is to physical ontology, then we get a Humean Mosaic, a global state of affairs constructed linearly out of a combination of local states, a simple summation of all local beables. Whether this can be done in quantum theory is the contentious issue to be discussed in the thesis.

### 1. 6 The research instrument and explanation

Though models of explanations abound in literature it is never straightforward to apply any of them to the particular scientific phenomena other than those they had been specifically designed for. It is sometimes said that we even need not fashion individual scientific explanations after general models. We shall have to take from each of the models that which is useful for the case-study instances and apply it in the present context. Precious little guidance can be gleaned from literature in that respect, as there is a scarcity of systematic accounts of the notion of explanatory depth, over and above proscriptive and descriptive delineations of the overarching explanatory models (Hitchcock & Woodward, 2003, p. 181). Explanations are often subjective beasts, when I consider something explained others might not. So one option would be to leave the issue out of the discussion altogether, we could just compare directly the two approaches presented in the thesis and see which one ‘clicks’ better. But that would be to give in too much to the subjectivity; I should in that case explain why I really like one of them so much over the other and hope the reader will like them too. Maudlin (2002) calls this choosing scientific theories on aesthetic grounds.

A more objective (and let’s leave ‘objective’ as implicitly understood here) route would be to try to explicitly devise the criteria upon which the value will be conferred to either of the approaches and then carefully collect the points of each on a scoreboard, using the final tally as an objective guide as to which one of them to ‘like’ more. What is needed is adjudication, over and above the descriptive account of the proposed explanations of the phenomena.

Upon such a strategy we need to try to box each of the approaches under a model as much as it will fit, in order to speed up the scoring, the more appealing the general model the more appealing will be the accounts subsumed under it.[28] In this respect we shall follow an instruction found in (Lipton, 2004)[29] to distinguish between, tentatively termed, epistemic and ontological (or metaphysical) explanations. Epistemic explanations cash in on satisfying our epistemic cravings alone: they provide us with good reasons to *believe* the phenomenon (explanandum) did actually occur or reduce the problematic phenomenon to what is already familiar. The ontological explanations, on the other hand, aim to present the phenomenon as a consequence of the way things really are in the world, regardless of how they may seem to us or how familiar they may be. As to how epistemology is connected to metaphysics, or more specifically ontology, in the simple transcendental strategy, we can follow Ruben's conclusions that explanations can and do have a virtue over a bare pragmatic satisfaction of 'explanatory hunger' (thus potentially making them mere narrative constructions).

> Explanations work only because things make things happen or make things have some feature ('things' should be taken in an anodyne sense, to include whatever the reader wishes to count as a denizen of reality). And making can be taken in a deterministic or in a nondeterministic (dependency) sense.
>
> And this, I think, is the ultimate basis for any reply to an explanation theorist who holds that full explanation is only and entirely a pragmatic or otherwise anthropomorphic conception. On my view, explanation is epistemic, but with a solid metaphysical basis. A realist theory of explanation that links the determinative (or dependency)

[28] They will also allow easier linking of explanatory strategies in individual instances into a wider reaching world-view.

[29] Lipton's account provides a useful starting point as he approaches the delineation of models from a utilitarian, not a purely descriptive, perspective. He asks what good an explanation is in science (and in sometimes related disciplines such as mathematics and philosophy) and sets up a simple 'three essential features of explanation' test that aims to respect these utilitarian goals. This test is not only useful in checking which models approach the utilitarian goal best, but also in alluding to the epistemic/ontological distinction. The three features test also appears to be applicable to the very instances that the explanations from the different theoretical approaches try to provide, and not just to the success of the models covering them. (Lipton, 2004, pp. 1-10).

> relations in the world with explanation gets at the intuitively acceptable idea that we explain something by showing what is responsible for it or what makes it as it is." (Ruben, 1990 , pp. 232-233)

As our transcendental strategy of section 1. 4 clearly requires ontological explanations to achieve realist conclusions, we shall focus on two such models to be applied to the case-study instances. These are the "unification conception of understanding" and the "causal conception of understanding" (Lipton, 2004, pp. 7-8).[30] As the unification model in general weavers between both epistemological and ontological explanations it will be interesting to investigate whether it can be pinned to the ontological side without being turned into a causal conception (with the pitfalls inherent in that from our 'troublesome' phenomena).[31] In that respect, as the historical analysis has illustrated (section 1.2.), causal conception can be seen as a subset of the unification conception; it provides unification through reduction of the wide range of phenomena to the universal causal mechanism. So the pure unification conception here will have to be what is outside that subset, the unificatory but not causal (or more precisely, causal-mechanical) segment of the model.

We will, thus, survey two conceptual approaches arguably aligned with the two types of explanatory models presented above. The aim is to investigate their explanatory content and scope, and especially to appraise the ontological characteristics of the explanatory narratives they provide for the 'troublesome' phenomena (as well as the wider scientific world-view). Each following chapter provides a more detailed introduction to the views of each of the conceptual and methodological approaches (the principle and the constructive one). A final tally is attempted in the last chapter where the explanatory success of the two approaches is directly compared.

### Comparative presentation of Lipton's models of explanation

Lipton (2004) devises makeshift criteria which help adjudicate explanatory worth (in the absence of a more lengthy analysis of 'understanding') based on a few simple insights about

[30] Lipton (2004) freely exchanges 'understanding' and 'explanation' in the text, as explanation is the means to achieve understanding. It would probably be clearer to call them conceptions of explanation, for understanding may be an unanalysable end-product of explanation. But it is the mystification of understanding that Lipton tries to avoid by, among other things, showing it to be something different than knowledge and practically available through the methods we use to explain things by.

[31] It is a mark of Kitcher's original advocacy (1989) of unificationist account, though not of Friedman's (1974) initial unificationist proposal, that in the realm of fundamental physics it is equated with the causal account, though in the special sciences it allows the divergence from the necessary construction of causal mechanisms.

the state we call understanding, both ‘phenomenological’ and comparative to other similar states. Thus understanding must be separated from bare knowledge by a gap that has to be additionally bridged, it must stop the endless why-regress at least until explicit further enticements (such as more detailed analysis or new phenomena) appear and it must have that wholesome character of all its elements obviously fitting into their places to form a uniform whole. These criteria Lipton terms, respectively, the

(i) Knowledge versus Understanding,

(ii) Why Regress, and

(iii) Self-evidencing Explanation[32].

In general, Lipton (2004) claims that casual-mechanical explanations fare better on the satisfaction of the three criteria and are on the whole best at satisfying the explanatory hunger. There is no need to quarrel with Lipton’s analysis here, nor to repeat it. What is more interesting is to apply the research instrument devised in this section, i.e. to show how the explicit instances of unification and causal explanations that we have chosen through our case-study instance actually satisfy the stated requirements.

But before that, it is worth summarising once more why Lipton deems the causal explanations as most successful in passing all the criteria and thus as the most desirable model of explanation in science. This is important also because it points to the direction our unification model of explanation should orientate itself in order to successfully compete with the general preference for the mechanical models (despite some of their failures that are to be discussed below). Lipton himself admits that the most tempting and succinct answer as to why causes provide better explanations than their effects, is that the causes have the power to confer understanding, at least in science. The idea would be: show the cause of a phenomenon and you have conferred understanding as to why the phenomenon occurs. But there are obvious problems with that, the first being that even though we could through counterfactual dependence show some event to be taken as the cause of the other, if there is no wider elaboration as to how it is its cause then understanding may still be missing. All we would have done is increased the stock of knowledge of facts, in this case that occurrence of the first phenomenon will under some circumstances lead to the occurrence of the other, that it will be the cause for it.

[32] A successful explanation not only conceptually unites the occurrence of the phenomenon into a wider conceptual scheme but shows just how the occurrence of the phenomenon is an essential part of our reason for believing that the explanation itself is correct (Lipton, 2004, p. 3). It ties the phenomenon and the explanation into a firm conceptual whole. It is hard to go deeper into structural analysis of this feature, and we take the lack of universal formal analysis of the syntactic structure of explanations to be a good indication that it needn’t be done here. Examples in this case seem to go a long way in replacing the formal analysis, such as Lipton’s illustration of the velocity of the recession of a galaxy as an explanation of its red shift even in the situation where the shift is an essential part of the evidence for the specified rate of recession (Lipton, 2004, p. 4).

His second attempt is to say that causes 'bring about' the occurrence of effects, but that might be taken as just synonymous for 'causes cause effects'. To avoid such a reading one has to look more closely at the temporal asymmetry of the phenomena deemed to be cause and effect, as well as abandon the Humean mosaic view of causation as entrenched but contingent conjunction. For at least one of our case-study instances that should not be a problem, as it relies heavily on just such a philosophical move. The other instance, should it make an attempt to move closer to the causation explanatory model will have to accommodate this distancing from Humeanism as well.

A much stronger support for causal explanation is provided by causes 'making a difference' between the phenomenon occurring and not occurring. In explaining a phenomenon, or more precisely its occurrence, that seems to be exactly what we are after i.e. showing what resides between the phenomenon occurring as it did and it not occurring at all. It is causes that often make a difference in this sense in science, whilst the phenomena we would deem their effects as rule do not (i.e. the asymmetry is not abolished). This kind of reading helps even in the situation where there are multiple possible causes or several of them contribute jointly. It is still the case that a better understanding is gained by selecting a cluster of causes that made the difference (preferably the crucial difference) to the phenomenon occurring, whilst at the same time having knowledge of their individual influences and joint interaction. Thus we come to another, often hidden value in explanation and that is not just showing that the event occurred but giving some detail (though not excessively) as to how it came about as well. A well structured causal explanation can do just this; provide a successful narrative of why and how our phenomenon occurred. Once that is done we may consider the phenomenon explained.

But there is a downside to this justification for the primacy of causal explanations, the use of contrastive explanations (Lipton, 2004, p. 16). Our desire to have the phenomenon explained often stems not from simple desire to learn why and how it came about, but from an implicit question why that particular phenomenon came about and not some other, similar phenomenon. Without going into further discussions of individuation of phenomena, it is clear that often in asking for an explanation of a phenomenon we are asking for an explanation of some crucial feature of the phenomenon, i.e. for explanation of why that feature obtains and not some other closely related feature. And causal explanations are not straightforwardly married with the 'contrastive requirement', as it is precisely the wider story and the more complex narrative construction that is needed to show how a particular cause, out of a cluster of closely related potential siblings, brought about a particular effect.[33] But on adding this criterion some causes can be shown to be weaker in providing explanations than the elements of other explanatory models, and this will be our concern in the section on depth of explanation. Of course those causes that surmount this hurdle will provide even better explanations. When explanations compete we want a 'deeper' one.

[33] This need not go to the extreme of denying chanciness and random outcomes even at the fundamental level. It is merely to claim that in competing explanations that which came closer to showing how a particular phenomenon came about from a particular cause will be considered a better explanation provided that both are equally empirically adequate.

## 2. PRINCIPLE APPROACHES

> Historically, much of fundamental physics has been concerned with discovering the fundamental particles of nature and the equations which describe their motions and interactions. It now appears that a different programme may be equally important: to discover the ways that nature allows, and prevents, *information* to be expressed and manipulated, rather than particles to move. (Steane, Quantum Computing, 1998, p. 119)

### 2. 1. Bohr and neo-Bohrianism

#### The founding father and his philosophy

Niels Bohr, a self-confessed non-philosopher, and one of the founding fathers of quantum theory, believed the "irrational element" (the Planck quantum of action) discovered through development of quantum theory has brought us against the insurmountable epistemic wall when it comes to the exploration and explanation of the physical world.[34] He expected philosophy to provide a 'band-aid' for the damage this wall has caused to the forehead of empirical research, but no more than that, as there is no way out of the dire predicament (Vukelja, 2004). Niels Bohr believed that quantum theory would have to adopt a radically different approach to investigation of physical reality, from the theories under the umbrella of classical physics.

In Bohr's eyes, due to the finite size of the Planck quantum of action, we can no longer perform experiments on objects that are elements of physical reality, without disturbing them 'beyond recognition'. The objects, independent physical entities, no longer exist in their own right, within the conceptual explanatory framework of the theory. This is not to say that there is no physical reality, or elements of physical reality, at the microscopic ('quantum') level (in a metaphysical sense), but that they have to stay forever epistemically inaccessible (or, epistemically insufficiently accessible) with respect to determination of individuality and physical characteristics. Thus we cannot construct a 'mechanical' conceptual scheme to describe the realm of the quantum.

[34] "There is an "irrational" element to nature: so stands the measurement problem on Bohr's philosophy" says Saunders in an updated version of (Saunders, 1994).

He took the major difference between the new language of quantum theory and that of the previous theories to be in that quantum theory's lacked the following four characteristics:

1. Determinism (or causality, Bohr finds the two terms almost synonymous, (Scheibe, 1973, p. 13)),
2. Terminology of pictorial description,
3. Independence of objects of observation from the experimental apparatus
4. Possibility of the combined use of the space-time concepts and dynamical conservation laws (Bohr, Atomic Physics and Human Knowledge , 1958, pp. 67-82).

The everyday (classical) language we use when discussing physical reality includes the above features, and is therefore not suitable to describe the reality as given by quantum theory. In Bohr's own words:

> All description of experiences has so far been based upon the assumption, already inherent in ordinary conventions of language, that it is possible to distinguish sharply between the behaviour of objects and the means of observation. This assumption is not only fully justified by all everyday experience but even constitutes the whole basis of classical physics. (Bohr, Atomic Physics and Human Knowledge , 1958, p. 25)

However we still have to use the classical terminology, the one we understand well from everyday use, to describe the results of the quantum measurement. This requirement is imposed so that those observations could be communicated, and made public, or even more precisely: the foundation of the realist explanatory conceptual scheme of physics is built on it.[35]

---

[35] We are treading over some fine notions here, most notably Bohr's understanding of 'objectivity'. Howard (1994) argues that Bohr made a break with a traditional concept of objectivity as independence of objects from observers, by defining it as "unambiguous communicability" of the scientist's descriptions of experiments and their results. Limitations of space preclude a wider discussion, though the notion will obviously be relevant to the expectations of ontology to be given by Bohr's 'interpretation of the formalism'. We can simply take this shift of definition to suggest similarities between Bohr's attitude to constructive ontology and the attitude of principle approaches to be presented below.

Bohr then considered that the chief aim of a consistent quantum theory is an unambiguous description of quantum phenomena, but obtained by including in their description the experimental conditions in which the phenomena occur (Scheibe, 1973, p. 18). Those experimental conditions are not to be clearly separated from the object, as in classical terminology.[36] But a problem arises because the apparatus is described by classical physics and the object by the quantum mechanical formalism, or in Bohr's words: "the essentially new feature in the analysis of quantum phenomena is…the introduction of a fundamental distinction between the measuring apparatus and the objects under investigation." (Bohr, 1963, p. 3). They no longer belong to the same language. Two different languages are required to describe what we expected is the same physical world on a continuous extension scale.

> From the above considerations, it should be clear that the whole situation in atomic physics *deprives of all meaning* such inherent attributes as the idealisations of classical physics would ascribe to the object. (Bohr, 1937, p. 293)

There is no room to enter into a detailed discussion of the route to Bohrian position, nor its eventual inadequacies from the present day vantage point. Insightful analyses can be found in (Vukelja, 2004); (Saunders, What is the Problem of Measurement?, 1994), (Saunders, 2005); (Barbour, 1999); (Hilgevoord & Uffink, 2006); (Bub, 2000); (Bub, 2004). What we really need here is an attempt to establish the outlines of his position with respect to methodology, metaphysics and explanations resulting from quantum theory, and how his views relate to the contemporary principle approaches which are often characterised as neo-Bohrian. Due to complexities of Bohr's own writing (Vukelja, 2004, p. 26) and extension of subsequent debate, the summation offered here serves the purposes of the wider positions outlined in the thesis without the luxury of argument and justification for such use (again due to limitation of space).

### *Treacherous metaphysics and limited explanatory potential*

As is outlined above, in perhaps too coy terms, Bohr advocated the agnosticism towards the constructive elements of reality at the quantum level due to the inadequacy of the mechanistic worldview in providing a description of them. But, as our existing, and culturally unchangeable, conceptual framework relies precisely on the mechanistic worldview, and is perfectly adequate for the description of the non-quantum experience[37], we are forced to use it to the best possible fit, even when describing 'quantum phenomena'. This is simply because of a contingent fact that it is the conceptual framework we have and one that we can't step out of when constructing another one anew.[38] This best fit is achieved by considering each measurement of the state of the inaccessible quantum object in isolation, but under internal holism. This is the uniqueness of individual phenomena. They become isolated from the wider context (e.g. physical history leading to the individual measurement) and thus do not allow formation of *unifiable knowledge* (Vukelja, 2004) about the individual elements of reality. On the other hand, the holistic element within each phenomenon precludes a clear-cut separation between the observer, the measuring apparatus and the object, so as to lead towards at least potential unification of the 'picture' of all of the object's properties.

[36] Bohr introduces a term 'phenomenon' to *replace* the object of observation, the apparatus used to observe the object and their mutual interaction that takes place during the process of measurement.

This implies that there is no possibility of providing a constructive-style theory of the elements of reality that interact with the measuring apparatus and the observer, assuming it subscribes to the causal-mechanical model of explanation. The language employed by quantum theory as a constructive theory cannot use the familiar concepts from the classical, everyday realm in the same sense that they are normally used in. The wholeness of the 'phenomenon' excludes the possibility of a clear delineation of new existents, their identification as objects traceable across different experimental contexts. Following on from that we cannot distil a unified picture of the object of observation, which is a telltale characteristic of the non-unifiable knowledge, and which, in turn, is the best we can achieve about 'microscopic/quantum phenomena'. Thus, in terms of epistemic access required for explanation we have to contend with wholesome phenomena, parcelled out from one another by the sea of standard mechanistically conceptualised experience.

Yet, this novel epistemology, rests on a metaphysical premise that is largely unacceptable today: the postulation of the existence of the "irrational element" that creates epistemic havoc in each instance of knowledge gathering in the quantum realm.[39] In each measurement interaction the "irrational element" disturbs the system, and this is *why* it is necessary to abandon hope of a 'phantasmal'[40] nature of observation that allows the observer to simply 'absorb' the state affairs, as it is in itself, unaffected by the act of observation. Thus, Bohr relies on a constructive step about the existence of an "irrational" element in order to avoid the discrepancy between the predictions of the theory and the observed outcomes (as contained in the measurement problem). In an ontological sense, we can almost picture the business as usual mechanics of the very small, treacherously disturbed by the unaccountable and unpredictable irrational element. However, the supposed "irrational" element does not feature in the quantum formalism, it is a purely interpretative philosophical addition (Saunders, 1994). But without the element, it is harder to accept the, almost metaphysical, necessity of limiting ourselves to non-unifiable knowledge of the 'quantum reality' however scarce that knowledge may be presently. In fact, Beller (1999, pp. 171-190; 197) cites opposition to Bohr's view from the likes of M. Born and W. Heisenberg, who held that there is no need to adopt such neo-Kantian view, and that a conceptual framework that includes quantum phenomena should be a correction of the inaccuracies discovered in the current everyday (classical) one.

[37] For presentation of Bohr's extensions of his 'quantum philosophy' to the realms of relativity theory, biology and psychology, see (Vukelja, 2004).

[38] That is, on a general level language contains a world-view and we cannot start constructing new private languages with altogether different world-views. Though we can correct the level of detail, in world-view construction we cannot start fundamentally from scratch, from some sort of non-linguistic starting point. Bohr thought that abandoning the mechanical view would require such a radical revision.

A more charitable reading of Bohr's approach, in Howard (1994) does not stress the reliance on the irrational element, but in fact sees Einstein's separability principle as the guiding idea behind Bohr's explanation of the phenomena. On Howard's account, the necessity of separability of elements of the universe is, according to Bohr, untenable in quantum theory. As the notion of objectivity as metaphysical independence of object and observer was also based on separability[41], it had to be redefined into 'unambiguous communicability' (see ftn. 35 above). On this reading Bohr's explanation of the phenomena rests on taking separability as the foundational presumption of our conceptual framework (i.e. language) and this is in perfect agreement with the theories of classical physics. In the quantum realm separability is violated and the language based on it cannot adequately describe the situation. Thus, we cannot have unifiable knowledge/explanation of the phenomena in that realm. With the separability broken, due to "irrational element" or something else, our conceptual framework has hit against the limit of understanding, and we must contend with agnosticism concerning the ontology at this level of reality.[42] One might also suggest that Bohr's acceptance of non-unifiable knowledge presents a criticism of the evidently limited mechanical-causal explanatory framework.

[39] This is a curious mixing of the principle and constructive methodology, as Bohr postulates a new existent of a special kind (the "irrational element") and uses that postulation as a constraining principle on the possibility of analysis and explanation of the experimental situation.

[40] Classical causal explanations of phenomena rest on the said 'phantasmal' nature of observation, i.e. possibility of detachment of the observer from the unfolding of the physical process (Vukelja, 2004).

[41] Namely, that the act of observation, a passive act by the observer, does not affect the outcome of the physical process as the whole process of observation consists of separable segments of physical process and a recording by the observer.

### *The methodological legacy*

Vukelja claims that it is Bohr's general position on the role of science that it should not aim at a conceptual mapping of reality, in a constructive sense of delineating existents and their interactions, but should instead aim to systematically unify human experience through objective presentation of the experienced phenomena (Vukelja, 2004). Hilgevoord and Uffink say that Bohr renounces "the idea that [conceptual] pictures refer, in a literal one-to-one correspondence, to physical reality" (Hilgevoord & Uffink, 2006). As will be discussed later, with reference to Einstein's explanation of his reasoning behind the use of principle approach, these can be seen as conforming with the principle methodology requirement of trying to abstract as much as possible from the assumptions and postulates about ontology, and formulating empirical generalisations expected to survive any foreseeable ontological clarification. Yet, this is not a full-blown principle approach based on achieving desired unification through one or more foundational generalising principles, and allows the use of alternative constructive conceptualisations (wave and particle mechanics) as useful fictions in individual contexts. Bub (2000) on the other hand, is not perturbed by this constructive misdemeanour, and claims that Bohr's position treats quantum theory as a principle theory with a Kantian twist (the necessity of using classical concepts).[43] As for the formalism, Bohr sees no reason to attribute credence to any of its particular demands as to the nature of reality.

---

[42] Of course, an important question of where exactly this cut between the levels is placed can be posed. Some commentators leave it as a weakness in Bohr's position to place it 'somewhere' between the scales of the macroscopic measuring instrument and the 'atomic' object. Hence, the metaphysical importance of the "irrational element" being the Planck quantum of action. Others, hold that the formalism should not permit 'quantum effects' to be amplified to the macroscopic size (though we do not observe that, and thus get the problem of measurement) and that the cut is not a matter of scale of material extension, but of choosing those parameters from the formalism that permit the accurate prediction of the desired experimental outcomes and the description that respects separability of object and apparatus. Such a description can be found in the formalism, at the expense of rendering unknowable other characteristics of the overall system. Thus, our description conforms to the classical conceptual framework but is irrevocably incomplete and does not allow construction of a unified explanation.

[43] Another similarity, presented in (Bub, 2004) is the *denial* of the measurement problem in Bohr's philosophy and the CBH principle approach. The former, according to Bub, simply placed the measuring instruments outside the domain of the theoretical description, however arbitrary the cut might seem. This way there was no problem to be solved (we were not to ask what happens to the measuring instruments between the

> The entire formalism is to be considered as a tool for deriving predictions, of definite or statistical character, as regards information obtainable under experimental conditions described in classical terms […]. These symbols themselves, […], are not susceptible to pictorial interpretation; and even [the formal predictions] are only to be regarded as expressing the probabilities for the occurrence of individual events observable under well-defined experimental conditions. (Bohr, 1948, p. 314)

On the other hand, what makes contemporary principle approaches of this chapter neo-Bohrian is their agreement that a constructive picture along the classical lines of the phenomena guided by the quantum formalism cannot be built. In fact both Fuchs and Bub admit Bohrian leanings, towards Bohr's position as they understand it to be ( (Fuchs, 2002b); (Bub, 2004)). We cannot construct metaphysical postulates that will satisfactorily fit into the existing overall conceptual scheme and provide a mechanical underpinning of the said phenomena. The quantum realm is conceptually radically different from the classical one, and we have to learn to respect that. Without any speculation as to the nature of ontology, we can ascertain that quantum formalism and separability are in conflict. Yet, the constructive approach of Chapter 3 is also willing to accept this, but build a modified mechanical picture of the processes 'producing' the experienced phenomena. Perhaps Bohr was simply wrong at the last step, and given some hindsight available to contemporary physicists he would have sided with the constructive picture and abandoned calls for non-unifiable knowledge (this would in effect be giving in to the criticism of Heisenberg and Born, as reported in (Beller, 1999)).

What can be seen as characteristic of the principle methodology in Bohr's position is the overall reduction in explanatory utility of the quantum formalism, whilst nevertheless holding on to some sort of determinism and realism. All principle approaches (which distinguishes them from pure unashamed instrumentalism) see the reduction as an indication of constraints on what can be known about the quantum-domain phenomena imposed by the reality itself (to a greater or lesser degree), thus not as a consequence of pure technical ignorance that

'ready'-state and the measurement interaction). The latter, purport to show that the measurement problem is a pseudo-problem that different interpretations waste time 'solving' (Bub, 2004, pp. 262-263).

further research might remove. Constructive approaches postulate entities that they hope will lead us out of such self-imposed ignorance with questions in the right direction.[44] The principle approaches and Bohr also treat the formalism as an instrumental tool and not an ontological pointer.

> Quantum [formalism] postulates a geometry of propositions because complete knowledge of the system is not possible; the geometry both guides and constrains the extent of our fragmentary knowledge of the properties associated with an instantaneous state. […] Our knowledge of the propositions true of the system is unstable and changing. It is so unstable that quantum mechanics proceeds by articulating only the exact fashion in which this instability is evidenced. (Demopoulos, 2004, pp. 103-104)

## 2. 2. Quantum information theory and principle approaches

### Step one and step two in principle-based explanations

Methodologically, the principle approaches of this chapter set out from the observation that formal theoretic accounts of the phenomena considered characteristic of quantum theory can be derived from a limited set of formalised principles about constrictions on the amount of knowledge an observer can have about reality, or similar principles about information acquisition and transmission when dealing with 'reality measuring' instruments. A common denominator for the approaches surveyed here is that they are explicitly in the state of development, i.e. that they do not offer complete explanatory accounts of the phenomena in question that are sufficiently couched in the wider explanatory framework concerning the physical world. We shall survey two such approaches, though most of the discussion in the end will be focused on one of them, a formally more complete one.

Yet, as their proposal is a deviation from the standard preference in physical explanation for causal-mechanical accounts the actual formal methodology of their derivation will have to be outlined to a greater extent. In that, the Fuchs programme can be seen as, conceptually, an intermediary step towards the more abstract CBH programme. As we shall see, though the more abstract programme is harder to fathom, it is less committed to 'sins' inherent in the principle approaches (cf. exposition of Einstein's principle derivation of Special Relativity in Chapter 1, section 1. 3. 1). Also, as both programmes are fresh and to a great extent still under development we can learn more for the purposes of assessing explanatory accounts based on them by considering two, rather than just the preferred one of those accounts. Likewise, with the non-constructive accounts being less common in philosophy of science, two are included here to help clarify matters. As a rule, at this stage they set the foundations and delineate questions to be addressed in future research. They are also not fully formally equivalent with 'standard' quantum theory, and seek to uncover 'metaphysical clues' from the ways of bridging the gap between their formalism and the 'standard' one.[45]

[44] Though, even they are aware of some serious obstacles on that route, as given by the in-principle unknowability of some important states of the universe.

These clues come from some formally describable situations (entanglement assisted communication, non-commutativity, dense coding, superdense coding, teleportation and the like), at least one of which we have introduced in the previous chapter. In attempt not to stray into too technical aspects of the discussion and lose sight of the primary aim of providing an explanatory framework of the phenomena intuitive enough to appeal to a wide enough audience, it suffices to say that the quantum information theory uses a well-known and tested classical information theory appropriated to the quantum context. Classical information theory concerns mathematical formalisation of quantification of transmittance and loss of information through classical communication channels (such as pieces of paper with pre-arranged code pushed through a boundary impenetrable to other information, or a standard telephone line, or a mobile phone radio frequency). The quantum context is provided by replacement of formal states of the communication devices expressed in terms of classical variables[46] by the formal states as expressed in terms of quantum variables.

### Information: classical and quantum

Before even introducing the two principle approaches, each of which has some unavoidable formal aspects associated, it is worth examining a general situation of dense coding (Clifton, 2004, pp. 431-432) in order to better illustrate why principle approaches are strongly oriented on the epistemic (and to a degree subjective) aspects of the situation. Namely, the classical analogue employs a system of codes inscribed onto blank cards and exchanged between agents. The situation is so set up that the receiving agent needs two cards (one initially taken with him, and the other received through the communication channel) to subsequently recover 2bits of information the sender is transmitting through the communication channel. The receiver, that is, needs both cards to make sense of the 2bit message, no relevant information is carried by either card in isolation. The codes on the cards are 'entangled' to provide a whole message.

[45] Though more explicit about this than the constructive approach surveyed in Chapter 3, this does not put them in a great disadvantage to the latter as those are also, at this stage, unable to complete the explanatory framework in every detail (as will be discussed in Chapter 3).

[46] These, of course, needn't and as a rule won't be the basic classical variables of a standard Newtonian phase space, unless one chooses to communicate through physically interacting point-particles, which is not the case in information theory. But classical variables are also other variables (such as orientation of an arrow or the amount of the electric current) codified in accordance with the mathematical formalism of classical physics.

If the analogy is perfectly appropriate, it seems to suggest that the information carried by the communication channel need not be parcelled out amongst the physical systems making up the channel, and thus that we need not invoke the metaphysical (even if we do it formally in terms of calculations) mysteries of entanglement to account for the dense coding phenomena. If the analogy is perfectly appropriate, there is no need to look for the ontology inherent in the quantum formalism over and above trying to fit that formalism with the classical ontology we are already happy with (and as has been repeatedly attempted for the past 100 years that quantum theory has been formulated). But the classical communication protocol Clifton describes is disanalogous to the quantum 'dense coding' situation in one important respect: for the sender to be able to choose the right sign (a piece of code) with which to convey the said 2bit message *she must know in advance what is already written on the receiver's first card* (the one he initially takes with him). And in the quantum versions of the protocol such foreknowledge is not envisaged, nor is it explicitly required (over and above whatever may be encoded in the formalism per se) for the protocol to be successfully completed. Thus, it seems at this tentative stage the quantum formalism somehow embodies the 'knowledge' required in the classical case. How the sender comes to acquire this knowledge remains a mystery (i.e. it is either a foreknowledge akin to common cause explanations, or it is a knowledge somehow acquired in the process akin to a holistic superluminal connection) that the principle approaches try to resolve (cf. Chapter 1, section 1. 5. 2 on teleportation, as well).[47]

Yet, we will not move sufficiently away from the 'troublesome' aspects of the phenomena if we attempt to explain away the mysteries by structural accounts of encoding large amounts of information directly into the material existents (this may also be a pointer in moving from

[47] Of course, the constructive approaches we shall consider later need not concern themselves with the mystery of foreknowledge as they have a metaphysical mechanism by which the non-local or holistic effects can be produced by local interventions, such as choice of signs to write on a card is.

Fuchs to CBH). We should not turn mystery of one kind (superluminal causal connection) to that of the other kind (instantaneous exchange of large amounts of information). Steane (2003) claims that processes involving quantum information transfer and manipulation, quantum computation, are not superior to classical computational processes in terms of efficiency. There is no mysterious transfer of large amounts of information.

What in fact happens in the quantum case is that the physical situation corresponding to entangled states, a physical entanglement, provides a sort of a 'physical shorthand' in information transmission and manipulation. That is, we get the appearance of efficiency in quantum information processes because "quantum entanglement offers a way to generate and manipulate a physical representation of the correlations between [entities represented by formal expressions of quantum states] without the need to completely represent the entities themselves" (Steane, 2003, p. 476).[48] What the characteristic of quantum entanglement provides is a way to represent and manipulate correlations directly, without having to go through a lengthier and computationally more expensive route of manipulation of the correlated entities. In conclusion, the principle approaches then try to present the 'troublesome' phenomena in a perspective that aims to remove from their description all that is metaphysically postulated but does not seem to do any work on conceptually linking the elements of the phenomena. And mechanical details of the physical systems might be just the thing if the phenomena are viewed in terms of outcome correlations on the black-box instruments. Thought this might be explanatorily 'efficient' in a sense of generating a wholesome narrative from a limited set of concepts, it faces the problem of satisfying explanatory depth (cf. section 4. 1) and conceptual connection with the simple transcendental strategy (as in section 1. 4).

### *Metaphysics: epistemic and ontic states*

The ontological characteristics of the principle and constructive approaches are most clearly seen in the interpretation of the '*quantum state*'. A quantum state is a part of the quantum formalism that purports to provide a formal description of the relevant characteristics of the physical system, thus a 'formal state'. It is useful here to introduce a dichotomy between states of reality and states of knowledge, following (Spekkens, 2007), as used in interpretations of formalism (thus, also of formal quantum states) of physical theories. Spekkens terms these *ontic* and *epistemic* states, respectively. From a classical and realist perspective, an ideal state in physics is an ontic state. An ontic state provides a complete specification of all the properties of the system.[49] An example of such state is a point in classical phase space.

[48] This means that the computational correlations can be so set up as to be able to produce desired results without the need to calculate a lot of 'spectator' results. For example, one can find the period of a function without calculating all the evaluations of the function, one can find a specific property of a quantum system (such as energy level) without also finding the complete quantum state, one can communicate some shared aspect of distributed information without transmitting as much of the information as one would otherwise need to. (Steane, 2003, p. 477)

But classical physics also provides examples of epistemic states, namely when the formal state specification expresses a probability distribution over phase space. In this case the formal state represents a relative likelihood (a probability distribution is a function, but this aspect need not concern us here) that some (human) agent assigns to the ontic states associated with the points of phase space 'covered' by the distribution. "The distribution [a formal state in this case] describes only what this agent knows about the system" (Spekkens, 2007, pp. 032110-2). Note that it is not claimed that there are no properties of the system, or that the system is not in some sense fully real (endowed with a full set of necessary physical properties).[50] It is rather that in the given experimental (physical) situation the agent is not in a position to know what ontic state corresponds to the true state of the system, but given some set of constraints is able to ascertain a probability distribution over some set of relevant ontic states. The metaphysical projection states that the system is in a state corresponding to one of the ontic states, but the agent cannot be sure which, though she can specify the difference in likelihood between those states.

Of course, the ideal situation is the one where the ontic and the epistemic states coincide, i.e. where the epistemic states encode complete knowledge and thus a complete specification of a system's properties. It is the claim of the principle approach that using epistemic states provides *conceptually* superior explanations of the 'troublesome' quantum phenomena (Spekkens, 2007, pp. 032110-2), even though constructive approaches are taken to provide equally valid demonstrations of the said phenomena as mathematical consequences of the theoretical formalism. Principle approaches concede that the explanations from the constructive approach (taking quantum states as *ontic* states) are conceptually equally well founded if one were to abandon certain preconceived notions about physical reality (such as the principle of separability, for example). But they argue that such abandonment does not make the phenomena sufficiently intuitive because, among other possible complications, it makes the construction of the overarching explanatory framework for the understanding of reality impossible (or at least too difficult to be worth the effort).

[49] These properties needn't all be explicitly listed in the specification of the state, i.e. some of them can be derived from the specification of the state and the overarching theoretical formalism. But the crucial point is that these 'implicit' properties can in principle be so derived at any stage with complete certainty. In other words, all the properties of the state are at all times in-principle epistemically accessible.

[50] One might interpret Bohrian metaphysics as claiming that there is no fact of the matter as to whether the system possesses all the properties, including those unknown or unknowable to the agent, but this is not what is claimed here.

Yet, at present the principle approaches have a recurrent explanatory pitfall of their own, one taken to be the plague of the pure unification-type explanations in general (Lipton, 2004, p. 7), in the lack of answer to what the epistemic state is knowledge about; what exactly in reality is the source of the knowledge codified in the epistemic state.[51] This is where a clarification of the analogy with the example from classical physics above may be useful. Whereas in the classical case the identification of the epistemic and ontic states was precluded on practical grounds (due to insensitivity of the measuring instruments or the practical limits of computational power), in the quantum case (i.e. according to our principle approach) it is precluded on theoretical grounds. The principle approach claims that it is not our lack of knowledge of some local and noncontextual hidden variables, or our ability to manipulate those computationally through the formalism, that prevents us from interpreting the quantum state as ontic state. They in fact take it (to a varying degree[52]) as the foundational principle of nature that the two states cannot be equated in interpretation of quantum theory, but as yet lack a further explanatory account as to why this is so.

> This is not to say that the question is not important. Rather, we see the epistemic approach as an unfinished project, and this question is the central obstacle to its completion. Nonetheless, we argue that even in the absence of an answer to this

[51] It is assumed here that having such knowledge would cure the unificationist type of many ills at once, most notable of the weakness in stopping the why-regress (Lipton, 2004) as description of material existents and their properties as a source of some phenomena observed about them is taken as a stronger explanatory foundation than the claim that a set of abstract principles holds about some phenomena we observe.

[52] Due to varying degrees of development and metaphysical commitment that the research programmes subsumed here under the umbrella of principle approach currently undertake it is difficult to provide a clearcut summary on this point, providing room for discussion to appear in this text as well. Some of the approaches considered take the most direct view that the epistemic states are just best rational guesstimates of the agents as to the instantaneous value of the relevant properties of the physical state (Fuchs, 2002). This is perhaps most akin to the classical analogue, only the preclusion of the identification of the epistemic and ontic states is seen as a ‘law of nature’ and not a technical difficulty. Others are much less direct and more explicit in claiming only an initial step in development of the satisfactory account, thus choosing to at this stage remain “agnostic about the nature of the reality to which the knowledge represented by quantum states pertains” (Spekkens, 2007, pp. 032110-2).

> question, a case can be made for the epistemic view. The key is that one can hope to identify phenomena that are characteristic of states of incomplete knowledge regardless of what this knowledge is about. (Spekkens, 2007, pp. 032110-3)

## C. A. Fuchs: constraining principles from a deep conviction

### *A simple method*

In Fuchs' programme we could view the claim, rephrased to suit the Spekkens terminology above, that 'quantum states are irreducibly epistemic states' as one of his foundational constraining principles. By respecting the nonlocal nature of the EPR situation Fuchs claims that quantum states cannot be ontic states, as if they were separability would be violated as a universal principle (we can thus take the expression of separability as another of his foundational principles). Fuchs further relies on the pre-communication segment of the teleportation phenomenon to argue that quantum states cannot be objective even in principle, and thus must be epistemic and uniquely tied to the individual experimenters that employ them. That is, before Alice in the teleportation protocol broadcasts her 2bit message no one can even begin to perform the operations that will complete the conversion of the distant state into the outcome of teleportation. And yet, the ontic interpretation of the state would expect the material for the conversion to already be in (the distant) place.

> If Alice fails to reveal her information to anyone else in the world, there is no one else who can predict [the final outcome of the teleportation] with certainty. More importantly, there is nothing in quantum mechanics that gives the [the power to reveal its ontic state out of possible a spectrum of epistemic states]. If Alice does not take the time to walk over to it and interact with it, there is no revelation. There is only the confidence in Alice's mind that, should she interact with it, she could predict the consequence of that interaction. (Fuchs, 2002a, p. 12)

However, Fuchs' programme still sees the 'troublesome' phenomena as outcomes of imperfect interaction between conscious observers and a strangely constructed reality. Though the quantum formalism is not a fully objective description of the physical system, it is somehow related to it, whilst containing many elements that are dependent on the individual observer. It is the aim of Fuchs's programme to wean the objective from the subjective elements.

> There is something about the world that keeps us from ever getting more information than can be captured through the formal structure of quantum mechanics. Einstein had wanted us to look further – to find out how the incomplete information could be completed – but perhaps the real question is, "Why can it not be completed?" (Fuchs, 2002a, p. 11)

Methodologically, the programme aims to re-derive the quantum formalism whilst ignoring all of its ad hoc metaphysical connections (such as what elements of the physical system have to correspond to which elements of the formalism, and what happens to the systems upon measurements) and respecting only his foundational principle that the states of the formalism are epistemic. In narrative terms Fuchs sees the formal states as individual conscious agents' guesstimates about the possible states of the physical system, guesstimates which then have to be updated upon each interaction with the system (the measurement intervention) in accordance with rational procedures of the Bayesian probability calculus.[53] Where the re-derived formalism differs from the historically developed 'standard' quantum formalism, we get a glimpse of the objective characteristics of interaction with 'quantum-level' reality different from what we have come across classically. One such glimpse says:

> The objective content of quantum mechanics (or at least part of it) is that if we subjectively set our probabilities for the outcomes of [any as-complete-as-possible measurement on some segment of the material reality], we are no longer free to set them arbitrarily for any other [outcome of same or

[53] It suffices to say here that Bayesian statistics is a formal mathematical for updating beliefs about future chancy outcomes based on the evidence previously gathered. It is the most rational form guesstimating available.

> different type of measurement]. (Fuchs, 2002b, p. 32)

### *Explanation: you mess up and you try to estimate the damage*

We may recall that one of the primary philosophical problems accompanying the development of quantum theory was the problem of measurement, i.e. the problem of explaining the collapse of the wavefunction during the measurement process. But more importantly, where explanation of the phenomena is based on the *outcomes of interaction with the physical systems*, measurement process plays an inexorable role. According to Fuchs, if we believe that the quantum state is rigidly bound to the elements of reality we "will never find a way out of the conundrum of "unreasonableness" associated with "state-vector collapse at a distance"", i.e. the nonlocal causal connection between the separated phenomena (Fuchs, 2002b, p. 164). Fuchs divides the measurement process into two parts, each of which is clearly illustrated by the limiting cases. The measurement process thus consists of (1) Bayesian conditionalisation and (2) further mental readjustment. (1) is the raw collapse of the wave function, the improvement of the 'guesstimate' of future measurements based on the outcome of the present one. It relies on Bayes' rule of 'factorising the fact' (the observed measurement outcome) out of the probability distribution. This is an entirely classical procedure that depends on the rational rules of Bayesian statistics and not some hidden characteristic of nature. Fuchs calls this the 'knowledge refinement'.

(2) is a further constriction, specific to quantum theory. It is a theoretical representation of the supposed intrinsic sensitivity of reality to experimental interventions. Fuchs calls it the 'back-action'[54] or 'feedback' that the measurement device inflicts on the system being measured, and that is dependent on the details of the measurement interaction, individual outcomes of measurement and the initial quantum state assigned by the observer. This 'back-action' is the specific quantum addition that is not found in the classical probability theory and that depends on the observer's rationalised subjective estimate of the consequences of her experimental intervention. Fuchs concludes:

> Quantum measurement is nothing more, and nothing less, than a refinement and a readjustment of one's initial state of belief. (Fuchs, 2002b, p. 34)

[54] The idea of back-action does not originate with (Fuchs, Quantum mechanics as quantum information (and only a little more), 2002). See (Valente, 2003) for further bibliography.

Thus Fuchs explains the basic interaction with a system in a state that one posses maximal possible information about as pure occurrence of the ‘back-action’ of the interaction with a reality sensitive to touch. Such a measurement does not provide the observer with any new information, but merely affects what she can predict due to the side effects of the experimental intervention. “That is to say, there is a sense in which the measurement is solely disturbance” (Fuchs, 2002b, p. 34)

But more interestingly, in the case of distant part of the system in the EPR situation, the experimenter has another limiting case of the two components of the measurement process, the refinement of beliefs without any disturbing interaction with the system. There is thus no violation of separability as no real change is induced in the system itself, but merely in the experimenters’ ascription of a state to the distant system. The change in the quantum state that is assigned to the distant system on the basis of such measurement corresponds to the pure (i.e. classical-like) Bayesian factorisation without any further ‘mental readjustment’.

It is these ‘mental readjustments’ that put Fuchs firmly on the Bohrian train, along with all the conceptual problems that may bring. But even before that, we have to note that in ascribing this intrinsic and insurmountable sensitivity to reality Fuchs breaks away from the principle approach into speculation about the nature of ontology at the quantum level. Yet, this speculation does not seem to be better supported than it was in Bohr’s day, i.e. exhibits great resemblance to the influence of the “irrational element”. According to Fuchs, the reality itself changes under invasive interaction (the measurement), thus we can never repeat the same type of measurement on the same system in order to achieve a fully complete description. It appears that in this case the constructive speculation does not rest on a satisfactorily complete principle-based explanatory account, but is in fact introduced to complete it. It is also not a formally negligible speculation, that might come about as a result of an oversight, as one could say about Einstein’s implicit assumption about the internal dynamics of measuring rods and clocks (cf. Chapter 1, section 1. 3. 1).

We thus do not get a sufficiently principle-based explanation of the troublesome phenomena. Though Fuchs claims no mysterious interaction between the separated segments of the entangled system taken place, he goes on to include a constructive postulate of ‘inherent sensitivity’ into the explanatory account. The correlations in the EPR-like situations are a product of the common cause that does not violate the separability, but that is, by some natural trickery, forever hidden from us. We shall never be able to gain complete knowledge about the initial state of the system, so are forced to surprising updates of it (such as the one in EPR situation) when the abstract formalism permits it. In the teleportation case, such updates are only possible with the assistance from other experimenters.

We conclude this section with the observation that though initially based on the intuitive generalisations from our 'troublesome' phenomena, namely that quantum states are inherently epistemic; this approach fails to show sufficient coherence to stand against the competing constructive approaches. This is largely due to its venture into the constructive domain where it bases the explanation of the phenomena at least in part on the changes of the physical systems themselves, which is conceptually on the same level as the constructive approaches. This in the end forces it into an explanatorily self-constrained position[55] akin to Bohr's and this is not solid enough ground to build explanatory accounts to compete against the causal mechanical account of Chapter 3. On an ontological side the explanation from Fuchs seems to rest on the narrative of changes to primary entities characterised by specific properties, but that only have a statistical existence, i.e. can never be claimed to exist (bearing the said properties) with certainty.

### The Clifton-Bub-Halvorson (CBH)[56] programme

A different sub-class of principle approaches is the route that does not start with the epistemic interpretation of the quantum state per se, but sets off by looking for more general principles of information reception and transfer (via microphysical material world, but not relying on any of its particular characteristics). Thus, on the question of nature of quantum states it remains as agnostic as possible, this way moving even further from the metaphysical projections (as the quantum state is probably closest one can come to the connecting point between the formalism and the existents supposedly behind it).

> […] the CBH [programme] should not be understood as providing a 'constructive' [*sic*] explanation of the quantum formalism, along the lines suggested by Chris Fuchs [ (Fuchs, 2002a) …],

---

[55] Recently, Timpson (Timpson, 2008) argues, rightly, that Fuchs is not a full blown instrumentalist. Fuchs remains agnostic about the details of the underlying reality, but is very much committed to its existence. Yet, with regards to the reality of the quantum state, i.e. interpretation of the quantum state (an element of the formalism) as a formal description of the physical state of the system, he is instrumentalist. That is, he denies the reality of the description and yet maintains the usefulness of the quantum state in making predictions about future interactions with reality. And that 'localized' instrumentalism is what we are concerned with here.

[56] Named after R. Clifton, J. Bub and H. Halvorson.

> but rather as a 'principled' reconstruction of the theory within a suitably general mathematical framework. (Bub, 2008, p. 15))

Thus adopting lessons from the pitfalls of the Fuchs programme (above), the CBH programme makes no use of the postulates about the nature of the physical systems employed in producing the 'troublesome' phenomena that result in the 'mysterious' correlations of measurement outcomes. In fact, it makes no *use* of the systems, measurements and outcomes in its derivation of the formalism, but focuses on constructing a mathematical description of the relationships between the formal expressions used as input and formal expressions for output of such procedures. Thus the phenomena to be explained on their view are mathematical structures that result from a coding game experimenters play with the lab instruments. No use of the structure of the instruments or their 'objects of observation' is made, in fact the CBH proponents prefer to call them '*black boxes*'. What they show is that if the game is played respecting certain principle constrictions on the moves (other than those restrictions that the formalism itself imposes, i.e. the internal mathematical rules) the resulting formal structure is sufficiently similar to the formal representation of quantum theory with the interpretative assumptions about the nonlocal interaction of the physical systems.

Here is a brief presentation of the principles in a language that avoids reference to complex algebras and connects the content of the principles to the more standard informal presentations of the 'theorems' of quantum theory, as given in (Timpson, 2004, pp. 199-205). The first principle forbids superluminal signalling via measurement, and corresponds to the more standard no-signalling via entanglement prohibition in standard quantum theory (to be explored in greater detail in section 4. 3). The principle mandates that *the state* assigned to the system at B, shall not be affected by any operation performed on the distant system A. The second principle in general forbids the 'broadcasting; of states, which can be seen as a generalisation of the 'no-cloning' restriction (applicable only to pure states). Simply put it forbids that a manipulating device takes a system to which we assign a certain mixed state and produces as an output two systems A and B, each of which will (through some further manipulations of the formalism) have a version of the initial state assigned to it. Though the exposition is more technical, intuitively we can understand the 'no-cloning' aspect of the prohibition as forbidding the systematic multiplication of the states assigned to systems through manipulations of material measuring instruments. The final principle, the no-bit-commitment, is even more technical, but following Timpson (2004, p. 203), we can understand it as a formal requirement that provides for the selection of formalisms that do allow entanglement, just as the standard (empirically adequate) quantum formalism does. This may be seen as a purely methodological move, for if we are trying to derive a formalism from the principles we want to hone in on the characteristics (however problematic) of the existing standard formalism, and exclude those formal constructions that deviate from it in significant respects.

In summary the methodological process strives to "derive the basic kinematic features of a quantum-theoretic description of physical systems – essentially noncommutativity and entanglement – from [the] three fundamental information-theoretic constraints" (Bub, 2004, p. 241), i.e. from the assumption that we live in the a world in which there are certain constraints on the acquisition, representation, and communication of information. Thus, it assumes that what defines any theoretical formalism as 'quantum formalism' is a certain characteristic algebraic (in other words abstract mathematical) structure of what the formalism takes to be observables and states. This structure is to be identical to the elements of the traditionally derived quantum formalism that are taken to exemplify noncommutativity and entanglement (as above). An example of these 'traditionally derived' formalisms is standard quantum mechanics of a system with a finite number of degrees of freedom represented on a single Hilbert space with a unitary dynamics defined by a given Hamiltonian, i.e. the standard university-course formalism of the quantum theory.

So, the methodological starting point in this case is twofold: on the one hand there is the abstract mathematical generalisation (some kind of constraint on what it takes for a chunk of formalism to be a quantum theoretical formalism), and on the other is the mathematical generalisation of the said information-theoretic principles. In terms of ontology, the former has more potential to smuggle in some metaphysics than the latter, though (as we shall see later) the latter on its own and in combination with the former carries some metaphysical assumptions about the world as well. In having to prove the similarities with the more constructive interpretations of the formalism, the CBH proponents have to keep going back to the conceptual framework of systems and properties.[57] Methodologically this is not a 'sin' in itself, provided the assumptions are explicated and we can keep an eye on them through the development of the formalism (i.e. through its approach to the structure of 'quantum formalism').

[57] At this stage the CBH projects relies heavily on the standard metaphysically burdened language of (at least) minimal interpretation of quantum formalism. Yet, as the presentation of the methodology above tried to outline this should not properly be the case, as the CBH approach professes to stay as far away as possible from the metaphysical speculations about the nature of the elements of reality behind the 'troublesome' (and other) phenomena. For the time being we can try to excuse this as an attempt to convince the standard

> […] if there is no minimum amount of mathematical structure shared by all theories, and if any fairy tale can count as a legitimate "toy theory" — then it would be hopeless to try to *derive* QM from information theoretic principles, or from any other sort of principles for that matter. (E.g., why assume that the results of measurements are real numbers? Why assume that measurements have single outcomes? Why assume that the laws of physics are the same from one moment to the next?) (Halvorson, On information-theoretic characterizations of physical theories, 2004, p. 292)

CBH do grant existence to physical systems, but it remains unclear just how much individuality (and in some respect: independent existence) these things have. The formalism, as derived by CBH, is only used to mathematically represent the statistical correlations between 'measurement' outcomes. Even 'measurement' is a problematic term here, for at this stage the measurement involves an epistemically rather agnostic situation of black boxes used to derive statistical correlations[58] (Bub, 2004). Yet, Bub also claims that the formalism constructed the CBH way excludes "haecceitist theories that associate a primitive 'thisness' with physical systems" (Bub, 2004, p. 253) in the description of the phenomena.

To associate the observables of the theoretical formalism with the properties of a physical system, as a system that is individuated and does have a primitive 'thisness', requires a metaphysical commitment to elements of reality with a 'mysterious' nonlocal connection,

---

practitioners, physicists using the formalism together with the language, of the worth of the newly derived formalism, i.e. of its equivalence with quantum formalism.

[58] Bub likens this situation to the one outlined at the beginning of Albert's (1992 ) book *Quantum Mechanics and Experience*, a familiar example in literature, in which the measurement simply takes "a system in an input state [… and produces] a system in one of two output states, with a certain probability that depends on the input state" (Bub, 2004, p. 253).

because of the appearance of entangled states. And this contradicts the authors' deep-seated expectations of explanatory ontology.[59] Bub quotes Einstein's letter to Born:

> […] whatever we regard as existing (real) should somehow be localized in time and space. That is, the real in part of space A should (in theory) somehow 'exist' independently of what is thought of as real in space B. When a system in physics extends over the parts of space A and B, then that which exists in B should somehow exist independently of that which exists in A. That which really exists in B should therefore not depend on what kind of measurement is carried out in part of space A; it should also be independent of whether or not any measurement at all is carried out in space A. […] However, if one abandons the assumption that what exists in different parts of space has its own, independent, real existence, then I simply cannot see what it is that physics is meant to describe. For what is thought to be a 'system' is, after all, just a convention, and I cannot see how one could divide the world objectively in such a way that one could make statements about parts of it. (Einstein, Letter to M. Born, 18th March 1948, 1971, pp. 164-165)

Furthermore, given teleportation and assignment of primitive 'thisness' to physical states it is possible to devise a hypothetical protocol, such that would allow the separated agents to send signals to each other, almost instantaneously and faster than the speed of light, relying on the measurement of the distant particle (the state of which will be steered by the operations on the proximal particles in the standard teleportation protocol) (Halvorson & Bub, Can quantum cryptography imply quantum mechanics? Reply to Smolin, 2008, p. 3). So respecting quantum information theory and the phenomenon of teleportation, along with classical (and empirical quantum) demands for no superluminal signalling, indicates that the physical states 'corresponding' to quantum states in the formalism do not have an individuating 'thisness'. In other words, 'teleportation is just a flashy name, but nothing material traverses the distance between the experimenters. The trick is then to explain what happens that enables the experimenters to know (and verify) what they do, and still try to respect the separable existence of material objects.

[59] Consider: "[…] an independence condition for distinct [spacelike separated] systems […] is taken for granted in both classical and quantum mechanics." (Halvorson & Bub, Can quantum cryptography imply quantum mechanics? Reply to Smolin, 2008, p. 1)

Given all this, the authors choose to remain in a precariously suspended state of denying a primitive 'thisness' to physical systems that are a part of the phenomena they aim to explain, and yet to use the concept of 'physical system' in providing a non-formal account of the phenomena. This brings us back to the track of neo-Bohrianism: denying the reference to the terms we are nonetheless forced to use in accounting for the phenomena. On the other hand, we lack a positive account of what it is that the structured regularities of the CBH formalism correspond to, what the phenomenal structure that is mapped by the algebraic structure of the formalism is. Despite the precariousness of their position, the CBH claim that the most rational position to take is one of cautious agnosticism about any metaphysical commitments (despite being forced to use a metaphysically richer language than, perhaps, they would like, in order to communicate to the physics community). This is because they see the ontic commitments and interpretations of the formalism beyond what is given above, as extensions of a quantum theory for the purposes of mechanical visualization, explanation and understanding. But as the formalisms associated with such extensions cannot produce additional empirical evidence[60] for the additional 'mechanical elements' over and above the evidence used to produce statistical correlations predicted by the bare formalism of a quantum theory, they see it as most rational to withhold metaphysical commitment in any direction.[61]

Still, one might wonder whether this is not putting things the wrong way round: surely we should use some theory about the structure of matter to show how the information-theoretic constraints arise. Not so, according to CBH, for that would not be respecting the full metaphysical implications of the principle methodology. The principle methodology, as explicated in the Bub (2008) article, does not seek to fight head to head with the constructive alternatives, but redefine the battleground altogether. This is not a difference between a top-down and a bottom-up approach, but one of radically different ontological world-views. Not just a list of what does and does not exist, but also how that which exists behaves and interacts (e.g. whether an electromagnetic field requires aether as a carrier, and whether all rods and clocks have ultimately fixable positions relative to the aether). This is what the CBH want by requesting that information be understood as a new physical primitive. The theoretical formalism then builds on this assumption:

> Quantum mechanics represents the discovery that there are new sorts of information sources and communication channels in nature (represented by quantum states), and the theory is about the properties of these information sources and communication channels. (Bub, 2008, p. 14)

> […]the claim that quantum mechanics is about quantum information—that quantum mechanics is a *principle theory* of information (in the sense in which Einstein regarded special relativity as a principle theory)—and that this physical notion of information is not reducible to the properties of particles or fields, is not to be construed as the claim that quantum mechanics is about observers and their epistemological concerns, […] nor that the basic stuff of the world is informational in an intentional sense. Rather, the claim is that the lesson of modern physics is that a principle theory is the best one can hope to achieve as an explanatory account of quantum phenomena. (Bub, 2008, pp. 15-16)

---

[60] Albert, 1992 also seems to imply that there are metaphysical commitments of different interpretations of the quantum formalism that cannot be decided amongst by experiment, i.e. that are empirically equivalent. (Albert, 1992 , p. 189)

[61] Though on the whole he finds a lot of problems in the CBH approach, and Bub's further elaboration of the philosophy behind it, even Duwell (shortly to be presented criticising the CBH approach) in the end expresses sympathy towards this metaphysical supposition. He says we can always prefer one theory over another (though, note, Bub is actually talking about theory extensions from the common core), but that it is not rational to have a cognitive state as extreme as belief that one theory is true and its *empirically equivalent* rivals false (Duwell, Re-conceiving quantum theories in terms of information-theoretic constraints, 2007, p. 198) [my emphasis].

## *Explanation*

According to the CBH programme we should not be trying to explain what happened mechanically in the 'troublesome phenomena', as we don't even have sufficient tools to properly account for the interaction between the object-systems, the instruments used to observe them and ourselves (or at least the hypothetical experimenters). We should instead view the situation as containing epistemological black boxes, which may in part be successfully described by some other physical theories, but not in terms that account for the 'troublesome' measurement outcomes. The black boxes in turn produce signals, that can be formally accounted for by the theory, and based on which establish signal correlations between different (and possibly distant) black boxes. The theoretical formalism allows the experimenters to attach some probabilities to certain signal correlations and not others. If the black boxes are 'real objects' (whatever they may be made of) it seems certain information transmission protocols are 'permitted' by reality and others are not (Bub, 2004).

But if the game of predicting signals is all that we can safely say to be doing in 'quantum experiments' then, Bub claims, the quantum formalism ('quantum theory' in Bub's terms) provides a theory about representation and manipulation of information, and not a theory about the ways in which non-classical waves of particles move, or the ways in which the universes split and recombine. As, on the CBH approach, the representation and manipulation of information is constrained by the information-theoretic principles, accounting for those principles becomes the fundamental explanatory aim of (this segment of) physics. This newly discovered aim has not produced many outcomes as yet, but the shift of focus marks an important departure for the provision of explanation from contemporary physical theories. Yet the primary focus in this thesis is on the ontological characteristics, so we will want to know what can be deduced about 'what is out there' from the constraints on the representation and manipulation of information that hold in our world.

> The explanation for the impossibility of a [description in terms of a classical conceptual framework] then lies in the constraints on the representation and manipulation of information that hold in our world. (Bub, 2004, p. 259)

Of course the ambitious aim lacks the sturdy output as yet, but Bub warns that we must "recognize information as a new sort of physical entity, not reducible to the motion of particles or fields" (Bub, 2004, p. 262). In principle this does satisfy the first question about the ontological characteristics of explanations from this type of contemporary quantum theory, namely that they should concern a new kind of 'stuff': information. Yet, this is a short-lived satisfaction for we are essentially changing categories altogether here. As will be discussed subsequently, it is disputable whether we can think of information as stuff at all. But even if we could, this is radically different stuff from our more familiar matter. And yet, we are by no means replacing matter with the new stuff (this would be a welcome and simple situation, then we could simply compare the explanatory success of the two). But as the above section on metaphysics indicates, the CBH are not suggesting that the world is 'made of information', or that material physical systems and the measuring instruments we use do not exist in material sense.[62] To explain what is going on in the world, what produces the phenomena we experience, we still need some account of the physical matter; or some account as to why we think there is a material world that produces phenomena in us in the first place.

But in the narrow domain of experience that is dominated by the prediction and measurement games of quantum physics, we have thus far been mistaken in thinking that the games we successfully played allowed us a glimpse of the structure of physical reality. We must now wake up to the fact, the CBH claim, that quantum physics was never about constituents of reality but about information manipulation. But information, that new stuff, is somehow linked to reality, and by investigating the link we can gain some understanding about the nature of reality, though probably (if the CBH theory derivation and assumptions are right) not about the mechanical aspects of its nature. There seem to be two possible routes to follow (which we shall investigate in greater detail in Chapter 4): (1) to sufficiently modify, or even replace the 'extended stuff' explanatory conceptualization (as perhaps presenting all 'extended stuff' as an illusion reducible to something else); or (2) to find ways to reduce the properties of the new stuff (information) to those of the primary qualities of the 'extended stuff'. There is as yet no suggestion in literature as to how the information-stuff and the extended stuff can coexist at the fundamental level. And as a way of introduction to discussion in Chapter 4, let us recall the notion of depth of explanation, briefly introduced in Chapter 1 (cf. (Hitchcock & Woodward, 2003)), that stresses the importance of the detailed account of the controllable variations in *objects* that the changes to be explained happen to. What replaces the objects in information-ontology will be important for discussion in Chapter 4.

[62] Though they do suggest a quibble with a 'primitive thisness' of those instruments, and their individuation may end up in some non-standard, albeit material, form. It remains an important open question how the proponents of this principle approach propose to connect the information-oriented research with the 'material foundation' of the common conceptual scheme.

## 2. 3. Principle approaches: problems and objections

The types of objections to different stages of the derivation of principle versions of quantum theory can be divided into those that object to the principle methodology (either that the adherents do not truly stick to it, or that principle methodology cannot be a valid road to explanation), to the metaphysical shyness (seen, perhaps, as deceit or trickery) and to explanatory robustness and lack of attention to detail. We shall try to survey all three types of them, though often the critique of one type is interconnected with another or they entail one another. The common point of most criticism can be summed up as the strong conviction that only constructive accounts can be sufficiently explanatory, and that no convincing explanation can stop at the principle stage without outlining the details of the metaphysics of the causal processes behind the phenomena.

### Methodology

Most vociferous criticism of the methodology of the principle approach, focused on the CBH version here unless explicitly stated otherwise, is that following Einstein's principle methodology of the 1905 Special Relativity derivation is unjustified in the current state of research in physics. Namely, Brown and Timpson (2006) claim that Einstein's own approach of 1905 represents a victory of pragmatism over explanatory depth that was only justified by the context of the chaotic state of physics at the start of the 20th century. They aim to stress that taking Einstein's 1905 approach as a role model fails to appreciate his own admission that such strategy was "a policy of despair, and represented a strategic retreat from the more desirable but, in his view temporarily unavailable *constructive* approach" (Brown & Timpson, 2006, p. 31). It seems Einstein never wanted to be followed in this respect, though it will take some further argument that he never should be (i.e. that his recommendation should be obeyed).

> It seems to me too that a physical theory can be satisfactory only when it builds up its structure from *elementary* foundations. […] If the Michelson-Morley experiment had not put us in the worst predicament, no one would have perceived the

[special] relativity theory as a **(half) salvation**. (Einstein, 1995, p. 50) (bold emphasis mine)

Brown and Timpson proceed to explicate just why the special theory of relativity is only a 'half salvation'. They illustrate how a much more satisfactory (though computationally more laborious) explanation of the workings of the single piston heat engine undergoing a Carnot cycle can be provided by statistical mechanics than by thermodynamics. Most notably they criticise the fact that the thermodynamical approach for failing to answer why the perpetual motion machines cannot exist, though it explicitly forbids them through its foundational principles. "What this theory gains in practicality and in the evident empirical solidity of its premises, it loses in providing physical insight" (Brown & Timpson, 2006, p. 32). And such theories are only acceptable in special circumstances, and then explicitly as temporary solutions until an overarching constructive theory is produced ("When we say that we have succeeded in understanding a group of natural processes, we invariably mean that a constructive theory has been found which covers the processes in question" (Einstein, 1954, p. 228)). So for our principle approach the proponents should demonstrate that the situation in the quantum theory and the explication of phenomena from quantum information theory is akin to the "worst predicament" of the Michelson-Morley experiment (cf. (Einstein, 1995), the quotation given above).

And Brown and Timpson rightly pick at segments of CBH's reading of history of Special Relativity (Clifton, Bub, & Halvorson, 2003) that are contentious in the philosophy of physics community today (cf. (Brown & Timpson, 2006, pp. 36-38), for a discussion of whether Minkowskian geometry should be seen as an algorithm for kinematic effects that require explanation through Einstein's theory or whether Minkowskian geometry is itself a constructive part of the special theory of relativity). Yet their own careful and thorough analysis seems to suggest that it was the impeding problems of quantum theory, namely the wave-particle duality which threatened to preclude a formation of a theoretically (i.e. precisely mandated by the mathematical formalism) sound conceptual framework for the electromagnetic and mechanical phenomena, that prompted Einstein to adopt the principle approach. Namely, he could not envisage the metaphysical conceptual framework that can 'reproduce' the phenomena in the climate of wave-particle duality, and these concerns over metaphysics pushed him to look for a solution in an unlikely place. He searched for a theory that could 'reproduce' the phenomena, even if all of the previously adopted metaphysics has to be abandoned (as it was undergoing a revision). If these difficulties with quantum theory metaphysics still give reason for concern today (as we have been trying to outline above), does that not give some impetus for a principle approach, despite Brown and Timpson's objections?

Thermodynamics was in Einstein's eyes the only theory to reproduce the phenomena without the troublesome metaphysics at the time, for whatever the speculations about the structure of matter, he could not envisage a situation in which its phenomenological principles were shown not to hold. So he opted for a methodological guidance from thermodynamics and searched for those aspects of the phenomena in electromagnetism (the domain of the 'troublesome' Michelson-Morley experiment) whose conceptual formulation could survive whatever structural hypothesis proposed for their constructive reduction.

> […]the speed of light is independent of the speed of the source and isotropic – something every ether theorist took for granted when the frame in question is taken to be the fundamental ether rest frame and something which remarkably Einstein felt would survive whatever the eventual quantum theory of radiation would reveal. (Brown & Timpson, 2006, p. 36)

Thus, it seems that despite the potentially erroneous "CBH historical fable" it is not entirely unjustified to treat the current situation in quantum theory as one where doubts about the metaphysical foundation for a unified conceptualisation of reality prompt for a principle speculation: for a search for those aspects of the phenomena that can survive any eventual construction of the explanatory metaphysics. It is of course, worth noting the warning from Brown and Timpson that even following the empirical success of Special Theory of Relativity, Einstein remained uneasy about the "sin" of the role that the ideal rods and clocks played in the theory (Brown & Timpson, 2006, p. 36). Special relativity, in Einstein's own words, divided the world into "(1) measuring rods and clocks, (2) all other things, e.g. the electromagnetic field, the material point etc." (Einstein, 1951, p. 59). Einstein admits that this is unacceptable in the long run, but also that it was a necessary, though unwanted, consequence of the derivation of the theory from the generalised phenomenological principles. In some sense, this is warning us that principle theories cannot in and of themselves (without further metaphysical speculation and theoretical construction) yield their own constructive replacements. But it is also giving us a historical example of how, despite the self-confessed conceptual shortfalls, principle theories can make advances in conceptualisation of the explanatory framework (even if, in places, pointing to its inherent explicit shortfalls).

A situation present in the quantum theory today, including the phenomena in the domain of quantum information theory, can be seen as justifying the return to the drawing board and a search for the foundational principles (as phenomenological generalisations). Namely it is difficult to see a conceptual framework for the theory that will combine the requirement of separability of physical systems and locality of physical processes, with the demonstrated phenomena such as teleportation. We might be prompted to search for those aspects of the phenomena that are best positioned to survive any future constructive speculation. Thus, a principle approach may be called for, though it is by no means clear which of the offered principle approaches it should be. But even if the CBH story (Clifton, Bub, & Halvorson, 2003) of Einstein's special theory of relativity furnishing an acceptable principle interpretation for the already existing empirically adequate Minkowskian formalism is not historically correct, it can be dismissed as an unsatisfactory analogy, without questioning the justification for the overall principle project.

Let us turn briefly to the criticism that in the proposed principle approaches (most notably that of CBH) the foundational principles do not correspond to the requirement of simple, intuitive generalisation of the key aspects of the phenomena. Most notably Duwell (2007) claims that the supposed natural empirically discovered constraints of natural processes (in this case information-theoretic processes) are not empirically discovered constraints at all. Namely, if the foundational principles are to be mathematically expressible generalisations of the phenomena, what exactly are the phenomena that the information-theoretic constraints generalise? Duwell claims that the evidence for the constraints is indirect and challengeable, for they are not unshakeable enduring straightforwardly *observable* characteristics of the phenomena, but are mere predictions of the standard theory. But then he goes on to say that such is the nature of any constraint, which constrains what is possible and cannot be tested directly (i.e. we cannot test how well we have recognised what is possible, as the impossible – the other side of the constraints on the possible – is not empirically/epistemically accessible at all, being physically impossible). Predictions, Duwell says, can be verified more straightforwardly, but constraints can't. However, in the light of the methodological discussion on whether to follow Einstein's example of the 1905 Special Relativity derivation, this criticism applies across the board for Einstein too had no means of testing whether the constraints he 'imposed' on the natural processes truly hold out in the world[63] (until they are demonstrably broken, that is). The situation can be taken as far back as Einstein's methodological role model, thermodynamics, for there too the fundamental principles of the theory are the constraints on the unfolding of natural processes, and this is precisely where the sturdiness of the theory lies.

Yet it is worth following this complaint a little further. The CBH constraints (unlike the 'Fuchs negative principles') do not appear to be empirical/phenomenal, nor straightforward. Perhaps the previous discussion showed, though, that their most remarkable characteristic should be their strength in the light of potential changes of the constructive structure that they might some day be reduced to. They must be the characteristics of the natural process that we expect will never disappear as 'unreal' from our explanations of the phenomena.[64] For example, the ban on superluminal information transfer via measurement is one such sturdy constraint, the one that seems to hide the deeply entrenched expectation of the separability behind it. But the ban on bit commitment would not seem even to many physicists as a physically necessary characteristic of reality (though it might be).[65]

Though this would be leaning away from the direct proscription by Einstein to look for *the unchangeable characteristics of the phenomena*, adoption of a mathematical formalism could help here, for we may find that some 'sturdy' characteristics are most economically expressed formally, even if this makes them less accessible to a wider audience. In a theory that aims to account not just for what people see (like maybe length contraction theory might be expected to do, be it a dynamical account through structure or a phenomenological account through principles), but also for what they get after manipulating instruments in accordance with their expectations of what they should have (could have?) gotten, it may not be so preposterous to introduce fundamental principles expressed in terms of some shorthand or mathematical formalism. But even if this were granted, Duwell objects to the choice of the mathematical framework that the constraints are situated in (Duwell, Re-conceiving quantum theories in terms of information-theoretic constraints, 2007, p. 184). He, rightly, expects the mathematical framework, the formalism, to be neutral regarding the choice of physical ontology to accompany the eventual quantum formalism. But, as Spekkens (2007) illustrates, the mathematical framework and the constraints are capable of yielding non-quantum theories, so the choice of formalism needs to be strengthened so as to exclude unwanted theories, such as the toy theory of Spekkens (2007). It remains an open problem of the programme whether all the possible, but unwanted, toy theories should be excluded by further modifications of the choice of mathematical framework (which will inevitably affect the choice of the metaphysical assumptions that go with it), or whether we should find what are reasonable constraints for the formulation of physical theories and rule them out on grounds of those.

[63] Think for example of the light postulate that has no direct verification, and where the debate about the conventionality of simultaneity (and thus the isotropic spread of the light wave) is still open in the philosophy of physics.

[64] We could for example liken them to primary qualities, namely the famous Cartesian derivation of the extension of the wax as its unchangeable quality (in *Meditations*). Unlike the secondary qualities that do not retain their phenomenal sturdiness when subjected to the explanation of what is really going on, i.e. secondary qualities as essentially dispositional and unreal.

[65] Duwell is in fact even more critical, he says the constraints only hold from the perspective of standard quantum mechanics, but not from that of other quantum theories (Duwell, Re-conceiving quantum theories in terms of information-theoretic constraints, 2007, p. 199). There is no room to enter that discussion here, but from the rest of the text it will become evident that this claim is tied to an erroneous assumption that the CBH argument starts from the 'standard quantum mechanics' and not the bare formalism common to all theories. Perhaps the warning by Halvorson and Bub (2008) that the CBH version of the theory is not developed in isolation from 'theoretical context' can be interpreted the Duwell way, but it needn't. The context can likewise be provided by the empirical results and the background assumptions about physical reality in general (such as is the one of separability) without subscribing to any particular interpretation of the quantum formalism beforehand.

Part of the answer to this question is given in Halvorson and Bub's response to Smolin's criticism of the CBH methodology (Halvorson & Bub, Can quantum cryptography imply quantum mechanics? Reply to Smolin, 2008). Smolin proposes to derive a mathematical formalism from the information-theoretic constraints that will not be the quantum formalism sought by CBH (Smolin, 2003). However, Halvorson and Bub swiftly respond that it was never the intention of the authors of the CBH approach to take the constraints in isolation from any theoretical (assuming this to mean physical, as well) context. Halvorson and Bub indeed acknowledge a whole host of explicit and implicit background assumptions (some of which have been considered here) that contribute to the particular derivation of the quantum formalism, and do not result in an altogether abstract mathematical game.[66]

### Metaphysics

As has been indicated above, and in the previous chapter, every principle approach carries with it some metaphysical assumptions that can point to the search for a more constructive conceptual framework, so it is worth investigating the objections to the metaphysics of the proposed principle approaches. But the approach of CBH has some even more provocative and explicit metaphysical commitments, namely the claim that a quantum theory should primarily be viewed as theory of information processing in 'the quantum world', and that information should be introduced as the new primitive in physics. This claim rests on a deeper principle that when mechanical theories (in this case theories of everything material and non-informational, from particles to waves) fail to show empirical supremacy over the metaphysically more conservative ones, then the latter should be preferred. A further step then requires that the representation and manipulation of information be recognised as the appropriate aim of physics (or the quantum segment of it).

[66] They, in fact, go further to find and point out more technical problems with Smolin's account, which make his mathematical formalism unacceptable as any sort of physical theory, but those details are beyond the scope of this discussion.

It is the deeper principle that is seen as problematic. Depending on different formulations, different readings of it in the literature, it either rests on the bare quantum formalism, or the more (though not much more) meaty 'standard theory'. If Bub's deeper methodological principle rests on the 'standard (quantum) theory' as the starting point for validating metaphysical speculations of other theoretical interpretations, then Duwell's (2007) criticism (explicitly credited to (Timpson, 2004)) that the CBH start from minimal metaphysical expectations of interpretative 'standard quantum mechanics' and not the bare mathematical formalism stands. Namely, what right do we have, other than historical contingency, for taking the 'standard (quantum) theory' as the basis for all metaphysical speculation; and without such right any other interpretation that is empirically adequate can be taken as the yardstick against which to measure the alternatives. On the other hand, if Bub did not have the whole package of the 'standard theory' in mind, but barely the formalism that is supposedly shared by (is common to) all the interpretations, then there is a clear reason to prefer it to the metaphysical speculations.

It is a categorical difference between the bare formalism, a mathematical tool, an abstraction, and all the other segments of particular interpretations (including the 'standard' one). The latter are not formalisms (or parts of the formalism), nor an abstraction, but are metaphysical conceptual frameworks built around the bare formalism in order to provide an explanation, or at least some sort of visualisation, of the physical processes corresponding to the mathematical representation. Fuchs' programme above also seems to rest on the assumption that there is a common formal mathematical core of all different quantum theories. CBH's search for a unique mathematical framework that would cover all the classical and quantum theories, and yield quantum ones through the introduction of the constraining principles, strongly suggests that there is in the background an expectation that a common mathematical formalism can be found in all quantum theories. Assumptions though do not amount to a proof, so it remains an explicitly open question whether principle approach authors recognise a common formalism in all quantum theories, whether such formalism can be separated from the theories so that the remainder can be compared between different interpretations, and whether Bub is relying on this assumption when using his deeper methodological principle.

Lacking the answer to the above, we can search Bub's writing to find whether he takes the 'bare formalism' or the whole 'standard theory' as the starting point. In the very same section that Duwell takes passages for his criticism from, Bub says:

> Note that the argument here is not that it is never rational to believe *a theory* over an *empirically equivalent rival*: the methodological principle I am appealing to is weaker than this. (Bub, 2004, p. 260), (my emphasis)

We can take this to be a strong indication, along with perhaps methodological errors[67] pointed out by Brown and Timpson (2006) above, that Bub does not imply that 'standard (quantum) theory' takes a privileged position as a starting point, but that it is the bare theoretical formalism (in itself insufficient to be taken as a theory, even a minimal one) that is common to all quantum theories/interpretations and thus worthy of the privileged position. Of course, Bub could be mistaken about there being such bare formalism, a distillate available from all theories/interpretations, but that, as is indicated above, even Duwell leaves as an open question.[68] However, even supposing that Bub is justified in holding on to agnosticism about the metaphysics behind the quantum phenomena, and preferring his own metaphysically agnostic theory over those that dare to speculate, there are problems with the ontological commitments of his approach.

[67] That CBH authors think Einstein starts with the ready made *formalism* provided by Minkowski, for which his special theory provides an interpretation.

[68] Duwell (2007, pp. 186-187) does actually recognise a problem vaguely along these lines, and sets off to rectify it by looking for conditions that might make one theory a foundation (or a common core segment) for the other, but does not open the discussion over common mathematical formalism. Timpson on the other hand is happy to accept the existence of the bare formalism and divides the quantum theories into three groups. Those that stick as closely as possible to the bare formalism (instrumentalist and (sic) Everett interpretations), those that appeal to non-unitary dynamics as modifications of the formalism (dynamical collapse *á-la*-GRW), and those that add extra metaphysical structure to the bare formalism (Bohm theory, hidden variable theories and some modal interpretations).

Namely, what is to be made of Bub's use of concepts of 'system' (or more precisely, 'physical system') and black box, in accounting for the troublesome quantum phenomena. On the subject of black boxes, Duwell is precise and devastatingly critical: these are not metaphysical black boxes, objects that we cannot now, but might be able to one day, take apart and come to know better. They are 'epistemological' black boxes, meaning we can observe and take them apart, just as physicists have been doing ever since they have been constructing the measuring instruments, but that (due to the guiding principle we have adopted) we cannot speculate the ultimate nature of (Duwell, 2007, p. 188). On a certain level our observation of the measuring instrument itself (not the physical system in the measuring process) will hit the 'quantum wall', will run into a constructive speculation, and because such speculations are banned, we will simply choose agnosticism about the whole thing full stop.

The metaphysical extent of this 'whole thing' is virtually endless, for there is no incontestable barrier between the measuring instrument, the rest of the world and myself, save for the implicit assumption of the mind-body dualism that allows me to escape the measuring instrument, at least at the level of consciousness. By epistemological black box, Bub seems to mean, that we can know what the thing does in terms of input and output, whilst remaining completely agnostic about where it is, how big it is, and what it is made of. Strong adherence to the methodological aim of dedicating physics to information manipulation and 'investigation' is all that is supposed to stop us from taking the objects at hand apart, nonetheless.[69]

But it is then only a minute step from accepting such view to committing to the CBH metaphysical speculation that the information is the newly discovered physical primitive and that quantum theory is our best theory about that. On the other hand, Duwell says that taking the environmental decoherence as the only joint segment of different quantum theories and thus not susceptible to agnosticism about metaphysics is not a safe road to take. He claims that there is evidence that environmental decoherence may not be sufficient to recover our experience of the world ( (Duwell, Re-conceiving quantum theories in terms of information-theoretic constraints, 2007, p. 188); referring to (Bacciagaluppi, 2004)). More recently

[69] This is no trivial matter for such adherence is attainable for those who accept from the start that taking the 'black boxes' apart has nothing to do with explaining the 'troublesome' correlation-based phenomena. But if the 'black boxes' are not to be taken apart, then either there is never to be an overall reduction of the information ontology to the ontology of extended matter, or the reduction should be directed towards some other segment of material reality (though it is hard to see what that would be).

Duwell (2008) suggests a technical account of how information can be analysed as an abstract entity, short of awarding it the status of substance, where substance is a general form of the material 'stuff'. Now that goes some way in helping the principle approach of CBH (though Duwell does not explicitly refer to that particular research programme) lay its ontological commitments in the open. It is not for us to evaluate the details of this proposal, but a few remarks that bear on its potential for provision of information are in order.

As abstract entity quantum information is not subject to change, or rather it possesses no durable changeable properties that would allow it to withstand identity under change. It must also not be regarded as a property (an abstract property akin to kinds or universals) of the underlying material ontology, as the Duwell analysis (relying on further technical distinctions in (Timpson, 2004)) explicitly shows it to fall short of the requirements for a substance or a property of material substance. Yet it goes some way to addressing the troublesome phenomena, by firstly disentangling them from the problems of separability violations by extended matter, and secondly showing that from a purely (and again technically) informational aspect their troublesome phenomena dissolve as they are in no way reliant on spatial extension or location of the information-entity. The latter in fact has no pretence to such grounding. Of course, as soon as we would try to treat the information as the new property of material substratum, the worries about separability violation would return. This difficulty in tying up (one such) proposed information ontology with material ontology points to a feature that is interesting from the perspective of explanation. Due to its resistance to alterations of properties, abstract entity information (if that were the paradigm we adopted) cannot feature in the explanations of the causal mechanical type. Duwell therefore advocates that "explanations of the quantum phenomena, if provided by the quantum information theory" (Duwell, 2008, p. 215), feature only in the unification type explanations. The unification of course should be provided by the phenomenological status of their constraining principles.

A brief comment, to be elaborated in greater detail in the discussion in the final chapter is in order here. Supposing we follow the suggested Duwell route, or a methodologically similar one, two objections arise, especially in comparison to the more candid constructive explanations of the next chapter. The unification explanations of the type proposed above would be extremely blunt about removing the troublesome aspects of the phenomena, not really making it clear how we came to see the phenomena as troublesome in the first place (except by simply saying we were constantly in error about what the object of physics at this level of reality should be). Tied in with that is the observation that they hardly even point towards the connection to objects whose changes in the material world lead to the appearance of the troublesome phenomena. And that is the truly interesting question: what is the link to the material foundation for the troublesome phenomena.

The question remains whether Einstein's guideline above: take only what you can be confident will not be affected by future metaphysical speculation, can help us out in this segment as well. Can we safely assume that whatever constructive explanation we may some day come up with for the behaviour of systems and apparatuses in the measuring process, they will always behave so as to have some input and some output?

On the face of it, it is not such a bad assumption, given that we are looking for something - anything - that survives the withdrawal from metaphysical speculation. We can never expect to 'see' directly into the measuring process at the extension level of the quantum phenomena, but there will always be some input and some output of the process. The only problem is, there is hardly any physical process that cannot be characterised under input-output principle, yet we have tried and have succeeded to find physical theories of greater precision than 'rubbish in, rubbish out' model. We have to postpone settling this discussion for the final chapter.

There is a further complication even for the assumption that the formalism is a mere calculational device, that the formalism is informationally incomplete (Maudlin, 2007a, p. 3155), as suggested in Fuchs's approach. In the troublesome phenomena, such as the EPR situation, the calculational device tells us that had things been different on the proximal side of the experiment, so they would have been on the other, distant side. If this is further coupled with recognition that the proximal outcome is a result of chance, an inherently unpredictable outcome of intervention in nature (or even, to strengthen the argument, a chancy choice of parameter to be measured), then we know things could have been different even with all the causal antecedents the same (i.e. our initial instrumental codification).[70] And, so Maudlin (Maudlin, 2002, pp. 146-148) argues, we get a counterfactual-supporting causal connection between the material outcomes on two sides of the experiment which cannot be explained by a common cause. Thus, even though the material existents are not described by the formalism of the theory, they do present a situation which cannot be explained by a common material cause for the two sides of the experiment. Our experimenters' guesstimates seem to still rest on the mysterious non-local connection between the material existents about which they have been formulated. More generally, it seems that any epistemic interpretation of the formalism that presupposes it has some direct links to the states of the world (however unpredictable and partial these links may be) will have to endow those states with non-local causal connections that violate the separability principle. If, on the other hand, holding on to material separability was one of the starting points of the particular principle approach (in this case Fuchs') that approach would appear to fail purely on the grounds of lack of internal consistency.[71]

[70] Of course, this assumes the analysis of causation based on support for the counterfactual situations, which there is no room to quarrel with here. Nonetheless, it serves as an indication of the difficulties that the Fuchs programme comes across, but that the CBH programme can hope to avoid.

More recently, Brown (private correspondence) objects to the notion of the evolution of the wavefunction (or state-ascription) in the long intervals between the measurement interactions. Why should the conscious agents expect their expectations (guesstimates) about the interactions with reality to change of their own accord in the intervals that they are not interacting (and not even planning to) with that very reality. Metaphysical commitments in Fuchs' response (private correspondence) clearly come about here again, strengthening the above criticism that separability violations cannot be avoided on this approach after all. In simple terms, the issue is why the state ascription, the guesstimate, changes with the formally calculated evolution of the wavefunction overnight whilst the experimenters are sleeping and are thus not likely to induce any unpredictable reactions into the super-sensitive reality. And Fuchs replies that something is, after all, changing about the material system overnight and the experimenters commitments are update in the morning to stay true to that commitment. He falls back on calling for the treatment of the quantum state as epistemic to be an unimpeachable dictum from which the further research programme should proceed, without at this stage providing an answer to worries about what in reality compels the experimenters to make the necessary overnight updates.

### Explanation

What is the explanation of the material (or otherwise) foundation for the 'troublesome' phenomena to be extracted from the principle approaches, individually and in general? Following Einstein's model principle theories do have embarrassing features (despite their empirical sturdiness), such as Einstein's privileged rods and clocks were. They are also only an interim step towards a more general constructive explanatory account. But for such an account to be possible, there has to be an empirically testable speculation about the limits of the principle theory, a constructive account has to provide a situation that needn't necessarily falsify the principle theory, but can show where to go beyond it. Bub, and Brown and Timpson agree that the theory of Brownian motion provided such superior metaphysical projection in the case of statistical mechanics: it allowed molecules to be directly counted and demonstrated the limits of validity of thermodynamics ((Bub, 2004); (Brown & Timpson, 2006)).

[71] It has recently been suggested in Timpson (2008) that Fuchs' programme makes no explicit (and formal) demands for the adherence to the principle of separability. Whilst this is strictly true, the narrative argumentation for the development of the new formalism from the principles, especially in the original proposals by Fuchs (2001)), relies on the unpalatability of explanations of phenomena that allow for the violations of separability.

Yet Bub seems to claim that there is no road beyond quantum theory, principle derived quantum theory that is agnostic about the mechanical structure behind the phenomena, that such advance is precluded *in principle* by all the empirically equivalent quantum theories (perhaps even by their common core, quantum formalism). For the case of the CBH programme Duwell concludes:

> [Though] no positive claims are made about what the quantum otology is, Bub thinks that it is not hidden variables, and no matter what it is, it is beyond the scope of physics to investigate it. Hence, quantum mechanics ought to be regarded as a principle theory of information. (Duwell, 2007, p. 194)

Yet there seems to be a missing step here: how come that a particular derivation of the bare formalism imposes any particular interpretation of that formalism? Given that CBH manage to derive what is some core formalism of all quantum theories, we must examine further steps that lead them to a particular interpretation. Of course, there is the deep principle of withholding judgement on metaphysical issues. And then there is the further claim that withholding judgment legitimises the hypothesis that (quantum) information is the new physical primitive.

It is worth reiterating that on the information-theoretic principle derivation of quantum theory, the objects of the theory whose behaviour is constrained by the fundamental principles are the macroscopic directly observable outcomes supported by the apparatuses (preparation and measurement instruments), whereas the apparatuses themselves are treated as unanalysed black boxes (as has been outlined above). The principles provide a derivation of relations between various preparations and measurements, and this is supposed to be the first-hand explanation of why the preparation and measuring apparatuses display the relations ("in terms of relative frequencies of various experimental outcomes", (Timpson, 2004, p. 216)) that they do. At the most basic level of interpretation of the formalism, the elements of the formalism are related with the observable physical quantities (the frequencies with which various outcomes of experiments may be expected). But in the principles themselves there is not much else that can help us go beyond this most basic level of interpretation (Timpson, 2004). Despite the nature of its derivation, such quantum theory would remain at best very similar to the 'minimal interpretation' (perhaps, 'instrumentalist interpretation', with the inherent pitfall concerning possible reliance on the metaphysical projections towards properties of the material background, summarised from (Maudlin, 2002), above), it would link the mathematical abstraction to the statistics of individual measurement outcomes, but it would not go much further in providing explanation of the material foundation of the phenomena that the outcomes are a part of.

What is needed to produce an explanation is the Redhead second-sense interpretation of the formalism, an account of the nature of the external world and/or our epistemological relation to it that serves to explain how it is that the statistical regularities of the formalism-outcomes relations come out the way they do (Redhead, 1987). This is not to say that a constructive account is necessary, though one such would obviously fit well with the 'nature of the external world and our epistemological link to it' requirement, but an extended principle account that goes beyond the minimal interpretation and is, preferably, explicit about any of its inherent 'sins'. Now we can see a further motivation for the employment of the deeper methodological principle, and the eventual road to metaphysical projections (the call for new physical primitives). But, according to Timpson (2004) the methodological principle involves a *petitio principii* argument and cannot be used against the rival constructive explanations (most notably, the Bohmian theory and the GRW dynamics) (Timpson, 2004, p. 220). To take the constraints as imposed natural laws is where the *petitio* lies according to Timpson: the constraints rule out the GRW interpretations for the latter can violate one of the principles, and the Bohmian interpretations because they cannot show additional empirical content over metaphysically more conservative interpretations. To simply state that the constraints hold as a matter of natural law (and thus physical necessity), is according to Timpson to beg the question against rival explanatory conceptions.

It is worth revisiting once more the way the constraints are introduced into the CBH (or, with alterations, some other principle approach). At least CBH are explicit about waiting for a demonstrable violation of any of the constraints. If such violation is to be found in practice, not suggested in principle, then the associated theory would falsify the constraints and the theory based on them. And the discussion would be over; the principle approach based on the violated constraints would fail. This is why the constraints are carefully chosen to be of the sturdy variety, to secure the best possible foundations for the principle theory. But no theory today is beyond the possibility of falsification, though we aim to build them to survive at least for some time. On the other hand, some of the contending theories, such as Bohmian mechanics, claim to be able to predict possible violations of the constraints, but cannot demonstrate them because we live in a particular universe in which all such violations are impossible (cf. the notion of the quantum equilibrium in Chapter 3).

Without going into details of this proposal at this stage (cf. Chapter 3), this appears to be a weak argument against taking the violations as outright forbidden. It has long been the case in the history of science that explanations based on universal conspiracy to conceal empirical support for explanatory frameworks have been considered unacceptable. Supposing that the CBH and similar programmes are open to empirical demonstrations of the violations of their constraining principles, but that no such demonstrations can, at this stage, be proposed, it is not circular to argue that the constraints hold in our universe as a rule (a phenomenological rule in some sense) and that additional metaphysical structures do not play an explanatory role. Of course, the constraints themselves can perhaps be *derived* (as lawlike, or merely approximate), rather than postulated, from the overall theories containing additional metaphysical structure, but that is a methodological issue of a different approach to the quantum formalism, not one of logical clarity. Given that there are reasons to consider a principle approach, adherence to principle methodology has to be respected. With hindsight we may correct the inadequacies of the robustness of principle approach (cf. (Bell, How to teach special relativity, 1987), concerning Einstein's derivation of Special Theory of Relativity), we may explicate its sins, but hindsight is not a luxury we have at the early stage of development of such theories.

Let us also briefly consider Timpson's objection that according to his grouping of the explanatory frameworks, even after the mechanical and dynamic-changing interpretations of the formalism have been discarded by the deep methodological principle, two further possible interpretations remain: the bare instrumentalism and the Everett interpretation (Timpson, 2004, p. 221). The former carries with it all the problems usually associated with instrumentalism in science, and for our purposes can be said to explain very little (and not to aspire to explain much more than that), and is therefore not a serious contender. The latter would take at least an additional chapter of its own to elaborate and analyse, but its greatest weakness in the present context is that it is just not as innocent of the metaphysical burden as Timpson portrays it. For present purposes we take it here to be a version of quantum theory with a heavy burden of (however fickle) existence of multitude of universes, through which the physical processes unfold, but where only the phenomena of one or relatively small group of them are epistemically available to us. But Timpson is right in calling for clearer explication of just how is it that quantum theory supersedes the bare instrumentalism (remember black box instruments) and becomes a theory about representation and manipulation of information (Timpson, 2004). For even if there is something special in quantum experiments, unlike in the more technically demanding classical ones, to suggest seeing the measuring apparatus as a source of signals, it is still sensible to ask what the signals signify or codify. In the context of search for explanations it is almost irresistible to ask what a particular measurement outcome is a signal of, given that it must not be a signal of something about pre-existing hidden variables (or some other details of mechanical structure of reality).

Until the notion of the 'new physical primitive' is further explicated, we can also take as strong criticism Timpson's complaint that it will not help turn an instrumentalist interpretation of quantum formalism into something more meaty by simply "[concluding] that information, or quantum information, is an entity" (Timpson, 2004, p. 222). A primitive, of course, does not have to be read as entity[72], in the same way that extension is not an entity (before or after Descartes). But we need to be told more about just what it is. It is certainly problematic for the CBH account to claim being open to falsification or some future clarification through a constructive theory (though not one of the kind available now and dismissed by Bub), whilst on the other hand changing the aim of physics in the quantum domain and claiming that the best we can achieve is a principle theory of information manipulation (where the measuring apparatuses remain essentially black boxes forever). And with the latter claim holding forth, Bub (and CBH in general) veer closer to the neo-Bohrian approach of Fuchs, by claiming that the reality is such that we will never be able to know the workings behind our measurement outcomes (a metaphysical claim of some sort). The principle approaches (that do not see themselves as mere unfortunate intermediate steps to a constructive theory) deny the implicit premise that a fundamental theory ought to apply to the workings of measurement devices that are constituted out of the very systems that the theory is meant to apply to. And yet, they have to think that on their account the measurement (or any other physical processes, but always those involving acquisition of new knowledge/beliefs) and the behaviour of directly observable devices in it is somehow explained (Duwell, 2007, p. 195).

[72] There is, in fact, no indication that it should be, and as frustrating as it might be for the title of this thesis, Bub does not explicitly commit to an ontological claim in (Bub, 2004). Duwell ascertains as much: "Bub does not out and out make an ontological claim" (Duwell, 2007, p. 193). In fact, a more charitable reading and one in greater accord with other texts, may be that Bub's explanatory framework is simply ontologically neutral regarding the underlying ontology of quantum physical processes involving interaction with conscious agents (i.e. measurement).

The only alternative Duwell sees to the hidden variables of the Bohmian mechanics type (to be presented in Chapter 3), is to go back to Bohr and state that the elements of reality that are represented by the quantum formalism "are simply not like classical definite valued properties" (Duwell, 2007, p. 196). This is a constructive approach of sorts, similar to the constructive elements in Fuchs, but it is too small a step towards a wholesome (mechanical) explanatory framework for the quantum phenomena. It may be linked to the 'sinful' status of reference-frame-defining rods and clocks in Special Relativity. Not that such a framework is impossible (which would be arguing in line with many who demand an outright constructive account for physical explanations, full stop), but we as yet do not seem to have enough of its structure to be able to take an explanation of the phenomena off the ground.

## 2. 4. Summary of the principle approaches

The principle approaches provide an explanation of the 'troublesome' phenomena, but the explanation struggles to provide sufficient features for the transcendental strategy as it struggles to connect to what we take to be ontological concepts therein. On the face of it it bridges the gap between knowing that a phenomenon occurs and understanding why it occurs, as in the conceptual framework of information-entities the occurrence of the phenomena is singled out of the sea of all possibilities by the constraining principles. But the caveat is that we just don't know enough about the information ontology to construct some story of how the 'information-entities' get into the state that evinces the observed correlations. From the perspective of exposition of pseudo-problems (cf. Kepler above and further discussion below), we might say that it achieves what it set out to do, it exposes the said gap as something different, a state of new entities rather than just a statistical correlation of macroscopic states of the material black-boxes. And it is true that the principle theories have little worries about the nature of the entities they take up. But this worry is more easily ignored only from the perspective of prediction, than the perspective of explanation.

For example, in thermodynamics we can predict the occurrence of certain observable states of properties of a wide range of objects (the ‘black boxes’) without any concern as to how those properties come to hold of those objects. We simply choose what to call an object and track the changes of a chosen set of its properties. This is a powerful predictive tool, but in terms of explanation it does not go far enough, as we can see different objects (these are macroscopic ‘black-bodies’ whose macroscopic constructs we can still see, there is no need to worry about unobservables as yet) being constructed in different ways. When certain external conditions can be satisfied about them (that they are in a thermal equilibrium with the surroundings – which we again do not deconstruct) we can predict a whole lot of their properties.

Yet, in terms of explanation, we know them not to be the same object. We took different objects to put them together. To put it bluntly, this type of explanation does not respect that we conceptualise the situations in terms of re-identifiable objects, the latter lose any meaning in the erasure of differences between complex objects in thermodynamical situations. We again jeopardise the conceptual starting point of the transcendental strategy. For explanation, if not for prediction, we would like to see some investigation of the conditions that lead to the same observable properties despite the differences in construction of objects. The real devil here is in the detail. Similarly, information ontology requires some further philosophical justification at the level of connection to the material substratum that is a part of the starting point of the transcendental strategy above. We can take the fact that several slightly different mathematical models can be used as toy-theory derivations of the formalism attached to the information-ontology as an illustration of this point.

Likewise, on the face of it the principle approach explanations stop the why-regress at the level of information ontology, simply by establishing that this is what this segment of reality is like. But even in taking the new ontology to be at the first stage of development of the Nersessian (1984) advocated route, the analogical stage, there is preciously little hooks to anchor the analogy on. For as soon as we start looking for the hooks, we are back at the common-sense conceptual scheme and the threat that ontological holism poses for it. This is in general how the principle approaches of Chapter 2 struggle with even complying with the unification-style explanation paradigm, as they cannot connect to the material ontology without threatening to make it non-separable. And the whole problem for the transcendental strategy goes back to the beginning again. Finally, Lipton's criterion of self-evidencing is easily satisfied in this case, as the occurrence of the 'troublesome' phenomena was methodologically an important point for the development of the whole new conceptual scheme. Yet this on its own does not go far enough.

Beyond the criteria, if we take the principle approaches' explanations as not of the ontological type, then they are of not much use for us here, seeking to compare the ontological characteristics of explanations. They are of not much use for the transcendental strategy either, as it aims to show how the non-problematic everyday ontology can be connected to the theoretical ontologies assumed to be fundamental. If we take it to be ontological, and trying to develop a novel ontology of its own, then we are back to the problems of connecting it to the common-sense conceptual scheme, as has been outlined above. A useful pointer to take at this stage, though, would be to look into how dissolving the danger of the non-separable (i.e. holistic) ontology can still be achieved, even without having to move to wholly novel ontological entities. This would mean taking some of the proscriptions of quantum formalism as incomplete, as guesstimates, whilst furnishing a sufficient generative mechanism behind such limited epistemology. In the vein of our transcendental strategy narrowed down to this special domain of experience we should look into what the world ought to be like so that we could know what we come to know about it through quantum formalism.[73]

So given that we are dealing with a unification-type explanation, it remains to show in Chapter 4 that it can be taken to fall under the ontological rather than the epistemic variety, and that it can stop the why-regress. For according to Lipton (2004, p. 7) this is the biggest problem for unification-type explanations in general. In some instances this can be satisfied by embedding those explanations into the "wider" pattern, but we will have to investigate in greater detail just how this is to be done. We should not forget that the constraining principles themselves carry with them some ontological characteristics, along with that carried by deeper metaphysical principles behind them. For example, the principle approaches are deeply committed to preservation of separability, an underlying principle that imports the individual existence of macroscopic objects and the like.

[73] One might suggest that this is precisely what the principle approaches tried to achieve with the identification of the constraining principles. But as yet they tell us nothing about what the world must be made of for the principles to hold as they do, and that is what is required for a deeper explanation: an account of the ontology that gives rise to these principles. What it certainly can teach us is to remove some deep conceptual expectations we may have had, by exposing them to be the root of our problems, and in this case one such expectation seems to be the account of the world whose fundamental structural feature is solely geometrical.

We should also bear in mind that the principle approaches are not aiming to replace the existence of material objects with information, but claim that the explanation of the 'troublesome phenomena' is essentially about information manipulation. Manipulation that is still performed with the aid of the material world, so we should be able to ask what the basic objects of such explanations (objects whose existence is invariable in counterfactual situations) are. This is to ask what is carrying the burden of explanatory work (most notably in the CBH programme). This is not a question about detailed nature of systems and instruments in the input-output manipulating process, but a request for clearer delineation of the existents suffering change at the level of information manipulation.

In summary, explanation, even of the unification type, will require a physical theory that steps away from bare instrumentalism, even if moving the whole debate to the level of macroscopic, directly observable, outcome manipulation, i.e. away from the mental processes. To set up an explanation of the unification type we need to explicitly state the segments that unify it with the rest of our (standard) conceptual framework.

# 3. CONSTRUCTIVE APPROACHES

## 3. 1. A quantum (mechanical) theory: 'Bohmian Mechanics'

### From epistemic restrictions to mechanical superstructure: historical and conceptual background

In the previous chapter the theories that follow Bohrian interpretation have been presented. They hold firm to some expectation of physical ontology, namely that it must be based on the familiar notion of macroscopic objects, some of whose properties must be directly perceived, whilst others are be derived from those. Directly perceived properties are spatial position and 'geometrical' extension, with existence independent from the surrounding environment. This way a body is conceptualised primarily in Cartesian-like primary qualities, with other perceivable properties reduced to further features resulting from primary qualities (such as e.g. colour). Of course, in classical physics further 'primary' properties must be attributed to such bodies, such as mass and charge, thus the picture is by no means perfect. But, it is assumed that such picture, modulo augmentations, is the fundamental conceptualisation of the physical world. Given that quantum theory contradicts such picture in certain aspects, it is taken not to provide a definite description of the micro-physical reality as this reality is also expected to conform to the general feature of the sketched conceptual scheme. It is thus taken that there must be some obstruction to acquiring the complete knowledge of the detailed nature of the physical objects at the micro-physical level, with the quantum theory providing the codification of the best of such knowledge that can be acquired.

The conclusion thus seems to be (though this will be further investigated in the final chapter) that we have to make the best of this limited knowledge, try to explain why we can't have it, but that we must not abandon this deep-seated expectation of what material reality must structurally be like. A parallel reasoning that runs alongside this is that there is not to be a hierarchy of physical theories associated with different 'levels of zooming in on reality'. That is, we should not have one theory describing the objects at one 'level of zoom' and another kicking in once we coarse grain the inspection. In fact, more than two such layers may be envisioned, and maybe even several entirely separate theories for different aspects of the reality at the same level of zoom, and things soon start running out of proportion. Given that the 'zooming' view is discarded it is taken that the theories that do not conform to the preferred conceptual structure (the one constituent of the preferred 'level of zoom') must be ‘epistemic’ and not ‘ontic’. Early precursor (though not altogether a prophet) of this view can be found in the philosophy behind Heisenberg’s derivation of matrix version of the quantum formalism (Lochak, 2007).

The theories presented in this chapter to a large degree share the convictions the above sketch starts with but take different conclusions. Generally, they agree with the denial of hierarchy of theories, i.e. do not accept the ‘level of zoom’ view, and aim to reduce all the phenomena to those of micro-physics as a realistic ontological foundation behind all others. One can note a certain agreement with the linearity of spatial zooming; the smaller things are expected to make up the bigger things, not the other way round. They take a somewhat diverse view as to the nature of objects at the ‘zoom-level’ of interest, as will be outlined in the below (section 3.1.2.). But even those give precedence to primary qualities (with some additions) over and above elevating traditionally secondary qualities (or inventing new ones) to a higher status. In that they seem to share the starting point with the theories of the previous chapter, but reach a different conclusion.

They say that we must do what we can with the primary qualities at this level, and treat the results as discoveries about the fundamental nature of matter, rather than project our expectations onto this level, and in resulting experimental disappointment give up on the project of delineation of quantum ontology altogether. We must, as de Broglie tried, explain the correlations and phenomena by reduction to deterministic objects and their standard and special properties (Lochak, 2007).

These theories, thus, reject more strongly than the ones from the previous chapter, the possibility of contending with ‘unsharp reality’ of objects at the level of micro-physics ( (Busch, Classical versus quantum ontology, 2002); for exposition of the alternative cf. (Busch, Grabowski, & Lahti, 1995)). Whereas the theories of the previous chapter could find a route to be reconciled with the ‘unsharp realities’ (though they did not set out to do so at the outset) through accepting ‘unsharp realities’ to be the ontological foundation behind their epistemic interpretations, the theories of this chapter stand firmly against ‘unsharp realities’ by delineating what some of the ‘sharp realities’ alternatives may be like. (There are, of course, other such alternative options that will not be considered here at all.)

### *Historical development of Bohmian Mechanics*

According to Lochak’s (2007) exposition, the historical development of quantum theory followed these two lines of reasoning from the outset (with a brief interlude of expecting them to be united through the Schrödinger wavefunction). The Heisenberg, Bohr, Pauli et al. camp advocated abandonment of ontological speculation about reality at the microscopic level, whilst Einstein, Planck, Schrödinger, de Broglie et al. aimed to supersede the theory as it was given at the time with a thorough ontological account. Historically, the Copenhagen camp won for some time, most of all, according to Lochak, due to easier formation of a unified camp ('there is nothing more to explain'). The anti-Copenhagen camp had trouble offering an alternative account as the difficulties in reducing the observed phenomena to the behaviour of simple ontological primitives were quite substantial and could not, at the time, be borne out in formalism. Thus, even de Broglie, the originator of the view that the particle like behaviour can be reduced to singularities of spatially extended waves, gave up for a while (Lochak, 2007, p. 78). Eventually, David Bohm resurrected (Bohm, A suggested interpretation of the quantum theory in terms of hidden variables, I and II, 1952) some of de Broglie's notions in his pilot-wave theory, which in the end gave rise to contemporary versions known as Bohmian Mechanics.

Even though there are different variants of Bohmian approach today, some of which we shall consider in more detail below, they all share a general conceptual dualism of particles and waves in existence. The particles (or 'the particle' in some cases) build up the macroscopic objects and behave in many ways as we expect from macroscopic objects themselves, i.e. they are finitely extended objects in space and time. Yet they are further guided in their behaviour by the wavefunction, a special and novel kind of entity that is not spatially localised and that provides 'the information' for the particles' nonlocal interactions. We thus have at the micro-level (this is now true only of one strand of Bohmian mechanics, the one that posits the existence of particles in ordinary spacetime, not in high dimensional phase space) objects similar to the objects familiar from everyday life and classical physics (i.e. characterised by primary qualities). Unlike in classical physics, alongside those objects there is/are also a novel and special kind of object: the wavefunction(s).

It is obvious that the status and the role of the wavefunction will prove to be the most contentious issue for our purposes. Again, versions of Bohm-style quantum theory differ on this issue and we shall focus on only one of them. The one to be discarded outright is the notion of the wavefunction as a physical potential field spread out in physical space or the configuration space. On such account the potential literally forces the particles along their trajectories. Though this would, at face value, be an appealing view from the perspective of search for the explanatory ontology that respects the traditional view of primary qualities, it faces technical and conceptual difficulties especially from the perspective of explanation. Namely, it presents the quantum theory as classical mechanics with a special metaphysical addition. This addition is responsible for all the non-classical phenomena but is itself highly obscure. It cannot be manipulated or investigated directly, but only through its influence on the particles. It is extremely nonlocal, but inert to any direct intervention (so can't be used for superluminal signalling). Philosophically, it can be seen as an *ad hoc* metaphysical addition with no other role but to carry the blame for all non-classical ('troublesome') phenomena encountered.

The other extreme is to make the particles equally unreal as the wavefunction, i.e. to claim that fundamentally reality corresponds to a highly abstract formal presentation of the observed phenomena in a high-dimensional configuration space. In that case there is a physically real universal and unique wavefunction for the entire universe and a single 'point-particle' in configuration space that is the summary of formal encoding of the position coordinates of all the supposedly observed particles in the three-dimensional physical space. The three-dimensional space and the multiple particles are not fundamental and must be reduced to the 'universal wavefunction + the marvellous particle' construction. This is highly speculative in terms of metaphysics and it is difficult to see how an isomorphism between the observed phenomena and their 'true' constructive explanation can be satisfactorily established (Monton, 2006). In terms of explanation it is a very expensive construct that generalises from 'the nonlocality of troublesome phenomena is an illusion' to 'the whole known world is an illusion'.[74]

Thus the approach to be elaborated in the rest of the chapter takes the middle ground. It claims that the micro-physical reality is irreducibly non-classical and that we should give up on trying to force it into a classical mould (particles moved exclusively by the force field). It acknowledges the need for a universal wavefunction (as there is no fundamental divide in the formalism between the wavefunctions associated with individual systems), but treads carefully in characterising its ontological features. It claims the material world is made out of particles, classically familiar objects embedded in space-time, but not that all of the properties we tend to ascribe to them are 'really true of them'. Notoriously, it acknowledges some Bohrian-like limits of knowledge through claiming that the world is fundamentally deterministic, but the details of this are forever obscured from us so that the best we can have is the stochasticity inherent in the quantum formalism. In that, it has to acknowledge the real influences of the wavefunction, but its unreality in the 'quantum potential' sense. Finally, it is openly nonlocal, allowing the wavefunction to coordinate behaviour (more precisely, motion) of the particles in synchrony that disregards the spatial separation.

[74] Monton (2006) specifies two main problems with this extreme view. The first is that such a view goes against the pragmatic rule that we should not accept theories which radically revise people's everyday understanding of the world when there are empirically equivalent theories on offer in which such revision is not as radical. In our case, the search for a deeper explanation, this pragmatic rule seems quite natural. Monton's other objection is that it is hard to see how our mental states, representationally supervening on some physical structure (i.e. relying on some isomorphism between the representational content and the physical structure of the world), would have the content that they do. That is, it would be hard to explain why we conceptualise the world in terms of three-dimensional objects evolving through time, given that the true reality consists of single high dimensional point-particle.

As we shall see in the final chapter there are contact points between this approach and the principle approaches of the previous chapter. Strangely, the wavefunction encodes important information about the world without corresponding to anything 'tangible' in that world. The most notable characteristic for our purposes of this approach is that it takes ontology as the starting point. It takes as given that the macroscopically observable world is made of something sufficiently similar at the microscopic level, namely particles, and then tries to reconcile this view with the observed 'troublesome' phenomena. The said particles are not classical, but they are endowed precisely with the primary qualities that have since the early modern era been so firmly established in our conceptual framework. This way a picture of the more complex phenomena is built out of the relatively simple formal scheme, just as Einstein (1954) required.[75] The key problem is that these particles do not enter into causal interactions in the way we classically expect them to, thus stretching to the limit the applicability of the preferred causal-mechanical model of explanation.

## Particle mechanics and the law-providing wavefunction

### *Introductory remarks*

Bohmian mechanics stipulates at the outset that the macroscopic objects familiar from classical physics are constructed out of particles. This is expected to hold as at least partially true, even if some more fundamental theory of fields or strings or some such eventually supersedes Bohmian mechanics. The particles will then be an intermediate stage, but conceptually clearly delineated and essentially populating the three-dimensional space. The macroscopic objects are reduced to particles, which themselves have to be further reduced to the more fundamental objects. But for the time being there are particles with definite positions and trajectories. These parameters are definite even when the formally assigned wavefunction is not an eigenstate of the position operator (Maudlin, 2002, p. 117). In general Bohmian mechanics takes a rather dim view of the naïve realistic interpretations of operators as formal representations of properties of real systems. The wavefunction (ascribed to the system, not the universal one mentioned above) itself evolves deterministically in accordance with the Schrödinger equation with no collapse occurring in the process of observation or measurement. The particles are guided by the wavefunction, but are not identical with it, thus there are no macroscopic superpositions (such as supposedly befall the Schrödinger's cat) even when the wavefunction represents a superposition of possible macroscopic states.

[75] There are claims that Einstein even expected this very route to be taken for quantum mechanics, i.e. that he expected something along the lines of Bohmian mechanics to play the role that statistical mechanics (as opposed to that other theory of gases: thermodynamics) does in the classical framework. This would make the Bohmian mechanics the constructive extension of the principle-style standard quantum formalism (Goldstein, 2006).

Alongside the particles as the constructive building blocks of matter, for any given system under consideration there is also the wavefunction. Its ontological status is more problematic, but let us not get into that yet. Formally the wavefunction provides a link between the Schrödinger equation as the fundamental formal encoding of the evolution of the system, and the derivative Bohm equation that specifies the temporal evolution of the positions of the particles. The Bohm equation is not formally sufficiently fundamental to simply incorporate the necessary elements of the Schrödinger equation and disregard any future talk of the wavefunction. Thus the Schrödinger equation remains the key element of the formalism, shared with other versions of quantum formalism, whilst the Bohm equation is a further step specific to the Bohmian Mechanics (as illustrated below).

$i\hbar(\partial\psi/\partial t) = H\psi$ (Schrödinger equation) ; $\psi$: the wavefunction

$d\mathbf{Q}_k/dt = (\hbar/m_k)\ \mathrm{Im}\ [\psi^*\partial_k\psi/\ \psi^*\psi]\ (\mathbf{Q}_1,\ldots,\mathbf{Q}_N)$ (Bohm equation); $\mathbf{Q}_\mathbf{k}$: position function for the $k^{th}$ particle

Of course, a question related to the formalism immediately arises: how come we still have to deal with probabilities in quantum formalism if this whole evolution is deterministic? Why can't we just investigate (as in observe, even if indirectly) how the particles behave and describe that through the formalism?

A simple answer to this question is that we don't know the exact initial positions of all individual particles, so cannot track their evolution formally and deterministically. We have to rely on ignorance probabilities, rational guesstimates of the possible overall configurations of particle positions. A more complex task is to explain why this is so, and for the moment we shall have to leave the precise exposition aside (cf. section on Quantum Equilibrium hypothesis). More importantly for us, Bohmian mechanics also precludes future determination of the particle positions to a degree of precision that removes this statistical guesstimate (Maudlin, 2002, p. 119). Thus it cannot empirically supersede the other interpretations of the bare quantum formalism in this respect. In this particular respect the constructive explanation along the Bohmian lines does not empirically offer more than the competing principle explanations. If this limitation to increase in precision of knowledge acquisition can be explained as a fundamental feature of nature, this can be a pardonable sin.

But there is another feature of Bohmian mechanics of crucial importance to us. It is manifestly nonlocal (Goldstein, 2006). The behaviour of the particles, i.e. their velocity (intensity and direction of motion), as codified by the Bohm equation, will typically depend upon the positions of other, possibly very distant, particles in situations (which are not at all rare) in which the wavefunction formally assigned to the system is entangled (i.e. is not a simple product of the single particle wavefunctions).[76] The wavefunction, whatever it is, is to be blamed for possible violations of separability, as we can have situations in which against our will (and even possibly against our knowledge, given the irreducible stochasticity) the distant objects affect the objects we are trying to investigate. The phenomena we are trying to explain can then not be simply reduced to the mechanical interactions of the constituent and nearby particles.

There is a partial escape from this dire situation, but only partial. Namely, in the multidimensional configuration space, in the arena for the abstract formal representation of the situation investigated, the 'troublesome' phenomena are not nonlocal, the trajectory of the abstract representation of the particles in configuration space is affected only by the value of the wavefunction around that point ( (Maudlin, 2002, p. 119). But unless we are to be pushed to the extreme view of reducing everything in the universe to the single multidimensional wavefunction and particle, we have to have a way of knowing when we have included enough information in our codification of the situation so that potential influences from higher dimension configuration spaces can be ignored. Moreover, though this helps with separability violation (by allowing us to sufficiently isolate our systems under observation from the rest of the universe) it does not remove the violation of locality in the three-dimensional space as observed in the EPR situations.

[76] In fact, in the extreme it can depend on the positions of all the particles in the universe, and we are back to the 'universal wavefunction + the marvellous particle' picture. But there are formal mechanisms of effectively decoupling the relevant systems from the rest of the universe so that we are not always forced to this picture.

It thus remains a task to specify in greater detail how the middle ground between the introductions of the unwanted 'quantum potential' situated in ordinary space and the all pervading wavefunction with a single multidimensional particle is to be constructed. Furthermore, this path has to offer viable models of explanation of the troublesome phenomena that violate locality and separability.

### *Methodology and metaphysics resting on explanatory constructs*

It is worth repeating once again the central methodological and metaphysical tenets of the Bohmian Mechanics constructive approach, those held by all versions. Methodology and metaphysics of this approach are straightforwardly linked, in that the proponents of Bohmian mechanics claim that one of the staring points for any theory must be to say what it is about. In this respect, the Bohmian approach starts with the metaphysical claim: quantum theory (or in this case its Bohmian alternative) must be about particles that build up the macroscopic objects. The secondary question is to determine what governs the particle behaviour, i.e. how their spatial positions evolve with time. It is at this step that the troubles begin, as the status of the wavefunction must then be elucidated.

Most of the criticism of the Bohmian approach is directed against the 'physical quantum potential pushes the well-defined particles about' view. As we shall not be focusing on that view, we can skim that issue here. What we have to assume (as there is no room to enter into the related debate here) is that the view that we shall focus on can overcome the problems generally levied against the Bohmian approach. Thus we shall assume that Bohmian Mechanics is indeed empirically equivalent to the bare quantum formalism. This is to simply disregard the criticism summarised in e.g. (Streater, 2007), most of which is directed more specifically against the 'quantum potential' view. The most potent criticism included in the given summary, that along the lines of Aharonov & Vaidman (1996), is primarily effective against the 'quantum potential' view. Modulo the discussion on the quantum equilibrium, below, we shall assume that the empirical equivalence between the bare formalism and its modification along the Bohmian lines stands, and that, therefore, Bohmian mechanics is a justified contender in providing explanation of the phenomena we are concerned with here.

Be that as it may, the 'quantum potential' view still gives us the most direct visualisation of the processes behind the 'troublesome' phenomena. In its absence we have to skim the technically demanding issue of the introduction of the quantum equilibrium and a philosophically more complex interpretation of the wavefunction as the fundamental dynamical law, which methodologically brings us closer to the principle approach. We shall elaborate on that further in the following section, but this early warning suffices to point towards the complexity of the problems addressed by our two approaches. When even the candidly constructive approach, the one that places the constructive methodology at the heart of its research programme, is forced to retort to principle-style steps, the initial unease (summarised in Chapter 1) about the general principle approach (of Chapter 2) is reduced.

Moreover, even the 'quantum potential' view, that is easy on visualisation, is forced to introduce some ontological oddities (beyond the unobservable potential) in dealing with the phenomenon of teleportation (cf. Chapter 1). In the explicit analysis of (Maroney & Hiley, 1999); and the subsequent criticism in (Timpson, 2006), strange information ontology is pasted on to the potential view. Namely, in the Bohmian case it is clear that no teleportation of the particle itself takes place, but that in fact some properties of a distant particle get (informationally-for-humans) assigned to the proximal one. In the Bohmian ontology the particles are the foundational existents and their trajectories through space are, at least in principle, traceable (they do not instantaneously jump from place to place).

What is supposed to happen is that some ways the quantum potential affects the particles get transferred through the classical communication channel (the telephone line) between distant and proximal locations (i.e. locations of experimenters Alice and Bob). When this 'information' is subsequently lodged into the quantum potential (through the operations Bob performs on his particle conditional on the message he receives from Alice) it enables the particle to behave in subsequent measurements as the distant particle would have (or at least it enables the experimenter to expect it to behave in that way, by relying on the formalism). The difficulty lies in explaining just what gets transferred between the separated locations, and how. In attempting to explain what goes on Maroney & Hiley (1999) edge ever closer to the law-like view of the quantum potential that will be developed in greater detail below. They take the potential to be holding 'information', alongside standard mechanical effect on the particles, but information in a special sense. The sense of "action of forming or bringing order into something" (Maroney & Hiley, 1999, p. 1408). This information is moved nonlocally through the potential, and is somehow available to the particle, but not to the experimenters.

That is, the experimenters can only work with what the formalism gives them, i.e. the probabilistic predictions of some future behaviour (position change) of the particles. In other words they deal with the 2bits of information exchanged classically, whilst the much larger quantity of information required to deterministically guide the particle is stored in the potential, and available to the particle only.[77] But, due to some other technical difficulties with the 'quantum potential' view, the authors are forced to introduce a further distinction into the 'information' inherent in the quantum potential, namely they distinguish between the active, passive and inactive forms of that information. These forms can be changed by action of the particles or their interaction with the measuring apparatus, and the picture becomes even more complicated.

Because of the non-classical nature of the potential itself we do not get a clear picture of what exactly is transferred and how, in the teleportation protocol. We are told by Maroney & Hiley (1999) that active information is moved through the potential, and coupled with further action of the experimenter Bob based on the message he receives from Alice, this information serves to make the particle at his possession behave just as desired. But how this 'active information transfer' process proceeds is left as a mystery.

> What we see clearly emerging here is that it is active information that has been transferred from particle 1 [Alice's particle, where teleportation originates] to particle 3 [Bob's particle, destination of teleportation] and that this transfer has been mediated by the nonlocal quantum potential. (Maroney & Hiley, 1999, p. 1413)

Timpson (2006, p. 609) objects to this understanding of information, as instead of making matters clearer (by supposedly defining a 'physical' rather than 'information-theoretic' sense of 'information') it requires ontology of 'action' such that it can be moved about as an object. That is, if active information is some property of the 'quantum potential' such that it performs an action on the proximal particle at the end of the protocol, as it did on the distant particle at the beginning, then the transferral of 'active information' in the protocol requires action to be moved about in space. For our purposes there is no need to claim, along with Timpson (2006,

[77] Available as guidance in future evolution of the trajectory, no one is attributing consciousness to the particles here.

p. 610) that this cannot be done, but suffices to say that this is not as straightforward as might initially have been expected of the constructive approach.

However, if the potential is not regarded as a physical field, then such difficulties need not arise. A more straightforward explanation of the teleportation process might involve the outright abandonment of any physical exchange in the protocol. The particle is not 'teleported' (in the sense of *trans*ported) nor are its properties transferred from one particle to another, as there were no properties (other than position; cf. objections to naïve realism about operators in (Goldstein, 2006)) to transport in the first place. What happens is that the wavefunction exhibits the nonlocal characteristics and *based* on the distant operations guides the local particle towards novel unexpected experimental outcomes. Yet an important question remains: how do the situations in which the protocol is enacted and those in which it is not differ; i.e. how is the proximal outcome 'based' on the distant one and not just contingently conveniently correlated ? Namely, how are the characteristics of the wavefunction *based on the distant operations*?[78] At this stage we have to postpone addressing this question (until Chapter 4), but I hope sufficient introduction is provided to take a closer look at the explanatory potential of the 'wavefunction as the universal law' view.

### *Other problems and objections to Bohmian ontology*

A powerful objection to the above solution-sketch turns the situation on its head. What if what is unreal, or less real, is not the wavefunction, but the particles? For however the particle ontology may seem appealing in terms of explanation of what is 'going on' in the 'troublesome phenomena', the whole picture rests on somewhat shaky legs empirically. In general we are barred from ever knowing the exact particle positions for any large enough collection of particles, and must work from some assumptions about the general characteristics of the entire collections of particles that we can never verify directly.

[78] Of course, one possible and rather simple (but for many non-physical reasons abhorrent) solution is that the wavefunction simply behaves universally as a prerecording of events, guiding all the particles through definite trajectories with no regard for their spatial location (in fact, in the 'marvellous particle + goo' view this is to be expected) or *inter*action. The particles simply dance according to the tune set from the beginning of time, and teleportation protocols are not enacted by the experimenters, but were simple coincidences of particle behaviour set out from the beginning to look like experimental outcomes. Though some of the major problematic consequences of such a solution (such as the question of free will) are outside the scope of this thesis, it does not score well as an *explanation* of what happens in the 'troublesome phenomena', as the latter presuppose a voluntary action on behalf of the experimenters, and this solution is simply a denial of these phenomena (as 'troublesome') altogether. It also disregards Bohmian Mechanics' respect for causal non-locality in the 'troublesome' phenomena.

That is, given that we don't know the exact values of all the parameters in the universal wavefunction, we have to work under the assumption that we are able to formulate effective wavefunctions, which help us describe the situation at hand whilst ignoring any effects the rest of the universe has on it. But to be able to form such effective wavefunctions in the first place, we must assume that (i) the universal wavefunction can be satisfactorily mathematically split into the 'relevant' and 'irrelevant' parts, and (ii) the actual particles of interest (those of the object system and those of the 'relevant' parts of the environment – even if distant) are guided by the 'relevant' parts of the wavefunction (for more precise technical exposition, cf. (Maudlin, 1995, pp. 480-482). So in describing the individual phenomena formally we are relying on the calculational, but really nonexistent, *effective* wavefunction and some assumptions about the particles that that can only be tested by the very occurrence of the phenomena themselves. In itself this is not a sin in terms of explanation, as laid out in Lipton's ( (2004, p. 3); Lipton further refers to (Hempel, 1965, pp. 370-374)) exposition of self-evidencing features of explanation. These account for situations in which what is explained provides an essential part of our reason for believing that the explanation itself is correct. They also are a part of Lipton's preference for both unification and causation types of explanation, over less popular reason and familiarity[79] types.

However, it seems that in trying to explain what goes on in the troublesome phenomena the wavefunction does most of the work, whilst the particles are there just because of their good relationship with the visualizable reality demand: they simply do a good job of playing the building blocks of material reality. In their survey of hypothetical and real neutron-interferometry experiments Brown, Dewdney and Horton (1995) show how many of the traditionally intrinsic properties of the neutron-particles, such as mass, spin and charge must be carried, in part, by the wavefunction-field rather than a particle with definite position. They are thus not purely intrinsic to the particles. Furthermore, it appears that in some situations such particles can even fool the specific detectors as to their position, again suggesting in reconstruction of the definite-path-for-the particles situation that even features of the phenomena related to the particle should more properly be attributed to the spread-out field and not the precisely localised position of the particle. This difficulty is more immediate for the view, not pursued here, that the wavefunctions correspond to real fields in space-time, as then we might be more tempted to pursue the general reduction of the re-identifiable objects of the common sense conceptual scheme to them,[80] than in the case where the wavefunction is taken to be more immaterial. From the perspective of competing interpretations of the quantum formalism, interpretations that we cannot go into here, this is simply not a good enough reason to admit them into the explanatory framework.[81]

[79] Neither of the latter two will be considered in greater depth in this thesis due to their theoretical weakness relative to unification and causation types.

Brown and Wallace (2005) stress other important features of the wavefunction that argue in favour of making it more than a mere law for the motion of particles. They see the wavefunction as a dynamical, as having degrees of freedom independent of the particles, and as being structurally very rich[82] (Brown & Wallace, 2005, p. 531). In other words, it may not be so straightforward to simply eliminate the wavefunction from the theory altogether, and formally recover it as "an effective, phenomenological object" (Brown & Wallace, 2005, p. 532). We shall devote the second part of this Chapter to grappling more closely with these issues, but it suffices to say at this stage that following this route Bohmian Mechanics is losing ever more of its explanatory head-start (gained initially by notionally subscribing to hardcore realism and the causal-mechanical type of explanation) over the principle approaches of Chapter 2.

In dealing with the *effective* wavefunction in the 'troublesome' phenomena we seem to be engaged in no more than knowledge updating (even when formally describing the situation, as the effective wavefunction has no direct real counterpart with particles being an indirect support). There are axiomatic conditions that have to be met for the formalism to be applicable to the phenomena in the first place, and (as we shall see in the section on quantum equilibrium) we have to postulate some general principles about the nature of reality and limits of knowledge acquisition for the whole approach to even get off the ground (Reutsche, 2002). From such perspective, treating the wavefunction as the only real and existing thing, out of which everything else arises, including the experimenters' consciousnesses, may not seem so strange.

[80] We might interpret Holland's (1993) warnings that without assigning energy, angular momentum etc. to the particles themselves serious problems arise in the classical limit, as arguing in this direction.

[81] It is simply too time consuming for us to go into a detailed elaboration of a further interpretation, the so-called Everett interpretation in this case. With its heavy ontological reliance on the wavefunction it complicates matters for the simple constructive-principle dichotomy, whilst at the same time introducing technical problems of its own. This is not a value judgement of its worth compared to the two case-study interpretational instances chosen, but a mere expression of limitations of this text. Nonetheless, the contemporary versions of the Everett-style quantum theory that take the single universal wavefunction to be the fundamental existing thing out of which the appearance of everything else arises, is a good starting point from which to address the wavefunction ontological denigration one senses in Bohmian Mechanics. In that we have to bear in mind that we have, above, been moving ever closer to the wavefunction-as-the-universal-law view of Goldstein and colleagues (Dürr, Goldstein, & Zanghi, 1996), and away from the wavefunction-as-the-potential-field-in-three-dimensional-space (e.g. (Holland, 1993); (Bohm & Hiley, 1993); (Maroney & Hiley, 1999)). In their criticism of the above view Brown and Wallace (2005) stress that it is at present a research programme and not a complete solution. From the perspective of the comparison to the Everett-style solutions this indeed is a valid point, but as the alternative approach we are considering here (cf. Chapter 2) is itself only a research programme, we needn't take that as a weakness.

[82] In fact, relatively richer than the mathematical field structures that can normally be 'argued-out' of physics by being shown to be functions on configurations space that are ontologically reduced to features of a more fundamental ontological elements (for example, point particles).

The greatest worry for the Bohmian Mechanics approach, from the perspective of constructing the simple transcendental argument (as in section 1. 4 above) is that what was taken to be fundamental material ontology almost entirely fails to feature in the causal explanatory account of the phenomena, except as a decoration added in by hand. As Brown, Elby and Weingard (1996) argue, there are situations where most interactions can be reduced to the quantum potential field, so as to lose even a mechanical account of how the corresponding field gets to distinguish the supposedly re-identifiable particles. That is, in some situations it is impossible to see how the interaction of the field and the particles takes place at all. As the particles were initially expected to perform the role of the re-identifiable objects in space and time, out of which the observable features of the phenomena are constructed, the tenability of the whole approach becomes questionable if the formal accounts of the phenomena need no reference (even in explanatory reconstructions, not just experimental predictions) of the particles' causal role. It appears they only stand in the place of 'space-fillers' for the geometrical construction of the macroscopically observable objects.

The reduction of properties to the wavefunction raises worries from a heuristic perspective as we have seen an increasing number of particles' intrinsic properties slipping away to the other entity of this dualist-ontology account (cf. (Brown, Elby, & Weingard, 1996) for this terminology), fearing for what eventually remains. But if the leakage of properties can be stopped so that the bare bones of the structural geometric isomorphism can be preserved, our initial aim for the transcendental argument will still be satisfied. From the perspective of everyday utilisation of the formalism, this may seem like decorative addition, large part of what we really need to predict and manipulate outcomes is in the wavefunction, so why as for more. From the perspective of construction of realist explanatory accounts that little more may still be needed, but even so must be seen to be very, very bare. By reducing the wavefunction (in either effective or the universal form) to a non-material law, a proscription for how the particles ought to behave without itself occupying space nor bearing properties, we appear to artificially recover some of the ontological explanatory justification for the particles' introduction.

### *The quantum equilibrium and the absolute uncertainty*

Adherents of the Bohmian mechanics view of the quantum theory repeatedly stress their commitment to constructive theories by putting the notions of ontology first in the construction and manipulation of theories. This, of course, suits the expectations of the research instrument, which aims to compare the principle and constructive approaches to the 'troublesome' phenomena. But it also asks of the Bohmian mechanics to account for the empirical equivalence with the competing extensions of the bare quantum formalism. Taking particles as primary existents should provide for alternative explanations of the 'troublesome' phenomena, but if those explanations are not to be of the classical kind (which they can't be, for the phenomena are indeed troublesome; cf. Sections 1. 5. 2 and 3. 1. 2 above ) we need to know the specifications of the difference between the classical particles and the quantum particles in Bohmian mechanics.

For the purpose of explanation-provision as set out in this thesis, we will first and foremost want to know what exactly happens to the particles in the troublesome phenomena. Yet, given empirical equivalence, Bohmian mechanics cannot help us with that, for even here there is a (neo-Bohrian) element of limits of knowability of the exact states of nature (Dürr, Goldstein, & Zanghi, 1996)). The exact exposition of notions summarised here is lengthy (Dürr, Goldstein, & Zanghi, 1992) and complex, and the brief sketch should suffice for the subsequent discussions concerning explanation and the comparison with the principle approaches in Chapter 4. The Bohmian approach we are to follow in the remainder of this chapter thus gives up on treating the system wavefunction as a real spatially extended object that (almost classically) guides the particles in their trajectories, with a caveat that it has no strong enough answer to the challenge that the wavefunction, the unreal calculational device is much more rich and descriptively complete than the bare particles ontology.[83]

So wherefrom the wavefunction for a system then? Let us not forget that whatever the ultimate speculation about the nature of reality turns out to be, if it is to be supported by science (even if it is not arrived at directly through empirical observation, but is a product of some delayed philosophical speculation) it has to agree with and explain the predictions made by the currently successful theory. That is, even extensions of the bare quantum formalism, such as Bohmian Mechanics is, must be able to tell us why the formalism works in the cases in which it does. If the whole universe is entangled in the single wavefunction how come we can get the non-local correlations and have them confirmed by experiment from a simple system wavefunction that does not explicitly include the formal description of particles in the Andromeda constellation? What is more, Bohmian mechanics itself is unable to go beyond the predictions for empirically observable phenomena made by the bare quantum formalism.

[83] Another reason to expect such abandonment is the expectation, also mentioned above, that through interactions the wavefunctions of larger chunks of matter, and eventually the whole universe should get entangled into an overall universal wavefunction. The nonlocality of wavefunctions also precludes the long-term isolation of the system wavefunctions.

How do restrictions of knowability come about from a theory that is decidedly deterministic, a theory in which the particles move along the trajectories that are set in stone for all eternity? Can we not, given enough effort, come to know at least some of these fixed trajectories, hopefully those of most significance for our everyday life? Bohmian mechanics is forced to explain wherefrom comes this limit on what can be learnt about the universe in a theory so precise, with precise motion of spatially located, almost tangible, particles. This is, so it seems, where the constructive approach leans close to the principle one, though the exact comparison will be left for the next chapter.

The proponents of the limited constructive approach have to postulate a universal constraining principle based on the simple phenomenological observation that the bare quantum formalism is the most we can know about the physical systems we are dealing with. They claim (cf. (Dürr, Goldstein, & Zanghi, 1992)) that we must assume that the set of initial distributions of the universe capable of yielding wavefunctions for individual systems that we in fact observe, out of the total set of all possible initial distributions, is itself very large. That is, given some universal wavefunction for the whole universe (the great universal 'goo') there are relatively many particular distributions of particle positions that accord with the given wavefunction and the ascription of individual system wavefunctions to many systems today ( (Dürr, Goldstein, & Zanghi, 1992), cf. also (Goldstein & Struyve, 2007)). So we can't know which particular particle distribution the universe started off in and has been evolving deterministically from ever since.

Further technical argument is then developed to show that we cannot in fact know more than the individual systems' wavefunctions tell us (and, remember those are stochastic and give rise to entanglement etc.) even for isolated systems today. The technical argument states that the individual system wavefunction can be thought of as a hypothetical part of the universal wavefunction. Hypothetical in that it does not represent a real object, but is an encoding of the best of human knowledge about what is going on. In order to work with systems at hand we can rely on such hypothetical separation of the world into the 'system at hand' and the ‘rest of the universe’ because, mathematically, such separation is complete modulo the wavefunction. Our best knowledge of the dynamical evolution of the configurations of interest will be given only by the individual system wavefunction. For that wavefunction provides the mathematical link between the abstract representations of the configurations of the system of interest and the rest of the universe. The configuration of the system of interest and the configuration of the environment are conditionally independent given the wavefunction $\psi$ of the system of interest (Dürr, Goldstein, & Zanghi, 1992)).

To summarise the above in even simpler terms. We can’t know the exact distribution of all the particles in the universe at some given point in time. Take that point to be the starting point. In order to derive the formalism that we use for the limited sets of particles today, we must assume that at the starting point the exact distribution of those particles was typical, i.e. that overall it was standard (that the particles, or the particle in many dimensional configuration space, were ‘randomly’ strewn about). That assumption then provides us with the mathematical tools to derive the individual system ‘hypothetical’ wavefunctions from the universal wavefunction (whose exact state is also unknown to us). Given that assumption we can relate our ordinary quantum formalism for the systems we play with in the lab and the universal wavefunction for the entire universe. The latter is unknown to us, but as long as the universe started in some typical state, we don’t even need to know it for we will be able to extract our ‘mini-wavefunctions’ for the systems of interest from the general outlines (the ‘typical’ features) of the ‘supreme global goo’.

But we, nonetheless, have to bear in mind the extreme nonlocality of the Bohmian Mechanics in which all the systems of interest are inextricably causally (though not mechanically) linked to all other particles in the universe through the universal wavefunction. So even when we extract our hypothetical wavefunction for the systems of interest, the predictions it is able to give us about the behaviour of the system are at best probabilistic, we only get a probability distribution of possible outcomes. So the world is made of particles that move in a unique manner through spacetime, but the exact manner of their movement (even their exact positions) is forever unknown to us. Unknown, because it is linked to all the other matter in the universe in a highly non-classical way (i.e. not linked through the causal mechanical interactions we are familiar with).

### *Troublesome phenomena as products of a global law*

So we end up with a strange world. The individual wavefunctions are not fields that spread through classical spacetime. The only such field is the universal wavefunction. But that wavefunction does not exist in the three-dimensional space with us and our everyday objects, it exists in the multidimensional configuration space and guides the universal particle, a queen bee of all the fundamental ontological entities in the universe. Somehow, through the universal particle (which itself is not real or fundamental, on this interpretation) the wavefunction affects all the universe's three-dimensional particles in a highly choreographed, nonlocal and deterministic 'dance'. Our troublesome phenomena are a product of the behaviour of the three-dimensional ontological primitives mysteriously instructed by something that is itself immaterial (by not being a part of the three-dimensional space of matter). Moreover, we have to postulate a constraining principle, namely the hypothesis of the (initial) quantum equilibrium, in order to reproduce the phenomenology of the bare quantum formalism and its role in the lives of the physicists. This principle is not unreasonable, it is more than a bare statement of the existing constraint, it aims to provide a rational justification for the constraint on the acquisition of knowledge about the precise current state of the particles in the reality, but it is nonetheless postulated as an *a priori* hypothesis in order to save appearances.

Another look at the situation described above opens up a perspective that we are dealing, unexpectedly for the constructive approach, with the in-principle limits of knowability situation again (just as in the neo-Bohrian approaches of the previous chapter). Yet following the historical precursor it is worth asking how it differs from the explanation of entropy through statistical mechanics rather than thermodynamics. Are the limits of knowability themselves explained or just posited as a theorem of the conceptual framework? Whatever the answer might be, the crucial difference for us is that statistical mechanics fitted well with the conceptual framework based on spatially extended particles in interaction, whilst Bohmian Mechanics has an extreme demand for separability violation. The saving grace lies in exploring the potentials for a conceptual framework without separability as its implicit foundational principle.

An account from Albert (1992 ) (and further modifications in (Maudlin, 2008) can help illustrate this problem visually. We consider a device that provides some 'measurement' of the particle, depending on the trajectory the particle takes through the device. More precisely, the particle can exit the 'measuring' device through an exit facing the ceiling, and in that case we say the particle has the value of some property 'up'; whilst if the particle exits the device through the floor-facing exit, we say its value of the given property is 'down'.[84] It can be shown that when a single particle is fed through this device (and because the trajectories in configuration space cannot cross) the initial details of the location of the particle affect its behaviour following the measurement. That is, the particle that entered the device via a route that is closer to the ceiling, ended up exiting it through the ceiling hole, and the one that enters closer to the floor ends up veering towards the floor-facing exit. In such case, even if it cannot be demonstrated experimentally because of the knowledge-gathering limitations of the quantum equilibrium, we would have a perfectly visualizable account of the physical phenomena formalised by quantum theory.

But in the entanglement situations things become more complicated. It turns out that if we set up two devices to 'measure' two such particles that are initially taken to be in the entangled anti-correlated state, then the outcomes of measurements of individual particles must be 'opposite' (i.e. one exits through the ceiling-facing exit and the other through the floor-facing exit) regardless of what their initial positions were. Or rather, the outcome of the second measurement to be performed must be opposite of that of the first, regardless of what the particle's position in the entrance hole of its measuring device was. Whether an individual particle exhibits a particular result cannot be determined simply by the initial location (in the range of positions allowed by the entrance hole), for if it could then there would be a completely local account of the EPR-style correlations, and those correlations would not be exhibited the way they are.

Suppose the two particles were both in the initial location ranges that would, had they not been in the entangled state, see them exit the device through the ceiling facing exit, and the devices are sufficiently separated in space. Albert (1992 ) shows that if the left-hand particle is 'measured' first, it will be found to exit the device through the ceiling-facing exit and the right-hand particle will be found to exit its device through the floor-facing exit. If the situation is reversed, and the right-hand particle is measured first then it will be found to exit through the ceiling-facing exit, and the left-hand particle through the floor-facing exit of its device.

[84] The original formulations of the example contain properties (such as spin) that make the situation more physical, but as the point is to demonstrate the importance of the location of the particle, and its dependence on the locations of other particles, I prefer not to introduce unnecessary technicalities here.

> "And this holds *no matter how far apart* the two [particles] are, and it holds without the action of any intermediary particles or fields traveling between the two sides of the experiment. So the behavior of the right-hand [particle] at some moment depends on what has happened (arbitrarily far away) to the left-hand [particle]. The dynamical non-locality of Bohm's theory is thereby manifest." (Maudlin, 2008, p. 162)

All of this is achieved, according to Maudlin (2008) by the way the wavefunction choreographs particle behaviour. We are not given a *mechanism* of how the effects of what happens to one particle can influence what happens to another (there are no particles or fields travelling between them), but rest on the simple summation that what one particle exhibits (in a 'measuring' interaction, for example) may depend on how a very distant particle is treated. In fact, when the real universal wavefunction is taken into account, instead of the conditional wavefunction for individual systems, then it may depend on how (indefinitely) many distant particles are treated.

What kind of explanation does this leave us with? Correlations in measurement outcomes on our separated particles cannot be attributed to a common cause (cf. the (Maudlin, 2007b) exposition of separability violation in the section 3. 2 below), but neither can they be attributed to the transmission of physical signals between the particles. They are taken to simply come about without a causal mechanism, but through a previously (prior to measurement) unknown nomic prescription (encoded in the universal law) that they should. A serious question arises: how does this explain them?

In Bohmian Mechanics the troublesome non-local phenomena are arrived at bluntly, even if the full justification of their 'explanation' is rather convoluted. The events whose 'outcomes' are mysteriously correlated over large distances in fact share a connection mediated by the wavefunction, rather than by some spatially localised physical conditions or particles that propagate faster than light (Maudlin, 2008). The distant correlations are thus explained by the dynamics that governs the total configuration of particle positions (the global wavefunction) by a global law rather than an effect of a local law on each individual particle. Of course, one problem with this notion is that it seems to require the absolute simultaneity, something that seems to be prohibited by Special and General Theories of Relativity.

The universal wavefunction, as some form of a universal dynamical (and causal) law must rely on some notion of absolute flow of time, in order to determine the instants of absolute simultaneity, and thus determine which particles enter their 'measuring' devices universe-wide. Though the latter is an interesting technical issue, it need not concern us here, as we are not arguing for locality from the technical position of conflict with Relativity, but from a more general position of universal application of the principle of separability. As far as our explanatory viewpoint is concerned, and especially its concurrence with the classical everyday conceptual framework, we can easily, taken at face value, accommodate the absolute simultaneity and the notion of flow of time.

What pushes us to consider the wavefunction in general, and most importantly the universal wavefunction, as the physical law rather than an element of the physical reality described by laws of nature? Two primary reasons are (1) the fact that although the wavefunction affects the behaviour of the particles, there is no formal account of particles affecting the wavefunction; and (2) for a system of many particles the formal expression of the wavefunction is not a field in physical space (such as, for example, electromagnetic field is) but on an abstract high-dimensional configuration space. However, formally, this is not a unique case as there are objects of formalism in classical physics which exhibit similar prediction-usefulness combined with abstraction, but are not considered to correspond to anything special in the real world. They are recognised as shortcuts in human descriptions of the real world, without accompanying ontological projections. They, though, are not dynamical.

The universal wavefunction does not itself change with time (though precise formulations are as yet insufficiently explored, according to (Goldstein, Bohmian Mechanics and Quantum Information, 2007)), but is just a nomological encoding of the changes of particles (which is what we observe in the end). In that case the derivative or system-wavefunction is just a phenomenological law, an instrumental ease of calculation device (similar ontologically to the suggestion of the principle approach from the previous chapter), whilst the (unknown) universal wavefunction is in fact the fundamental dynamical law governing the behaviour of all the particles in the universe.

What is more interesting for us, and is related to the discussion about simultaneity in Bohmian mechanics in literature (cf. (Albert, 1992 ); (Maudlin, 2008)), is the empirical inaccessibility of the planes of absolute simultaneity, i.e. the precise global dynamics of the particles as governed by the wavefunction. That is, we cannot, for reasons sketched above, experimentally determine the exact position of the particles in Bohmian Mechanics (Maudlin, 2008). With each attempt to physically determine the exact positions of the particles we disturb the wavefunction and thus those very positions of the particles.[85] History of science notes a strong dislike (in part due to the tradition of logical empiricism, but only in part) for the explanations based on the postulation of empirically indeterminable facts. In this case we have the perfect determinism of the distantly correlated events, precise constructions of macroscopic objects out of unique positions of constituent particles, and the unique temporal evolution of the wavefunction governing them; but all of them forever inaccessible to empirical observation. The best we can contend with are the probabilistic 'guesstimates' as encoded in the standard quantum formalism.

Maudlin (2008) offers two lines of reasoning in defence of such obscurantism. Firstly, the posited structure is not physically superfluous, it does some explanatory work and is not merely introduced into the theory as a decoration. He sees the Newtonian Absolute Space as such a decoration, because not only can it not be physically detected (or rather the position within it cannot be physically ascertained) but also its postulation has no physical consequence (unlike that of the Neo-Newtonian, or Galilean, space). But all the ontological elements of the Bohmian scheme are not physically superfluous; they cannot be subtracted from the conceptual framework without physical consequence.

The second line of reasoning aims to show that there is no extra work being done to cover-up the existence of the empirically inaccessible structure. That is, we do not add new elements in the Bohmian theory that do no other work but obscure some elements of its ontology from empirical observation.[86] Maudlin claims that the inaccessibility of some of the ontological elements is an involuntary (maybe even unwanted) consequence of the simplest dynamical solutions to the explanatory problems we are facing. Take the world made of particles, take the information about its behaviour as given by the formalism (or its important element, the wavefunction) and you get a mechanism (supposedly) explaining how the troublesome phenomena arise, but not permitting the direct accessibility of the ontological elements the said explanation depends on. This does not have to be direct observation, it can be some form of empirical testing designed to tease out the precise characteristics of the ontological element. Though such explanatory mechanism may not be popular, Maudlin claims it is not devastating for the viability of the Bohmian conceptual framework, as the empirical inaccessibility of the said ontology is a consequence of the physics, but not of the physics designed or motivated to produce that inaccessibly. Though the latter line of reasoning seems shakier than the former, we can temporarily accept both as defence of the viability of the Bohmian framework. They will both prove to be a relative weakness of that framework, though, if the explanatory models it is compared against can do without them.

[85] Notice the functional similarity here, that is at the moment only to be noted and taken at face value, between the Bohmian inaccessibly of the exact particle positions, Bohrian and neo-Bohrian sensitivity of the real systems to observer-intervention and the structural 'black boxes' of the CBH programme. Yet, we can still expect to get different explanations of the troublesome phenomena from these varying theoretical programmes based on the role the empirical inaccessibility plays within each account.

[86] One might argue, though, that postulating the quantum equilibrium hypothesis achieves exactly this, but as has been argued above there are additional reasons for its introduction.

Perhaps unnecessarily repeating what has been stated above, it turns out on this account that explaining the 'troublesome' phenomena rests on an instance of knowledge-updating so it would accord with the pre-determined universal 'choreography'. On such view even the separability loss is not so crucial as the supposed fundamental principle behind our ordinary conceptual framework was just an illusion arising from ignorance, anyway. So on extreme reading even influences can be sent to achieve change from proximal to distant measurement (and vice versa), only we are in-principle not in a position to learn about them directly. The following half of this chapter examines once more, from various philosophical angles, how we could learn to live being forced with such a predicament.

## 3. 2. Laws as part of fundamental explanatory ontology

### Metaphysical problems related to realism about unobservable microstructural concepts

Leaving aside, for the moment, quibbles over the role of fundamental laws it is worth briefly considering the philosophical problem of resting scientific explanations on properties of entities that are not directly observable, such as the Bohmian particles are. Unlike the principle approaches of the previous chapter, the constructive approach of Bohmian mechanics must be able to account for the classical criticism, most notable from the extreme empiricist camp, against relying on speculations about the properties of unobservable entities in producing scientific explanations.

Most philosophers of physics would agree that novel predictions in science provide a good reason to believe the theoretical constituents they rely on. A simplified version of the ‘miracle’ argument (Putnam, 1979) could say that it would be a miracle for a novel prediction to come out right and the theoretical construction preceding it to be wrong (or at least wrong in more than inessential details). Were we to be given such a prediction, which resulted in confirmation, and for which one of our approaches above had a ready made explanatory account whilst the other struggled to even incorporate it into its world-view, the case would be next to decided. This is in fact what the traditional accounts in philosophy of science expect from the competing theories. However, to the best of my knowledge our ‘troublesome’ phenomena still lack such predictions, not to mention their confirmations.[87] Thus novel predictions remain excluded as the deciding factor between the explanatory successes of our two approaches. As has been sketched in the introduction most of our preferred, successful explanations rely on the mechanisms that contain unobservable entities. We might even say that the preferred explanations in contemporary science consist of reductions to unobservable entities. Add to those the causation and laws, the possibility of manipulation as exemplified in the counterfactual situations and the predictive success evidenced in contemporary science, and we see that alternative models should only be sought for *in situations which make the causal mechanism utterly unpalatable*. But there are more general arguments that work against resting explanatory success on unobservables whose essential function is to produce the observed phenomena.

For example, van Fraassen (1980) argues against using the explanatory virtues (in this case the adherence to a widely popular model of explanation) as reasons for believing a given theory. He distinguishes between epistemic and pragmatic virtues of theories. A pragmatic virtue might be the property of a given theoretical framework to make quick and easy calculations. Though this would count in favour of using the framework when dealing with the phenomena covered by the theory, it cannot be the reason for considering the given framework to be closer to truth than its alternatives.

Thus, van Fraassen claims that the only epistemic virtues of theories are the empirical virtues of getting more observable consequences right and fewer of them wrong.[88] On such a view

[87] This is not strictly true. The constructive approach is able to offer some predictions, but they deal with much deeper theoretical generalizations than the narrow group of phenomena under consideration (separability violations). The principle approach also has some theoretical expectations closer to the ‘zone of observation’ but currently out of reach of verification. The caveat is that the theoretical constructions characterised by the two approaches prevent the empirical verification of the said predictions, i.e. they have built-in a priori constraints to empirical verifications of the differentiating predictions, cf. (Albert, 1992 , pp. 183-189).

[88] We shall not enter the discussion of the pros and cons of empiricism, nor whether van Fraassen’s view sketched here is an instance of excessively strong empiricism. Let us just assume that empirical adequacy is

explanatory virtues of our two approaches constitute a pragmatic virtue, and as such cannot decide between them, given that both are empirically adequate. But this objection is ignorant of the special situation we are in given the ever increasing closeness among our two opposing approaches, as well as the admitted reaching for purely philosophical tools outside the realm of good empirical scientific practice. We are, thus, choosing to simply overstep van Fraassen's concerns in order to move out of the stalemate of empirical adequacy of both approaches, even if van Fraassen were to declare the choice a purely aesthetic one. An upshot of further and more detailed criticism of using explanatory power as virtue in defending mechanistic accounts (cf. (Boyd, 2002)), or similar realist accounts, is the requirement that the findings of the relevant background sciences should be relevantly approximately accurate. Now, such justifications can indeed be provided, but not *a priori* as the reliance on the explanatory virtue requires. Furthermore, it will not be an easy task to provide them in the light of alterations to the conceptual scheme required by the failure of separability. This seems to be another respect in which the mechanistic approach of this chapter is comparable to the speculative elements of the previous one.

### *Properties*

Yet in the troublesome phenomena it is not just correlations within the entangled states that are the problem, but the actual swapping of properties in the phenomena such as teleportation. Is it at all possible to explain such processes by the 'action' of a law? What form would a law of regulated property swap need to adopt and how would it fit with the wider worldview?

> The fundamental ontological tradeoff reflects the perennial tension between explanatory power and epistemic risk, between a rich, lavish ontology that promises to explain a great deal and a more modest ontology that promises epistemological security. The more machinery we postulate, the more we might hope to explain – but the harder it is to believe in the existence of all the machinery. (Swoyer, 2000) guaranteed for both of our approaches and that it constitutes the bedrock below which neither of them can go.

We are here interested in determining properties (in the traditional sense) that withstand the loss of separability and are affected by the laws imposed on the world as fundamental. A most pressing issue is to survey the choice of properties traditionally (in classical physics) assumed fundamental and investigate any possible changes to them by acceptance of the Bohmian strategy for addressing the troublesome phenomena.

Swoyer (2000) claims that properties are usually introduced into ontology in order to help "*explain* or *account* for phenomena of philosophical interest." They are usually taken to be the ground of phenomena in a manner that some phenomenon holds in virtue of some properties. We can then play the game of investigating the conditions imposed on the property by its explanatory role: investigate what properties would have to be like in order to play the roles of explaining the phenomena. It is, of course, possible to claim that this is a vacuous game, that properties have no explanatory power and are a mere fig leaf to cover our lack of understanding of what a given phenomenon is. In our case, if that were really so, a strife to settle for any sort of explanation of the troublesome phenomena should help us decide how much, if at all, we really need to rely on the properties proposed.

The philosophical topics surrounding properties are wide in scope and not always empirically grounded. What we are concerned with specifically is how the classical properties characteristic of physical objects fare in the physical interactions of a novel kind, such as those presented as the troublesome phenomena. These phenomena themselves do not directly dispute that the objects participating in them have a position is space or even some discrete extension, but it is the nature of changes of those properties that is troublesome from the viewpoint of classical physics, the physics that introduced those properties to explain its own phenomena of interest. We are thus more concerned with how a property of a particle can change or be undetermined (metaphysically, as well as epistemically) without *observable* physical interaction with other particles or fields, rather than whether a given property, a universal, can simultaneously exist in more than one place (which is a popular problem related to properties). Moreover, what kind of a world is inhabited by objects that seem to interchange properties as if they were coats without us being able to keep a precise record of the details of those exchanges and what governs them.[89] Finally, what can we hold fixed in such a world, so as to recognise a change as a different state against a background of things that do not change?

[89] Of course, even coats are exchanged along some traceable spacetime route, so we are really stuck for analogies here.

Such firm foundation was provided by primary properties, made most famous by Locke, though the notion goes back to the Greek atomists. The primary properties are the directly recognisable objective features of the world, the most straightforward exemplifications of the isomorphism between the structure of reality and the formal elements of our physical theories (and accompanying conceptual frameworks) describing that reality. They are often so fundamental that they are used to explain why things have the other properties that that they do. Traditionally these have most famously included shape, size (features of extension) and some variants of mass and charge/force field. The secondary properties, on the other hand, are the reflections of powers inherent in objects to produce certain responses in humans, but are primarily rooted in primary properties (cf. Descartes' rules for understanding complex phenomena in terms of primitives, in *Rules for the Direction of the Mind, Rule XII*, (Descartes, 1931)).

What kind of properties can we expect the constructive approach of this chapter to rely on? With the particles moving in the physical space, extension remains a fundamental property. But what other properties are there and what role does extension play if it is not sufficiently/significantly contributing to the changes in those other properties given that they can change instantaneously at a distance? If we cannot account for the systematic attainment and alteration of properties by the token objects (in this case particles) what sort of realism can we cling to with regard to the physical reality described by quantum theory? Abandoning realism, even if of some weaker kind, would put this approach to the troublesome phenomena in the same metaphysical boat (if not even worse) as the principle approaches of the previous chapter, at least when it comes to accounting for the real changes in the world that stand behind the observation of the troublesome phenomena.

In Devitt's (2006) account (relying on his detailed exposition in (Devitt, 1997)) realism assents to existence of the most common-sense and scientific physical types as objective and independent of the mental. Opposed to it is the view that the independent reality cannot be epistemically accessed and correctly conceptually described and that the phenomena we are concerned with are partly constructed by our forced imposition of concepts onto the manifold of the bare perception. In Devitt's view one of the appeals of realism, other than its intuitive acceptance outside the intellectual circles (2006, p. 6) is the rational rejection of the alternatives as unsatisfactory. From our perspective, the downside of the alternatives to realism is Devitt's claim that they are *explanatorily useless*. That is, accepting that there might be some kind of world out there that is behind all the phenomena we are struggling with, but that that world cannot be known for what it is, leaves us with very little else to turn to in order to provide the sought for explanations. If constructive approaches of this chapter were characterised as such their stake in provision of explanation sought would instantaneously vanish.

Although particles would still have an extension, it seems it would no longer be fundamental in their interactions, as they can alter their properties (to the extreme point represented in teleportation phenomena) without respect for the constraints spatial extension imposes on physical interaction. In the causal accounts relying on the primacy of the physical state, such as Harré (1996) advocates in the next section, laws of temporal evolution take a back seat. If the laws of temporal evolution are to be made primary, with the property possession and exchange depending entirely on them, are we threatened with a slide into anti-realism with respect to how we come to explain the phenomena we observe?

Suppose concepts such as ‘redness’ (a colour concept) do not have a direct ‘isomorphic’ connection to some real feature of the world, but designate “a disposition to produce a certain sort of response in normal humans under normal conditions” (Devitt, 2006, p. 11). Globalising the argument to all properties runs as follows: all property concepts, not just those of secondary properties, are response-dependent. So all that we take to be properties in the real world are in fact response-dependent dispositions to produce certain sort of responses in normal humans under normal conditions. The abhorrence of world-making along these lines lies, according to Devitt, in the need to posit something even more wildly speculative than the realist metaphysics: the noumenal things-in-themselves which are really behind the observable phenomena painted by concepts. This way, of course, antirealists (world-makers) expect to limit irrational speculation, put some material constraints on what we can actually do with words and concepts. But, as Devitt points out, these noumenal things only present an illusion of a constraint, we can *ex hypothesi* know nothing of the ‘mechanisms’ by which they exercise their constraint, we can not explain or predict any of the constraints, nor can ever hope to be able to do so. For if latter was the case we would be overstepping the bounds of world-making and venturing into speculative scientific metaphysics proper. Furthermore, causality is part of the existing scheme of concepts and cannot be extended to the link between the noumenal world and the conceptual scheme, thus we don’t even have a notion through which to connect the world-in-itself and our supposedly constrained view of it.

To slide into world-making is to subscribe to the view that our concepts make up the world, that the structure of the world is dependent on our classificatory activity and not vice versa. Then the conceptual requirements of the theories would not be a discovery of what there is in the world (however primitive and coarse a discovery), but an act of literally recreating the world. The only way to avoid this is to say that somehow, by blunt fact, the facts of the world impose constraints on how our concepts are created and interlinked but that nothing more can in-principle be said about that. But even here, if there had been no conceptualizers, us, there just would have been no macroscopic material objects as well as the microscopic speculative metaphysics. That must be a claim that the anti-realist of the world-making camp must be committed to (Devitt, 2006, p. 8). Whatever the outcome of this debate in philosophy and linguistics in general, it is clear that world-making explanations will not fare well in our case, even if confronted on the other side by mere agnosticism about entities and structures at the micro-level.

As the laws governing the temporal evolution are primary existents, whilst the observable properties of objects are just a temporary product of their operation, those properties are not in the traditional sense fundamental and in the objects. We are just disposed to observe them as such under the influence of the laws, whilst they are not really there in the world, in the same way that redness as experienced by us is not in-the-world. In this way, through the abandonment of separability (to be argued for below), among other things, the undulating-high-dimensional-goo view and the laws-are-primary view of quantum theory become two sides of the same coin.

The objects that we either directly observe, or geometrically project as isomorphic sustainers of what we observe (cf. (Sellars, 1963) and section 4. 1), the atomistic construction of the observable world out of the unobservable fundamental particles, are to a great extent our projections arising from the dispositions of the true elements of reality behind them (the goo or the bare particles choreographed by the fundamental laws of temporal evolutions) to produce a certain response in us under normal conditions. This is because even the primary properties, such as extension, or consequently physical separation, are not metaphysically fundamental, existing in their own right and in direct isomorphism to how we conceptualise them.

Of course, it is the issue of what is fundamental that is important here. It is not that we have not had response-dependent concepts in the explanatory schemes before, such as the colour concepts introduced above might be. But that did not pose any problems because there have always been some fundamental property concepts that these could be drawn from, such as surface texture or microphysical interaction with surface particles. The latter were fundamental concepts rooted in the conceptual and theoretical isomorphism with material reality based on the primacy of the concept of extension.[90] With denying such fundamental status to all properties or to all formally describable physical interactions (which depended on the participating objects having certain fundamental properties that further predisposed them for a certain interaction) we lose the firm footing for a realist explanation of the observed phenomena.

### *Causation*

If the troublesome phenomena are better explained by the reliance on operation of laws and causal processes than on unification of the phenomena into a wider world-picture held together by shared properties across phenomena and degrees of magnitude of extension, then we can regard momentary properties employed in *descriptions* of the phenomena as fleeting shadows of a classically constructed language. The onus is then to show how the non-unifying account will work and also that it will not *rely* on the classically introduced properties, at least not in crucial instances. Of course, properties have been known to feature in causal accounts, especially where reduction of causation to causal powers is introduced (cf. (Harre, 1996), but in that case we must lay them on a more firm account that does not slide into anti-realism proper ('worldmaking' in (Devitt, 2006)).

Though causation per se is outside the scope of this thesis, it is inextricably linked with issues of ontology in physics, both in historical development of the mechanical explanations (cf. section 1.2.), and in conceptual characteristics of contemporary desirable forms of explanations (cf. introductory sections of Chapter 4). Essentially there are two conflicting overarching accounts of causation: one affirming the key role it plays in explanatory conceptualisation of reality and the other denying a fundamental role for the notion of causation in explanation (reducing it to either a psychological error or a merely heuristically useful device). In a nutshell, we expect the real ontology to account for causal processes, but can never strictly *observe* anything other than a concurrent regularity of physical phenomena with no inherent mark of what makes them causal. We get a pro- and anti- realism views of causation.

[90] And isomorphism should be taken seriously here; it designates an easy or natural correspondence with a basic everyday conceptual scheme. Something a multidimensional configuration space may not be able to achieve.

The segment of history of the role of causation in science relevant here starts in the seventeenth century with the abandonment of the Aristotelian final causes and focus on the search for efficient causes through mechanical philosophy. Yet the problems immediately arose with those segments of reality that could not be modelled by strict mechanical contact, such as gravity. And it is here that divine will was often invoked in place of the mechanical essence: "Gravity must be caused by an agent acting constantly according to certain laws" (Newton, 1957).[91] Even though such mystical explanations had to be grudgingly accepted, the majority of science was expected to move towards the ideal of a Laplacian demon where causation and the deterministic nature of laws of physics provided an exact mechanical description/explanation of all physical processes, past, present and future. This attempt for overall regularity in science, and physics in particular lost its general appeal with the advent of non-deterministic theories early in the twentieth century.[92]

But, of course, philosophy was not to be swayed by such scientific strivings and we thus have the great Humean analysis of the fictional nature of causation as a mere psychological erroneous projection of human expectations onto the physical processes. Hume argues from the epistemic atomism of individual (perceptible) states of physical reality to the conclusion that all that can be learnt from observation alone is the concurrence of certain types of states (e.g. stone hitting, glass breaking), but not their necessary or physical connection through some causal process. Causes and effects are held to be absolutely independent in reality, and consequently must be held to be so in concept too (Harre, 1996, p. 311). We thus have the Humean Mosaic.

The reliance of the Humean doctrine on 'epistemic atomism' is of importance for our purposes. Atoms of experience are held to be the experienced sensory elements that are both the ultimate components of perception as well as of the world-as-experienced (the phenomenal world). As opposed to the explicit denial of the Humean Mosaic later in this chapter, if the approaches of the previous chapter are shown to rely on just such experiential atoms in their analysis of the troublesome phenomena, then they can straightforwardly be expected to be deniers of the reality of causation, and with it, of the causal explanations. Narrowing this down solely to their chosen field, namely information manipulation, could help them escape explanatory zeal for causal accounts. The theories presented in the two central chapters could be viewed as either affirming or denying what Norton (Norton, 2007) terms the 'causal fundamentalism': nature is governed by cause and effect (in the case of this chapter: a primitive law) and the burden of individual theories is to find the particular expressions of the general notion in the realm of their specialised subject matter.

[91] Of course, this historical problem can be resolved in the same sweep as our current ones by taking the fundamental laws of temporal evolutions as ontologically primitive and simply attributing all gravitational interaction to obedience of gravitational laws regardless of the media and details of interaction.

[92] This is not just a case of quantum theory, the popular champion of indeterminism, but even in Relativity theory processes on a larger scale cannot be uniquely specified through a single causal process. In fact, Harré (1996, pp. 304-307) charts attempts parallel with development of mechanical philosophy that either argue for or against dynamism, a view that sensibly inaccessible forces (similar to contemporary field theory) produce and sustain causal processes across the universe, and even replace material ontology of spatially extended entities (in a Cartesian sense).

Harré, on the other hand, proposes arguments from psychology and epistemology to show that the sensory invariant in the experience of phenomena (though, material not informational phenomena) is not the wholeness of the phenomenon itself, but the general things, the fundamental units of realist ontology out of which the experience is constructed. This opens a way for him to argue for the reality of causal processes at least in some cases, and our two approaches can then be compared on the types of causal processes they propose. Harré (1996, p. 321) sees the fundamental ontology of science as constructed out of entities whose essential natures are given by their causal powers, and whose causal agency[93] is well delineated. This chides well with general preference for causal explanations. In this case the constructive ontology and the account of causal processes are inextricably linked, as is suggested by the nature of causal explanation and the adherence to the primitive role of laws of temporal evolution. This does require, however, that the constructive approach recognises the equally fundamental role played by the general things, the token-type objects, alongside the law of temporal evolution. If such material ontological components of the 'troublesome' phenomena can be found, or if the 'information ontology' (of Chapter 2) can successfully replace them, then we might be able to compare the outlines of causal pictures suggested by either approach.

Functionally, then, we want causation to be understood through knowing what would happen to the object central to our phenomenon to be explained had the relevant surrounding circumstances been different. The ontology that takes laws as primary and yet epistemically inaccessible might struggle to give a workable solution along these lines, as it would lack the details of the *manipulable* mechanism that leads to observable changes in objects. Yet quantum theory overcomes this obstacle surprisingly well, with the effective wavefunctions (though, necessarily statistical in nature), thus providing a workable manipulable mechanism. The problem is that our notion of object and its durability through changes is slightly altered, and we shall have to bear that in mind in the next chapter (sections 4.4. and 4.5.).

[93] This is a more detailed aspect of Harré's exposition that need not concern us here. He distinguishes agents and patients in causal interactions, where patients must be stimulated to produce actions, whilst agents need only be released to act (Harre, 1996, p. 322).

### *Chanciness*

Our everyday (non-technical) conceptual framework views causation as part of regulated (i.e. not completely chaotic) behaviour of ordinary objects. This behaviour is determined by a small set of conditions: the object's dispositions to respond to various sorts of interference and the listing of the sorts of interference the objects of that kind in fact encounters. Speaking plainly, we know for most everyday objects when they will break and when they will fly, and what local situations will arouse either behaviour. But in non-local physical theories, no small set of conditions suffices to determine an ordinary object's behaviour. We need to specify the entire state of the world at one time in order to determine the state of even a small region at some future time (Elga, 2007). This is the nightmare of non-local theories, such as quantum theories examined here are.

To wake up out of the nightmare, we might suggest, as Norton (2007) does, that the specification of the entire state of the universe is a task only extreme pedants ever need fulfilled. He sees the everyday view of causality as the approximately correct model in certain limited domains, and that the physicists need not ever venture into something more except when extreme precision of prediction or description is required. This would mean that the pedantically formalised laws are such that in certain domains they can make the everyday view true, this must be their formal feature. But Elga (2007) argues that some laws (and these are our 'troublesome' ones), whilst formally perfectly respectable, are nonetheless such that they do not make the everyday view even approximately true in any domains at all.

In the case of some of such laws Elga claims we are warranted by appropriate statistical assumptions to treat the law-following behaviour as intrinsically chancy. This allows us to treat the objects susceptible to such laws as mostly isolated and feign to hand over their supersensitive causal connectedness to the intrinsic indeterminacy of the physical reality. Of course, as sketched above, this not a very prudent position to take, and neither of the approaches presented in this thesis will ever fully embrace it. That would mean secretly committing to the super-connected ontology, one that utterly removes separability as real constraining principle, and yet develop a formal theoretical approach that chooses never to tackle this characteristic of reality formally. We would then have to claim instead that the indeterminacy predicted is not epistemic (i.e. is not an ignorance interpretation) but is a formal expression of the deep chanciness of nature. It is hard to imagine an extreme abstraction where the two extremes are one and the same thing: where chanciness just is the supersensitive connection of everything in the world.

Furthermore, quantum theory makes probabilistic predictions about the chances of different phenomenal experiences, which are extremely well confirmed on the aggregate level. But at the individual level, when each phenomenon or macroscopic event is viewed in isolation, we must also make sense of the probabilities that the formalism assigns to each particular event (or, to be precise, to a set of possible events). Formally, our explanations must also account for the completeness of Schrödinger dynamics, as well as the quantum state. That is, we have to say whether the probabilities ascribed by the formalism are real chances in nature, or are a product of our ignorance about the true laws of nature (and then also explain how that ignorance comes about and what is to be done about removing it).

And this seems to be the point where the roles of our two approaches are reversed. It is Bohmian Mechanics that takes the probabilities as merely epistemic, and states that the laws of nature are actually deterministic. It is only a calculational opportunism that leads to describing the processes as chancy (Maudlin, 2002, p. 146). The principle approaches that take the non-realist route to the quantum state ascription, are now pressed against a wall of taking an even stronger non-realist stance (claiming that the formal evolution of states is also a result of human ignorance) or accepting that something in reality, whatever it may be like, justifies the ascription of probabilities for each individual phenomenon. The latter requires taking the stochastic laws seriously at the ontological level, and thus taking the probability ascription equally seriously. This in turn means that the result of admitting a basic indeterminism in reality is the acceptance of probability ascription for particular event as a basic physical fact (Maudlin, 2002, p. 147). One that then must somehow be a part of the explanatory ontology.

### Abandon separability in favour of a radical causal mechanism

The radical proposal of Bohmian Mechanics is worth recapitulating once again through a slightly different formulation of the EPR situation. The quantum formalism differentiates between the m=0 triplet state and the singlet state (just technical terms for formally different states of particle pairs). But the statistics for the outcomes of measurement on the separated components are the same, which in combination with the separability principle suggests that these formally different states are in fact one and the same physical state of the particles (and the wavefunction or some such accompanying item). The problem once again stems from the troublesome correlations we are pressing to explain. Although no local measurements on the individual parts of the composite states (either the triplet or the singlet) can yield differences between the two, a global measurement on the overall pair can. Namely, if we decide to measure a property closely formally related to the property originally used to describe the state of the composite system, the formal differences between the states take on a more important role empirically. What we in fact get is quite different expectation statistics for the two composite states, statistics that is empirically confirmed upon measurement. More precisely, if the original composite state was the singlet state, upon separation the measurement of the related property on each of the particles will yield directly opposite results (say 'up' and 'down') with a 50% chance of either combination (i.e. first 'up' – second 'down', and vice versa). In the triplet composite state though, the results will also have a 50% chance to come out either way, but this time with identical outcomes for distant separated particles, i.e. either both 'up' or both 'down'.

The conclusion is that we cannot identify the singlet and the triplet states. But in that case we cannot have a sensible definition of separability either, for separability requires that either the states be identified or that we can tell what the difference between them is.[94] But neither composite state can be expressed as a combination of individual particle states and the spatiotemporal relations between, for we cannot specify the individual states of the particles with certainty, except as a part of a composite system. And separability required, in summation, that the whole is no more than the sum of the parts (including spatiotemporal relations). Maudlin (2007b, p. 61) concludes that no physical theory that takes the wavefunction seriously (i.e. that considers the formalism to be a complete veristic description of the physical system) can be a separable theory. In the language used in the paragraphs above we may say that considering the formalism (with the Bohmian additions included) to provide a complete description of the composite systems requires that we do not see the systems as separable entities that can be described by the momentary state of the component particles and the spatiotemporal relations between them alone, at any given instant of time. The wavefunction seems to be doing serious work that violates the separability of the states involved in the troublesome phenomena. Of course, and Maudlin refers to a similar proposal in (Loewer, 1996) here, the state may be considered separable in the configuration space rather than the three-dimensional physical space, but that is a further metaphysical step we have chosen not to follow in this thesis (cf. the initial parts of Ch3).

[94] A brief recap why this is so, in the current terminology. As no detectable signals are passed between the states, nor are they formally expected to, we expect the differences to be borne out of the initial formally indistinguishable states. This is because separability permits differences to be observed experimentally only if there is some detectable (or at least predictable) interaction (or signaling) between them. As there is no such distinguishable difference between the initial states, yet the global difference is distinguishable upon locally performed measurements, separability is violated.

The central tenet of the constructive approach states that there is no other way out but to abandon separability.[95] Everything else is deemed instrumentalist (the quantum formalism is incomplete and needs more work), or idealist (the states change under the conscious intervention or do not correspond to real physical changes), or demanding the alteration of logic (quantum logic) to accommodate a metaphysical principle (separability). The Bohmian approach is, of course, not the only viable such constructive solution and not the only one to abandon separability, though the only one with initial interest in mechanical structure as required by our two explanatory models. Given that one of the approaches that does respect separability is presented in the previous chapter, what we would like to know here is what the world without separability is like. But an interesting caveat opens in the preceding paragraphs, and even Maudlin (2007b) points to it: separability has something to do with 'knowability', more so than with necessity. What leads to the conclusion that it must be abandoned, as presented above, is not so much that the physical reality as described by the identification of the singlet and the triplet states would be a priori impossible, but that it would be strangely closed to epistemic access.

That is, something, and we can't say exactly what, would preclude us from ever determining what state the particles in the composite states are really in. We would assume that they are in some definite state, that the state of the composite overall is a combination of their states and the spatiotemporal relations between them, but we could never tell what the initial definite states are.[96] We seem to be forced to choose between two evils: limiting how much we can learn (empirically) about the material world, or abandoning the comfortable epistemological and causal apparatus we relied upon to hitherto successfully gather the knowledge about that same world. If this dogmatic issue can be at all deconstructed and evaluated, that will not be attempted in this chapter. Let us first turn our attention to what else separability abandonment, and with it the supposed impossibility of "the postulation of laws which can be checked empirically in the accepted sense" (Einstein, 1948, p. 322) requires.[97]

[95] Maudlin (2007b, p. 62) warns that abandonment of separability is not the same as the abandonment of locality, for separability can be maintained by non-local theories with superluminal or temporally reversed causal connections.

[96] And thus we are almost pushed into neo-Bohrian conclusions that the meaning of state of the particles can only be given in their relation of the system as a whole, i.e. the system and the measuring apparatus and the measurement required to determine the states afterwards. All of these include the operation of the irrational

What Maudlin (2007b, pp. 61-62) alludes to is that separability is an important ingredient in the Humean Supervenience (Lewis, 1986), and that when forced to abandon separability we might also be forced to abandon the Humean Supervenience. This means abandoning the position that "all there is to the world is a vast mosaic of local matters of fact, just one little thing and then another (Lewis, 1986, p. x)

> (But it is no part of the thesis that these local matters of fact are mental.) […] we have local qualities: perfectly natural intrinsic properties which need nothing bigger than a point at which to be instantiated. For short: we have an arrangement of qualities. And that is all. All else supervenes on that." (Lewis, 1986, p. x)

That makes the physical state of every space/time point independent of the laws that supposedly govern the evolution of phenomena, and thus suggests that laws are unreal, a mere human projection on the sequence of total factual states. All our explanations of the observed phenomena had an implicit reliance on the supervenience, the completeness of the description in the state of material existents. Of course, in explaining a process we had to include some projections of the causal relations between existents, but the description of an outcome was contained in the momentary physical state (primarily). And sufficiently distant states could not be essentially connected. With the supervenience abandoned as well, we have cleared the way for the introduction of laws as primary ontological entities alongside material existents.

---

element and thus prevent us from inferring more than momentary outcomes of measurement and global relative states (cf. Chapter 2 sections on Bohr).

[97] Of course, Maudlin (2007b) leaves some room for the middle ground as well, interpreting Einstein as demanding that theories be built on some minimum set of separable states, but not that all properties that are empirically ascertainable must be separable or depend on separable states. Presumably, Bohmian particles would provide such separable entities, whilst the wavefunction provides the inseparable ingredients.

What would a law as a primary ontological existent be like? Conceptually, this means that "the idea of a law of nature is not logically defined from, and cannot be derived in terms of other notions" (Maudlin, 2007b, p. 15). This is to say that laws are the patterns that reality necessarily exhibits, an essential part of an overall structure (whether we can observe them or not).[98] Thus, what is physically possible is what is constrained within those patterns. But such a status, in Maudlin's analysis still gets us no further to determining which of the regularities (such as the correlations between distant events) that we observe are fundamental laws. We may be, he says, living in an unlucky universe, or part of one, in which random stochastic processes produce perfect correlation between distant measurements without any underlying fundamental law (Maudlin, 2007b, p. 17). This would be a stroke of extremely bad luck, but it is a possibility we shouldn't lose sight of when fitting the explanation of the troublesome phenomena into the overall world-view.

But supposing our luck serves us, we may take the Schrödinger equation as a fundamental law of temporal evolution of the universe, and thus the mysterious 'activities' of the wavefunction are just a consequence of the operation of that fundamental law on the primary existents, the particles. As shorthand, we may then call this fundamental law (mathematically formalised in the Schrödinger equation) the action of the wavefunction, but have no need for the wavefunction as the actual existent that somehow 'pushes the particles about' (akin to a potential field or some such). Of course, logically, conceptually and formally the law and the wavefunction cannot be identified, but we might (in admitted sloppiness) call this underlying law: the wavefunction.

Maudlin (2007b, p. 49) admits that this still does not provide an easy (or straightforward) explanation of all the troublesome phenomena in quantum theory. The entangled states of multiple particles cannot be understood as the sum of local physical states of each particle, with fundamental laws governing only the epistemically accessible interactions between particles. Moreover, as has been indicated previously, the evolution governed by the supposedly fundamental law behind the Schrödinger equation proceeds in Hilbert space, and not the ordinary physical space in which the particles sit. But he is more prone to revise our concepts of counterfactuals, locality and causality based on classical physics, than the empirically confirmed quantum theory. As the concepts of law, possibility, counterfactual, causality and explanation are deeply connected we could infer from quantum theory the direction the revision should take in providing the desired explanation.

[98] Maudlin (2007b) also requires that the passage of time be considered as an ontological primitive, accounting for the basic distinction between the past and the future of an event. There is no space to enter that aspect of the problem here, but it neither detracts from nor adds to the problems of explanation we have considered in this and the preceding chapter.

To begin with, it is intuitively clear that laws (if correctly identified) carry more explanatory power than mere truth-statements (be they accidental generalisations or not). In the first instance it is not difficult to provide explanations of individual instances of a phenomenon by subsumption under a law, but such explanation cannot be achieved by subsumption under an admittedly accidental generalisation. But we might, and often do, seek a further explanation for the law, or at least some further differentiation between a law and an accidental generalisation, other than claim that it just *is* a fundamental law. Note that a request to provide explanation places a serious requirement before a law (and a theory it forms a conceptual and formal part of). An 'anything goes' law would logically satisfy the subsumption of all the observed phenomena, but could hardly be said to explain any of them. Thus, Maudlin (2007b) concludes in criticism of van Fraassen (1989a), that science has to aim at true theories (in his view construed round true fundamental laws) rather than just empirically adequate ones that need not bother with the ontological (and hierarchical) status of their formal statements.

Supposing that we seek theories with greater explanatory power, what should we be looking for? Metaphysically adequate theories, claims Maudlin. Theories whose model constructs stand in one-to-one correspondence with the physically possible states of affairs. And the limitations of this physical possibility will be provided by the laws of nature. And, the stronger the limitations the more explanatory the theory will be, i.e. it will have fewer model constructs that correspond to possible states of affairs and be close to the list of observed/actual states of affairs. But there is a hidden danger here of multiplying restrictions until we get a simple description of the current state of affairs, which would be metaphysically adequate on the above account, but would not really be explanatory ('the world just is as it is in every detail and it is the only way it could have been'). Our troublesome phenomena then need no more explanation than any other phenomenon, or indeed any fact, in the world.

But, as Maudlin correctly points out, this does not describe the scientific practice. Scientists, even quantum physicists, do not work on producing an unchangeable and minutely detailed *description* of the current state of affairs, but a shorthand way of *understanding* what states of affairs are possible and where the current/observed one fits in. Thus, Maudlin claims, the contents of the model constructs are determined by three factors: "the laws, the boundary values, and the results of stochastic processes" (2007, p. 50); where the boundary values presumably allow for some determination of participating objects and states of their properties. The regularities we observe as patterns in model constructs can be entirely explained by subsumption under laws, whilst the regularities stemming from the other two factors may just not have an explanation at all within a given model (and the 'final' ones among them may not have any explanation at all if we admit fundamental chanciness in the physical reality).

Adding laws as ontologically primitive allows us to better select for the theories with greater explanatory power, than mere objects-only-are-primitive theories can allow for. To borrow Lewis' terminology, theories with fewer world-models give better explanations. By specifying laws as ontological existents we narrow down the availability of the world-models compared to the multitude available in the only-objects-are-primitive situation. On the other hand, in the Humean Mosaic, laws cannot be used to explain its particular features because they are nothing more than generic features of the very same mosaic themselves. What they can do is contribute towards a unification type explanation by showing commonalities of structure among various distinct regions of space-time (Loewer, 1996, p. 113). In this way they can provide explanations of some phenomena (isolated segments of the mosaic) through unification with a larger class of the phenomena based on multiple snapshots of the mosaic, but there is certainly no explanation of the entire state of the mosaic at any given time. Primitive as it is, in its entirety the mosaic just is. Through their connection with the mosaic, from this perspective at least, adherence to separability and unification model of explanation go hand in hand.

What those types of explanations cannot do, on the other hand, is provide an account of how some phenomenon was produced for they lack the causal mechanisms between different mosaic snapshots. But with the laws as primitive existents we can connect a structure in one snapshot with causally related structures in further snapshots. In this way we could provide an explanation of the occurrence of some structure in those further snapshots. In our case-study instances, the 'production' of the later-state structures (the narrowing down of the class of possible world-models) is achieved by the introduction of a fundamental law as an in-itself-unexplainable primitive behind the troublesome phenomena. The correlations between the object-existents cannot be further explained than be specifying the law that governs the correlations regardless of how far in physical space the objects are and what further barriers may separate them. This strategy shares some similarities with the principle approaches of the previous chapter in seeking to functionally reify the boundaries on behaviour of objects or updating of knowledge about those objects. Maudlin does not provide a recipe to decide between the two types of explanation available, other than to argue that neither Ockham's razor nor the standard Inference to the Best Explanation can be used as arbiters in this case (Maudlin, 2007b, p. 181).[99]

### 3. 3. Summary of the constructive approaches

The constructive approaches of Chapter 3 fare better according to the Lipton criteria. They bridge the gap between knowing that a phenomenon occurs and understanding the circumstances that lead to its occurrence through relying on the concepts of generative mechanism consisting of the particle-objects and the law governing their behaviour that is capable of inducing changes in the objects non-locally. The problem is that the details of the actions of the law are in-principle inaccessible, so the best we can have is again the guesstimate encoded in the quantum formalism. The details are inaccessible due to a peculiar state the whole universe is in, the quantum equilibrium. So when the phenomena are considered globally a radical cut in the generative story must be accepted so that the effects of the law on the particles is not uncovered through piecing together local states of the particles only, but considering the holistic elements that arise from the glimpses of the global law, as well. This is not damaging for the separable conceptualisation of the world as the holistic elements are relegated wholly to the non-spatial law, and the deterministically incomplete predictions of the local behaviour of objects cannot be improved on due to epistemic limitations of the quantum equilibrium state.

The why regress is successfully blocked by providing a description of what the material world is like, including the acceptance of the universal law that plays a part in its changes. The problem is that we have no genuine explanation of why the quantum equilibrium constrictions hold, except for formal statistical considerations, it must be entered as a postulate that blocks the why-regress bluntly. This exposes the weaknesses of the constructive approach that do not allow it to escape much further than the principle approach gets. Finally the explanations have a self-evidencing characteristic in that the introduction of the universal law was motivated by the problems caused on purely separable view of 'troublesome' phenomena. It is also clearly an ontological explanation, though we are justified in wondering what the ontology of laws, once they are taken as primitive existents and not supervening on the states of the material ontology, is like in greater detail.

[99] In fact, he says that in this case the two amount to the same principle, and again one type of explanation is preferred over the other on individual aesthetic grounds.

The explanation also has characteristics of a deeper explanation, at least notionally if not in practice, as we can construct a story of how we can change the relevant aspects of the phenomena by manipulating the particles, given their subjection to the law (which is unknown, but some aspects of its action can be derived formally, as given in the effective wavefunction). Providing we have an independent account of how the interactions of the particles select which distant particles they create effects on (and we can assume a further technical notion of decoherence provides us with this), we can claim the knowledge of the law we have through the effective wavefunctions allows us to alleviate worries about unexpected effects on the state of the material ontology globally, i.e. that we can hone in on the 'troublesome' effects when they arise in reality.

This allows the transcendental strategy to be given through reliance on the concepts of enduring objects and non-local laws. Yet, this seems to require that in the transcendental strategy we change the starting point from objects being defined in terms of primary qualities alone into objects conceptualised as enduring individuals subject to the universal law. This way we would be 'cutting nature at its joints' not through the selection of structure across space, but in selection of structure across law-permitted changes across space and time. The laws would enter our initial concept of objects essentially. The final problem to address though, remains in justifying the fundamental role of the material ontology at all, given such a structurally essential role for the universal law (or more of them). In terms of quantum formalism, we may ask ourselves why we need to shy away from the (epistemically) inaccessible universal wavefunction, if the essential properties of the objects are going to be dispositional on it. Are we not merely enslaved by the expectations of realist structure imposed by the transcendental strategy and depth-of-explanation as we know them (but which are both somehow anthropocentric)? The real challenge might be to reconstruct the transcendental strategy and deeper explanations in terms of the law alone. The latter though, is beyond the scope of this thesis.

As for the 'troublesome' phenomena specifically, they arise out of the changes that directly observable objects (measurement instruments in this case) undergo. These, in turn, are reducible in their structure to the microscopically fundamental ontology of the particles continuously enduring in space and time. This structural link directly connects the continuous endurance of the macroscopic objects and their observable parts with the extension-based segment of the fundamental ontology. Yet, not all of the properties of the fundamental ontology are in this way reducible to their positions and space-time relations, though some sufficient segment of them is. As for the rest, and those are interesting properties in our 'troublesome' phenomena, they are dispositional on the nomological local proscriptions of the epistemically inaccessible fundamental universal law of temporal evolution. That is to say, some properties (those not reducible to position and spatial relations between particles) do not continuously hold of the particles at every instant.

To us, with our limited epistemic access to the universal law governing the particles, it appears as if they do not have the particular property at the time (making at least some aspects of them seeming dispositional and subject to world-making hypotheses of the antirealists). But on the retrodictive explanatory account all we have to permit to accommodate the 'troublesome' phenomena is that the properties can change as dictated by the law, without the change being induced by a spatially continuous signal as the cause of change. Though the causes can in explanatory accounts be traced back to the activities of the agents and their particular interactions with other particles, they are locally induced in the 'distant' particle by the nomic proscription of the universal law. Upon gathering more information concerning the global aspects of the situation we come to form conclusions about general correlations between the distant and proximal aspects of the phenomena. It is important to stress, though, that the constructive approaches do not permit doubts about the chaotic and haphazard 'jumps' in the intrinsic non-relational properties of the fundamental ontology. But this is where our problems arise: many of the traditionally intrinsic properties of particle ontology turn out to be dispositional in relation to the universal law, and not truly intrinsic to the ontological constituents themselves. As one of the hypothetical cases examined above even the position of the fundamental material existents is dependent on the proscriptions of the wavefunction-law, making them vulnerable to charges of ultimate dispositionalism. Of course, these charges need not be accepted and can be carefully defended against: the position of the particles and their spatial extension (their 'being' in space) is not unreal nor explicitly denied by the theory. It is, in fact merely taken to be less permanent and less informative on its own. To give an account of the world (even its local segment such as the constitution of some directly observable macroscopic instrument) it is not sufficient to specify solely the arrangement of the fundamental material existents and the physically significant relations between them. We have to also specify the instantaneous local proscriptions of the universal law.

The ‘troublesome’ phenomena then consist of special situations in which the non-local action of the universal law becomes acutely visible even from the macroscopic perspective. This is where the law orders the fundamental existents to behave in way unexpected in the macroscopic realm. But, crucially, their identity and potential for independent re-identification are not denied, once the proscriptions of the primitive and universal law are taken into account. Without those proscriptions the situations seemed paradoxical, but the paradoxes arose from our erroneous expectation to reduce all physical accounts to the intrinsic and relational properties of material (extended) ontology only, disregarding the fundamental role laws play in the understanding of the world. The ‘transcendental’ argument can then rest on the irreducible role of the extension in the construction of objects constitutive of the phenomena, provided that fundamental role played by laws is duly appreciated.

Methodologically this is a constructive account, as it shows the constructive mechanism behind the phenomena. But it is a radical constructive account that requires that we revise some deep-seated expectations of physical theories and explanatory generalisations, so as to abandon the fundamental status of the Humean Mosaic, and admit extension to be structurally important though not wholly sufficient for the explanatory connection between the standardly and regularly experienced and physically foundational. This is not an impossible move to make and one that still does not permit the antirealist to claim that simple realist strategies are bogus nor that objects in lawfully constrained interaction cannot be identified in the experience. Only what will identify the objects will no longer be their shape and spatial position, along with some other aspects of geometrical structure, such as texture, but also the relation the objects hold to the fundamental law of temporal evolution. Immediately we must ask though: what use are the objects we cannot directly observe in explaining the phenomena when all their identifying features are dependent on the proscriptions of this fundamental law? May we not explain the phenomena as consequence of the fundamental law at directly observable level, without having to construct the narrative of objects? These are important objections to be addressed in the final chapter. Finally, metaphysically it is clear that constructive approaches of this chapter argue for a dichotomy of the fundamental role of extended material ontology (just as preferred by the ‘transcendental’ argument) and fundamental though non-material laws of temporal evolution. But it is also clear that they place ontology on a high position methodologically, and a particular type of primary-qualities-come-first ontology at that. Here is what Albert says of chances of uniting Bohmian Mechanics with more general field theories:

> "Bohm's theory (as it presently stands) is quite deeply bound up with a very particular sort of ontology; the trouble [is that this sort of theory is not a replacement for the bare formalism in general, like the Everett –style theories, but only for those interpretations of the formalism] which happen to be theories of persistent particles;". (Albert, 1992 , p. 161)

We come to wonder whether this staunch adherence to persistent particle-objects is too high a price to pay in order to save the simple transcendental strategy of the section 1. 4.

# 4. COMMON-SENSE CONCEPTS AND EXPLANATION

## 4. 1 In search of the structure of a deeper explanation

> [...] explanation is not a logical structure, [...] it cannot be characterised in syntactic terms, but it is rather an epistemological structure, and, more specifically, a structure organising conceptual content." (Hansson, 2007, p. 3)

Setting the issues of realism and deeper attitudes to the methodologies in sciences aside (or laying them to rest having discussed them in the previous chapters), in this chapter we turn to precisely the selected problematic concepts introduced in each of the approaches and assess how well they can be organised into the overall conceptual scheme of our language, so as to achieve the goals of explanation as Hansson (2007) lies them down. Hansson shows that some degree of complexity is required in order to make the explanations better, and thus the critics of the scientific endeavour cannot rely solely on the fact that some of the introduced concepts are hard and not straightforward as the 'tables' and 'chairs' seem. The most general structure of the explanation will contain, in the most general Hempelian style a list of properties an 'object' before us has, and the laws connecting those properties to the environment/context. But when choosing the level of depth and the complexity of interconnection of these concepts within an explanation, we must bear in mind that the essential function of explanation (both unification- and causal-style) is to gain understanding by connecting the previously disjointed knowledge of 'facts' into a unified whole of a world-view.

Usually this is achieved by connecting the observation, an experienced phenomenon to be explained (though this need not prejudice the choice of language or be limited solely to supposed bare 'observation statements'), with the highly general law known to be directing the acceptable variations covered by the concepts appearing in the phenomenon. The number of steps required will depend on the previous knowledge and understanding the explainee has, whereas the link between the steps is provided by the conceptual framework inherent in language. As at least one of our approaches to the explanation of the 'troublesome' phenomena contains limits to overall unification of knowledge (at least temporarily), does it follow that it is immediately precluded from providing an explanation on these grounds? The answer is yes only for staunch Bohrians, who insist on not modifying the conceptual scheme in any way, whilst showing it to be insufficient to provide a full unification of the phenomena with the well-understood theoretical terms.

In the cases where we have to introduce new concepts in explanation, as will be further discussed below (Nersessian, 1984) the new concepts have to fit with the exiting conceptual framework so as to help 'cut nature at its joints' (Hansson, 2007, p. 9), i.e. allow a better empirical (and manipulative) access to the phenomena they cover. This is a precursor to a more detailed debate on depth of explanation, but the main idea is that the new conceptual framework, consisting of the insertion of new concepts into the old framework, should provide explanations of the phenomena that allow more variability (even if all of it is not empirically confirmed) as part of the understanding of particular occurrence of the phenomena. In other words, they should allow wider spectrum of counterfactual situations involving the said concepts, but differing in the relevant way from the phenomena actually observed.

Even before we look in more detail into the requirements of depth of explanation (Hitchcock & Woodward, 2003), it is easy to see that the explanations in terms of objects, their properties and causal processes they are subjected to generally fit this requirement well. Explanations of the causal–mechanical type are then just a more extreme example such general scheme, providing detailed specifications of the nature of objects and the relevant interactions they can undergo. And yet, Hansson warns, following exclusively this prescription, and not falling back on the idea of conceptual unification and organisation, would lead to us to extreme and absurd lengths in providing explanations of even the simplest phenomena. Where the operations of nature are complex, and they more often are than aren't, conceptual economy goes a long way in providing explanation and thus allowing meaningful interaction with the world without having to adopt a God's eye view. Variable depth is required of every explanation, and "concepts are more flexible than properties" for this purpose (Hansson, 2007, p. 10).

Thus, Hansson concludes, good explanation is like an exercise of proof in mathematics, an epistemological exercise of linking concepts under objective constraints. Of course, such internally consistent, but somehow irrelevant conceptual networks can be created ad lib (cf. for example the works of Duhem, Quine and Feyerabend) and accepting such a strong linguistic turn will play well into the hands of the critics of the explanatory potential of science. In order to avoid that it is advisable to rely on the conceptual networks that already exist, that form the well connected global system of orientation in the material environment and function well in a variety of contexts. But in this position, which is a kind of unificationism, unification should not be sought for in and of itself, but as a consequence of other goals on the conceptual level. “While the classical unificationist is right in asking for intellectual and epistemological economy, she is wrong if she identifies this with having as few premises or beliefs as possible. Rather, global economy concerning what concepts are needed to make the world intelligible is more basic than either global or local economy of assumptions or premises” (Hansson, 2007, pp. 10-11). From the perspective of provision of explanation this seems to agree with the starting point of the transcendental strategy of section 1. 4, as much as possible rely on the readily available concepts.

In the troublesome cases under consideration here appearance of some properties of objects or general characteristics of situations are seen as at first glance improbable, or are at least unexpected on the straightforward account of the phenomenon. We thus have to do extra work to connect them to what is ‘expected’ in the conceptual scheme that we start with. Weber and Van Bouwel (2007) argue that explanatory depth has intrinsic value in such instances and those explanations that can provide the required depth will be considered better explanations in such cases. Explicitly, the contexts in which explanatory depth is seen as useful are those in which we ask whether the occurrence of some property or event is a predictable consequence of some other, more familiar or more widely expected events. More generally, Weber and Van Bouwel (2007) argue that contexts of asking explanatory questions have to be taken into account in assessing explanatory worth. What is important for us is that on their analysis, given that the troublesome phenomena we are concerned with fall under the right context, explanatory depth (to be explicated in the following section) will be of intrinsic value. Troublesome phenomena are seen as anomalies from the perspective of the plausibility of the transcendental strategy of section 1. 4 and explanatory depth will prove as a useful heuristic in comparing the explanatory potential (i.e. their potential for deeper explanations) of our two approaches.

Sellars (1963) reminds us how we needn’t view the claim that behind the perceptible appearances of objects and phenomena there lie fundamental explanatory physical ontology, as a claim that ‘everyday objects’ don’t exist. He claims that by reducing the perceptible to the physical explanatory ontology we are not challenging the claims about tables and chairs within a framework, but are trying to replace the whole framework with one that can support and explain it, but goes further in providing understanding of the wide range of perceived phenomena (Sellars, 1963, p. 27). This is in line with the strategy traced back to Descartes in section 1. 4, the ontological projection should not only provide an explanation of how the phenomena arise but also how our appearance of them has the peculiar features (including those that lead to *prima faciae* erroneous ontological projections) that it does. This was his famed replacement of the manifest image by the scientific image which both supports and explains our use of the conceptual framework of the manifest image (as it was ideally posited by Sellars).[100] Swoyer elucidates that we are using the analogy of length measurement formalisation, where "an isomorphism of an appropriate sort explains the applicability of mathematics [i.e. mathematical formalism] to reality" (1987, p. 284) to outline the way that conceptual frameworks (though not nearly as formally coherent as Swoyer's formalised measurement theory) when seen as somehow isomorphic or homeomorphic to relevant aspects of the world can provide an explanation of the applicability of thought to reality (which is just what we needed in section 1. 4 of the introductory chapter).

But for the said replacement to go through the manifest image must already possess 'the germ of the solution' of how the two images are linked and can conceptually coexist. It is suggested here (with special reference to Descartes in section 1. 2) that the wanted germ is given by the geometrical regularities based on the foundational role of extension. If our 'scientific' image, i.e. the explanatory frameworks stemming from our two approaches, are forced to somehow deny that foundation, i.e. if the replacement of the images goes so far as to deny the very link of the replacement-route can we still use Sellars' programme? This is a question we come to pose in light of the conclusions of Chapter 3, where the details of the law seem to be more informative than the bare positioning of the particles. Alternatively we may ask whether the notion of laws contains enough conceptual stability to be the sole new provider of the link with the geometric isomorphism of primary qualities taking a back seat. Addressing these problems will have to await some further stage-setting.

[100] Immediately this might invite the question of replacing one paradigm with another (cf. Chapter 1), however the two supposed paradigms here do not compete but rather one encompasses the other. For this to present an effective criticism a further charge of incommensurability of the two supposed paradigms would have to be levied. Sadly, Sellars is difficult to pinpoint on this matter (DeVries & Triplett, 2000), and for the sake of brevity we will have to work on intuitive understanding of the proposal here. The scientific image grows out of and replaces (though this is not strict reductionism) the manifest image, and has to be able to "deal with the questions raised in the manifest image and the phenomena familiar to it" (DeVries & Triplett, 2000, p. 114). What is clear though is the permanent request in Sellars for the continuation of postulational reasoning with ordinary modes of explaining and understanding our world.

One route left open is to criticise Sellars' view in the context of this thesis as simply presupposing the predominance of the mechanistic views (in fact we might accuse Sellars himself of helping establish such a dominance in the philosophy of science), and thus trying to show that the approaches which are aware of a link between preference for causal mechanical explanations and the conceptual primacy of the geometrical isomorphism will not be threatened by the consequences of the 'troublesome' phenomena for the passage from manifest to scientific images. They might either claim that the separability violation is an illusion, an error, in the case of 'troublesome phenomena', or might claim that for the route from the manifest to the scientific that they are building separability violations do not present as much of a threat as is portrayed above.

## Metaphysics of deeper explanations

Hitchcock and Woodward (2003) note the paucity of literature on systematic account of this notion. In such a context it is worth merely surveying their own cited attempt for the ontological features that might provide pointers in the desired direction, with the proviso that the previous chapters were supposed to point towards the depth-providing characteristics of the specific case-study instances. On their account (Hitchcock & Woodward, 2003) greater depth is achieved by explanations which *depend* on more variables changes of which lead to more *significant* interventions in the phenomenal outcome. But they have to be those variables interventions on which can produce variation in the observable effects within the explanandum, and not some related concepts. Thus, deeper explanations depend on (not just contain) more elements which can observably alter the key segments of the observed phenomena, that can pander to the greater range of the relevant what-if questions. But, and this is the key point Hitchcock and Woodward are trying to make, this does not mean taking the most general account of the situation to be explained, inclusion of the widest possible set of background conditions, *but selecting those features of the situation that can be identified as possible properties of the very object or system that is the focus of explanation*. To avoid going round in circles here as to what really carves nature at joints, and how to recognise, it is worth reminding ourselves of the purpose of the transcendental strategy connecting the everyday conceptual scheme with the specific one employed in the explanatory account. To avoid the dangers of general syntactical game-playing that wreaked havoc of the general deductive-nomological explanatory account, clear conceptual unification with the wider conceptual scheme is required. As Psillos (2007) warns the counterfactual variations can be superficially achieved by any law abiding account, and the real mettle of some explanatory construction is proven through the unification with the wider conceptual scheme. We must thus aim to identify the object that is the focus of explanation and see how its properties relate to what is more directly experienced.

This may seem an obvious point, but one that is not readily adopted in great many scientific explanations, for it is precisely the difficulty of identifying those possible properties of the system which make the focus of explanation that proves most difficult. It also reintroduces the chicken and egg problem of what is to guide our selection of those properties, i.e. is the explanation the prerequisite or the consequence of the featured ontological entities. And it is here that we can see firmer foundation for preference for the causal-mechanical model of explanation over unificatory and other models.[101] That is, the ontological primitives, explicitly named as such, of the causal-mechanical model are postulated as the very objects whose properties (or their changes) lead to the desirable observable variations in the phenomena that are the focus of explanation. Namely, according to Hitchcock and Woodward (2003) generalizations (which is what all explanations based on theoretical framework come to be) provide deeper explanations than others if they provide the resources for answering a greater range of what-if-things-had-been-different questions, i.e. are invariant under a wider range of interventions. But, crucially, the interventions must be of the kind that focus on the hypothetical changes in the "system at hand" (Hitchcock & Woodward, 2003, p. 198), and not the changes in the systems adjacent to the one whose features are to be explained. This, again, stresses the importance of appropriately hypothesising the ontology in advance.

Though this confirms the popular preference for the causal mechanical explanations, it does not preclude further investigations in our case-study instances, as there are considerations of conceptual unity and efficacy to be taken into account also (cf. (Hansson, 2007) above). But it does point towards what the minimal ontological requirement for greater explanatory depth is, namely the identification of variant properties of the system/object that is identified as the element of reality under investigation, the subject of explanation. What we must bear in mind then is that our case-study instances of explanation must be able to at least name the elements of reality (objects or systems) that are the focus of explanation[102], and attain explanatory depth through explication of interventions on those that produce effects that can be conceptually accounted for (i.e. described or predicted).

[101] It is worth bearing in mind here that Woodward and Hitchcock present their account as part of a wider scheme to provide a model of explanation that is nether the standard causal nor unificatory model, and that can satisfy the requirements of explanatory depth better than the two traditional rivals. This need not concern us here, though, as their account of explanatory depth still provides criteria of evaluation (that need to be further explicated when we encounter individual instances of attempted explanation), without necessitating adoption of their model in particular instances. In other words, they are searching for a general model of explanation in science, which may be insensitive to the particular difficulties we are trying to respect here.

Psillos (2007) criticises the above account of depth of explanation for failing to provide clearer guidelines about the truth-conditions of the counterfactual situations, whilst distinguishing them from the relevant evidence-conditions. The latter distinction is important for the counterfactual musings to be explanatory, i.e. to be able to tell how phenomena would have turned out differently due to counterfactual interventions on them. Of course, the interventions can be, and in the interesting cases are, hypothetical, i.e. we can provide explanations of this sort even in the situations in which the direct evidence conditions for the counterfactuals are empirically inaccessible. This is the weakness of the depth-of-explanation account of Woodward and Hitchcock (2003b) in the situations that are far removed from the simple past events or simple accounts involving unobservables. As Psillos (2007, p. 99) notes in the latter situations there are well-known stories to be told as to what the difference between truth- and evidence-conditions in counterfactual situations is. This taps into the important psychological underpinning of the satisfaction with deeper explanation: we want to know what it is that *makes* the explanatory account true, not just how we verify its truth; what the conceptual structure that generates truth of the explanatory account is. The safest route to provision (at least notional) of the required truth-conditions is, in Psillos' (2007) view, to rely on the laws of nature. That is to include the laws of nature in the truth-conditions for the relevant counterfactuals. Laws have to be in place before we construct, by relying on counterfactual interventions, an account of what is and what is not invariant under relevant interventions on the objects.

> "Without independent account of what laws are, there is no clear way in which we can deem some (interventionist) counterfactual assertions true or false. Which interventions are physically possible and which interventions leave certain relations invariant depends on what laws there are. The latter cannot be fully understood as relations that remain

[102] Though, crucially, not necessarily the primary existents, the fundamental ontological entities.

> invariant under interventions since they specify what interventions are possible." (Psillos, 2007, p. 105)

This is important for us in two ways. Firstly, it suggests that our 'transcendental strategy', coupled with desire for deeper explanations from the two case-study instances, will not go far enough in providing the conceptual link through the selection of ontological elements and the 'geometrical' structural isomorphism between the fundamental ontology and the everyday material objects. What it needs to have added is the minimum set of laws of nature that are expected to hold between the fundamental and everyday account of the phenomena. In most cases this is not a problem, and largely the minimum set consists of the fundamental logical connections, and in many other cases we have enough uncontroversial information about the conceptually supportive causal structure. Thus, Psillos (2007, p. 106) says that when we are dealing with stable causal or nomological structures interventionist counterfactuals are meaningful and have truth values. The problem is that our 'troublesome' phenomena may not be supported by enough of such structure to let us construct a convincingly deeper explanation, and thus provide for the comparison of our two case-study approaches. In any case, it calls for an explicit justification of the stability of whatever nomological (if not always causal) structure the approaches can rely on, alongside the material ontology they employ, in order to provide them with sufficient grounds for the construction of the 'transcendental' argument ( (Luntley, 1995); cf. section 1. 4 above).

Secondly, the account which provides a separate account of laws relevant to the situation will be better prepared for the task of providing a deeper explanation. Psillos worries, though, that the depth-of-explanation account as constructed by Woodward and Hitchcock (2003b) above highlights and employs the symptoms of good explanations (in particular of good causal explanations) without being able to provide a fully fledged theory of what an (causal) explanation consists in. "Invariance-under-interventions is a symptom of causal relations and laws. It is not what causation and lawhood consists in" (Psillos, 2007, p. 106). In our case both accounts can use the pragmatic virtues of the depth-of-explanation account provided they are explicit about how they will overcome the problem Psillos raises. The principle approaches can claim not to aim for a causal account at all, and search for deeper explanations through supplying the relevant laws as directly observable empirical generalisations not justified, nor accounted for, through their role in the 'troublesome' phenomena themselves. Due to their supposed simplicity these can then be easily linked with the wider unchallenged set of laws governing the behaviour of material reality. The problems arise, though, when the phenomena are interpreted as constitutive of ontology that is not easily linked with the material ontology of the everyday conceptual scheme. The constructive approaches, on the other hand, have (cf. Chapter 3) provided an independent account of relevant laws, most notably the universal law governing the behaviour of the ontology, through abolishing the Humean Mosaic and making laws primitive existents alongside the ontology. Each of the accounts then has to show that this general model can be applied to the 'troublesome' phenomena and the potential consequences they can have for the 'transcendental' argument.

Yet, one might object that on this reading preference is given to the causal-mechanical model of explanation right from the start. How could a unification model satisfy the requirements for hypothetical manipulations on system at hand, accompanied by a network of stable laws that provide the truth-conditions for the counterfactual situations? The answer is simple, if not directly applicable to our principle approaches: take the uncontroversial objects that feature in the phenomena and show the limits of manipulations possible (the hypothetical situations where only the relevant aspects of the central objects are changed or affected). In the troublesome phenomena this would involve showing how the objects central to the phenomena would have been different had relevant changes in them been instigated, whilst the remainder of the context (this includes the laws and the other objects) had been kept unchanged. It is hoped that the principle approaches can in this way provide sufficiently deep explanations (though not expose the 'mechanism' that gives rise to the phenomena) without having to construct awkward connections between the central tenets of the explanatory account and the everyday conceptual scheme. They would gain the upper hand over the causal mechanical accounts if the latter were forced to do just that, to add entities and change qualities of the core conceptual scheme in order to satisfy the construction of the explanatory account. But we must bear in mind the impermanence of objects on principle approaches, where precedence is given to universal applicability of generalised principles to all and any 'thing' featured in the phenomena.

### 4. 2 Principle approaches and the depth of explanation

It remains an open methodological problem for the principle approaches, one that ties in well with the overall overview of the role of physics and the requirements of arguments for simple realism, whether all the possible formal models that the principle approaches can derive (and that agree with the constraining principles) should be excluded from considerations by further modifications of the choice of the suitable mathematical framework for quantum theory (along with the implicit metaphysical assumptions that might come along with them), or whether we should find reasonable general methodological constraints (these are not our constraining principles) for the formulations of the physical theories and rule them out on the grounds of those. We shall proceed in the following sections on the latter assumption, i.e. that the provision of deeper explanation suitable for the 'transcendental' argument is a reasonable criterion to adopt. We are no longer worried about the details of possible common formalism inherent in all quantum theories as those with deeper explanations will be preferred overall. It is another issue whether this equivalence is the very assumption that Bub is relying on when using his deep methodological principle.

In light of the above it remains to be seen how Bub's 'deep methodological principle' ( (Bub, 2004); cf. Chapter 2 as well) aligns with the requirements of provision of deeper explanation and upholding of the 'transcendental' argument. Following Bub's principle we must refuse to venture further than macroscopic *'records'* of the inputs and outputs of the measuring processes, effectively making their conceptual framework reliant on the epistemic atomism of the momentary states of the input and output status of the macroscopic apparatuses. Even when given in terms of the information concepts this remains a Humean Mosaic view of the phenomena as causally independent sequences. But, though all physical processes can be given in such terms, in the past we have been able to move beyond this timid generalisation of the world.

It seems historically heuristic to view the principle explanations as a step towards novel, as hitherto unexpected constructive explanations. Explanations from principle approaches (principle explanations) are primarily concerned with exposing the competing explanatory approaches as focusing on a pseudo-problem, striving to explain something that essentially does not require an explanation over and above that it stems from an erroneous perspective on the phenomenon to be explained. In that they have to stay away from the thin line of slide into a full blown instrumentalism, whereby no steps towards a future new explanation are offered but every route to explanation through ontology is effectively closed. Fine (1989) suggests as much in denying that we are forced to accept "the explanationist challenge" (Fine, 1989, p. 191) and speculate about the hidden hands and propensities that guide the 'troublesome' correlations. Fine claims that the demands of explanatory adequacy come *a priori* from the outside the quantum theory, and are a remnant of a different kind of physical thinking. As much as this would rid us of the struggle to provide an explanation from the principle approaches, it lands us squarely in the neo-Bohrian (but what is worse neo-Bohrian with an extreme slant that even Bohr is likely to shy away from) camp characterised by abandoning all hope of understanding the processes that give rise to the troublesome correlations in material terms, as well as all hope for the unified knowledge of the macro- and the micro-physical realm.

Fine shies away from constructive steps and advocates firm adherence to the establishment of principles that expose what is prohibited in the correlations, whilst quoting a statistician Moses when accounting for the non-local influences, mysterious background guidance, mutual dependencies and passions: "Much less is true" (Fine, 1989, p. 194). Hughes (1989), in the same volume, is supportive of this view. It is his argument that if the elements of our standard conceptual scheme cannot find a suitable home in the explanations of the troublesome phenomena, and yet the phenomena are taken as real, empirically verified, then we must abandon the use of the conceptual scheme or seriously modify the key elements of the conceptual framework. Hughes wants the elements of the new conceptual framework to be clearly identified within the mathematical structures used by the theory. An obvious problem for our principle approaches is to show how the new conceptual scheme unifies conceptually with the standard one of extended matter, so as to achieve our goal of avoiding anti-realist criticism of the possibility of scientific explanation in general. This would in fact be a route of making the entire material conceptual framework dispositional, emergent from the new ontology (such as information-ontology might be). Though Hughes argues that the new metaphysics would have stronger resistance from refutation by emerging 'naturally' from quantum theory itself rather than being artificially tacked onto it by metaphysical demands external to construction of physical theories, he admits that may not be able to do any useful *explanatory* work. "However, it is not clear what useful explanatory work this interpretation would perform over and above that provided by a full articulation of the models the theory presents" (Hughes, 1989, p. 207).

The CBH programme (of Chapter 2) can then shift the explanatory focus to a different realm, that of information manipulation. This is admittedly a risky route to take in provision of explanation, as it explicitly shies away from providing the explanatory account in terms of the conceptual framework that we initially required for the transcendental strategy.[103] Though risky, here is how the route might proceed nonetheless. When asked to provide an explanation of the 'troublesome' phenomena, the principle approach advocates might proceed by pointing out that nothing is neither exchanged nor travels, and no explicit mysterious connection is established between the material existents characterised by the primary qualities. We have come to have an erroneous view of the situation and have thus entangled ourselves in a pseudo-problem. We must, fully and truthfully, suspend all speculative expectations and return to the conceptual scheme of material existents at hand in the situation. Alongside, we must dissolve the troublesome characterisation of the phenomena and relieve any worries that the ontological separability of material objects is threatened. The principle approaches are asking us to take a step back: leave the material existents as they are in the standard conceptual scheme, connected only by the physical interactions that respect the space-time extension and separation. That part of the conceptual framework remains intact. And that part of the conceptual framework plays no role in the establishment of the phenomena. What does then? Here we have to be presented with the phenomena in the new light. Bear in mind though, that a small but important constructive step has implicitly been made: separability has materially been upheld, i.e. whatever the appearances nothing is expected to characterise the macroscopic material existents over and above what characterises them locally in their space-time region. Likewise, all the changes they can be expected to endure must be understood as local phenomena, requiring no knowledge of distant states or some global set-up.

[103] As Timpson (private correspondence, but cf. also (Timpson, 2008)) puts it, we want to know what the physical processes behind the phenomena are, not what the experimenters can know about them or in what ways we can interpret the supposedly correlated signals from the 'epistemological black boxes'.

According to Sklar (1990) the greatest contribution of the principle approach in physics is to remove the need to adjudicate between the equally empirically adequate, but metaphysically divergent, explanatory constructions. When the difference between such constructions cannot be adjudicated empirically, it has sometimes been useful to present the difference as a pseudo-problem, to show us how we could account for the phenomena (again, without the explicit constructive mechanism) by ignoring the constructive conflict and looking elsewhere whilst holding on to what is phenomenologically unalterable: the constraining principles. Again, drawing on Einstein's derivation of Special Theory of Relativity, Sklar claims that the latter exposes what were considered rival but empirically equivalent descriptions of the universe as equivalent descriptions of the same state of affairs (as the search for absolute motion is abandoned). Again, we must bear in mind the warning of the sinful constructive step Einstein makes (cf. Chapter 2 above), but also that it is not damaging for the kinematical considerations of the theory. But what are the explicit advantages of the principle approach over the rival constructive approaches? According to Sklar they are supposed to be speculatively more cautious through abandoning the metaphysical expectations that cannot be directly verified. They are also supposed to unify a greater range of phenomena under one explanation, rather than requiring a range of respective different explanations (i.e. not just instances of one basic explanatory conceptualisation tailored to individual situations). Finally, the explanatory power of the principle approaches is supposed to be greater by avoiding what would otherwise be mere coincidences in agreement of different explanatory constructions.

In other words, the greatest power of the principle approach should come from telling us how come the phenomena consist of the same appearances even when we approach them along different constructive schemes. This goes further than strictly explaining the phenomena, but aims to explain the occurrence of the illusion. Of course this can immediately be charged with criticism of pragmatic shiftiness in the choice of observables. We open up to the possibility of re-examination of the fundamental concepts we implicitly take for granted in the transcendental strategy. Everything is suddenly thrown into doubt, and the principle approach takes liberty in choosing what to call observable and non-speculative. And Sklar says as much in his analysis. He says that our theory, however conservative on speculation, must carry with it some metaphysical baggage that does the explanatory work. Rather than being per se simple, the supposedly sturdy conceptual structure must do extra work to explain how it fits with that which can still be held as well-understood and free from illusion. He sees the spacetime structure of Special Relativity to be such minimal baggage, a replacement for the aether and the absolute velocity. "[A mere set of observational consequences taken as a theory], unlike the special theory with its theoretical space-time structure, fails to offer genuine *explanations* of the observable phenomena." (Sklar, 1990, p. 155) Principle theories have to supply that extra weight that distinguishes them from bare phenomenalism and instrumentalism, so as to provide explanations. That is the most important lesson for our principle approaches of Chapter 2. But a serious caveat is immediately put forth by Sklar: this is increasingly difficult to follow in the cases where the considerations strike at the very foundation of our conceptual schemes.[104] Again, Einstein's sinful constructive step can be seen as just the required avoidance of the tinkering with the foundations of the conceptual scheme. Though principle approaches drag with them a constant risk of sliding into excessive instrumentalism or phenomenalism by their adherence to almost primitive empiricism, the risk is worth taking when our pragmatist (explanatorily too weak) and realist (conceptually threatening and observationally underdetermined) accounts strike at the very heart of our well-entrenched conceptual scheme.

[104] In Duhem's (1991) insightful criticism of the declarations of methodological superiority of the principle-like approaches advocated by Ampere, we see that even in the less conceptually troubled domains, researchers are forced to make implicit (and in a way operational) constructive hypotheses by borrowing analogies from existing constructive disciplines and operationally objectifying hypothetical entities. This is a declaration even before Einstein's qualms about the structures behind Special Theory of Relativity of the implicit constructive theorizing in the declaratively simple principle approaches. I am grateful to Simon Saunders for pointing out this case.

On the metaphysical side there is no clear suggestion in the literature as to how the 'information-stuff' (provided we can argue there is such a thing) and the extended material stuff can coexist at the fundamental level of reality. The notion of depth-of-explanation above stresses the importance of the detailed account of the controllable variations in *objects* that the changes to be explained happen to. This is the most serious of weaknesses attributed to the principle approaches and one that can only be avoided if we can somehow show that the 'transcendental strategy' can be more effectively constructed with principle approach concepts even without the *prima faciae* concerns for the depth of explanation. This is to show either that:

1. the 'extended stuff' can be modified or replaced in the explanatory conceptualisation required for the 'transcendental' argument (perhaps by presenting the 'extended stuff' as an illusion reducible to something else); or
2. there are ways to reduce the properties of the new stuff (presumably, information entities) to those of the primary qualities of the 'extended stuff' making the former a dispositional illusion to be removed from the conceptualisation of the 'transcendental' argument that respects the occurrence of the 'troublesome' phenomena.[105]

But if the latter strategy was adopted we might ask ourselves what the contribution of the principle approaches is, other than providing an alternative way of looking at things, i.e. we would be at loss to identify the exact pseudo-problem that principle approaches have helped us out of. We must constantly be asking ourselves what it is that the principle approaches can hope to achieve (other than satisfy Bub's methodological principle) given that we already have empirically adequate constructive attempts. If we are to go beyond all of them, what is the direction that the principle approaches are suggesting? On such reading the preferred direction seems to be to establish the novel ontology that does not threaten separability violations, but the question is then how to combine it with the material ontology of the 'transcendental' argument.

[105] The latter seems to be exactly the strategy that the constructive approaches follow. In this way the principle approaches would in the end be reduced to the constructive ones in terms of explanatory ontology, and would thus be making that step towards the more explanatory constructive theories, as Einstein required (cf. Chapter 2, above).

The principle approaches then can rely on diffusing the threat of teleportation phenomena, along the lines advocated in Timpson (2004), where it is claimed that the conceptual puzzles arise when information is mistakenly taken to be a substantive, rather than an abstract term.[106] What is in fact phenomenologically the case, is that Bob can extract only one bit of information from his black-box, upon the successful run of the protocol in which Alice has sent him 2 classical bits. If there is no material substratum to the phenomenon assumed, or at least none is speculated about, then there is no great quantity of information (which was physically meant to be stored in the material referent of the quantum state) transmitted in the protocol. For if things had been otherwise the no-signalling theorem would be violated. What remains puzzling is the role of the quantum formalism in the whole situation. It seems to allow for some counterfactual situations involving the distant experimenter, Bob, which would not be possible in the pure-black-boxes case.

What role do the general constraining principles play then, in an overall understanding of reality as required by starting point of the simple transcendental strategy? The principles must stand for something explanatorily, even if just to say that Bob cannot in reality obtain more than 1bit of information from his black box. To be a constraining principle, no-signalling theorem tells us that things could have been different and that the fact that they are not is significant for our situation. But unless we assume that the situation is characterised by the potential for a larger information extraction, the 1bit (without the constraining principle) is not the least surprising nor 'troublesome'. As soon as we bring the principles in, we are assuming something more about the ontology behind them, an ontology that does refer to the potential for large quantities of information to become available to Bob conditional on the distant actions Alice takes.

The constraining principles must constrain something, and the interesting question immediately becomes what it is. One option is to follow Timpson's suggestion (2004, p. 72) and to rephrase the question in terms of the material ontology behind the phenomenon (thus abandoning the black boxes, and falling prey to the traps of non-separability). The other is to simply admit that when manipulating the black boxes we are constrained by the general principles, and then seek an explanation of those principles in terms of the structural familiarisation with the new entity. Of course, that is just moving the game to a different playing field, but it still remains a hot task to link the information ontology to the material ontology that is the major supplier of our experience. Now this needn't be an entirely obsolete route, as the investigation of new entities, even if abstract and non-material can still tell us something about the world we inhabit. For if we were to take information to be an abstract entity, such as a mathematical triangle might be taken to be (cf. suggestions in (Duwell, 2008)), we can still learn something about the 'geometry' of our world even if we do not talk directly about the material objects affected by that geometry. Suppose information should not be understood in either of the Timpson (2004) senses[107], but as an entirely new entity. Nersessian's (1984) analysis investigates a precursor for such an approach from history of physics.

[106] Timpson (2004) is adamant that information cannot be understood as any kind of entity (even an abstract one) at all, and that this is where the error of the principle approaches lies. They should instead turn to the material foundations of the concept of information.

In (Faraday to Einstein: Constructing Meaning in Scientific Theories , 1984) Nersessian aims to present how a new concept of a 'field' was introduced into scientific parlance (with respect to theory and observation) (Nersessian, 1984, p. 27). She suggests that new concepts change from being a heuristic guide to other ends, through a stage of elaboration, into being full-blown philosophically justifiable concepts capable of sustaining rigorous analysis. This does not mean we can form a clear definition through a set of sufficient and necessary conditions for some phenomena to be characterised by our chosen concepts, but that they feature a set of family resemblances where each instance varies in the degree of qualitative conformity to the lot. First a primitive qualitative concept is introduced, with no clear mathematical structural unification into formalism, as an operational alternative to the existing explanatory view. Further development through a series of analogies to furnish additional detail to the new concepts, with analogies serving as explanations (or in the Hansson view above: conceptual links into the wider explanation) of the newly discovered details. Nersessian's 'analogies' provide a function similar to Cao's metaphors (cf. Chapter 1) of carrying over understanding from a familiar domain (most notably that covered by the everyday conceptual scheme) to the 'troublesome' one containing the explicit descriptions of the phenomena under consideration. In other words, they carry the transcendental step, through sufficient structural isomorphism.

[107] The two sense of the term information, supposedly confounded in the principle approaches are the common-sense 'type information' sense and the technical (in terms of Shannon's (1949)) communication theory) 'quantity information' sense.

Finally the new concept can adopt the role of substance (the practice Nersessian bases her analysis carries over more easily to the case of principle approach's information, than the constructive approach's fundamental laws in this case, but that needn't concern us at this stage) in the conceptual scheme. At this stage it is possible to consider a wide range of problems and objections, to address them and to clarify the links of the new concept to the existing conceptual scheme (which may have been partially changing alongside it, or even with it). Now a clear understanding of the new concept is achieved and it is successfully unified with the prevailing conceptual scheme.

Signs of that understanding are provided precisely by the ease with which it plays the explanatory role and addresses the questions such as: "What does it do? How does it do it? What is its function? What effects does it produce? What kind of 'stuff' is it? How can it be [(sic)] located?" (Nersessian, 1984, p. 156). Some of these questions our candidates will have to start grappling with, other may not be applicable to them. What is important is that we can start building explanations from them, and comparing them to each other and existing explanations even at the early stage, working all the way to complete the steps towards the next stage or opening up new questions. Thus we do not have to have a demonstrable reference bearer at the outset for each concept we introduce, nor do we need to be clear about all aspects of its connection into the conceptual scheme in order to work on an explanation of the 'troublesome' phenomena.

But the principle approaches of Chapter 2 are a long way away from understanding the new concept in this way. Moreover, Nersessian's paradigm concept of a 'field' relies on the same essential qualities of extension as does the common-sense concept of an extended object (though there are important differences as well) and interaction 'by contact', unlike the entities of information ontology. Other elements of our ordinary conceptual scheme are also present in the defining questions that Nersessian poses: such as "What does it do?", "What effects does it produce?" Those causation-related elements are not even hinted at in metaphysical extensions of the principle approaches of Chapter 2. Thus such alternation of conceptual framework has a long way to go, and as yet there are no clear indications that it is going in the right direction.

Duwell (2008) attempts to construct a starting point for the novel ontology behind the 'troublesome' phenomena. To a degree it relies on partially dissolving the 'troublesome' nature of the phenomena, but also strongly argues for the existence of information not as substance (which must be spatially located, and then subjected to generation of 'troublesome' aspects of the phenomena), but an abstract entity outside the constrictions of the material ontology. The details of this account need not concern us, but the general potential for explanation, and most importantly for the 'transcendental' argument, will be of interest. Duwell (2008, p. 215) advocates seeing the explanations resultant from this metaphysical extension of the principle approaches as those of a specific unification type: the deductive-nomological explanations. That the latter have been severely criticised in the philosophy of science, and often in the end amended through addition of causal aspects, should be a sufficient pointer of their explanatory worth for our purposes. Yet the criticism often centred on their overly syntactic aspects, and what we are primarily concerned with is the conceptual explanatory potential Duwell can generate from their content.

Unfortunately, Duwell's account is abruptly cut short here, and beyond advocating the "unificatory view of explanation" (Duwell, 2008, p. 215), he fails to tell us how the experienced phenomena will be explained in terms of lawful behaviour of quantum-type information distribution. There is a legitimate suspicion that two plains of being will be introduced, one of material ontology and one of quantum-information ontology, with all the supposed 'troublesome' aspects of the phenomena relegated to the latter. If this allowed adherence to the principle of separability at the level of material ontology then our 'transcendental' argument may still be able to survive the antirealist charge, but there is no indication that this is so. The original 'troublesome' aspects will be generated in the conceptual scheme of the quantum-information ontology, but we are told nothing about how they connect to the material ontology. The issue is simply swept under the carpet. The legitimate worry then remains that to produce the phenomena, wherever we consigned their 'troublesome' aspects to, violations of separability must be accepted at the level of material ontology (not that material ontology is dispositionally reliant on the quantum-information ontology, the two simply exist side-by-side). Yet the transcendental strategy, that aims to include the 'troublesome' phenomena too, needs the account in terms of material ontology also, as it forms the grounds of our epistemic access to the quantum-information realm.

So what happens when direct consequences are drawn from the principle level generalisations to the material ontology? The first of our principle approaches (tentatively abandoned even in Chapter 2) does not fare well in this respect. As Timpson (2008) shows, Fuchs' approach faces a severe explanatory deficit: "it is unclear how what *is* explanatory could be so" (2008, p. 607). This poses problems for our transcendental strategy of section 1. 4, also. The extreme sensitivity of the fundamental ontological realm delineated in Fuchs' principle approach denies that there are any "facts about the world, prior to the measurement outcome actually obtaining, which determine what the outcome would be, or even provide a probability distribution over different possible outcomes" (Timpson, 2008, p. 595). In that we lose the structural connection providing for re-identification of objects at the fundamental ontological level (this is not an epistemic, but a metaphysical deficiency now). When constructing the full-blown dispositional account of the fundamental ontology, we cannot provide a stable foundation for the repeatable, regular behaviour of objects in interaction, "the rules of composition of the powers are too loose (or are non-existent) [...], giving rise to the lawless pattern of events" (Timpson, 2008, p. 597). Our transcendental account of section 1.4 not only loses the ground of separable ontology, but an altogether greater one of anything that can be said about how things are "occurrently" (Timpson, ibid.). This plays into the hands of the postmodern critic, when Timpson (2008) recalls Wittgenstein's claim that nothing would do as well as something about which nothing could be said. Any hope of the depth of explanation is likewise lost.

However, even Bub, as one of the proponents of the CBH programme, seems intent to follow some way down Fuchs' route in suggesting the possible metaphysical glimpses beyond the principle approach. In Bub and Pitowsky's (2008) exposition a principle theory is the best epistemic account of a metaphysically fundamentally indeterministic universe. In that they block the route to any deeper explanation beyond what can be given by the acceptance of the constraining principles of information manipulation. This we take to be the meaning of their claim that there is no explanation of the series of observed events through real change in the correlations between separated events at the micro-level, as opposed to other possible observed events in a quantum measurement process – the occurrence is constrained by the generalised principles of information manipulation, and only by those. Even if this does not directly damage neither the separability expectations for the fundamental ontology nor the structure that is meant to connect it to the observable ontology, it nonetheless denies any possibility of a deeper explanation by making senseless any truth-conditions for conceivable counterfactual situations. The consequences for our transcendental strategy are simply that even if the supposed damaging separability violations are an illusion, we await to be told what the connection is between that which is constrained by the generalised principles and the fundamental ontology of the world. In historical terms we must have at least a possibility of finding the Lorentz-style constructive explanation of length contraction, for it is the conceptual prerequisite of a framework (of Special Theory of Relativity) in which the rods demonstrably contract and clocks slow down.

### 4. 3. Non-separability and the derivation of fundamental physical laws

One possible route to be taken as a lesson from the principle approaches to enlighten us any deeper on the potential separability-violation issues is to try and find the ways of holding on to the transcendental strategy whilst admitting non-separability as an explanatorily benign feature of the material world. This marries the principle approaches' attempts to dismiss threatening separability violation as an illusion with the (hard-core) realist approach of the constructive approaches in assigning the essential characteristics of the 'troublesome' phenomena to material ontology. That is to argue against Howard's contention (1989) presented in section 1. 4., that separability violations threaten the very core of our foundational conceptual scheme, the isomorphic connection between the physically fundamental ontology and the objects of everyday experience, through the primary qualities of material existents. In fact, if Newtonian physics could (albeit uneasily) live with the non-local laws and yet account for the everyday experience, maybe quantum theory can find ontological elements to bear the brunt of the separability-violation without denying the realist firmament of the stable extended material existents.

This is precisely the position Dickson (1998) advocates, arguing against the problems put forth by Howard (1994) in section 1. 4. Dickson claims that it is ontological holism that is threatening to our core conceptual scheme (the latter featuring in Luntley's transcendental strategy) and not simple action at a distance. He proposes to align (if not identify) what Howard calls separability-violation with holism (i.e. to claim that holism implies violations of separability and vice versa), and what Howard calls locality-violation with action at a distance (again: violations of locality imply action at a distance and vice versa). His conclusion is then that quantum formalism in the troublesome phenomena requires accepting action at a distance (i.e. violations of locality), and that that in itself is not damaging to our conceptual scheme as it can be accommodated in a way similar to accommodating existence of gravitational influences in the conceptual scheme of Newtonian physics. Maudlin (2007a) was also presented above as arguing for a similar point, by requiring the ontology of beables to be local whilst the laws (also a part of ontology, or at least the physical world-view) could be non-local. Yet the initially plausible analogy has to be further addressed before the end of this chapter.

So here is a possible middle ground to be extracted from the multiple presentations of the problem above, and cast in the light of our second case-study instance: the constructive approach of Bohmian Mechanics (below, cf. also Chapter 3 above). We are in fact looking for a way to show that though notionally separability is violated, the violation is not such as to threaten our entire conceptual scheme based on the extended matter (as suggested in section 1. 4 above). This is effectively arguing for the violation of locality, i.e. showing that separability as a deeper principle can be conserved if we allow only some aspects of our foundational ontology, and not those central to the suggested transcendental strategy (of (Luntley, 1995) and section 1. 4 above), to display action-at-a-distance and thus violate a weaker principle of locality.

In their analysis of the issue Timpson and Brown (2003) claim that separability in Einstein's works takes the form of a transcendental strategy, with somewhat different purpose of the one we had been considering above, argument for the possibility of framing empirical laws. This can be seen as part of the Luntley's transcendental strategy sketched above, as along with the stable ontology the argument implicitly requires a possibility of grasping the laws that govern the changes of the material ontology. Again the primary qualities of ontology can be said to give the laws their understandable form, i.e. when referred to those features the laws can be seen as contributing to the isomorphism between the 'real' processes and the experienced phenomena. But Timpson and Brown (2003, p. 7) go on to push for a distinction within Einstein's 'original invocation of separability' in (Einstein, 1948) into separability-proper ("requirement that separated objects have their own independent real states (in order that physics can have a subject matter, the world be divided up into pieces about which statements can be made)" (Timpson & Brown, 2003, p. 7)) and locality ("requirement that the real state of one system remain unaffected by changes to a distant system" (Timpson and Brown, ibid.)). The transcendental strategy can then go through, and not have to adopt empirical adequacy of the quantum formalism as its scientifically derived counter-example, if we take the 'troublesome' phenomena to be *violating locality*, but *not violating separability*.

For as quantum formalism (with its generalising principle of no-signalling) shows we can formulate empirically adequate laws governing the locally observed phenomena without having to take into account the state of affairs at a set of unspecified distant locations. This does imply that the formalism of the theory will not be as precise as we might have wished it to be (though this need not imply that it is formally incomplete), but it does not imply that in order to make it more precise we must take into account the state of affairs at various distant locations. When we use the laws to predict the occurrence of phenomena the general prohibition of superluminal signalling (respected in both our principle and constructive approaches) guarantees that whatever phenomena occur at distant locations, our predictions concerning our local phenomena cannot be improved. Of course, if we include the classical signal improvements can be achieved, but that very signal is not even a locality violating process. Thus if we take the empirical testing of laws to be achieved through correct predictions of the phenomena, then "[it] is established by the no-signalling theorem [that] the *probabilities* for the outcomes of any measurement on a given sub-system, as opposed to the state of that system, cannot be affected by operations performed on a distant system, even in the presence of entanglement. Thus the no-signalling theorem entails that quantum theory would remain empirically testable, despite violating locality" (Timpson & Brown, 2003, p. 8).

Due to no-signalling we can then not predict locally the changes our system is supposed to have undergone on the subsequent explanatory account, but that also allows us to rest all local explanatory accounts on what we can predict without fear that they will be falsified by such changes. In other words we need not open the possibility of ontological holism. We do import from the principle methodology the acceptance of the generalised no-signalling prohibition (modified so as not to be expressed in terms of information-ontology) which of necessity skips over the contentious issue Einstein raised: how come we can reliably formulate laws when we cannot satisfactorily conceptually isolate our objects of experiment from the rest of the universe. And admittedly this element remains mysterious, though the constructive approaches' notion of quantum equilibrium aims to give some account of it. If we consider Einstein's stronger version of separability as an epistemic condition on formulation of laws then blunt acceptance of the no-signalling theorem (regardless of its subsequent constructive account through the complex notion of quantum equilibrium) provides us with effective epistemic separability as required. Bub's deep methodological principle suggests we should not go further than that, but in search of the explanation that can be united with the transcendental strategy we have already forgone that prohibition. Now we view one of the CBH generalising principles as an epistemic, not metaphysical limitation. This certainly weakens the ideal account that the realism of the transcendental strategy would want, but as we shall aim to illustrate below it does not prohibit all possibility of ontological explanatory connection between everyday experience and the troublesome phenomena.

Prediction should not be directly equated with explanation, and in fact some of the grounds for Luntley's transcendental strategy against the antirealist criticism is provided precisely by that asymmetry (this is in the cases where the prediction is imprecise). So one might object that when giving an explanatory account of the micro-physical phenomena violations of locality will still provide difficulties for provision of a complete and precise account. But Timpson and Brown (2003) claim that this is a problem of a different kind, a problem that may be resolved by appeal to different measures, from the objections that empirical laws cannot be determined due to doubts about underlying ontological holism. In our case, a defender of Luntley's transcendental strategy would claim that though troublesome phenomena require additions to the conceptual scheme that encompasses the common sense core and classical physics, the very conceptual scheme is not throwing up inconsistencies between requirement of primitive individuation of the segments of material reality (the basic ontology of objects) and the ontological holism of the same material substratum. It is the empirical generalisation of no-signalling, or its deeper constructive explanations in terms of quantum equilibrium, that secure the viability of the quantum formalism alongside our common-sense understanding of the world. In other words, we can argue that neither the formalism itself, nor the constructive renderings of it, force us to a view of ontological holism (and it is worth bearing in mind that this is a stronger threat than the notion of an all permeating field, for the latter still allows for a local individuation of characteristics of the foundational 'element' of reality) that forbids the individuation of objects in local regions of space-time.[108]

But it does impose some demands on the explanatory conceptual framework of our case-study constructive approach. Most notably, though predictively this was not required if we stay at the level of quantum formalism and its statistical character, in terms of explanation it must account for the violation of locality (i.e. account for the no-signal action-at-a-distance), whilst showing how ontologically separability is maintained in the 'troublesome' phenomena. We can permit that separability to be of the weaker form (out of several possible forms considered in (Healey, 2004)): physical processes behind phenomena in a spacetime region R supervene on an assignment of intrinsic physical properties to extended objects (again this can include fields as well) and the local proscriptions of the universal law governing the changes of the intrinsic properties at points of R and/or in arbitrarily small neighbourhoods of those points. But we must give some account of how changes to the overall entangled system (as implied in what Maudlin (2007b), in Chapter 3 above, termed abandonment of separability) are communicated into changes (even if locally unpredictable and imperceptible, as in the case of symmetric and anti-symmetric triplet and singlet states) of the local separated extended material ontology. In essence we have to show whether, and if yes how, Luntley's transcendental strategy can survive the 'troublesome' phenomena on constructive account. Though separability is not violated in a sense that we can't formulate any laws governing the behaviour of a localised group of objects, that law itself cannot be taken to supervene at all times solely on the structured arrangement of the intrinsic state of those objects alone. Our empirical equivalence then results in alternative views of the problem of whether a primitive thisness of objects or supervening generalisations should be maintained.

[108] It is worth bearing in mind here, and this is also further explored in Timpson and Brown (2003), that Everett interpretation, missing from this analysis, is not forced to adopt ontological holism either as it outright excludes the notion of collapse which Einstein used alongside that of entanglement in exposing the tension between the completeness of quantum theory and principle of separability.

### Constructive approaches in the light of non-separable laws

How does the universal law 'transmit' (or even record) the local mechanical interactions that the proximal particle undergoes to the distant one? Correlations in measurement outcomes on our separated particles cannot be attributed to a common cause (cf. the (Maudlin, 2007b) exposition of separability violation in 3. 2. 2 above), but neither can they be attributed to the transmission of directly detectable signals between the particles. They are taken to simply come about without a contact-interaction causal mechanism, through an unknowable nomic prescription (encoded in the universal law) that they should. A serious question arises: how does this explain them? Is this not simply hiding the lack of separability-respecting explanation under a carpet, a carpet imprinted with a neo-Bohrian pattern similar to the epistemic limitations of principle explanations?

In terms of comparison with the principle approaches of Chapter 2, we have to ask whether allocating the occurrence of the 'troublesome' phenomena to the universal law capable of affecting locally the distant particle, based on changes the proximal particle undergoes at a separated location, is not just a return to the 'black box' explanatory agnosticism about material processes as given by the principle approaches. Bohmian mechanics is forced to explain wherefrom comes this limit on what can be learnt about the universe in a theory so precise, with precise motion of spatially located, almost tangible, particles. This is, so it seems, where the constructive approach leans close to the principle one.

Finally, when explanation of the 'troublesome' phenomena is offered by the constructive approaches how well does it tie in with the requirements for durability of individuation of spatial entities seemingly behind the 'transcendental' argument? Most notably, if all the relevant information for dissolving the 'troublesome' aspect of the said phenomena as required by the transcendental strategy comes from the universal law (wavefunction) alone, how fundamental is the extended material ontology?

The constructive approaches of the previous chapter, most notably the ultimate suggestion to treat the wavefunction in Bohmian Mechanics as the universal law of temporal evolution governing the behaviour of the fundamental primitive ontological entities, the particles, aim to outdo the principle approaches in the provision of explanation compatible with the 'transcendental' argument by specifying how empirical adequacy of the theoretical formalism is achieved in terms of the material existents, the very same entities that physically construct the objects that our everyday concepts refer to (Goldstein, 2007). They specify what ontological elements of the real world make the quantum formalism empirically successful. They aim to not only uphold the same constraining principles that the principle approaches put accent on, but to show how those principles arise in the world of ontology that is supposed to support explanations required by the 'transcendental' argument. In that they encounter problems of their own, most of which we shall try to address in the following sections, but more importantly they shed light on the nature of explanation required to accommodate the 'transcendental' argument and the 'troublesome' phenomena.

With the postulation of the primitive ontology of particles and the kinematic guidance they receive from the wavefunction, elaborate arithmetical proofs are employed (as surveyed in the previous chapter) to show that the slightly modified formalism of Bohmian Mechanics is empirically equivalent to the bare quantum formalism. Now if the latter is capable of generating empirically testable situations, as suggested in the previous section, despite not being able to guarantee ontological separability of all elements of nature, then we have a way of dissolving the worries about potential consequences of the implications of violations of separability for the 'transcendental' argument. Bohmian Mechanics, just like the minimal versions of quantum theory focused on the predictive manipulations of the formalism alone, can support enough stable empirical structure for the postulation of existence (and tentative guesstimates of) laws of nature.

The problem is, though, that it must treat the fundamental element of the bare formalism, the 'system' wavefunction featured in the Schrödinger equation, in the same way as principle approaches do: as a rational guesstimate of the state that the particles of interest are in conditional on the state of the remaining particles in the universe and the universal wavefunction. Great deal of mathematical derivation is employed to show that this can be done (cf. references in Chapter 3), but even more is required to show why this must be so: i.e. why we cannot simply directly read off the state of the universal law and its effects on the local particles (the 'objects at hand' required for the depth-of-explanation). The latter is enshrined in the assumption of the quantum equilibrium hypothesis, a constructive version of the constraining principles. Given this hypothesis which limits in principle what we can epistemically access concerning the fundamental ontological elements, the ordinary system wavefunction is the best information[109] we can have about the system at hand (Goldstein, 2007, p. 13). Though this seems to play into the hands of the principle approaches of Chapter 2, it needn't be seen as such. The further step is provided by allowing us to draw inferences about the nature of reality that operates in a regulated and understandable manner even with this epistemic limitation.

Goldstein (2007) claims that this is by no means putting the cart before the horse, because the analyses cited in Chapter 3 (most notably (Dürr, Goldstein, & Zanghi, 1992)) show that we are justified in treating these rational guesstimates from the postulations of fundamental ontology as genuine probability statements about real-world events, statements that are relevant to characterisation of what phenomena we actually expect to experience (and can experientially verify). Our survey of the issue above should also convince us of the conceivability of this claim. Moreover, Goldstein draws on formal analyses that show that no more detailed information can be available about the changes in the fundamental material ontology than is given by the system wavefunction which respects the constraint of the quantum equilibrium hypothesis.

[109] It is important to note here, though, that the marriage between the principle approaches and this particular constructive approach is not as straightforward as suggested by Goldstein (2007) in the light of the lengthy discussion above. Most notably, what Goldstein and our answer to questions posed above are referring to is the qualitative sense of information ('type information' along the lines suggested by Timpson (2004) above), and it remains to be seen what its relationship to the quantitative sense that the principle approaches employ is. It is a further task for the constructive approach along these lines to show how the principle approaches can methodologically arise, given the nature of reality as suggested by this particular constructive approach. This is not necessarily an impossible task, but is one requiring further elaboration than is given in simple equating of the effective system wavefunction with as-complete-as-possible information about the state of the particles of interest.

Thus the gap between the knowledge of the occurrence of the phenomena and its understanding is bridged by claiming that the phenomena arise (through a structural isomorphism) out of the spatial configurations of the fundamental ontological elements, when governed by the universal law of temporal evolution. As the law itself is not directly epistemically accessible to display this governing, we rely on the informationally as-complete-as-possible guesstimate of its proscriptions given by the system wavefunction that is formally conditional on the state of all the particles in the universe and the universal law governing them. The further why regress, as to why the universal law proscribes what it does, is stopped by its fundamental ontological status: the (epistemically unattainable) universal wavefunction just is the formal expression of the universal law governing the spatiotemporal changes of all fundamental building blocks of material objects.

Despite the objections of the anti-realist critic we peacefully accept a certain form of ontological holism when we employ the equilibrium-conditioned guesstimation of the 'universal law plus particles' mechanism (i.e. information codified in the universal wavefunction) in our experimental situations. But the said holism is not threatening as it still allows us to formulate empirically adequate rules (though not themselves the fundamental laws of nature but conditional on them) regulating the occurrence of the phenomena. Goldstein claims that there are further mathematical guarantees that "the observed deterministic regularities would be classical" (Goldstein, 2007, p. 16). That is to say, formally we can expect all the observed regularities to be obeyed, just as our core conceptual framework requires. Through the notion of "local beables", the fundamental extended entities of the material reality, we have a direct structural and conceptual isomorphism with the core concepts of the everyday conceptual framework, such as are given by the primary qualities of directly identifiable objects. But, our explanation of the troublesome phenomena, requires that we admit into the ontology another essential element: the universal law that allows (in fact instigates) the elements of the material ontology to behave in a non-separable way the macroscopic effects of which we observe in the troublesome phenomena. Nonetheless, given the constraining principles (and the constructive account of their origin) the foundation of the conceptual framework is not jeopardised as its elements are not shown to be illusory: we can use it as the starting point of the transcendental strategy. Moreover, the beables give us a straightforward way to identify the object that undergoes real and counterfactual changes in the situations that we aim to explain. Still we will have to say more below about how exactly this proceeds, i.e. what kind of explanation is required to marry the partially non-separable ontology with the seeming expectations of separability from the everyday conceptual scheme.

Yet, does this legitimise us saying that we understand the interactions between separated formally entangled objects, any more than the establishment of limiting principles for information manipulation does? We have to be careful not to use this question to slide back into the view of universal wavefunction as the all-permeating field that takes on to itself the mechanical influences from the local particles and transmits them in a non-separable way to the distant set of particles (and vice versa). This was shown to be an erroneous view from the beginning as even the formalism does not encode any influences from the particles to be 'recorded' in the effective wavefunction. We might, though, expect them to be recorded in the fundamental one, only not expressed in its derivative – the manipulable effective wavefunction. Goldstein is adamant that we must never confuse the effective with the universal wavefunction, although the former is dynamical and manipulable, the latter is not even expected to be.

> "But for Bohmian mechanics, that the [universal] wave function does not change is, far from being a problem, just what the doctor ordered for a law, one that governs the changes that really matter in a Bohmian universe; of the variables Q describing the fundamental objects in the theory, including the 3-geometry and matter." (Goldstein, 2007, p. 18)

Yet, the universal law itself, upon which so much hinges in this explanation, is in-principle unknowable and directly susceptible to be ostracised by Bub's 'deep methodological principle' for example. The answer to this requires drawing on the realist traditions that claim that we can know a law exists even if we don't know exactly what it is (Bhaskar, 1978). This is to widen our transcendental strategy to include the causal ascriptions of reality alongside structural durability of objects. This, in turn, was argued for by the abandonment of the epistemic atomism of the structural state of matter in phenomena, and shifting the focus on the atomism of an enduring object that undergoes changes in the phenomena (Harre, 1996). That is, we have to permit ascriptions of reality that result from a causal natural (not logical) necessity, as well as what is mediated by the bare structure of extension alone.

Our second worry might be that in explanatory sense elements of this constructive approach are pushing us back to the disregarded world-view of pre-established harmony. Namely, if there is no mechanism through which the material ontology (the particles) affects the *effective* influence transmitter, the universal wavefunction, are we not consigned to the blunt view of individual particles locked in a monadic dance choreographed by the universal law? The picture is one of perfect clockwork, but clockwork where no hands can be stopped as there are no influences actually transmitted between different elements of the mechanism. But to accept this criticism is to be too attached to the mechanistic world-view as the only form of causal world-view. As the discussion in Chapter 1 showed, this can be historical mistake also, and there are precursors even in classical mechanics where we have been forced to accept the action-at-a-distance without the mediating mechanism. The question is how we were not pushed to considerations of pre-established harmony then. And the answer is, some, like Leibniz, were, but the rest of us just took it to be a non-explicable (and thus foundational) fact about the world that objects with mass will affect each other at a distance. The effective regularity was there (even in the absence of the mechanism) and that was enough. It had to be.

Likewise, we can allow that the universal law *specifies* (but not transmits) how the fundamental objects will affect each other in interactions. Moreover it tells us how the relations established between the objects will reflect in their local states, by having some glimpse of the law we can learn more about the states of the objects than we can simply from observing each of them in isolation (i.e. locally), because the law provides a rule by which such inference is legitimised. The law, or what we can derive from it, will also tell us what to predictively expect of the objects, but due to limitations of derivation, will not tell us exactly what will become of them in the future. The derivation on the other hand should be sufficiently formally regulated to alleviate fears of chaotic modification (most notably those affecting the possibility of reidentification) of the structured state of the primary qualities of the said objects. Thus in our troublesome phenomena we can generate counterfactual situations in which we can show how the interventions on one of the particles produced 'regulated' effects on the distant one (for example by 'providing' the *local conditions* required for informationally rich future measurements on a distant particle in teleportation – without actually instantaneously moving the particle itself; as well as allowing the proximal experimenter to predict the results of potential measurements on the distant one by 'reading off' the conditions set up in the universal law and the states of the local particle – previously coupled with the distant one).

The fundamental material objects will undergo changes in such circumstances that cannot be *predicted* from the state of their local environment (even from the limited local derivation of the limited conditional wavefunction – the mini-law), but that can be predicted – and what is most important for us, can be explained – when a more global perspective is adopted, better conditionalized on the universal law specifying their changes in time. Just as in Newtonian mechanics taking increasing number of distant masses into consideration (but under a rational guidance of what is sensible for the given situation) increases the predictive capabilities of the change in non-inertial motion of the local mass. We gain better understanding of the mass' behaviour when seeing it as a system of 'gravitationally' (i.e. 'regulatedly') interacting objects, then when trying to account for potentials for re-identification of a single isolated massive objects seemingly irregularly undergoing changes of inertial motion (most notably the changes of the rate of change of position). The depth of explanation is provided by showing how potentially varying the state of the objects crucially involved in the interaction changes the phenomena in a regulated way, by showing how adding or removing the masses and altering their relative positions affects the phenomena in the way that simply altering the position of a single mass in isolation cannot. Yet we have to see how this explanation is better for the 'transcendental' argument than the one in which no significance is attributed to the mass of the objects themselves, but rather to the general constraining principles governing their interplay.

Of course, just as is the case with the principle approaches, there are further technical difficulties to be resolved, most notably those of rigorously showing that given all that we can rationally infer about the universal wavefunction we are justified in holding the conditional wavefunction to be behave just as expected from the empirically successful bare quantum formalism. That is we need a formal demonstration how the system can be for the purposes of many versions of the transcendental strategy suitably decoupled from the totality of the universe and that most complete Schrödinger equation for the totality of the ontology in the universe can at least have an appropriate form (given that it can't be specified exactly) (Goldstein, 2007, p. 19). Still the arm-waving information provided above should allow us enough insight to compare the potential for deeper explanations concordant with the transcendental strategy of section 1. 4 and the occurrence of the troublesome phenomena. We are also interested in addressing the general structural components of such explanations to be applied to the common-sense conceptual framework. The starting point, though, should by now be clear; we should be able to see what Bohmian Mechanics on the final rendering from the end of Chapter 3 says about the nature of independently existing reality.

## 4. 4 Explanation and the two approaches

From the perspective that accepts Bub's 'deep methodological principle', the perspective of empiricism, even though the local beables enable us to make an easy and intuitive connection with the fundamental structural features of the direct experience they cannot furnish a deeper explanation than the explanatory models that don't contain the right sort of beables at all (the principle approaches in our case). Though on the face of it, the constructive approach seems appealing because of its structural similarity to the much preferred causal mechanical model, at the present stage of development the appeal is a result of an illusion. The reason for that is that with committing to metaphysical postulates the constructive approach cannot avoid the dangers of the separability violation in the right way. If it consigns them to an action of non-spatial entity, such as the universal law, it is merely hiding behind another cloak the bare phenomenological generalisation of no-signalling prohibition: we cannot know the exact mechanism by which the action-at-a-distance phenomena come about. The mechanism is there, it does not involve transmission of influences along space-time paths, but we are forever prohibited from knowing exactly how it works (how the proscriptions of the law in a limited region relate to all the relevant proscriptions in other regions, i.e. what the global wavefunction is).[110] What they effectively say is that though the ultimate universal wavefunction is informationally complete (though, crucially, not ontologically complete in the terminology of (Maudlin, 2007a)) in-principle limits of knowability prevent us from making explanatory use of the completeness.

So we end up in the same mess as those who claim that effective wavefunctions are informationally incomplete (e.g. the Fuchs principle approach as presented above) and then have to search for the ontological account of the limits of knowability. What constructive explanatory account in fact presents us with is the pre-established harmony situation, where distant elements of reality sometimes affect each other without any (epistemically accessible) intervening mechanism established between them in space. The effect is 'transmitted' (and that term has to be taken with great caution here) through the causal action of a fundamental law, the universal wavefunction, so as to allow for some visible correlations between the states of the separated and separable elements of reality. And this, warns van Fraassen ((1989b, p. 112), original emphasis) can only accomplish two functions: to postulate an entity that has either predetermined all our supposedly free interactions, or simply coordinates what we call an interaction 'externally' to both parties; or "*to admit that we have no explanation but to refuse to consider the correlation mysterious nonetheless*". Pre-established harmony is just not a token causal-mechanical type of explanation, and cannot pride itself on having its traditional virtues. Yet as we have seen, the principle approaches struggle to even get a deeper explanation off the ground as they refuse to rely on any causal structure that is not a product (and not a pre-requisite) of our explanatory conceptualisations of the phenomena. Due to 'troublesome' nature of our phenomena of interest they cannot find any such stable structure and are forced to relegate explanation-stumps to the unfamiliar territory of abstract entities that are strongly mind-dependent.

[110] This, of course, holds for the case-study instance as presented here for the specific purpose of comparison. There are in fact suggestions in the literature (cf. (Valentini & Westman, 2004) for recent suggestions) how the limit of knowability may be circumnavigated or removed, and suggestions for empirical verification should certainly be explored. In the present case, however, we take them to still be lacking and that the in-principle unknowability as resulting from the quantum equilibrium state holds.

The difference between the approaches in the end lies in the philosophical position adopted, as might have been expected from the initial empirical equivalence of the different 'quantum theories'. From the perspective of trying to provide an explanation sufficient for the transcendental strategy but *limited to the epistemic accessibility* of the ontic concepts employed, i.e. the perspective where ontology is largely reducible to epistemology, the two approaches come strikingly close together, despite explicit methodological differences. The principle approaches are forced to admit a dispositional aspect of the properties they venture to ascribe to the elements of material reality (the *'be-ables'* instead of 'beables'; cf. (Howard, 1989)). These properties, despite quantum theories' success in providing predictive laws, cannot supply sufficiently firm grounds for the 'transcendental' argument. Nothing can be known about reality-in-itself, as it is so unpredictably sensitive to observation-intervention. The principle approaches are forced to this position because they are unable to say what the ultimate nature of ontological elements is, beyond their dispositions to exhibit certain properties *when prompted to do so by experimenters' actions.* They start off with what is directly epistemically accessible and happens to be enshrined in the conceptualisation of persisting objects, but there is no possibility of linking (in the present state of the development of the programme) the potentially dispositional concepts employed in the explanation of the 'troublesome' phenomena to this conceptual framework as conformant with minimal realist structure in the transcendental strategy. Their explanation of the phenomena (including the 'troublesome' ones) in terms of fundamental ontology cannot provide truth-conditions for counterfactual situations centred on the ontological elements to provide an explanation that goes beyond the regularities predicted and observed. They leave the changes the ontology undergoes as mysteriously holistic and essentially indescribable. In that they struggle to both bridge the gap between knowing that the phenomena occur and why they occur as they do. Furthermore, the why-regress cannot be easily stopped.

The constructive approaches overcome this problem, by speculating on the nature of the universal law that governs the changes (thus making the changes be real alterations from one state of the exiting property to another such state, not from a collapse from a spectrum of dispositions into a concrete state). They view the phenomena as a wholesome process of an operation of a really existing law on a really existing ontology, with occasional disregard for the spatial and mechanical structure relating the existents. Yet they are forced to admit an in-principle epistemic inaccessibility of this law and rely on elaborate mathematical speculation as to how we can gain incomplete glimpses of the requirements of the law. By philosophical commitment to viewing the phenomena as more than a series of events, they venture to offer grounds for the transcendental strategy of direct structural connections between the fundamental ontology and the common-sense conceptualisation of the world. But for the argument to succeed they have to modify the starting point of the transcendental strategy to recognise more structure in the common-sense conception than was originally envisaged. Through this loophole they can make the requirement acceptance of what is in-principle forever epistemically inaccessible as real. For it to be acceptable, the anti-realist critics would have to be convinced of sound reasons to abandon empirical realism in general analyses of the conceptual framework, to replace it by so-called transcendental realism.

Instead of seeing the principle approaches as conservatively limiting speculation to the merely epistemically accessible, we could view them as committed to the 'epistemic fallacy', when shying away from the ontological investigations behind the apparent phenomena (especially, 'troublesome' phenomena). In that way they commit to the Humean Mosaic of series of appearances, such as the informational relations that are established between the separated 'black-box' instruments, but that do not inform us of the laws governing the objective non-instantaneous behaviour of the black boxes. The laws that they do establish as the principle generalisations cannot be understood as causal laws in the material domain at all. They can at best concern the information-ontology. And this is to be expected of efforts to stick to the metaphysics that is always reducible to epistemology: it creates its own 'implicit ontology' and 'implicit realism' (Bhaskar, 1978) only in special domains. So we get the ontology based on the category of experience and a realism based on the presumed characteristics of the objects of experiences, in this case the expected informational content of the formal quantum states. Bhaskar claims that such strategy leads to the generation of methodology that is either irrelevant to science, or relevant to science but inconsistent with epistemology. We can see this well in the information-ontology speculation, where we are either waiting to be told of what the relevant connection to the material objects of science is, or we have to establish the new science of information-manipulation but one that is difficult to connect to the conceptual scheme of our everyday (but also experimental in this case) experience of objective laboratories and 'black-box' machines.

The greatest weakness of the principle approaches, despite the expectations we might have had of them at the outset is that they struggle to connect the abstract novel ontology to the foundational elements of the core-conceptual framework. The ease of unification of concepts was meant to be their greatest strength, but in the light of the conceptually challenging 'troublesome' phenomena its advantages have been lost and the principle approaches have been left with inadequate resources to connect the structures of the two ontological realms (the informational and the material). On the other hand, when they venture to establish this connection they jeopardise either then separable 'individuality' of elements of material objects (as in instrumentalism) which their particular methodology aimed to preserve, or the requirement of the transcendental stem that there are epistemically accessible facts about the world. In the light of the world-making charge, they don't provide a way to successfully conceptually connect the novel ontology (at this stage this can be entities or properties) with the foundational aspect of the common-sense conceptual scheme required for the transcendental strategy for simple realism. The novel ontology on its own though, is not sufficiently well understood to be able to enter the common-sense conceptual scheme (and there are claims that it never will be able to aspire to such status; cf. Timpson, 2004; 2008) without this mediation via matter as extended substance.

In the light of explanation of the 'troublesome' phenomena' required for the successful universal application of the transcendental strategy, the principle approaches lack a foundation for a formulation of a causal law that can abandon the perspective of the Humean Mosaic. Such mosaic, when featuring elements otherwise included in the 'troublesome phenomena', does not on its own provide enough structure to extract concepts of enduring objects that play essential role in the starting point of the transcendental strategy. The general constraining principles of information manipulation do not provide sufficient conceptual ground for a causal law 'limiting' changes in matter, and a parallel notion is not offered in the information ontology, so the transcendental strategy cannot be constructed. That is their weakness as compared to the constructive approaches when providing the explanation needed for the transcendental strategy to succeed given the apparent separability violations in the 'troublesome' phenomena.

Humean Mosaic as presented above fails to account for the necessity and universality of laws which in turn, given our empirical experience of the 'troublesome' phenomena, requires an explanatory ontology that cannot sustain the transcendental strategy for its very existence. The 'troublesome' phenomena teach us that we can come to know more, and manipulate more, of the real world when we take it to be a structure of material ontology (with essential features related to separability) coupled with (at least partially epistemically accessible) causal laws that govern the changes under which it still maintains identity as the fundamental object of experience. Reality is on this strategy attributed to (initially) speculative ontology on the basis of causal lawfulness as well as direct perception of states of objects.

Explanations aim at global economy of concepts, but such that provides for a greater variety of changes in characterising the object enduring through the phenomenon. In fact, the phenomenon is to be set in the conceptual network as the regulated change of the enduring object. Through this object-concept link such explanations connect to the transcendental strategy which rests on the universal acceptance of central role of the concepts of individual enduring material objects. If the concepts taken as central in explanation of the phenomenon are not characterised as enduring identifiable entities (and re-identifiable through further development of the situation) then such explanations struggle to connect to the realist attitude developed through the transcendental strategy that is seen as the background securing superiority of such explanations over provisionally constructed narratives. Traditionally, primacy of extension was seen as a possible straightforward connector of the speculative elements of explanatory narratives and the segment that is directly epistemically accessible. That notion of extension carried with it some constrictions on behaviour attributed to separation. Quantum theories deny that constricting role of separation. Our case-study explanatory constructs are challenged to give an account of what replaces it.

When we construct explanations we come to rely on more than on what can be predicted, we retrodict to an account that makes sense, that unifies a particular experimental experience (which is, crucially, more than just an observation of the local state of material existents upon completion of the experiment) with the core elements of the conceptual scheme. The constructive approaches that deny the Humean Mosaic can, at least notionally, achieve this by showing the phenomena to be a product of extended objects and specific (classically unheard of) causal laws, the latter giving rise to the changes in material ontology that cannot be accounted for solely from its local powers. It remains to be seen what the price for this 'achievement' is.

### Fundamental ontology and the acceptance of the universal-law-Bohmian worldview

The constructive approaches overcome all these obstacles, when they venture beyond the limits of direct epistemic accessibility, but we must address the question whether the ontological (potentially non-separable) features can be distinguished from the epistemological features of the explanation they provide. The aim is to guarantee the isomorphism between the gross structure of directly experienced reality, as the prerequisite of the transcendental strategy, and the fundamental structure postulated by contemporary quantum theory. Maudlin (2007a) seems to suggest that we can, as Bohmian Mechanics marries the "local beables" (entities characterised by the primitive constrictions of extension, including related spatial separation) and the non-local, but also non-material wavefunction. So, from the perspective of section 1. 4 we seem to be on safe ground: our experience is connected to our projections about fundamental ontology through an enduring stability of the essential function of the spatial extension (this can be enhanced to have a temporal element in the relativistic sense, as well) as given by the theoretical fundamental entities, the spatially located material particles with finite extension, separated by finitely extended spatial regions, in a word: the local beables. We say to the anti-realist that although tables and chairs are not the fundamental elements of our ontology, we can reconstruct them out of the fundamental elements that have an irreducible property of spatial extension. We can reconstruct them out of *essentially* similar components.

Yet, the extended objects we were expecting to provide a conceptual foundation on which to unify the scientific and the everyday accounts, seem to harbour a threat for our transcendental strategy. Though we cannot consciously and willingly subject them to unpalatable changes, we must accept that they can in the end undergo just such changes. If the phenomena elucidated by the theory involve an abandonment of locally specifiable intrinsic 'thisness'[111] of objects, such as seems to be the case with teleportation on a constructive approach interpretation, then we seem to lose the desired connection to the common-sense conceptual scheme. And the latter we required to get the transcendental strategy off the ground. Though maintaining some aspect of the explanatory conceptual framework as primitive and fundamental prevents us from a rapid slide into anti-realism of the world-making type (cf. (Devitt, 2006); (Devitt, 1997)), there is still a worry that if the concepts of properties we use (all of them) have a mere dispositional basis in the real world and no direct structural correspondence with the realistic interpretations of contemporary science the common-sense realism will not have a sufficient conceptual foundation (in the 'geometrical' structural isomorphism alone). The problematic dispositional basis lies in the proscription of the 'thisness'-bearing properties by the universal law, effectively calling for their reduction to the law. This is, in a sense, saying that fundamentally there is just a law, but that expectation is difficult to connect with the starting point we need to diffuse the world-making challenge. Especially, as we also have to admit the universal law is in-principle unknowable. This is where taking quantum theory (at least in the case-study instances considered here) seriously exposes constructive scientific explanations to difficulties similar to those in the more ontologically economical principle approaches, and threatens the tenability of the 'transcendental' argument of section 1. 4. Now, the crucial thing in the fundamental status of some properties for the realist explanation, was the constriction (independent of humans) on what could be said about the world. This was the realism's upper hand over unrestricted world-making, as the latter could not account at all for why some conceptualisations work better than others. But, unlike the case of unrestricted world-making (cf. for example (Putnam, 1981), (Pettit, 1991)), in the case of fundamental laws we do once again have constrictions imposed by the real world: constrictions on what experiences can be expected of the phenomena.

In applying the transcendental strategy we accept that there are two ways in which our concepts of dispositional (or secondary quality) properties are restricted: by how they depend on us and how they depend on the real world. The later provides a structure-characterised base for the transcendental strategy; our conceptual frameworks already contain concepts that can be identified as this base. Explanations of the phenomena can then be built through account of causes (which are often elucidated through the explanations of the mechanistic style) of the activation of our disposition to judge the situation as characterised by a particular secondary quality. Thus we explain away the illusion. But this causal dependence can be relied on only if it is open to empirical investigations of the constraints it imposes on our thinking and concept formation. And traditionally, again, here we employ the connection of theorising with material reality and eventually the everyday material objects (the starting point of the transcendental strategy). And those, in the end, rely on a transcendental strategy of accepting that some of the concepts essentially associated with them are those 'natures' of those objects and as such *are not dependent on our judgement*. The question we now face is whether the law alone can provide the required natures.

[111] Again, as a reminder, this is an arm-waving intuitive thisness, not the technical terms of R. M. Adams and others.

But first, it might be objected that the law we want to rely on, the potential fundamental law of temporal evolution, is in-principle unknowable. So we seem to be saying that, unlike the world-makers, we can know how it is that some conceptual categorisations are better suited to explaining experiences than others, and this knowledge relies on the in-principle unknowable ontological element: the universal law. Let us pause to carefully unravel this conundrum. To start with, the universal law is not entirely unknowable, as we have useful 'effective' glimpses of it through the effective wave-functions of the quantum formalism. It is, on the other hand, unknowable in sufficient detail to make the formalised 'glimpses' any more than epistemic prediction tools. But to argue from that that it is entirely 'unreal' is to commit to an 'epistemic fallacy' (Bhaskar, 1978), to expect the ontological claims to be confined to the same limits as the epistemic ones (in our case the limits of knowability). There is no urgency to accept this limit (which is essentially Bub's 'deep methodological principle') if there is hope of providing a deeper explanation of the experienced phenomena than the principle approaches can hope for. We can direct our explanations to answer ontological questions without having to transpose them into epistemological terms first. We can accept Bhaskar's claim that causal laws are ontologically distinct from patters of events that are epistemically accessible to us. This would allow our transcendental realist of section 1. 4 to argue that given that we have the science we have (i.e. a functioning quantum formalism) the independent reality must exist and be of a certain type. But a further problem to resolve is how this type can be unified with a common-sense conceptual scheme so as to avoid the charge of world-making.

The problematic task is then to show how constructive approach explanations can be united with the common conceptual scheme, in the way that that unification is easily achieved when the questions are reduced to the epistemological realm. We must also show what further benefits an ontological speculation can provide, other than chiding well with the structure of the depth of explanation. It is, perhaps, important at this stage to include one final step towards the connection with the 'troublesome' phenomenon of teleportation. We have already seen its greatest mystery lies in supposedly infinite availability of information to the distant experimenter, which has in the end dissolved as a characteristic of a global not a local set-up. In other words, that experimenter needn't worry that his account of the phenomena generated locally will be factually incorrect concerning the local features, only that it can be improved by updates from a special person – the holder of the other half of the entangled pair. Again, the no-signalling prohibition (enshrined in the structure of the formalism already), an epistemic consequence of the quantum equilibrium postulate, guarantees that with respect to local predictions (and even subsequent explanations) the distant experimenter Bob will not be able to tell whether anything metaphysically significant has happened to the object in his possession. However, once he takes Alice's manipulations into account, he will come to know that the object in his possession has indeed undergone important changes to its 'thisness'. Here is where we tread a fine line separating this account from utter ontological holism. Though individual experimenters working locally cannot gather enough information to be certain that they can successfully re-identify objects they are working with, the full-blown world-making is still restricted by the supposed existence of a generative mechanism that allows certain changes to the local object, and only those. The worry to address is whether this is sufficient to allow the transcendental strategy to get off the ground.

Namely, if in the teleportation process the particles, the local beables, do actually instantaneously traverse the required distance (effectively instantaneously swap places across any distance) and carry with them all the interactions with the universal law then our simple strategy of reducing composite objects with intrinsic 'thisness' to equally primitive constituent local beables fails. This is because the identity of the constituent beables can be changed at will from any location in the universe (provided some special operations in the past light cone, potentially very deep into the past), and the occurrence of change is not open to objective investigation. Some experimenters (Alice) have a unique epistemic position in the universe concerning the relevant beables. Given such a situation we have no guarantee of locally detectable endurance of the fundamental objects, against which to construct deeper explanatory accounts.

If on the other hand the beables themselves do not traverse the distance, but merely serve as placeholder for different property ascription by the universal law, then we may wonder as to their utility in the first place. And here the greatest weakness of the realist project on the constructive account is exposed. Detailed analysis of it would take us too far away from the limited project of comparison of the explanatory potential of the case-study instances, so only a brief sketch will have to do. On one hand it seems that we could replace the constructive account as given above by one that removes the very bare place-holder particles and converts the informationally rich universal law into the only existential primitive, either as a holistic single field in geometrical space or even as s more complex object in higher dimensional spaces. This is a return to the Humean Mosaic viewed in its entirety as the variations in structure produced by the ontologically holistic wavefunction, all of which structures are non-separable in a metaphysical sense. We could, as a sketchy illustration, imagine this as a sea (that maybe consists of individual 'water-particles' and maybe doesn't) where all the structures interesting from the perspective of the starting point of the transcendental strategy are further structures created by the sea such as waves, and whose identity is not necessarily tied to individual component 'water-particles'. A realist might immediately object that this illustration also commits us to belief in space and in which both the water-particles and the emergent wave-structures reside and endure through changes, which might again give some explanatory primacy of individuation to the water-particles occupying specific positions in space. But less sketchy structures of this kind can be devised, perhaps along the lines of super-substantivalism (cf. (Sider, 2001), (Schaffer, 2007)) which reduces all emergent structures to property ascription to space alone.

Yet we might worry that generation of common-sense concepts along those lines mimics that of anti-realist constructivism that effectively leaves the account of how concepts depend on the real world unexplainable and empirically untestable, making the explanations offered a mere empty facade of constrictions on our thinking and concept generation. Following this route is helped by the non-separability inherent in all aspects of the universal wavefunction, with no conceptual reliance on the characteristics of concepts from the framework that rely on the inherently separable elements of the world. Even with the (principally given and unexplainable) no-signalling constriction on prediction, in explanatory retrodiction we admit of possibility of dispositional responses that generate object-concepts to be systematically thwarted so even those concepts become inherently vague.[112] Effectively this in the end makes all our concepts inherently vague, barring further explanatory account, which should in principle be empirically testable at least in part, of how the concepts of common-sense depend on the real world. "Reality may be indeterminate, and the cognition of reality may be subject-involving, in certain surprising ways." (Pettit, 1991, p. 623) In the ascription of the independent reality to the bare particle-objects in constructive approach above we tried to cling to the notion that something at least can be pointed to as the real constraint on our judgements as to the character of the independently existing material reality. In a realistic account with some aspects understood as essential the constraints were provided by the typings of objects that are not dependent on us to explain the conceptual frameworks they provided. "A little bit of world-making is alright against a background of a world that we did not make and that influences our little effort." (Devitt, 1997, p. 255) What we have to bear in mind is that the universal wavefunction in this materialised form is still in-principle unknowable, so we are short of constrictions for the explanatory account of the difference in Alice's and Bob's local accounts of the, for example, teleportation phenomenon.

But if the above sketch is a convincing exposition of the slide into anti-realism, such as the transcendental strategy tries to avoid and that the explanatory accounts of the 'troublesome phenomena' of the constructive approaches aim to outdo, it is one could argue that the view advocated by Bohmian Mechanics above cannot avoid collapsing into them. Even on that view the distant experimenter Bob cannot ever come to know the important changes that occur to the objects in his region of space, until Alice broadcasts the true account of her actions on her particle, even though his local object has really changed under the influence of the universal law. We don't need anything to travel between the distant locations, but the important changes to his objects are at some instant prior to his investigations hidden from him and anyone else, but not from the other experimenter Alice, until she announces the results of her local actions. Now this is not to say that the stability of all objects is forever thrown into doubt, for Bob has a reason to be careful of what he assigns to his local object given that it is one half of the original entangled pair (and not just some object picked at random), but however hard he tries he will never come to know fully what its local state is. Though this needn't immediately put the possibility of constructing the transcendental strategy into jeopardy, it does place a great onus of what we importantly need to know about the world onto the epistemically inaccessible law. Effectively, without knowing what the law proscribes, for Bob there has never been any teleportation at all, and yet the explanation of the phenomenon requires that the local object has been altered in a dramatic fashion (which is just short of saying that it has been entirely replaced by a different object).

[112] An intuition behind this is the classical consideration of the Ship of Theseus, all of whose parts (boards, beams, masts etc.) get replaced with time by the new wooden elements of the same shape (new boards, beams etc.). Though relying on the form alone we can say that it is still Theseus' ship, particularly as we can account for the history of changes of its elements. If 'another' ship is reconstructed again from the original boards and beams (say they have been cleaned and the rotting has been stopped), when the two ships are compared side by side we are still tempted to call the *reconstructed* one the real Theseus' ship. In this re-identification the actual history (from being a part of the ship to being taken out and cleaned) of the particular constructive elements (the boards and beams) plays an intuitively important role. This is not to say that replacement of constructive elements automatically destroys the identity of objects, nor that individual humans become new people when their cells are replaced, but that the account of the history of these materially fundamental elements is somehow important in the common-sense accounts of individuation.

The minimal realist constriction provided by the constructive approach as given in the previous chapter, relied on the separability and durability of the material existents, the spatially located particles. However, that may not be enough to allow for the explanations of the troublesome phenomena that respect the transcendental strategy. For the strategy itself is considerably weakened the more of its starting concepts we take to be dispositional (or response-dependent in the sense of (Pettit, 1991)). In simplest of terms, the bare durability of the extended stuff in space may not be sufficient to explain all the appearances readily found in the common-sense conceptual scheme. That is why the most comprehensive, blunt, forms of the transcendental strategy, as renewed in for today's purposes outside considerations of quantum theory, take more of the elements of common-sense framework as directly related to the nature of things in the real world, minimising the world-making as much as possible (cf. (Luntley, 1995, pp. 118-119; 235 note 6)). This is abandonment of our strategy above to select from the conceptual framework in which we present the immediate experience that which is directly related to nature of things in the real world, and using it to explain that which is illusory. Through the considerations of the 'troublesome phenomena' as presented in Chapter 3 we have come to leave only the barest of spatial position as directly characteristic of objects of the common-sense framework, and reduced all other aspects to the 'illusion' structurally dependent on the universal law.

As these considerations take us further away from the investigation of the explanatory ontologies of the actual case-study instances we shall stop short here with a few remarks. Explanation features strongly in our strategy, and it requires a conceptual unification of the diversity of phenomena through primitive concepts. Traditionally (cf. illustrations from Descartes in Chapter 1), extension is one of those primitive concepts and it has strongly featured in the traditional versions of the transcendental strategy. The more blunt of those versions as suggested by Luntley's later remarks (1995, p. 235) are "contentious and still poorly understood". This is not to say that they are wrong, but only that they require further deeper investigation as to how they differ significantly from the anti-realist world-making accounts (cf. just for illustration (Rorty, 1980), (Putnam, 1981), (Pettit, 1991)). In the end our explanations will require reliance on structures that restrict the world-making, however liberating acceptance of some world-making might be. The holistic material structure does not provide enough of those restrictions as the typification it provides for the generation of concepts is fuzzy due to effective *dependence on our judgments to interpret* the structure emergent from the holistic ontological substratum as such.

### 4. 5. Playing the constructive game, retypifying common sense

This seems to be the precarious situation we are in. To discourage anti-realist criticism we had to show the possibility and explanatory utility of the transcendental strategy from the basic structures of the common-sense conceptual framework to the fundamental ontology of all phenomena experienced in an *interaction* with the material world. Those phenomena included some 'troublesome' instances generated in the domain covered by quantum theory. Those instances appeared troublesome for they seemed to provide an experiential basis for the denial of the realist-style validity of the elements of the common-sense conceptual scheme we take as the starting point. The latter is most notable as an individual 'thisness', given by the constrictions of extension taken as primitive and isomorphic in both the fundamental ontology and the objects of common sense experience, including the role of spatial separation in the conceptualisation of identity ('thisness'). So as not to block the possibility of the unified explanation of the everyday experience and the 'troublesome' phenomena in terms of fundamental ontology we had to add further non-separable elements to it. Yet that very element, the universal non-separable law seems to be more problematic than expected as it is outright characterised by ontological holism and potentially more important for the desired explanation than the extended material ontology taken to be the fundamental connector between the 'troublesome' and the non-'troublesome' in phenomena.

This is the lesson for explanatory accounts to take from the struggles of constructive approaches to provide deeper explanations than principle approaches can (though, for the time being, there is still no verdict whether in fact they can achieve that): neither the bare surface structure of the phenomena nor the human constructs imposed on the interpretation of them are sufficient for deeper explanation. A deep explanation that can still serve the transcendental strategy is concerned with the structural constraints which endure despite not being directly epistemically accessible. That is, in the above account the phenomenon is not given by the bare fact of the appearance of the correlations between distant measurements, it is given by the whole account of the experimenters' production of the correlations with manipulations of macroscopic equipment as objects in space and time. This seems to require also that our transcendental account starts not only with the conceptual framework of objects with certain essential structure (in our simple case, the geometrical structure of spatial extension) but with a wider framework of the interactions and changes those objects can endure (and still be re-identified as the same objects) and the effects we as human agents (and not pure observers) can have on them. This is asking for a slightly higher price for our transcendental strategy, but not a price that must be unacceptable to the antirealist critic. After all, our experience of interaction with objects is as much as part of our everyday conceptual scheme as is the bare experience of perceiving those objects. If so much is admitted we can add to the essential requirements of isomorphism not just the durability of extended objects but also a notion of regularities of the changes they undergo.

That is, it seems that we have to be careful not to presuppose in the starting point of the transcendental strategy that at any instance *a total description* of the situation is embodied in the purely empirical descriptive concepts employed. Those concepts are ones of objects not bare geometrical structure, and the former include an implicit understanding of the causal/lawful properties as well as the spatial ones. These properties must also be understood as primitive, and not dispositional. The essential structure is given by the objects' shape and the existent laws that can act on it in the right circumstances. These laws are not observable to us in the same way as individual material entities, but are inferentially no less real than material structure, and cannot be reduced-away in terms of locally (i.e. not a total description) specifiable concurrence of events (though, this is how we ate first come to speculate about their existence, to form the required metaphysical projections).

### The price, in terms of conceptualisation, of the constructive depth of explanation

The constructive approach of Bohmian mechanics, outlined in the previous chapter, denies metaphysical separability, whilst nonetheless trying to avoid the threat of ultimate full and complete ontological holism. The latter would provide a non-starter for our defence from antirealist criticism from section 1. 4. as we take it to invite response-dependency for all concepts of the common-sense conceptual framework. In the light of the previous section, the constructive approach argues that in the retrodictive explanation of phenomena we must contend with the violation of separability as we come to know that the physical processes in some spatiotemporal region are not wholly supervenient on assignment of qualitative and quantitative physical properties at the points of the said region and their arbitrarily small neighbourhood (cf. (Healey, 2004)). Yet, our limits of knowability, enshrined in the no-signalling theorem, assure us that even if we could know of the non-separable change of properties, the physical laws we can empirically deduce for our region would not have been different. There are non-separable changes taking place, but they (due to no-signalling prohibition) do not crucially affect the limited predictions we can make about the behaviour of objects in the said region, do not affect the possibility of performing manipulative science from which to derive the truth-conditions for the relevant object manipulation on the extended material ontology in the local region. In other words, though our explanatory conceptual framework must not contain total separability, we can still do science; to the extent that we do in experimental and descriptive employment of the quantum formalism.

The problem is that once we come to put things this way we can legitimately ask whether we really have a deeper realist explanation of the phenomena, than we have been offered on our principle approaches with an instrumentalist slant. Pause just for a moment: the fact that the change of properties in the separated region is governed by a well structured law prevents us from having to fear the ultimate ontological holism, taking the entire material universe to be definable only as an indivisible whole with all partial definitions together summing up to insufficient global understanding. Our constructive approach in fact assures us that in any given region we can formulate the laws of physics and reconstruct experience of the material world on the basis of the properties of local objects (as they are formalised in the bare quantum formalism) and infer the existence of empirically inaccessible universal law governing their behaviour (in which all the non-separable effects are codified). So there is no need for metaphysical holism couched in the non-separable connection of properties of objects, the apparent violation of separability is achieved through the dictates of the universal law, which is itself immaterial. The central character of the role of extension in our conceptualisation of the real 'mechanisms' behind the phenomena does not lose its ontological significance: small things still add up to the big everyday things, and only these local small things add up to this here local big thing.[113]

[113] But, and this is crucial, our phenomena do not consist only of what is added there but also of what the things added are expected to do and to know what that is we can't simply summarise all the properties and propensities of the small things making up the big one.

Well, knowing the universal law then would allow us to regain the strong separability in the sense of Healey (2004). But, and here is the snag, the limits of knowability prevent us from ever knowing the exact details of proscriptions of the law for our given region, though they make them stable enough to allow correct probabilistic predictions of the future phenomena, and law-abiding accounts of the past ones. But predictions are not explanations. And our explanation explicitly involves action at a distance: in the 'troublesome' phenomena (we come to know once we take a more global view) a change in the intrinsic properties of one system induces a change in the intrinsic properties of a distant system without there being any process that carries the influence contiguously in space and time (Berkovitz, 2007).

This seems to be the consequence for a conceptual scheme to be employed in explanations of the troublesome phenomena and the construction of the transcendental strategy. As the universal law is in in-principle epistemically inaccessible, save for some details, to fend off the slide into excessive dispositionalism (where everything is reduced to the dispositions of the law, but those are unknowable) we must employ the tried and tested technique of relying on the 'geometrical' isomorphism between then common-sense conceptual scheme of re-identifiable objects and the fundamental ontology of spatially situated particles (the local beables). Yet to justify the existence of an external criterion of correctness of explanatory conceptualisations of this reduction of the empirically accessible to the empirically inaccessible, especially with respect to the 'troublesome' phenomena, we must postulate the existence of the non-local universal law that affects the conditions of re-identification of the fundamental ontology. In that, as we struggle to conceptualise the details of a causal connections between separated elements of the fundamental ontology, we must make the universal law primitive and modify the starting conceptualisation of the empirically accessible in phenomena to include both the spatial extension of objects and their subscription to (unknown) law. Our starting point in the transcendental strategy must also include the objective nomological structure of the world.

Otherwise we face the problem of not being able to account for the external constraints on our explanatory conceptualisation, we are again threatened by the excessive dispositionalism charge which we cannot dispel as our transcendental strategy cannot get off the ground. This is because the initial conceptualisation of the separate re-identifiable objects in space is just an illusion imposed by us onto the essentially holistic fundamental ontology of forever inaccessible world-stuff. Our typification, our carving of the world-stuff into manageable concepts is just an illusion, and any such carving is as good as another: a game of freely constructing the facade before the noumenal world. But on such account all explanations are equally vacuous, as there is no matter of fact as to what explain what. The price to pay for this (in the absence of a satisfyingly primitive account of causation) is to view the world from the outset (the very simplest starting point of the conceptual scheme employed in everyday conduct) as characterised not just by momentary spatial relations, but also by the mind-independent (primitively characterised) nomological structure. This mysterious guiding-hand-behind-events requirement may be too much of a price to pay on some worldviews. Especially as the theory itself demands that the universal law behind quantum phenomena (and fundamentally behind most physical phenomena) remains in-principle epistemically inaccessible. Furthermore, the role of the law at times becomes so fundamental as to affect the very individuation of the materially fundamental ontology, the particles, inviting a question whether those are again illusory projections included to save appearances, most notably the starting point of the transcendental strategy.

What this leaves us with is a road to modification of the starting point conceptual scheme, but not a modification that is outright unacceptable. We start from arguing for the necessary minimal typification of experience into that of enduring objects. This is an uncontroversial route the starting point of which is forged by Devitt ( (1997); and as presented in Chapters 1 and 3 above). To produce any explanations of the experience, and particularly deeper explanations of experience it is desirable to have some account as to how the real world affects our formation of concepts (rather than leaving us to freely dream up conceptual schemes of our choosing, even permitting they have *internal* consistency). In the latter case all our conceptual connections depend on our judgements (or even unwilling dispositions) that something is the case or that a set of concepts is in some way interrelated. But we cannot call upon the external world to account for a causal influence on how these judgements come to be formed, and why some of them might be more appropriate accounts of our experience than others (this may be appropriate to a particular purpose, even fulfilment of a pragmatic aim like acquiring more experiences *significantly like some given experience*).

As we cannot take an external position and view the world as it is, it is prudent to start from a shared ground, that of the common conceptual framework. As noted by Devitt, above, anti-realist interpretations of the experience as presented through the common-sense conceptual framework (or any similar conceptual framework, for that matter) cannot explain our experience. Even simple realism of the most basic kind has the tools to start producing explanations of the experiences given the common sense conceptual framework. The idea is that the basic germs of the realist accounts, which may grow to be extremely complex in the case of explication of formal contemporary theories, are already present in the said conceptual framework. We can then construct increasingly deeper explanations of an increasingly wider range of phenomena. But for the explanations to be possible in the first place, we need a transcendental step: a necessary condition for breaking the anti-realist explanatory impasse. Again (cf. Chapter 1), this is not a strict necessity of the form usually employed in the transcendental strategy, but an explication of the sensible conceptual commitment the possibility of explanation of experience as encoded in the common sense conceptual framework.

From here we rapidly proceed from accepting that we all have thoughts about material objects to 'necessitation' of the commitment to the conceptual scheme that sees the objects as existing independently of us in an objective framework of space and time. This commitment can further be distinguished from a sensorily similar commitment that there appear to be objects existing independently of us by further investigation of how the notion of those objects participates in our objective accounts of the world, including the intersubjective communication. The said commonsense conceptual scheme with the prior commitment sees the material objects (which are also *in* space and time, in some way that needn't be precisely specified at this stage) ontologically basic. In this way the persons engaged in the communication can identify and re-identify the particulars that are being spoken *about*. Other than demonstrative pointing to the objects, they can also be identified (given a 'thisness' as suggested in Chapter 1) by providing a description which, in the given circumstances, applies uniquely to the particular elements of reality concerned. Being ontologically basic within the common sense conceptual scheme, material objects do not need further reference to particulars of a different sort (Strawson, 1959).

As we investigate the nature of material reality in greater depth we come to uncover a number of illusions inherent in the above conceptual scheme, which must be removed from the scheme of the ontologically basic. Many of the identifying properties of material objects are dispensed with, but the germ of structure immediately evident and independent of our judgment remains, that of the necessary primary quality of extension in space. The identity of objects remains founded in the combination of identities of smaller objects that make them up, all related to each other through definite relations in space. Though our explanations no longer take the material objects as we perceive them as fundamental, they tell us how the appearance of the objects arises out of their fundamental structure, and the typification that does not slip away long this route is the extended structure of objects as constructed out their constituents. When the structure is subject to change, the details of the change can be tracked along the change of positions and shape in space. The germ of the connection between the Manifest and the Scientific images (Sellars, 1963) is given in the shared nature of extension in both the account of fundamental physical ontology and the directly perceivable material objects. Of course, there are other fundamental properties as well, but those can be added as attachments to the objects identified through their extended structure.

Yet, this kind of image might still lead us down the wrong path, and in some cases it seems to have done so for centuries. For sometimes it appears as if we have not taken on board the lessons required for a starting point for our transcendental strategy above. Namely, though we have argued for the conceptual primacy of the common-sense conceptual framework and the search for the realist metaphysics out of its ontological commitments, the commitments have strayed to one side only. With excessive focus on the spatial (geometric) structure, we have again allowed too great a reduction of the elements of what were supposed to be ontologically basic concepts. The focus on spatial structure alone allows for a return of the anti-realist suspicions through the back door. For the macroscopic spatial structures are again nothing but an illusion, and though there is an account of how the common-sense conceptualisation of experience arises and the required germ of connection is in place, we can allow for judgements that *reduce* the supposed ontologically basic concepts to products of an illusion. The world may exist independently and be made of the fundamentally extended things, but the structures that we see as arising from those things are nothing but castles in the sky. The generalised thing, the supposed fundamental unit of a realist ontology is an illusion, a human projection onto the real external world in the same way that a visible image is a projection onto a structure of pixels. When quantum theories threaten to deny the individuating characteristics to supposed fundamental elements, our entire house of cards threatens to collapse. If the spatial arrangements of the fundamental elements are not stable, then the structures we see as arising from them are not stable either. The anti-realist says once more that the transcendental step cannot be legitimately made, that by committing to the illusory structures we are not thereby committing to any further beliefs about the origin of the shared experience. If the directly re-identifiable material objects are nothing but provisional spatial (and even the significance of that condition can now be questioned) arrangements of the even in-principle non-individuatable fundamental elements, then we cannot explain our experiences as they are given even in the common sense conceptual scheme. That is, we cannot explain them in a better way than the anti-realist accounts can describe the same experience. It is important to note that this further difficulty arises only when we accept that the fundamental elements of the realist structure do not have an individuating identity, even in principle, regardless of their position in the overall spatial (or geometrical) structure.

And as Harré ((1996), and elaborated in Chapter 3 above) reminds us, the fundamental unit of the realist ontology is not the totality of the directly perceptible situation (the instantaneous state of the extended structure), but a generalised thing. "Things and other invariants through change are ineliminable fundamental elements of experience" (Harre, 1996, p. 312). The common sense conceptualisation of experience relies on more than the geometric structure and relation between illusory constructions, it includes at every step the notion of invariance through change. And the generalised objects, those fundamental referents for re-identification, are not a conjunction of structure statements, but something more. The further element can be provided by the notion of primitive laws governing the changes that the said objects can undergo. The laws account for the external limitations of the changes that the objects can undergo, thus participating in the very notion of the definition of an object (though, admittedly, not in the same way as the geometric structure or some other materially fundamental element might). They also provide limitations that provide for deeper explanations given as conceptual connections between the experienced phenomenon featuring the said object and the counterfactual situations it can be conceptually envisaged in. The same notion of laws allows us to account for the changes that the fundamental elements of extended ontology undergo at the 'ontologically deeper' level, providing explanations even for the 'troublesome' phenomena that arise in the domain of quantum theory. So, even in the cases where it seems that the individuating 'thisness' cannot be attributed to the particulars of fundamental ontology, the universal law governing their behaviour allows for their individuation and re-identification when required. Joining those phenomena to the common-sense conceptual framework does not then commit us to the ontological holism, which would eventually invalidate the possibility of identification and individuation of material objects within the common-sense conceptual scheme. In summary we are philosophically permitted a commitment to the conceptual individuation of material objects within a commonsense conceptual scheme, and further ontological commitments as required by the simple transcendental strategy.

As Worall  (1989) notes, realism in general has been pronounced dead before, but has successfully resurfaced. What the above discussion teaches us is that the worth of realism in explanation should not be easily abandoned, even at the price of modifying what is considered primitive and constituent of the common-sense conceptual scheme. That is not to say that we can and should go changing the basics of the everyday conceptualisation of the world as we please every time a slightly troublesome physical theory needs to be accommodated. But it does permit that we look hard at the elements of the conceptual scheme and reason about possibilities of seeing them in a different light so as to accept new primitives which we were previously hoping to reduce to some others. In our case universal laws of temporal evolution have to be admitted as primitive and recognised as such in the common-sense conceptual framework. There is no a priori reason why good-natured anti-realists would not accept this move, provided that appearances of the phenomena are saved as they are, and that we can still talk of those phenomena in the way that we ordinarily do. Accepting laws as primitive, along the lines that Harre and Bhaskar suggest, seems to allow for all this. There is, nonetheless, a high price to pay in admitting that there are foundational elements that we must accept as epistemically inaccessible and open only to inferential guesstimates that do not show signs of empirical improvement as yet. If that is the price, so be it, say those intent on commitment to realism of some sort. There are of course those for whom this may be a step too far to make, but in abandoning ship at this stage they must go back over the ground covered from those first tentative steps of the transcendental strategy. They must ponder the potential for explanations of the phenomena, including the troublesome ones, and the general worth of explanations. They must also be prepared to address additional problems that plague our principle approaches, which initially wanted to avoid any tinkering with the common-sense conceptual scheme, but then struggle to connect their account of the phenomena with even the most basic elements of the realist outlook (that there are material objects behind the troublesome phenomena in the first place). The only other alternative is to embrace the ontological holism and search for some kind of reconstruction of experience along those lines. Though they are not impossible, the above argument aims to suggest that they cannot follow the route of the simple transcendental strategy traversed above, but must start from scratch in accounting for the conceptualisation of experience as an error arising from historical misconception or sensory deception. Whilst this is by no means an impossible route to take its struggles with the anti-realist criticisms along the lines of dispositionalism seem much greater than those attempted here.

Perhaps ‘rejecting the grammar which tries to force itself on us’ (Wittgenstein, 1967) is to accept that ordinary, everyday concepts of objects in spatial framework and temporal duration and interaction presuppose inclusion of lawful, entirely externally conditioned, behaviour of those objects over and above the external limitations of their structure as identified through space and time. A chair is then more than certain spatial structure before us, it is a durable object whose temporal structure is, just as the spatial one, limited by what primitive laws of nature allow its material constituents to do and suffer. What the direct comparison of our approaches teaches us is that perhaps we looked in the wrong place from the start. Given the empirical equivalence of the two approaches perhaps the secret of their differentiation is not in which can axiomatically construct a better explanation of the world that contains the 'troublesome' phenomena, but what our expectations of the understanding of the world must be in the light of the troublesome phenomena. Both our approaches would agree that we can't get to the nature of the fundamental entities in a direct empirical way, that we cannot distinguish between them empirically (which just is to restate the empirical equivalence). To break the equivalence we must look into the starting position of the search to see how the equivalence has arisen in the first place and how the 'troublesome' phenomena have come to be seen as troublesome. The idea is that saying that we must start with objects that can be successfully reidentified is not problematic in itself, but simply relegates the problematic aspects to another domain as yet to be addressed. Whether it is the new ontology of abstract information-entities or the more classical one of local beables, the interesting question is how the phenomena that display the non-local connections between separated directly empirically accessible (macroscopic) objects can be generated.

And for that we need a different conceptual starting point. What our transcendental requirement must recognise is that the starting point cannot be the conceptualisation of individual objects solely on their intrinsic properties reducible to extension. Instead, we must conceptualise the objects as elements of generative mechanisms that contain both their spatial location and the universal laws that contribute to their local changes, but are themselves not bound by the requirement of locality or separability. The idea is not to identify things by the stability of their spatial position but by the stability of the role they play in the generation of processes. One may wonder whether this is not just making the processes ultimately fundamental, with the object-entities as their more or less enduringly recognisable features. This is certainly one avenue to explore, but it is not of necessity the only route left to take. For one thing it would make the construction of the transcendental strategy difficult, as we would have to not just modify, but fully replace its starting point, one of the world characterised in part by the concepts of macroscopic objects. To alleviate that difficulty we can hold on to the concept of objects but claim that the concept is not completely adequate when understood in terms of primary qualities alone.

The objects are not just what exists in terms of certain permanence of extension. The objects exist in a sense that they can be re-identified through the changes of a certain type. The key to the type in question is that there is a recognisable natural law governing the change, rather than a combination of such laws or a haphazard string of changes that cannot be understood as a law. It is the role of the law in interaction with objects that has to be better understood and investigated, and it is the recognition of conceptual foundation rooted in both laws and objects that distinguishes the principle and constructive approaches above. For the former turn to be inadequate in providing an explanation, primarily a conceptual connection between knowing that a phenomenon occurs and understanding why and how it does, for they lack any tool for identification of relevant (and then eventually shown to be conceptually fundamental) objects. The latter play up to this requirement, but must provide extra work in showing how this is not just a trick to fit the ready-made mould of the explanatory model. To do that they must look into ways to break the limits of knowability, finding ways to suggest how this might be done. Alternatively, we could try to rebuild explanatory ontology in terms of the structure emergent from the fundamental holistic entity, following the empiricist line (including the Humean Mosaic of the momentary state matter) and avoiding search for deeper causal mechanisms. Even when ignoring the attendant technical difficulties (such as the preferred choice of the formal basis for the decoherence that makes the emergence of the desired structures possible) such explanatory constructions cannot rely on our simple transcendental strategy as they lack the ‘germ of the solution’ for the connection of the directly observable experience and the fundamental physical ontology. There are other possible emergent and stable structures that are not in correspondence with our conceptual framework, but might be in good correspondence with some other possible such framework (making the existing one contingent in the fundamental structure, not just details). As the transcendental strategy starts with the preference for the essential features of the existing conceptual framework, explanatory ontology along the lines of the emergent structure would struggle to fend off the worldmaking charges and ontological relativity.

## CONCLUSION

The primary issue addressed in this thesis was the comparison of two case-study instances of methodological and explanatory approaches to 'troublesome' phenomena in the domain of quantum physics. In light of anti-realist criticism, which claims that beyond a limited network of concepts related to direct experience objective competition of explanatory narratives is not possible, as well as that different explanations are as good as each other, the thesis investigates the different strengths and weaknesses of two explanatory approaches: the principle and constructive one. Such anti-realist criticism is potent not only from a purely philosophical perspective, but also for a wider-reaching conclusion that contemporary science in general cannot offer convincing explanations, and that it is sometimes not even in the business of doing so, beyond the limited perspective of direct experience. If in fact we require that the directly experienced phenomena be seen as part of a unified whole of material reality then even in this sturdy everyday domain we lack objective explanations due to explanatory deficiencies in its foundational ontology, in the hypothesized primary constituents of all things material.

The two methodological approaches were initially chosen for their clear opposition in conceptualisation of the problem. The principle approaches were expected to overcome the said explanatory deficiencies of the hypothesised primary constituents by relying on the concepts familiar from everyday discourse and explaining the phenomena in the framework of generalised constrictions on natural processes, without reference to the ontological elements inaccessible to direct experience (i.e. concepts outside the scope of the common-sense conceptual framework). Though at first glance this might seem a strained strategy, it has been shown to work in well-known instances in the history of physics, such as thermodynamics and special theory of relativity. The explicit advantages expected were the unification of the phenomena that would otherwise require separate explanations and stronger explanatory potency by elimination of brute coincidences between competing explanatory narratives. The explanatory model such approaches were expected to fit in was one of unification, covering a wide range of phenomena within a strictly delineated conceptual framework. The constructive approaches, on the other hand, opted to openly rely on the hypothetical elements inaccessible to direct experience, focusing on the elimination of the supposed deficiencies or their rebuttal. Their general explanatory model was a widely popular one of causal-mechanical interaction, explanation of phenomena as causal processes arising from the physically deducible (though not always directly perceptible) interaction of the fundamental ontological elements, interaction characteristic of the said elements' properties and propensities. Through the evaluation of the way these general models dealt with the specific issues arising in the domain of quantum theory the aims was to distil conclusions for a generalised explanatory strategy concerning material reality.

In addressing these issues the thesis opens with a survey of the role and nature of explanations in modern and contemporary physical science. It proceeds to argue for the importance of explanation in scientific discourse, rather than the separation of the two as has repeatedly been suggested throughout the history of modern physics, and commitment of 'pure' science to descriptions useful for prediction and technological development. Moreover the opening chapter argues that although explanatory narratives are essentially epistemological constructions, they require a general metaphysical backing through the explainer's and explainee's commitments to take the concepts and higher structures composed of them as directly referential. Part of the success of the explanatory constructions examined in this thesis will be evaluated on the acceptability of the commitments that stand behind (as a 'backing') the concepts we employ in everyday communication, the ontological characteristics from the title.

Thus the opening chapter outlines the 'transcendental strategy' to be employed in comparing the ontological worth of the opposing explanatory strategies. Though a partial misnomer, the said strategy requires of all speakers of a given language, in our case any natural language used to provide the required explanations supplemented by the minimum necessary formalism of quantum theory, to accept that limitations to our acting and thinking rationally commit us to a conceptual framework that contains objects existing independently of us in objective space and time. It is then a further task for our explanatory approaches to try to fit in the explanatory narratives constructed to provide understanding of the troublesome phenomena with this general strategy. A historical overview outlines how this was achieved through the development of physics from early modern times to the occurrence of 'troublesome' phenomena with the rise of quantum theory in early 20th century. To permit the increase in knowledge through detailed empirical investigations the transcendental strategy is forced to select between the more and less fundamental elements of the conceptual scheme. The latter are then subjected to change under increased empirical investigation and the former provide a permanent and stable connection between the old and the new, between the directly experienced and the hypothetically explanatory. Historically, physical spatial extension and the geometric properties provided this desired connection.

When quantum theory appears on the scene it introduces some phenomena that require a careful selection of the agreed upon set of characteristics used to construct explanations that respect the essential elements of the common-sense conceptual framework. By violating separability, these phenomena seem to call for explanations that do not share the widely accepted minimal conceptual framework of objects in space and time. For the latter requires that these objects can claim an existence independent of one another insofar as they occupy different parts of space. The objects may be in discernible interaction, but they ought to have separate intrinsic states that can be altered through such interaction. Moreover, composite objects should acquire all their properties from the constituents' intrinsic states and locally intrinsic interactions.

If the phenomena in the domain of quantum theory violate separability the transcendental strategy for realism is threatened by denial of the possibility of spatiotemporal separation as the primary objective criterion of individuation of the elements of foundational ontology, elements which play a foundational role in the most universal conceptual scheme. In other words, they form the core element of every conceptual scheme as they are particulars that can be identified and re-identified without reference to the particulars of a different sort. The 'troublesome' phenomena from the domain of quantum theory seem to invite a holism that denies the possibility of the application of a transcendental strategy, and thus pose a challenge for the conceptual connection between explanatory narratives suited to quantum theory and the simple and sturdy common-sense conceptual scheme, a starting point of the transcendental strategy.

It is further outlined in Chapter 1 how the 'troublesome' phenomena such as the EPR correlations and the more novel 'teleportation' raise questions about the continuous existence of individual particulars in a systematic way predicted and confirmed by the theory. Chapter 1 concludes by acknowledging the general preference in literature for causal-mechanical type (constructive approaches) of explanations over those of the unificatory type (principle approaches). The following two chapters are devoted to examining the strengths and weakness of the case-study instances of the two approaches in the light of the general problems each of the explanation types encounters in the specific situations, most notably the requirements of contrastive explanations which cannot be easily cast into the causal-mechanical mould.

The second chapter surveys the epistemological position of one of the founding fathers of quantum theory, Niels H. Bohr, as an introduction to the principle approaches, presented as neo-Bohrian in methodology. They accept the necessary limits to epistemic accessibility and adopt an overall agnosticism about the structure of material reality out of which the perceived phenomena arise. They focus on the general limitations to knowledge gathering and information transmission between conscious subjects as sufficiently clear foundations upon which to build the explanations of the troublesome phenomena, without having to connect them to concepts of individual material objects (of any particular size or type) that are threatened by the non-separable aspects of the phenomena. In other words the Fuchs and the CBH methodological programmes call for a change of perspective that would eliminate the need for the jeopardised connection between the 'troublesome' phenomena and the common sense conceptual framework.

Following a classification of Bohr's philosophical position as principle theory (shying away from even the possibility of mechanical conceptualisation of matter of 'microscopic' size) with a Kantian twist (the necessity of classical concepts for objective description of the physical realm), Fuchs' programme sees the supposed quantum descriptions of matter as codified epistemic guesses about the future macroscopic outcomes of measurement. Yet to avoid the pitfalls of instrumentalism Fuchs ventures into constructive domain, but on a weak footing of 'inherent sensitivity' of reality to all empirical observation. The CBH programme (named after R. Clifton, J. Bub and H. Halvorson), explored in much greater detail on account of methodology, metaphysics and explanatory potential, aims to reconstruct the theory within a suitable mathematical framework with minimal ontological commitments (the epistemic 'black boxes'). They propose to see the macroscopic objects in physical interaction as mere displays of output and input states from the perspective of 'information transmission', with no epistemic access to their material structure (nor any need for such access). Unlike Fuchs, the proponents of the CBH programme claim that they do not show that the theory deals with the epistemological concerns of the observers nor that the basic stuff of the world is informational, but that the principle-style explanatory account is the best that can be achieved about the 'troublesome' phenomena.

However, the explanations so constructed struggle to provide sufficient features for the transcendental strategy to remove the criticism of a vacuous narrative. Though such explanations satisfy key segments of the unification-type explanations in general, they leave a gaping hole in the connection between the conceptual parts of the universal constraining principles they rely on and the successful connection they achieve between knowing that a phenomenon occurs and why it occurs. As in the case of thermodynamics, this can be an extremely useful predictive tool and even goes some way to providing an explanation, but when deeper explanations appear as contenders it is left wanting. The principle explanatory strategy, though nominally respecting the existence of material objects and their necessary separability, in the very provision of explanation does not respect that we conceptualise situations in terms of re-identifiable objects. Though elaborately avoiding the separability-violating threats to the transcendental strategy, the principle approaches must commit at least to some novel ontology of their own. The latter, on the other hand, is taken to be the first stage of its development, difficult to connect with the common-sense conceptual framework which was the starting point of the transcendental strategy.

The third chapter outlines the history of one particular type of the constructive approaches, the one following the work of D. Bohm and well suited to the particulars of the constructive strategy through insistence on the point particle as the fundamental ontological element. Such particle nominally satisfies the requirements of a re-identifiable object in space and time, though the accompanying element of the quantum field or potential (required to reproduce the specifically quantum 'troublesome' phenomena) is presented as marred with explanatory inadequacies, especially in the light of the novel phenomenon of 'teleportation'. Even without teleportation, the 'particles plus the real field' view struggles to maintain sufficient 'intrinsic thisness' of the particles in certain situations, and thus to prevent the slide into a fundamentally field-based holistically non-separable ontology.

Following further introductory presentations of the philosophical notions of causes, properties and deterministic realism, an ontology of equally real (but ontologically of distinct type) particles and universal laws governing their behaviour is presented. Such ontology accepts non-separability through abandonment of the Humean mosaic that sees only momentary arrangements of material objects as really existing at any given time (and thus making the laws governing their changes a mere human projection onto the real state of affairs). The non-separable aspect of the phenomena is relegated to universal law though, and thus not attributed to the material constituents. A summary of the technical arguments connecting quantum theory with such a worldview is presented and the 'troublesome' phenomena are recast in new light. A new problem arises though, for the said explanatory construction requires not only an abandonment of the Humean mosaic as the foundational conceptual commitment, but also an acceptance of the epistemic barrier to access to the said universal law due to an axiomatically attributed state the whole of the universe is in. Thus the ontological elements are all named, but a barrier to their direct empirical investigation is once again raised.

Chapter 3 concludes with a survey of satisfaction of the criteria for explanation set out in Chapter 1 by the final constructive approach based on the point particles and the universal (though epistemically obscured) law. It is shown that the constructive approaches fare better in satisfying the Lipton criteria of explanation, and as is to be expected of the ontological, causal-mechanical explanations they show potential for providing deeper explanations (something that is discussed in greater depth in the subsequent chapter) than the principle, unification-style explanations can. This allows the transcendental strategy to be given through reliance on the concepts of enduring objects and non-local universal laws.

Yet this seems to require that in the transcendental strategy we change the starting point from objects being defined in terms of primary qualities alone into objects conceptualised as enduring individuals subject to the universal law. The nature is now 'cut at the joints' not along the lines of instantaneous structure in space, but through the selection of structure across law-permitted changes in space and time. But a final caveat opens here, especially in the light of the 'troublesome' phenomena introduced above: how can we justify the fundamental role given to material ontology, the point particles, if so much of their contribution to the overall structure is dispositional on the proscriptions of the universal law and is not intrinsic to the given ontological elements themselves? Such questions open up the validity of adherence to the transcendental strategy at all, given the conceptual obstacles raised by contemporary quantum theory.

The final Chapter turns to a presentation of the general characteristics of deeper explanations. It is shown that deeper explanations do require some notion of laws as conceptual background against which the permissible alternatives to the experienced phenomena are evaluated. Furthermore, deeper explanations focus on the explanatory narrative that has an object, a system undergoing a regulated change, at its centre. This object has to be re-identifiable in its own right, not just as a structural feature of the phenomenon to be explained. On such an account of the depth of explanation, especially when metaphysical projections beyond, but related to, the constraining principles are sought after, the principle approaches are found wanting. As they generally shy away from any specification of the metaphysics of the elements of reality responsible for the experience of the phenomena it becomes unclear how what is supposed to be explanatory on their account can actually be so. Furthermore, even the transcendental strategy that the sturdy and non-specific principle approaches were expected to connect well to, becomes problematic when they aim to clearly separate from straightforward instrumentalism. Both the Fuchs and the final Bub (representing CBH) approaches display inclinations toward a metaphysically fundamentally indeterministic universe (in the present, not just the future sense), one that cannot be isomorphically related to the common-sense conceptual framework. The structures available on such a view then struggle to give rise to re-identifiable objects that endure through change.

Finally, lessons from the principle and the constructive approaches are combined to diffuse the threat of separability violations for the very core of the foundational conceptual scheme, the isomorphic connection between the physically fundamental ontology and the objects of everyday experience through the primary qualities of material existents. Observed regularities in the separability violations themselves are employed to show how they affect the predictive and explanatory aspects of the interpretations of quantum formalism respectively. As the separability violations can never be used for superluminal signalling, epistemic and metaphysical restrictions can be combined to allow for the construction of the explanatory models which respect the formal requirements of the theory and the realism-supporting aims of the transcendental strategy.

They do have to lean on neo-Bohrianism to some extent though, in admitting explicit limitations to our epistemic access, but dare venture beyond it in asking for metaphysical projections that can account for the core features of the basic conceptual scheme and suggest areas of investigations where the said limitations might be experimentally removed (cf. the breaking of the quantum equilibrium). The constructive approaches suggest respecting the structurally important role that point particles play in connection between the common-sense conceptual framework and the vagaries of contemporary physics (by playing the role of local be-ables, however flimsical), and the non-material nature of the universal law of temporal evolution not subjected to limitations of separability. But this brings forth consequences for the starting point of the transcendental strategy, the core conceptual scheme. It requires that that very scheme admits as the fundamental ontological unit not the totality of the directly perceptible situation (the instantaneous state of the extended structure including fields), but a generalised thing.

The common sense conceptualisation of experience must rely on more than the geometric structure and relation between illusory constructions, it must include at every step the notion of an object enduring through regulated change. As metaphysical limiters of the changes the objects can undergo, the universal (and ontologically non-material) laws participate in the very definition of an object (though admittedly not in the same way as the geometric structure or some other directly observable feature). Thus the 'troublesome' nature of some phenomena in quantum physics need not just draw consequences for the axiomatic structure of the explanatory conceptualisations specifically constructed for them, but can influence what our expectations of the understanding of the world that contains the troublesome phenomena ought to be like. A transcendental strategy of arguing for realism must start not with the conceptualisation of individual objects solely on their intrinsic properties reducible to extension, but use also the generative mechanisms that contain both their spatial location and the universal laws that contribute to their local changes across time. These laws themselves are not, though, bound by the requirements of locality and separability. The idea is to identify things by the stability of the role they play in the generation of processes. Simply put, to understand a 'chair' we must see it as capable of smashing a window as well as having four legs and a backrest.

An avenue for further research opens up in connection between the above discussion and the so-called Everettian conceptual approach to contemporary physics. The latter makes the spatially extended, but holistically conceived structure as absolutely primary, though conceptually disassociated from the everyday objects of experience. Though this might be a natural route to take in response to the impermanence of the intrinsic properties of the point particles in some experimental situations (crucially involving precise position as well) it is metaphysically demanding (through calling for parallelly existing but unperceivable universe-duplicates) and difficult to connect to the starting point of our transcendental strategy for realism.

## Curriculum vitae

**Mladen Domazet**

Born on 14th September 1978 in Sisak, Croatia. Completed primary school in Sisak, as well as the first two years of the grammar school. Transferred to Uppingham School in the United Kingdom and completed A-level exams in Biology, Physics, Mathematics and Further Mathematics. In 1997 enrolled in and successfully completed the first year of Mathematics and Physics course at the Faculty of Science, Zagreb University. In 1998 enrolled in the Physics and Philosophy course at the University of Oxford, United Kingdom. Graduated in 2002, having completed an undergraduate thesis on methodological parallels between Special Theory of Relativity and Quantum Mechanics. In 2002 enrolled in a postgraduate course in Philosophy, specialising in Philosophy of Science, at the Studia Croatica, University of Zagreb. Since 2003 employed at the Institute for Social Research in Zagreb. Professionally specialising in natural sciences education, scientific explanation and popularisation of science through the projects of the Institute's unit, Centre for Educational Research and Development. Since 2007 participates in undergraduate teaching in philosophy at the University of Zagreb. Published a number of articles in philosophy of science, philosophy of physics, educational philosophy and comparative education, in English and Croatian. Presented papers at several philosophical and educational research conferences in Croatia and abroad. In 2006 completed a summer programme in philosophy of physics at the Central European University in Budapest, and in 2008 a three month study visit with the University of Oxford Philosophy of Physics research group. Member of professional associations such as the Society for Advancement of Philosophy, Croatian Society for Analytical Philosophy and Society znanost.org.

## Životopis

**Mladen Domazet**

Rođen 14. Rujna 1978.g u Sisku. Osnovnu školu i prva dva razreda opće gimnazije završio u Sisku. Nakon toga prešao u Uppingham School u Ujedinjenom kraljevstvu i završio A-level programe iz Biologije, Fizike, Matematike i Više matematike. 1997.g upisao i položio prvu godinu na smjeru prof. matematike i fizike na Prirodoslovno-matematičkom fakultetu Sveučilišta u Zagrebu. 1998.g upisao dodiplomski studij Filozofije i fizike na Sveučilištu u Oxfordu, Ujedinjeno kraljevstvo. Diplomirao 2002.g, uz završnu radnju o metodološkim paralelama između Specijalne relativnosti i Kvantne mehanike. Iste godine upisao poslijediplomski studij filozofije, sa specijalizacijom iz filozofije znanosti, na Hrvatskim studijima Sveučilišta u Zagrebu. Od 2003.g zaposlen kao asistent na Institutu za društvena istraživanja u Zagrebu. Istraživačkim radom u Institutu specijalizirao se u području obrazovanja iz prirodnih znanosti, prirodoznanstvenih objašnjenja i popularizacije znanosti, kroz znanstvene projekte jedinice Centar za istraživanje i razvoj obrazovanja. Od 2007.g sudjeluje u dodiplomskoj nastavi filozofije na Sveučilištu u Zagrebu. Objavio stručne, znanstvene i pregledne radove iz filozofije znanosti, filozofije fizike, obrazovne filozofije i komparativnog obrazovanja, na hrvatskom i engleskom jeziku. Izlagao i na desetak međunarodnih i domaćih znanstvenih skupova. 2006.g završio ljetni sveučilišni program iz filozofije fizike (pod vodstvom istraživačke skupine s Rutgers University, SAD) na Central European University u Budimpešti, a 2008.g boravio u tromjesečnom studijskom posjetu istraživačkoj skupini za Filozofiju fizike Sveučilišta u Oxfordu. Član Udruge za promicanje filozofije, Hrvatskog društva za analitičku filozofiju te Društva znanost.org.